\documentclass[letterpaper,useAMS,usenatbib]{aa}
\usepackage{graphics}
\usepackage{natbib}
\bibpunct{(}{)}{;}{x}{}{,}
\usepackage{graphicx}
\usepackage{amssymb}
\newcommand{\micron}{\hbox{$\mu$m}}
\begin{document}
\title{A SCUBA survey of bright-rimmed clouds}

\author{L.K.Morgan
	\inst{1,2}
	\and
	M.A.Thompson
	\inst{3}
	\and
	J.S.Urquhart
	\inst{4}
	\and
	G.J.White
	\inst{1,5,6}
	}
	
\offprints{L.K.Morgan}

\institute{CAPS, The University of Kent, Canterbury, Kent CT2 7NR\\
	\email{lmorgan@nrao.edu}
	\and
	Present address: NRAO, Green Bank Telescope, P.O.Box 2, Green Bank, WV 24944
	\and
	Centre for Astrophysics Research, Science and Technology Research Institute, University of Hertfordshire, College Lane, Hatfield AL10 9AB
	\and
	School of Physics and Astronomy, University of Leeds, Leeds LS2 9JT
	\and
	Space Physics Division, Room 1.71, Space Science \& Technology Division, CCLRC Rutherford Appleton Laboratory, Chilton, Didcot, Oxfordshire OX11 0QX
	\and
	Dept. of Physics \& Astronomy, The Open University, Walton Hall, Milton Keynes MK7 6AA
	}

   \date{}

\abstract{Bright-rimmed clouds (BRCs) are potential examples of triggered star formation regions, in which photoionisation driven shocks caused by the expansion of HII regions induce protostellar collapse within the clouds.}
{The main purpose of the paper is to establish the level of star formation occuring within a known set of BRCs. A secondary aim is to determine the extent, if any, to which this star formation has been promulgated by the process of photoionisation triggering.}
{A primary set of observations is presented obtained with submillimeter SCUBA observations and archival data from near-IR and mid- to far-IR have been explored for relevant observations and incorporated where appropriate.}
{SCUBA observations show a total of 47 dense cores within the heads of 44 observed BRCs drawn from a catalogue of IRAS sources embedded within HII regions, supportive of the scenario proposed by RDI models. The physical properties of these cores indicate star formation across the majority of our sample. This star formation appears to be predominately in the regime of intermediate to high mass and may indicate the formation of clusters. IR observations indicate the association of early star forming sources with our sample.
  A fundamental difference appears to exist between different morphological types of BRC, which may indicate a different evolutionary pathway toward star formation in the different types of BRC.}
{Bright-rimmed clouds are found to harbour star formation in its early stages. Different evolutionary scenarios are found to exist for different morphological types of BRC. The morphology of a BRC is described as type `A', moderately curved rims, type `B', tightly curved rims, and `C', cometary rims. `B' and `C' morphological types show a clear link between their associated star formation and the strength of the ionisation field within which they are embedded.

An analysis of the mass function of potentially induced star-forming regions indicate that radiatively-driven implosion of molecular clouds may contribute significantly toward the intermediate to high-mass stellar mass function.} 

\keywords{stars -- formation: HII regions: Clouds: Submillimetre: Dust}

\maketitle

\bibliographystyle{aa}

\label{firstpage}
\section{Introduction} 

Bright-rimmed clouds (BRCs) are small molecular clouds found at the edges of large HII regions. These
clouds are potential examples of triggered star-forming regions, whereby shocks driven into the BRC by
the photoionisation of the cloud exterior results in the collapse (and perhaps even the formation) of
sub-critical molecular cores within the cloud (e.g.~\citealt{Deharveng2005}; \citealt{Elmegreen1998};
\citealt{Sugitani1991}). The process of photoionisation-induced collapse is usually known as
radiative-driven implosion or RDI  \citep{Bertoldi1989,Lefloch1995,Lefloch1994}.  The RDI of molecular cores at the periphery of HII regions may thus be responsible for subsequent
generation of star formation, amounting to a possible cumulative total of several hundred new stars per
HII region \citep{Ogura2002} and perhaps 15\% or more of the low-to-intermediate
mass stellar mass function \citep{Sugitani1991}. 
IRAS-selected protostars associated with HII regions are systematically more luminous than those that are not \citep{Dobashi2001}. This highlights a trend toward higher mass star formation and greater star formation efficiency.
Confirming bright-rimmed clouds as star-forming and quantifying the nature and location of their star
formation thus provides important insights into the clustered mode of star formation and the overall
star formation efficiencies of molecular clouds.  

\citet{Sugitani1991}, hereafter referred to as SFO91, searched the Sharpless HII region catalogue
\citep{Sharpless1959} for bright-rimmed clouds associated with IRAS point sources, their selection was based upon the FIR colours thought to identify YSOs/protostars. Later, \citet{Sugitani1994} -- SO94 -- extended their search to  include bright-rimmed clouds from the ESO(R) Southern Hemisphere Atlas. A total of 89 optically identified bright-rimmed clouds have been found to be associated with IRAS point sources. For brevity (and
consistency with SIMBAD), we will refer to the combined  SFO91 and SO94 catalogues as the SFO catalogue. 
Whilst a few individual clouds from the SFO catalogue have been studied in detail 
\citep[e.g.~][]{Lefloch1997,Megeath1997,Codella2001,Thompson2004a,Thompson2004b,Urquhart2004,Urquhart2006,Urquhart2007} and shown to harbour
protostellar cores, the question of whether star formation is a common occurence within BRCs (irrespective of the formation mechanism) is still unresolved.

We are in the process of carrying out a star formation census of SFO bright-rimmed clouds in order to
determine the current level of star formation within these sources and to relate it to their physical properties and morphologies with respect to their ionising star(s). In two previous publications we reported observations of the ionised gas surrounding the
bright-rimmed clouds \citep{Thompson2004,Morgan2004}. Here we report a Submillimetre Common User Array (SCUBA) imaging survey of the
submillimetre continuum emission from the clouds, in order to reveal the presence of  compact, potentially
star-forming, dust cores within the clouds and to contrast the properties of these cores with those
found in other star-forming regions. We use NIR archival data from the 2 millimetre all-sky survey (2MASS) database to construct colour-colour diagrams to investigate the current level of star formation within these clouds and to identify YSOs.  We examine the kinematics and temperature
of the dust cores revealed by our SCUBA observations to show that a number of the SFO bright-rimmed
clouds are associated with active star formation.

\section{Observations}
\label{sec:obs}

\subsection{SCUBA continuum maps}

Simultaneous 450 and \mbox{850 \micron ~} images were obtained using SCUBA \citep{Holland1999}
on the James Clerk Maxwell telescope (JCMT\footnote{ The JCMT is operated by the Joint Astronomy Centre
on behalf of PPARC for the United Kingdom, the Netherlands Organisation of Scientific Research and the
National Research Council of Canada.}). SCUBA is a dual-camera system comprised of two bolometer arrays
which simultaneously sample a similar field of view ($\sim$ 2$^{\prime}$ square). Each bolometer array
is arranged in a hexagonal pattern and both arrays are observed using a dichroic
beamsplitter. The 91 pixel short-wave array is optimised for operation at \mbox{450 \micron ~} and the 37
pixel long-wave array of 37 pixels is optimised for operation at \mbox{850 \micron ~}. The arrays do not
fully spatially sample the field-of-view and in order to provide fully-sampled images the secondary
mirror of the JCMT is moved in a 64-point pattern (``jiggling'') whilst also chopping at a frequency of
1 Hz to remove sky emission. This procedure is commonly known as a ``jiggle-map''. 

Our initial source list was formed from those BRCs in the SFO catalogue lying within the declination
limits visible from the JCMT, which gives a total number of 50 clouds (45 from the mainly northern
hemisphere SFO91 catalogue and 5 from the southern hemisphere SO94 catalogue). We also included the two
bright-rimmed clouds SFO 11NE and SFO 11E in our sample. These two clouds lie in close proximity to the
bright-rimmed cloud SFO 11 \citep{Ogura2002}. SCUBA observations of these three clouds have
already been reported in \citet{Thompson2004a}. Thus our initial source list comprised 52
bright-rimmed clouds.  However, observations of only 47 clouds were completed, due to the fact that our
observations were performed in the flexibly-scheduled mode used at the JCMT. As observations of three of these clouds have been previously reported by \citet{Thompson2004a} we present here observations of the remaining 44 clouds. Flexibly scheduled observations
are not scheduled over a pre-defined period but are carried out according to the appropriate weather
(atmospheric opacity) band, the visibility of the sources from the telescope and the scientific priority
of the observations. The observations of the BRCs in our sample took place over several days from
November 2001 to April 2002.

As the angular diameters of most of the clouds in the SFO catalogue are comparable to the field-of-view (FOV)
of SCUBA we obtained single jiggle-maps for each cloud. A small number of clouds in the sample either
have larger angular diameters or lie in associations that are larger than the SCUBA FOV. In these cases
multiple overlapping jiggle-maps were obtained and mosaiced together to cover the cloud or association.
Varying integration times were used for  each cloud, depending upon the atmospheric opacity and the
brightness of the submillimetre emission that was detected. A chop throw of 120\arcsec\ was initially used for
each jiggle-map but after inspection of the first images this was changed to 180\arcsec ~to avoid
chopping onto extended emission. The atmospheric opacity was measured using the CSO 225 GHz radiometer
and by performing hourly skydips with SCUBA at the approximate azimuth of each observation.
Typical sky opacities were 2.270 at 450 \micron ~and 0.326 at 850 \micron.

The data were reduced and calibrated in the standard manner using the SCUBA User Reduction Facility SURF \citep{Surf2000} and
the Starlink image analysis package KAPPA \citep{Kappa2002}. Absolute flux calibration was performed
using calibration maps of the primary calibrators, Uranus and Mars, or the secondary calibrators CRL
618, CRL 12688, 16293-2422, IRC 10216 and OH 231.8 depending upon the observation date, time and sky
position. Predicted fluxes for Uranus and Mars were estimated using the values given by the Starlink
package FLUXES \citep{Fluxes1998} and predicted fluxes for the secondary calibrators were taken from the
JCMT calibrator webpage\footnote{http://www.jach.hawaii.edu/JACpublic/JCMT/Continuum
\_observing/SCUBA/astronomy/calibration}.\\

The JCMT has a well known error beam contribution that may be modelled by a combination of two
circularly symmetric gaussian functions representing the main and error beams. We removed the error-beam
contribution to the flux by deconvolving the calibrated images with a model two-component beam determined
from azimuthally averaged maps of the primary calibrators  \citep[using the method outlined
in][]{Hogerheijde2000,Thompson2004a}. It is well known that CLEAN processes do poorly with extended structure and therefore as many of our sources are extended on the scale of the JCMT primary beam CLEANing was performed only until negative components were encountered. Images were then restored to a resolution of 14\arcsec. This is the native
resolution at 850 $\mu$m and hence smooths the 450 $\mu$m data from its native resolution of 8\arcsec.
The smoothing has the advantages of facilitating the comparison of the 450 $\mu$m and 850 $\mu$m images
and improving the signal-to-noise ratio of the 450 $\mu$m data to extended emission by approximately a
factor of 3.

\subsection{Archival data}

\subsubsection{IRAS HiRes Data}
\label{sec:IRASHiRes}

IRAS HiRes images at 12, 25, 60 and 100 $\mu$m were obtained from the NASA/IPAC Infrared Science
Archive\footnote{http://irsa.ipac.caltech.edu} to complement the SCUBA data, extend the FIR wavelength coverage of each source and
enable the spectral energy distribution (SED) to be measured. The IRAS HiRes image construction algorithm utilises the maximum correlation method (MCM) \citep{Aumann1990}
in the enhancement of the original all-sky survey data. The angular resolution and signal-to-noise
ratio of HiRes data varies depending upon the number of deconvolution iterations used and the position
within the image. HiRes images are also subject to a number of processing artefacts, most notable of
which are the negative `bowls' surrounding bright sources that are known as ringing. This effect is much
more prevalent at higher wavelengths, we found that 60 and 100 \micron ~images may need significant correction, while 12 and 25 \micron ~images needed very little. Corrections to source fluxes were estimated through measuring any negative offset present close to the source and adjusting the source flux accordingly.

For IRAS data sets with typical coverages, correction factor variance (CFV) values reach 0.01-0.001 at strong, converged sources \citep{Levine1993}. It was found that the default processing parameters for HiRes images were not sufficient to achieve this in the majority of our images. Visual inspection of the produced maps allowed us to discount the possibility that the regions were saturated, overly noisy or contained `bad' scans.
  Coverage maps of all of our sources allowed us to ascertain the level of coverage per source, as well as the uniformity of the coverage over the beam area. This was good for all sources and so we iterated all sources until we reached the recommended CFV values of 0.01-0.001, this was generally found to be after $\sim$150-200 iterations, similar to the number of iterations used in the analysis of the Serpens star-forming cloud core \citep{Hurt1996}

Typical (mean) angular resolutions were 36\arcsec, 34\arcsec, 49\arcsec ~and 73\arcsec ~at 12, 25, 60 and 100 \micron ~respectively, which are generally only sufficient to identify the strongest cores in each SCUBA field  as point sources. The absolute calibration uncertainties of the HiRes data are estimated to
be around 20\%, which is similar to the original IRAS survey data.

Individual fluxes for each of the SCUBA cores were measured from the HiRes images by aperture photometry and are listed in Table
\ref{tbl:HiRes}. It should be noted that the fluxes tabulated here differ from those given for each source in the IRAS point source catalogue. This is, no doubt, due to the nearly five-fold enhanced resolution of the HiRes images over the original IRAS survey. HiRes fluxes are typically 2-3 times lower than those listed in the IRAS PSC with the largest differences seen in the 12 and 25 \micron ~bands. A comparison of the luminosities as determined from HiRes fluxes to those derived by SFO91 will be addressed in Section \ref{sec:Classification}.
 This is likely to have a large effect upon the conclusions of SFO91 who concluded that the IRAS sources embedded within these bright-rims had large luminosities in comparison to IRAS sources embedded within dark globules and other dense cores. With our improved IR resolution and extended coverage of the sources SEDs we aim to investigate the true luminosity function of the Sugitani catalogue.

\subsubsection{2MASS}
\label{Sec:2MASS}
Data from the near-infrared $J$, $H$ and $K_{\rm{s}}$ 2MASS catalogue \citep{Cutri2003} were obtained
to search for protostars and embedded young stellar objects (YSOs) associated with the BRCs.  The
catalogue data were obtained from the  2MASS  Point Source catalogue held at the NASA/IPAC Infrared
Science Archive\footnote{http://irsa.ipac.caltech.edu}.  The accuracy of the photometric measurements in
the 2MASS Point Source Catalogue is between 1 and 2\%, though can be much larger for weak sources. The
limiting magnitudes at $J$, $H$ and $K_{\rm{s}}$ in the 2MASS catalogue are 15.8, 15.1 and 14.3 respectively.

\begin{center}
\begin{table*}
\caption{Fluxes of each source as measured in IRAS HiRes images.}
\begin{center}
\begin{tabular}{ccccc}
\hline
 
 {Source} &\multicolumn{4}{c} {IRAS HiRes Flux(Jy)} \\ 
 {} & {12 \micron} & {25 \micron} & {60 \micron} & {100 \micron}\\
 \hline
SFO 01  & 1.3  & 2.9   & 46.8   & 173.2  \\
SFO 02  & 0.9  & 1.5   & 13.6   & 51.9  \\
SFO 03  & 0.2  & 1.5   & 12.2   & 52.1  \\
SFO 04  & 0.5  & 0.9   & 11.9   & 25.2  \\
SFO 05  & 1.8  & 30.2  & 95.0   & 155.4  \\
SFO 06  & 0.2  & 0.3   & 3.4    & 15.5  \\
SFO 07  & 0.5  & 1.1   & 18.0   & 74.9  \\
SFO 09  & 0.4  & 0.6   & 15.5   & 47.6  \\
SFO 10  & 1.0  & 1.9   & 25.6   & 72.8  \\
SFO 12  & 0.4  & 1.1   & 16.7   & 36.6  \\
SFO 13  & 1.2  & 2.3   & 35.9   & 84.5  \\
SFO 14  & 5.2  & 49.6  & 420.2  & 534.7  \\
SFO 15  & 0.2  & 1.0   & 7.3    & 23.6  \\
SFO 16  & 0.1  & 1.3   & 14.6   & 23.0  \\
SFO 17  & 0.1  & 0.7   & 3.3    & 16.0  \\
SFO 18  & 0.1  & 1.0   & 14.0   & 22.2  \\
SFO 23  & 0.1  & 0.4   & 4.1    & 9.4  \\
SFO 24  & 0.3  & 0.6   & 8.7    & 26.0  \\
SFO 25  & 0.3  & 1.4   & 17.9   & 44.1  \\
SFO 26  & 0.2  & 0.3   & 4.7    & 17.7  \\
SFO 27  & 0.5  & 1.8   & 15.8   & 97.2  \\
SFO 28  & 1.4  & 1.6   & 37.4   & 126.6  \\
SFO 29  & 0.3  & 0.3   & 5.7    & 29.5  \\
SFO 30  & 9.1  & 69.0  & 665.0  & 901.2  \\
SFO 31  & 1.0  & 3.8   & 26.2   & 127.5  \\
SFO 32  & 0.4  & 0.2   & 6.8    & 25.7  \\
SFO 33  & 0.2  & 0.8   & 7.0    & 43.4  \\
SFO 34  & 0.6  & 1.0   & 6.8    & 24.7  \\
SFO 35  & 0.3  & 0.6   & 4.2    & 22.7  \\
SFO 36  & 0.3  & 1.7   & 14.9   & 54.8  \\
SFO 37  & 1.8  & 11.2  & 35.6   & 45.2  \\
SFO 38  & 0.9  & 3.6   & 62.2   & 172.2  \\
SFO 39  & 0.7  & 3.8   & 17.4   & 33.7  \\
SFO 40  & 0.5  & 0.4   & 5.6    & 25.8  \\
SFO 41  & 0.5  & 0.7   & 5.7    & 21.6  \\
SFO 42  & 0.3  & 0.6   & 5.6    & 20.1  \\
SFO 43  & 2.8  & 12.1  & 95.4   & 144.0  \\
SFO 44  & 0.6  & 8.7   & 124.4  & 300.2  \\
SFO 45  & 0.1  & 0.3   & 1.2    & 7.7  \\
SFO 77  & 0.5  & 5.1   & 55.7   & 67.1  \\
SFO 78  & 0.4  & 1.1   & 9.7    & 37.2  \\
SFO 87  & 3.3  & 4.2   & 138.5  & 430.6  \\
SFO 88  & 1.6  & 3.8   & 120.7  & 396.4  \\
SFO 89  & 3.8  & 5.5   & 111.9  & 330.2  \\
\hline  	    
			   	    
\end{tabular}\\		   							      
\end{center}								        
\label{tbl:HiRes}									         
\end{table*}								      
\end{center}								        
  
\section{Results \& Analysis}
\label{sec:results}

Of the forty-four BRCs that were observed, 39 yielded a detection of at least one submillimetre core and five showed no evidence of submillimetre emission. We define a core
as a closed contour of submillimetre emission at or above the 3$\sigma$ level and equal to or greater than the size of the FWHM of the JCMT beam. Individual cores were identified through a visual inspection of the submillimetre images and while most of the BRCs were found to be associated with a single submillimetre core, a small
number of clouds were found to possess multiple cores (SFO 25, SFO 39 and SFO 87).

 Clouds that did not exhibit any detectable submillimetre emission greater than 3 times the r.m.s.~flux level were automatically classified as non-detections and all sources were assessed visually to determine if the map showed emission consistent with an embedded core and if that emission was spatially correlated between the two wavelengths. If the source failed these two criteria then it was also classified as a non-detection. At 850 \micron ~SFO 06, 17, 28, 77 and 78 failed these criteria and 19 clouds failed these criteria at 450 \micron. Upper limits to the fluxes are given in Table \ref{tbl:corepos}.

We present the SCUBA images as contour plots overlaid on DSS R images in the online appendix. The
properties of the individual cores are described in Section~\ref{subsect:cores}.

\subsection{Core Positions, Fluxes and Sizes}
\label{subsect:cores}

The position of each core, their measured fluxes and effective diameters are given in Table
\ref{tbl:corepos}. The positions given for each core are those determined from the peak of the 850 \micron ~emission.
Multiple cores have been designated with an SMM number to indicate each component. 

The Starlink package GAIA \citep{Gaia2004} was used to measure the peak and integrated fluxes of each core, where integrated fluxes represent a summation over a 30\arcsec ~diameter aperture centred at the position of peak flux. a 30\arcsec~aperture was chosen as it represents a width of 5$\sigma$ of a 14\arcsec~beam. Background levels were estimated from emission-free regions of each map and
subtracted from the measured flux values. We estimate that the systematic errors in measuring the fluxes
of the cloud cores  are no more than 30\% in the case of the 450 $\mu$m measurements and 10\% for the
850 $\mu$m measurements (including errors in the absolute flux calibration).

Because of the higher signal-to-noise ratio of the 850 \micron ~maps in comparison to those at 450 \micron ~the effective angular diameter of each source was estimated from the longer wavelength maps. A Gaussian function was fitted to the
azimuthally averaged 850 \micron ~flux. The beam size was taken into
account through the assumption of a simple Gaussian source convolved with a Gaussian beam of
14$^{\prime\prime}$ ($\Theta^{2}_{\rm{obs}}$=$\Theta^{2}_{\rm{beam}}$+$\Theta^{2}_{\rm{source}}$). Three sources (SFO 33, 41 and 45) were found to be marginally resolved or unresolved at 850 \micron ~.

The distances to each cloud given in SFO91 and SO94 allowed a calculation of the physical effective diameter of each core (see
Table \ref{tbl:corepos}). These spatial diameters range from 0.01--0.71 pc with a mean of 0.17 pc.

  The size of the cores may be placed in the context of star formation by considering turbulent motions within molecular clouds. Observed turbulent motions become subsonic on the smallest scales creating smooth, regular structures. The size of the smallest star-forming cores may then be determined by the scale at which turbulence transits from supersonic to subsonic. This is of the order 0.05 - 0.1 pc \citep{Larson2003} (the resolution of the JCMT, 14\arcsec, corresponds to 0.09 pc at the mean distance of the sample, 1.29 kpc). The cores in this sample are evenly distributed in terms of size around this turbulent transitional scale. The correlation of the core sizes with the transitional size scale is supportive of the identification of the cores as being star-forming or, having the potential to be such.

\begin{center}
\begin{table*}
\begin{center}
\caption{Core positions, fluxes and effective diameters. Upper peak flux limits for non-detections are given as the three-sigma value. The integrated detection limit is taken as the three-sigma flux value integrated over our 30\arcsec~aperture. Where there are no detections at either 450 or 850 \micron ~the superscript $^{a}$ indicates that the recorded position is that of the map centre.}
\label{tbl:corepos}
\scriptsize
\begin{tabular}{ccccccc}
\hline
 \multicolumn{1}{c}{} &  &  &  & {Integrated Flux} &  & \\ 
 {}& {$\alpha_{\rm{2000}}$} & {$\delta_{\rm{2000}}$} & {Peak Flux(Jy/beam)}                   &(Jy)                                   &D$_{\rm{eff}}$ & Rim Type\\
 {Source      }& {       (Peak)  } & {    (Peak)     } & \mbox{450 \micron}  ~~\mbox{850 \micron} &\mbox{450 \micron}  ~~\mbox{850 \micron} & (pc)  &	(Section \ref{Sec:CC})\\
  \hline
SFO 01			& 23:59:32.3  &   +67:24:03 &  2.8$\pm$0.2~	0.69$\pm$0.07	& 8.4$\pm$2.3   	0.99$\pm$0.19	& 0.03    & B \\
SFO 02			& 00:03:58.6  &   +68:35:12 &  2.8$\pm$0.2~	0.60$\pm$0.03	& 15.6$\pm$2.0  	1.03$\pm$0.07	& 0.10    & A \\
SFO 03			& 00:05:22.7  &   +67:17:56 &  $<$1.5~~~~~ 	0.29$\pm$0.03	& $<$13.7~~~~~  	0.47$\pm$0.09	& 0.08    & C \\
SFO 04			& 00:59:01.0  &   +60:53:29 &  2.7$\pm$0.4~	0.16$\pm$0.02	& 10.3$\pm$3.0  	0.33$\pm$0.06	& 0.06    & B \\
SFO 05			& 02:29:02.2  &   +61:33:33 &  7.8$\pm$2.0~	1.44$\pm$0.04	& 13.1$\pm$18.2 	2.04$\pm$0.11	& 0.16    & B \\
SFO 06			& 02:34:45.1  &   +60:47:48 &  $<$5.4~~~~~ 	$<$0.21~~~~~~	& $<$49.1~~~~~  	$<$0.57~~~~~~	& $^{a}$  & A \\
SFO 07			& 02:34:47.8  &   +61:46:29 &  6.6$\pm$0.4~	1.24$\pm$0.05	& 27.4$\pm$3.3  	1.77$\pm$0.14	& 0.19    & B \\
SFO 09			& 02:36:27.6  &   +61:24:02 &  2.4$\pm$0.4~	0.18$\pm$0.02	& 6.6$\pm$3.1   	0.30$\pm$0.05	& 0.33    & A \\
SFO 10			& 02:48:12.3  &   +60:24:32 &  $<$2.6~~~~~ 	0.16$\pm$0.02	& $<$19.3~~~~~  	0.21$\pm$0.06	& 0.35    & A \\
SFO 12			& 02:55:01.5  &   +60:35:43 &  3.1$\pm$0.7~	0.68$\pm$0.02	& 1.9$\pm$4.9   	1.10$\pm$0.06	& 0.31    & B \\ 
SFO 13			& 03:00:56.5  &   +60:40:24 &  11.6$\pm$3.0	0.69$\pm$0.03	& 15.6$\pm$22.3 	1.33$\pm$0.09	& 0.40    & B \\ 
SFO 14			& 03:01:31.3  &   +60:29:20 &  12.1$\pm$1.2	2.68$\pm$0.04	& 50.8$\pm$9.0  	4.54$\pm$0.11	& 0.40    & A \\ 
SFO 15			& 05:23:28.3  &   +33:11:48 &  $<$1.5~~~~~ 	0.10$\pm$0.01	& $<$11.3~~~~~  	0.12$\pm$0.03	& 0.71    & B \\ 
SFO 16			& 05:19:48.4  &   -05:52:04 &  $\pm$1.2~   	0.85$\pm$0.03	& 23.0$\pm$9.2  	1.54$\pm$0.07	& 0.06    & A \\ 
SFO 17			& 05:31:28.1  &   +12:05:24 &  $<$0.7~~~~~ 	$<$0.20~~~~~~	& $<$5.4~~~~~~  	$<$0.52~~~~~~	& $^{a}$  & A \\ 
SFO 18			& 05:44:29.7  &   +09:08:55 &  2.0$\pm$0.2~	0.72$\pm$0.02	& 6.3$\pm$1.5   	1.25$\pm$0.05	& 0.09    & A \\ 
SFO 23			& 06:22:57.8  &   +23:10:15 &  2.5$\pm$0.5~	0.23$\pm$0.02	& 3.8$\pm$3.7   	0.40$\pm$0.06	& 0.27    & A \\ 
SFO 24			& 06:34:52.2  &   +04:25:38 &  $<$3.3~~~~~ 	0.19$\pm$0.02	& $<$21.8~~~~~  	0.12$\pm$0.04	& 0.25    & B \\ 
SFO 25 SMM1		& 06:41:03.2  &   +10:15:09 &  5.1$\pm$0.8~	1.16$\pm$0.03	& 10.0$\pm$5.2  	1.66$\pm$0.08	& 0.06    & B \\ 
SFO 25 SMM2		& 06:41:04.8  &   +10:15:01 &  3.7$\pm$0.8~	0.90$\pm$0.03	& 10.2$\pm$5.2  	1.70$\pm$0.08	& 0.06    & B \\ 
SFO 26			& 07:03:46.7  &   -11:45:52 &  $<$1.4~~~~~ 	0.14$\pm$0.02	& $<$9.1~~~~~~  	0.23$\pm$0.05	& 0.13    & A \\ 
SFO 27			& 07:03:58.3  &   -11:23:04 &  $<$1.5~~~~~ 	0.32$\pm$0.02	& $<$9.8~~~~~~  	0.59$\pm$0.06	& 0.48    & A \\ 
SFO 28			& 07:04:44.4  &   -10:21:41 &  $<$1.6~~~~~ 	$<$0.14~~~~~~	& $<$15.7~~~~~  	$<$0.39~~~~~~	& $^{a}$  & A \\ 
SFO 29			& 07:04:52.4  &   -12:09:43 &  $<$2.7~~~~~ 	0.45$\pm$0.06	& $<$25.3~~~~~  	0.60$\pm$0.18	& 0.04    & A \\ 
SFO 30			& 18:18:46.8  &   -13:44:28 &  5.0$\pm$0.3~	1.89$\pm$0.03	& 20.0$\pm$2.5  	2.42$\pm$0.07	& 0.33    & B \\ 
SFO 31			& 20:50:43.1  &   +44:21:56 &  2.6$\pm$0.2~	0.62$\pm$0.04	& 5.6$\pm$2.3   	0.72$\pm$0.12	& 0.04    & A \\ 
SFO 32			& 21:32:29.5  &   +57:24:33 &  $<$0.8~~~~~ 	0.22$\pm$0.01	& $<$7.9~~~~~~  	0.13$\pm$0.03	& 0.20    & A \\ 
SFO 33			& 21:33:11.9  &   +57:30:07 &  $<$1.0~~~~~ 	0.14$\pm$0.02	& $<$8.2~~~~~~  	0.10$\pm$0.06	& 0.01    & A \\ 
SFO 34			& 21:33:32.1  &   +58:03:32 &  2.0$\pm$0.1~	0.48$\pm$0.01	& 5.7$\pm$0.6   	0.60$\pm$0.03	& 0.04    & A \\ 
SFO 35			& 21:36:03.4  &   +58:31:25 &  $<$1.6~~~~~ 	0.20$\pm$0.02	& $<$10.3~~~~~  	0.36$\pm$0.05	& 0.11    & A \\ 
SFO 36			& 21:36:07.4  &   +57:26:41 &  14.9$\pm$0.3	1.73$\pm$0.03	& 38.7$\pm$1.7  	2.12$\pm$0.07	& 0.07    & A \\ 
SFO 37			& 21:40:28.9  &   +56:35:53 &  $<$1.5~~~~~ 	0.98$\pm$0.04	& $<$7.7~~~~~~  	1.40$\pm$0.10	& 0.06    & C \\ 
SFO 38			& 21:40:41.8  &   +58:16:14 &  6.2$\pm$0.3~	4.02$\pm$0.05	& 24.0$\pm$2.0  	6.68$\pm$0.14	& 0.11    & B \\ 
SFO 39 SMM1		& 21:46:01.2  &   +57:27:42 &  3.8$\pm$0.6~	0.66$\pm$0.03	& 13.5$\pm$4.0  	1.12$\pm$0.09	& 0.06    & B \\ 
SFO 39 SMM2		& 21:46:06.9  &   +57:26:36 &  3.4$\pm$0.6~	0.64$\pm$0.03	& 8.7$\pm$4.0   	0.94$\pm$0.09	& 0.03    & B \\ 
SFO 40			& 21:46:13.4  &   +57:09:42 &  3.4$\pm$0.6~	0.30$\pm$0.04	& 9.4$\pm$4.4   	0.44$\pm$0.10	& 0.10    & A \\ 
SFO 41			& 21:46:28.6  &   +57:18:35 &  $<$1.1~~~~~ 	0.20$\pm$0.02	& $<$10.4~~~~~  	0.38$\pm$0.05	& 0.04    & B \\ 
SFO 42			& 21:46:34.7  &   +57:12:31 &  $<$1.3~~~~~ 	0.25$\pm$0.03	& $<$12.0~~~~~  	0.50$\pm$0.09	& 0.09    & A \\ 
SFO 43			& 22:47:49.3  &   +58:02:51 &  2.0$\pm$0.5~	0.60$\pm$0.06	& 7.6$\pm$4.4   	1.19$\pm$0.16	& 0.31    & B \\ 
SFO 44			& 22:28:51.1  &   +64:13:40 &  18.7$\pm$0.6	2.93$\pm$0.06	& 63.2$\pm$5.4  	3.88$\pm$0.16	& 0.06    & A \\ 
SFO 45			& 07:18:26.6  &   -22:05:56 &  $<$1.1~~~~~ 	0.14$\pm$0.01	& $<$8.8~~~~~~  	0.04$\pm$0.04	& 0.05    & A \\ 
SFO 77			& 16:19:53.9  &   -25:33:39 &  $<$4.1~~~~~ 	$<$0.09~~~~~~	& $<$40.6~~~~~  	$<$0.24~~~~~~	& $^{a}$  & A \\ 
SFO 78			& 16:20:52.9  &   -25:08:07 &  $<$3.6~~~~~ 	$<$0.10~~~~~~	& $<$35.7~~~~~  	$<$0.28~~~~~~	& $^{a}$  & A \\ 
SFO 87 SMM1		& 18:02:49.8  &   -24:22:26 &  3.2$\pm$0.7~	0.64$\pm$0.04	& 8.0$\pm$7.0   	1.22$\pm$0.11	& 0.12    & B \\ 
SFO 87 SMM2		& 18:02:27.3  &   -24:22:57 &  2.2$\pm$0.6~	0.44$\pm$0.02	& 5.3$\pm$6.2   	0.79$\pm$0.07	& 0.15    & B \\ 
SFO 88			& 18:04:11.5  &   -24:06:41 &  1.6$\pm$0.2~	0.90$\pm$0.03	& 7.8$\pm$2.3   	1.96$\pm$0.09	& 0.23    & A \\ 
SFO 89			& 18:09:56.6  &   -24:04:21 &  2.5$\pm$0.3~	0.52$\pm$0.02	& 14.2$\pm$2.6  	0.97$\pm$0.05	& 0.21    & A \\
\hline

\end{tabular}\\
\end{center}		
\end{table*}								      
\end{center}

\subsection{Temperature, Mass and Density}
\label{sec:DC}  

It has been shown (e.g., \citealp{Sridharan2002,Beuther2002}) that the SED of luminous YSOs can be modelled by a two component modified blackbody fit. The two components consist of a compact, hot component which dominates the IRAS 12 \micron ~and 25 \micron ~fluxes and a component which represents the cooler, more extended, dust. The cooler component of the SED dominates the IRAS 60 \micron ~and 100 \micron ~fluxes in addition to fluxes at submillimetre wavelengths.
  The cooler component arises from re-radiated emission, absorbed and then emitted by the outer envelope of the protostellar source. The temperature that is then associated with this component is referred to as the `dust' temperature of the cloud. Measured fluxes classified as non-detections have been discarded when fitting SEDs, thus some SEDs have been created using only the IRAS fluxes and the 850 \micron ~flux.

Certain sources were not subjected to SED fitting analysis, even if they have been classified as detections. This was due to high levels of error in flux measurements or poor morphological agreement in the source structure at different wavelengths. These sources were SFO 24, 32, 35, 36, 38, 44, 45 and 88.

\begin{center}
\begin{figure*}
\begin{center}
\includegraphics*[scale=0.50]{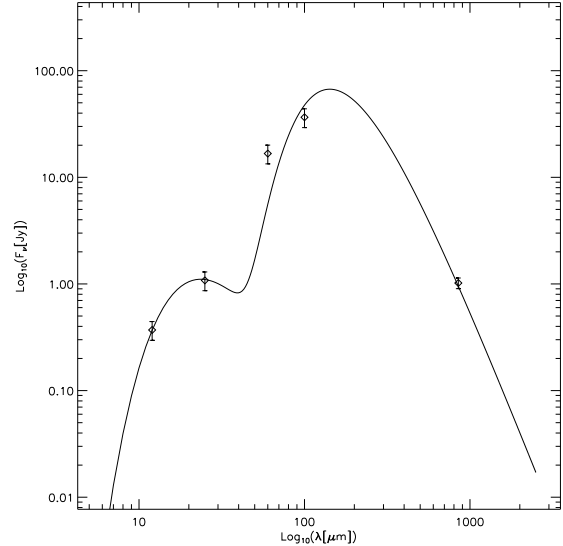}\\
\caption{SED fit to fluxes observed toward the object SFO 12}
\label{fig:sedplot12}
\end{center}
\end{figure*}
\end{center}

Fits to the IRAS \& SCUBA fluxes were made using the approach outlined in
\citep{Dent1998}, specifically,~the integrated flux at each frequency ($F_{\rm{\nu}}$) was determined from the following equation:

\begin{equation} F_\nu
= \Omega B_\nu (T_{\rm{d}})(1-e^{-\tau_\nu}), \end{equation}

where $\Omega$ is the solid angle subtended by the aperture used when summing fluxes, $B_\nu (T_d)$ is the Planck function evaluated at a dust temperature, T$_{\rm{d}}$, and
frequency, $\nu$, and $\tau_\nu$ is the optical depth at frequency  $\nu$. The optical depth of the dust was assumed
to follow the form  $\tau_{\rm{\nu}} = \tau_{\rm{ref}}\left(\frac{\nu}{\nu_{\rm{ref}}}\right)^{\beta}$, where $\nu_{\rm{ref}}$
is a reference frequency and $\beta$ is the dust emissivity. As the SEDs of the cores are defined by only five or six points we chose to reduce the number of free parameters
in our model by fixing $\beta$. For the small, hot component $\beta$ was fixed at a value of 1 in following with previous works (e.g.\citealp{Faundez2004,Sridharan2002}) and at its canonical value of 2 for the cooler, more extended gas (e.g.\citealp{Thompson2004a}). An example SED is shown in Fig.\ref{fig:sedplot12}.

The mass of each core was determined using the method of  \citet{Hildebrand1983}, here we measure the mass of the cold components of our SEDs, the dominant component of mass in the cloud. For an optically thin cloud with a uniform temperature the total ($M_{\rm{dust}+}M_{\rm{gas}}$) mass of the cloud $M$ is given by

  \begin{equation}
  M = \frac{d^{2}F_{\rm{\nu}}C_{\rm{\nu}}}{B_{\rm{\nu}}(T_{\rm{d}})},
 \label{eq:mass}
  \end{equation}
  
   The parameter $C_{\rm{\nu}}$ is a mass conversion factor combining both the gas-to-dust ratio (typically assumed to be 100)
and the frequency dependent dust opacity, $\kappa_{\rm{\nu}}$. Quoted values for $C_{\rm{\nu}}$ range from 21.4 g
cm$^{-2}$ \citep{Kruegel1994} to 286 g cm$^{-2}$ \citep{Draine1984}. A value of $C_{\rm{\nu}}$ = 50 g
cm$^{-2}$ at $\nu$ = \mbox{850 \micron ~} has been adopted here following \citet{Thompson2004a}. Number densities of
the clouds were determined from the mass by assuming a spherical geometry for the core. The dust
temperatures resulting from the modified blackbody fits, along with core masses and H$_{2}$ number densities are
presented in Table \ref{tbl:sp}. Upper limits for the masses of potential cores in clouds labelled as `non-detections' are presented in table \ref{tbl:corepropsnondets} along with upper limits for density and extinction values.

The uncertainties in the mass (and hence density) of the clouds are generally dominated by temperature
effects in the non-linear Planck function in Eq.\ref{eq:mass}. Typical values of observed 850
\micron ~flux and SED determined dust temperature have uncertainties of $\sim$ 10\% and $\pm$ 1-2 K
respectively leading to an factor of $\sim$ $\sqrt{2}$ uncertainty in mass and density. In addition, it should be noted that the value of $C_{\rm{\nu}}$ is a factor of two smaller than that typically assumed. This was to maintain consistency with other works based on the same set of observations \citep{Thompson2004a}. This uncertainty will also propagate into our derived values of mass and density and so these results should be scaled by the best known value of $C_{\rm{\nu}}$. In addition, there will be some uncertainty in mass and density values introduced by our uncertainty in $\beta$. This is commonly assumed to be 2 and can be fairly well constrained in certain circumstances, note that only the cooler component of $\beta$ is relevant in this case as we are only deriving the mass associated with this component of our sources, the majority of the mass of the system.

Values of the visual and near-infrared extinction ($A_{\rm{V}}$ \& $A_{\rm{K}}$) toward each core were 
derived from their submillimetre fluxes following the
method of \citet{Mitchell2001}. The submillimetre flux $F_{\rm{\nu}}$ may be related to the visual extinction by:

\begin{equation}
F_{\rm{\nu}} = \Omega B_{\rm{\nu}} (T_{\rm{d}})\kappa_{\rm{\nu}}m_{\rm{H}}\frac{N_{\rm{H}}}{E(B-V)}\frac{1}{R}A_{\rm{v}},
\end{equation}
  
  where the total opacity of gas and dust is represented by $\kappa_{\rm{\nu}}$, which is functionally
equivalent to the reciprocal of the mass conversion factor $C_{\rm{\nu}}$, $m_{\rm{H}}$ is the molecular mass of
interstellar material, ${N_{\rm{H}}/{}E(B-V)}$ is the conversion factor between column density of hydrogen
nuclei and the selective absorption at the relevant wavelength. Following \citet{Mitchell2001} (and
references therein) a value of ${N_{\rm{H}}/{}E(B-V)}$ = 5.8 x 10$^{21}$ cm$^{-2}$ mag$^{-1}$ has been assumed. A value of R = 5 has been assumed in line with predictions of the extinction of the inner regions of molecular clouds (e.g.\citet{Campeggio2007}). The K-band extinction A$_{\rm{K}}$ of each source was found by multiplying by the ratio
A$_{\rm{V}}$/A$_{\rm{K}}$ = 8.9 \citep{Rieke1985}. Values of A$_{\rm{V}}$ and A$_{\rm{K}}$ are listed in Table
\ref{tbl:sp}, along with the dust temperatures and masses, etc. derived from the submillimetre
observations.

\subsection{2MASS colour-selected YSOs associated with the BRCs}
\label{sec:2MASS}

\begin{center}
\begin{figure*}
\begin{center}
\includegraphics*[scale=0.50]{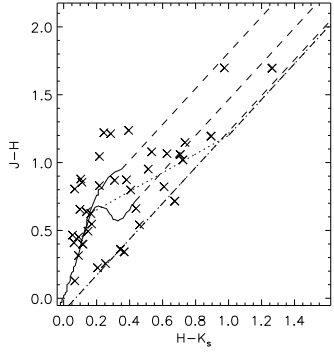}
\caption{J-H versus H-K$_{\rm{s}}$ diagrams of the 2MASS sources associated with the cloud SFO 12. The solid lines represent the unreddened loci of main sequence and giant stars from \citet{Koornneef1983} The dotted line represents the classical T-Tauri locus of \citet{Meyer1997}. Reddening tracks are shown by dashed lines and the dash-dot line represents the reddening track of the earliest main-sequence stars.}
\label{fig:jhkplot12}
\end{center}
\end{figure*}
\end{center}

The 2MASS catalogue was searched for near-infrared objects within the optical bounds of the surveyed
clouds to identify any YSOs that may be associated with the BRCs. YSOs occupy a distinctive region in a 
$J-H$ and $H-K_{\rm{s}}$ colour diagram that is intrinsically redder than the main-sequence population
\citep{Lada1992}. An example of this type of diagram is shown in Fig. \ref{fig:jhkplot12}, with similar
diagrams for the rest of the clouds in the survey presented in the online appendix. The solid
lines represent the colours of main sequence and giant stars \citep{Koornneef1983}, while the dotted
line shows the Classical T-Tauri (CTTS) locus \citep{Meyer1997}. 

YSOs may be differentiated from older stars that have had their radiation extincted by large amounts of
intervening dust through their infrared colours. As a source is subjected to higher levels of extinction
(larger intervening columns of dust) it moves towards the top right corner of the diagram, following the
reddening tracks shown by dashed lines in Fig. \ref{fig:jhkplot12}. Thus, sources found to lie to the
left of the reddening track associated with the latest main-sequence stars are likely to be main-sequence or giant stars, while those sources
that lie between this track and the rightmost track associated with the CTTS locus and above the locus itself may be identified as CTTS. Sources to the right of the
rightmost track are likely to represent Class I protostars hidden by a large, extended dusty envelope. The numbers of CTTS and Class I protostar sources were determined using these constraints and are listed in Table \ref{tbl:ysonos}.

The regions occupied by main sequence or giant stars,
CTTS Class II protostars and class I protostars in these diagrams are not completely discrete; each population overlaps with the others 
\citep{Lada1992}. Nevertheless the colour-colour diagrams indicate the general population of recently
formed stars and protostars 
associated with the clouds and allow us to identify \emph{candidate} CTTS and class I protostars for
later followup and confirmation. When identifying potential CTTS sources we have discounted sources that may simply be extincted main-sequence objects, thus, the number of identified CTTS sources is probably an underestimate.

Investigation of the position of 2MASS sources in the plane of each DSS image fails to reveal any consistent trend of source position with cloud morphology, i.e. the sources are not consistently clustered towards either the submillimetre cores or the optically bright rims. Data from the Spitzer telescope will yield much information upon the IR content of each cloud and will be presented in future work.
 
 We indicate the numbers of candidate CTTS and class I protostars associated with each
BRC in Table \ref{tbl:ysonos}. The majority of BRCs are associated with at least one YSO and in many cases multiple YSOs; this is consistent with recent and/or ongoing star formation activity: out of our sample of 47 detected cores there are 10 that do not show any candidate CTTS
or class I protostars (SFO 1, 24, 26, 28, 34, 78, 87a, 87b, 88 and 89).  Protostellar sources in the early stages of evolution are likely to be highly embedded within their host clouds behind large degrees of extinction. These sources are also likely to have low fluxes at NIR wavelengths, thus the number of associated 2MASS sources may be considered as a lower limit. This indicates that the majority of observed BRCs are still in an early phase of their star formation history. The positions of each source identified as either CTTS or protostars are overlaid on the images in the online appendix. The 2MASS sources do not appear consistently related to either the position of peak submillimeter flux or the position of the optically bright rim.

 As mentioned in Section \ref{Sec:2MASS}, the limiting magnitudes of the 2MASS are 15.8 in the $J$ band, 15.1 at the $H$ band frequency and 14.3
in the $K_{\rm{s}}$ band. The average $K$-band extinction of the sample set ranges from 0.2 to 4.5 with a
mean of 1.3 (the determination of $K$-band extinction is discussed in Section \ref{sec:DC}). The maximum observable $K_{\rm{s}}$-band magnitude should then be 9.8, equivalent to a 3.2 M$_{\odot}$ protostar with an age of 0.3 Myr at the mean distance of the sample (1.29 kpc) \citep{Zinnecker1993}.
The numbers of identified YSOs found should then accurately trace the star formation within the clouds, at least for protostars more massive than $\sim$3.2 M$_{\odot}$.

\begin{center}
\begin{table*}
\begin{center}
\caption{Numbers of class I protostellar candidates (CI) or candidate T-Tauri stars (CTTS) identified from the JHK$_{\rm{s}}$ diagrams (online appendix) for all sources}
\begin{tabular}{ccc|ccc}
\hline
{Source} & Number of      & Number of         & {Source} & Number of      & Number of  \\ 
{}    & associated CTTS & associated CI &    {}    & associated CTTS & associated CI \\
\hline
01        & 0 &  0 & 30        & 2  & 1 \\
02        & 3 & 10 & 31        & 1  & 0 \\
03        & 0 &  1 & 32        & 0  & 2 \\
04        & 0 &  2 & 33        & 0  & 2 \\
05        & 2 &  1 & 34        & 0  & 0 \\
06$^{a}$  & 0 &  6 & 35        & 0  & 3 \\
07        & 0 &  4 & 36        & 0  & 3 \\
09        & 0 &  1 & 37        & 0  & 4 \\
10        & 0 &  2 & 38        & 1  & 3 \\
12        & 4 &  3 & 39        & 0  & 8 \\
13        & 1 &  3 & 40        & 0  & 2 \\
14        & 1 &  9 & 41        & 0  & 2 \\
15        & 0 &  1 & 42        & 0  & 1 \\
16        & 0 &  3 & 43        & 1  & 2 \\
17$^{a}$  & 1 &  0 & 44        & 1  & 2 \\
18        & 0 &  2 & 45        & 1  & 3 \\
23        & 0 &  2 & 77$^{a}$  & 0  & 1 \\
24        & 0 &  0 & 78$^{a}$  & 0  & 0 \\
25        & 2 &  3 & 87a       & 0  & 0 \\
26        & 0 &  0 & 87b       & 0  & 0 \\
27        & 4 &  2 & 88        & 0  & 0 \\
28$^{a}$  & 0 &  0 & 89        & 0  & 0 \\	
29        & 1 &  1 & & &  \\
\hline		
		
\label{tbl:ysonos}
\end{tabular}									      
\end{center}		
\scriptsize$^{a}$Non-detections at SCUBA wavelengths
\end{table*}								      
\end{center}		

\subsection{Additional IR Information}
\label{sec:AddIR}
In addition to the IRAS data presented in our SED analyses and the 2MASS fluxes utilised in Section \ref{sec:2MASS}, infrared observations have been performed of large areas of the sky by both the Midcourse Space Experiment (MSX) and the Spitzer telescope. Searches were performed for observational data relevant to our sources in both the MSX and the publically available Galactic Legacy Infrared midplane Survey Extraordinaire (GLIMPSE) catalogues. Only one source from our sample was found to lie within the field observed by GLIMPSE, which covered only the central two degrees in latitude of the Galaxy. This source was SFO 30 and will be the subject of a more detailed study by the authors in the future. Many of our sources were found to have been observed at 24 \micron ~in proprietary Spitzer programmes and these data were examined for evidence of protostellar content within our sample. A total of 37 multiband imaging photometer for Spitzer (MIPS) maps were obtained from the Spitzer archive, covering sources from the SFO catalogue. Of the sources covered by these maps four (SFO 8, 19, 21 and 22) were not observed by the reported SCUBA observations and an additional two (SFO 6 and 17) were non-detections in the SCUBA survey. Of the remaining 31 sources, 25 show 24 \micron ~emission clearly associated with the SCUBA cores identified in Section.\ref{subsect:cores}. The other six clouds appear to have non-correlated 24 \micron ~and 850 \micron ~emission.\\

  The correlation of 24 \micron ~emission with the positions of the reported 850 \micron ~cores highly supports the identification of these cores as star-forming. This includes cores in which no 2MASS YSO sources were identified (Section.\ref{Sec:2MASS}), these sources are SFO 01, 24, 34 and 88. The non-detection of these cores by the 2MASS is most likely due to these sources being at early stages of their evolution as described in Section.\ref{Sec:2MASS}. A more detailed analysis of the publicly available Spitzer archival data is underway by the authors and will be reported in a future work.

\begin{center}
\begin{figure*}
\begin{center}
\includegraphics*[scale=0.40]{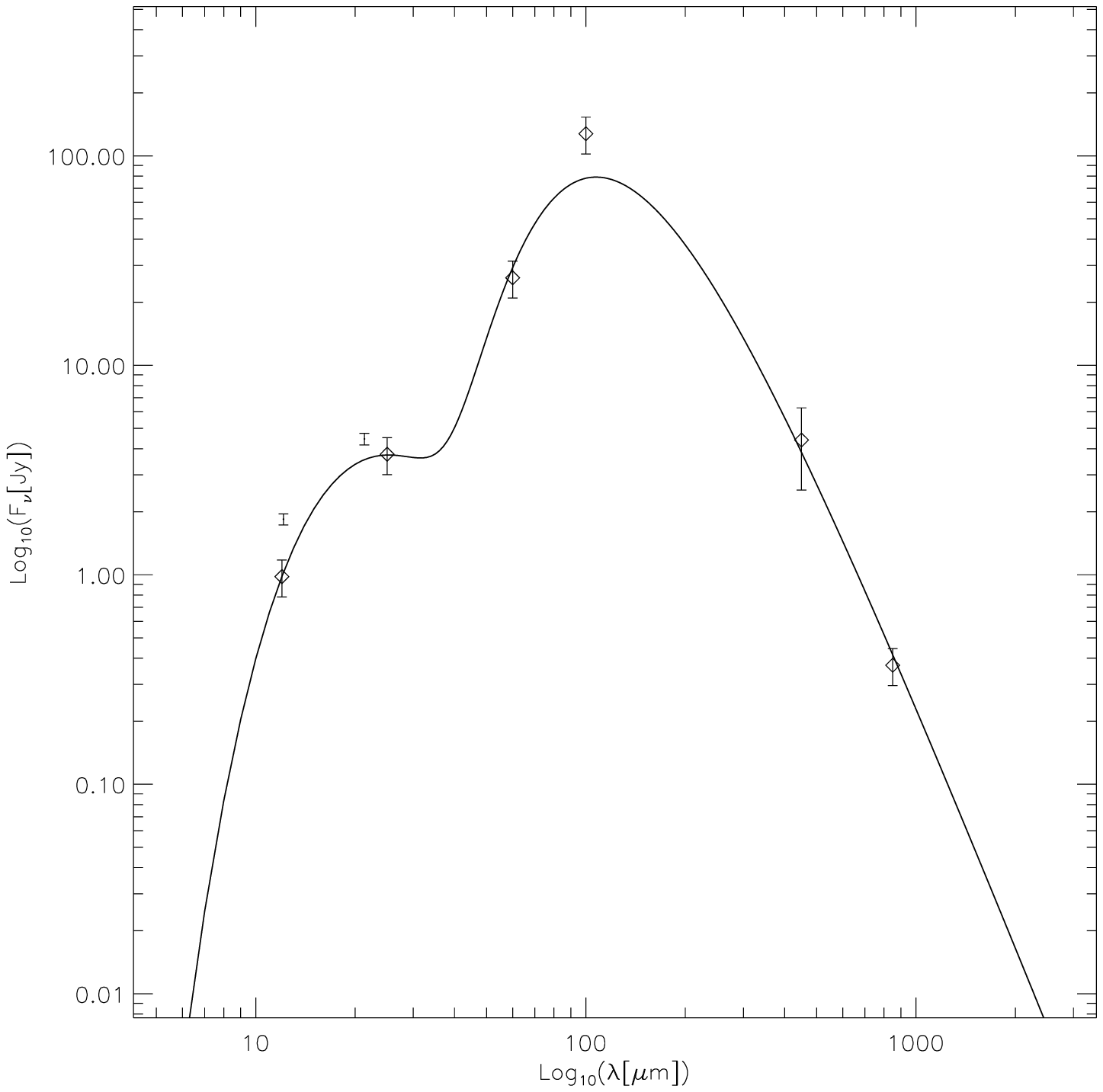}
\includegraphics*[scale=0.40]{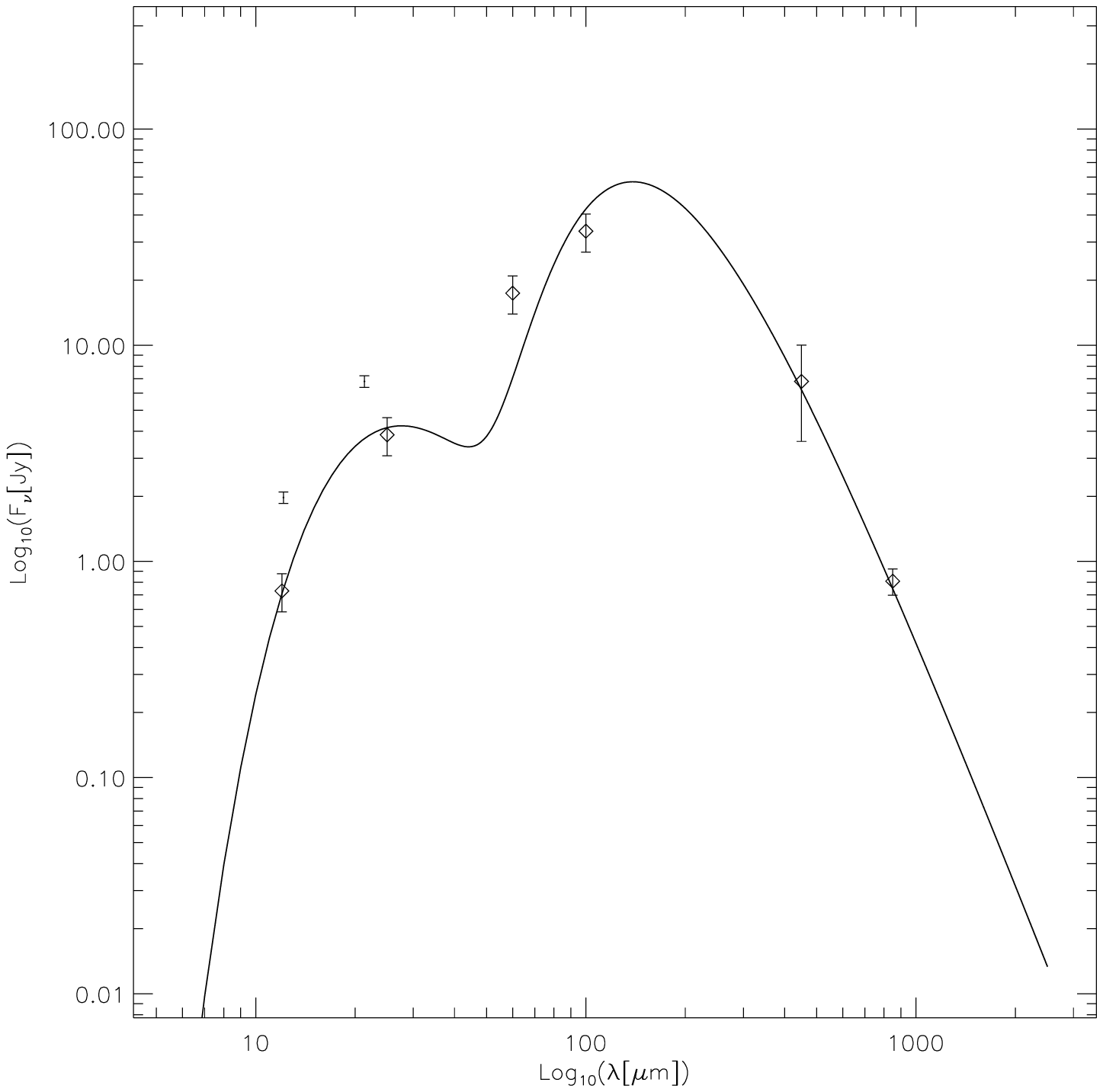}
\caption{SEDs of the sources SFO 31 (left) and SFO 39b (right) including fluxes from the MSX survey. MSX fluxes are shown as error bars only, IRAS and SCUBA fluxes are plotted as diamonds.}
\label{fig:msxSEDs}
\end{center}
\end{figure*}
\end{center}

Several of our sources were found to contain objects identified in the MSX survey. However, a very limited number of these sources were found to be closely associated with the submillimeter cores identified in this work (the criterion being a position within one IRAS beam-width). An even smaller number of these sources were found to have reliable fluxes within the MSX catalogue. Sources with flux qualities of lower than a 5-sigma detection were rejected. After applying these criteria, sources were found associated with SFO 5, SFO 31, SFO 39 and SFO 43.

Fluxes from the C and E bands of the MSX survey were plotted upon the SEDs of the relevant sources (SEDs for SFO 31 and SFO 39b are shown in Fig.\ref{fig:msxSEDs}). These fluxes (found at wavelengths of 12.13 and 21.34 \micron) serve as a useful comparison to our IRAS HiRes fluxes, found in Section. \ref{sec:IRASHiRes}. The SEDs show that MSX fluxes largely agree with the IRAS HiRes fluxes found at similar wavelengths for three of the five sources found to have relevant detections in the MSX survey, given the errors inherent in comparing the two surveys. This similarity is exemplified by the source SFO 31, shown in Fig.\ref{fig:msxSEDs}. Attention may be drawn to the two sources within which fluxes were not in such good agreement. SFO 39a and 39b show rather large disagreement between the MSX-derived fluxes and the line of the SED as drawn by our best-fit method. It should be noted that these objects are fairly poorly fit in our SED analysis and that the IRAS HiRes 60 \micron ~point is also discrepant with the line. A possible explanation for this is that the source is known to comprise at least two distinct sources and so there may be `blending' of the source emission in at least the IRAS beam if not the SCUBA beam. As further Spitzer data is released, the quality of NIR measurements of these sources will allow a great improvement in the fitting of these sources' SEDs at shorter wavelengths.

\section{Analysis}

\subsection{Cloud Morphology}
  The SCUBA images of the cloud sample show submillimetre cores generally located immediately behind the optically bright rim (online appendix). The submillimetre core positions are generally consistent with the IRAS source positions presented in SFO91 and SO94 (i.e. the submillimetre cores lie within the IRAS beam `footprint'), though some discrepancy may be due to the IRAS tracing the more diffuse dust surrounding the cores.\\

\begin{center}
\begin{figure*}
\begin{center}
\includegraphics*[scale=0.45]{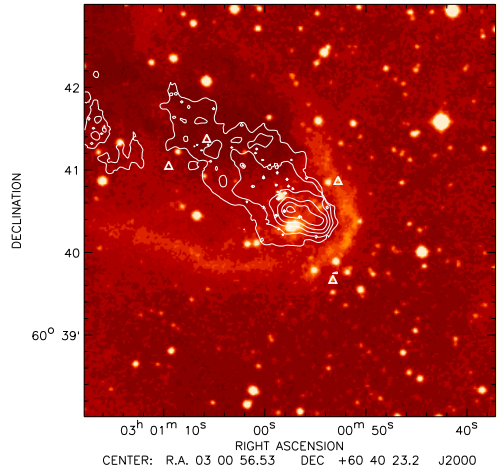}
\caption{SCUBA 850\micron ~contours overlaid on DSS images of SFO 13. Triangles indicate infrared sources from the 2MASS Point Source Catalogue \citep{Cutri2003} that have been identified as YSOs. Contours start at 6$\sigma$ and increase in increments of 20\% of the peak flux. }
\label{fig:scudss13}
\end{center}
\end{figure*}
\end{center}

\begin{center}
\begin{figure*}
\begin{center}
\includegraphics*[scale=0.45]{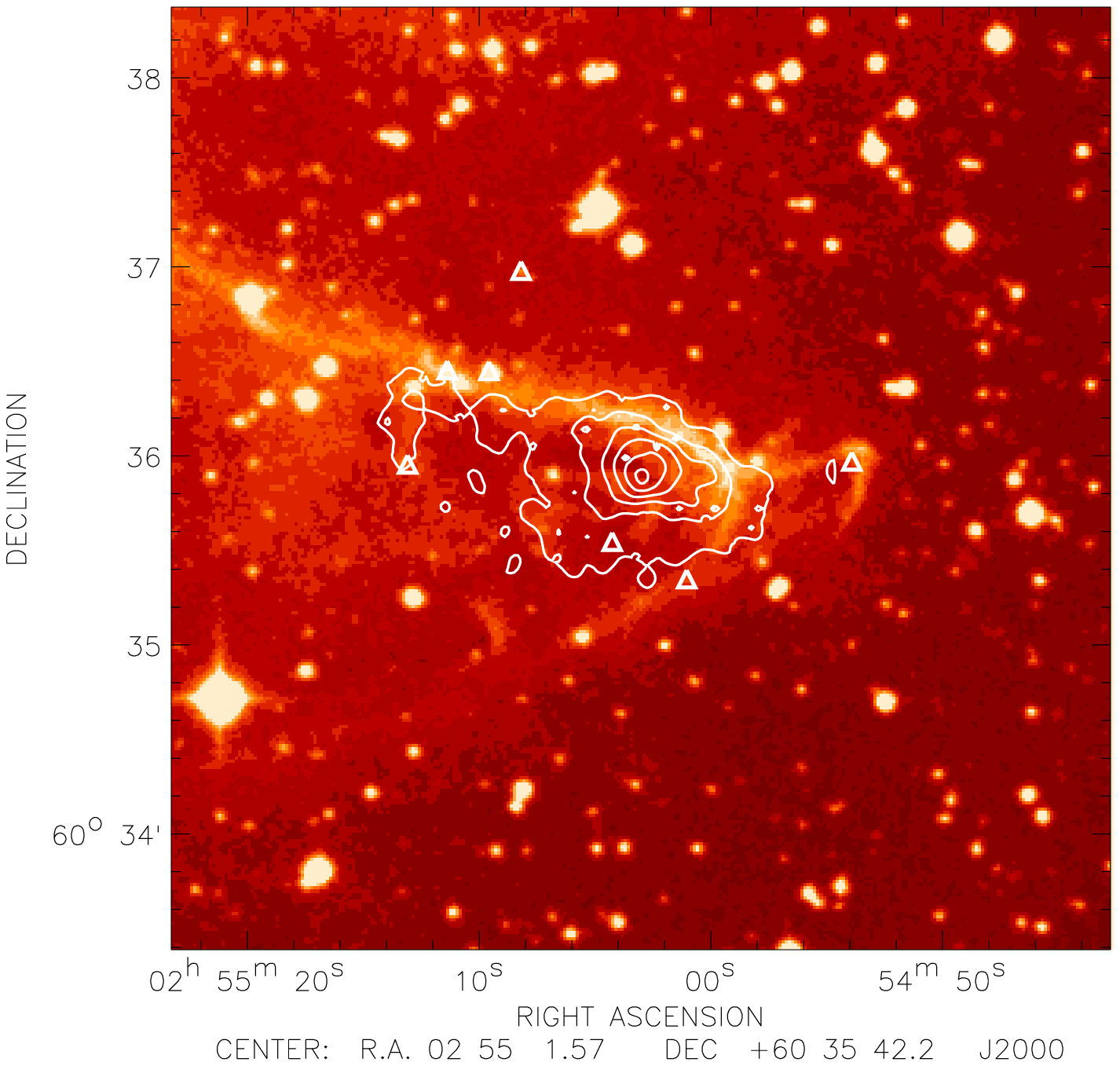}
\caption{SCUBA 850\micron ~contours overlaid on DSS images of SFO 12. Triangles indicate infrared sources from the 2MASS Point Source Catalogue \citep{Cutri2003} that have been identified as YSOs. Contours start at 4$\sigma$ and increase in increments of 20\% of the peak flux. }
\label{fig:scudss12}
\end{center}
\end{figure*}
\end{center}

The submillimetre emission from each core is approximately radially symmetric at the 50\% level. Below the 50\% flux level the weaker, more diffuse, submillimetre emission from each source generally follows the optical boundaries of the bright rimmed clouds and thus may show a cometary morphology (e.g. SFO 13, Fig. \ref{fig:scudss13}). A general summary of morphology is then: radially symmetric contours surrounded by diffuse emission following the optical cloud morphology and often elongated along the direction of the ionising source, exemplified by SFO 12 in Fig. \ref{fig:scudss12}.

\subsection{Luminosities of Embedded Sources}
\label{sec:int}

Use of the far infrared (FIR) and submillimetre fluxes of the sources allows the estimation of the spectral class of any YSOs that may be present in the cloud cores. The luminosities of each source were estimated by integrating under the modified blackbody curve fitted to each cloud's observed fluxes, as presented in Table \ref{tbl:sp}. Distances to each source were taken from SFO91 and SO94. The derived properties of the internal sources are presented in Table \ref{tbl:sp}, upper flux limits from those sources that were not detected by the survey have been utilised to determine upper limits for the masses of potential cores in those clouds, along with upper limits for density and extinction values. These values are presented in Table.\ref{tbl:corepropsnondets}.

\begin{center}
\begin{table*}
\begin{center}
\caption{Core properties determined from modified blackbody fits.}
\tiny
\begin{tabular}{ccccccccccc}
\hline
                           & {}&{}&{}&{}&{}&{}&{}&{}&{Distance}&{}\\ 
\multicolumn{1}{c}{Source} & {T$_{\rm{Hot}}$} & {T$_{\rm{Cold}}$} & {M}		  & {Log$_{\rm{10}}$(n)} & {A$_{\rm{V}}$} & {A$_{\rm{K}}$} & Luminosity	&   {Spectral } & To Cloud$^{b}$ & L$_{\rm{submm}}$/L$_{\rm{bol}}$\\
                           & {(K)}       & {(K)}        & {(M$_{\odot}$)} & {cm$^{-3}$}	    &	        & {}	    			   & (L$_{\odot}$) &  {Type$^{a}$} & (kpc) 		  &  (x 10$^{-3}$)\\ 
\hline
01   & 169  & 26  &  2.18  & 4.6  &  8.5 & 1.0 & 122   &  B8	& 0.85  &  6.9  \\
02   & 182  & 22  &  3.37  & 4.8  & 13.2 & 1.5 & 57    &  B9	& 0.85  & 15.5  \\
03   & 126  & 24  &  1.15  & 4.3  &  4.5 & 0.5 & 39    &  A0	& 0.85  &  9.2  \\
04   & 182  & 26  &  0.03  & 4.6  &  2.1 & 0.2 & 1     &  G2	& 0.19  &  6.7  \\
05   & 107  & 24  & 25.85  & 4.6  & 20.2 & 2.3 & 1154  &  B4	& 1.90  &  6.5  \\
07   & 170  & 21  & 31.26  & 4.7  & 24.4 & 2.7 & 400   &  B5	& 1.90  & 19.0  \\
09   & 185  & 27  &  3.03  & 3.7  &  2.4 & 0.3 & 181   &  B7	& 1.90  &  6.5  \\
10   & 181  & 29  &  2.22  & 3.5  &  1.7 & 0.2 & 268   &  B6	& 1.90  &  4.0  \\
12   & 157  & 21  & 16.87  & 4.4  & 13.2 & 1.5 & 201   &  B7	& 1.90  & 18.5  \\
13   & 177  & 23  & 19.60  & 4.5  & 15.3 & 1.7 & 444   &  B5	& 1.90  & 11.5  \\
14   & 121  & 27  & 51.79  & 4.9  & 40.5 & 4.5 & 3205  &  B2	& 1.90  &  5.8  \\
15   & 130  & 25  &  7.04  & 3.3  &  1.7 & 0.2 & 317   &  B6	& 3.40  &  7.3  \\
16   & 60   & 17  &  1.25  & 5.3  & 22.1 & 2.5 & 8     &  A8	& 0.40  & 24.4  \\
18   & 108  & 18  &  1.15  & 5.3  & 20.3 & 2.3 & 6     &  F0	& 0.40  & 30.6  \\
23   & 125  & 19  &  6.26  & 4.2  &  6.9 & 0.8 & 40    &  A0	& 1.60  & 27.1  \\
25a  & 139  & 19  &  5.59  & 5.1  & 25.9 & 2.9 & 43    &  A0	& 0.78  & 24.3  \\
25b  & 139  & 19  &  5.69  & 5.1  & 26.4 & 3.0 & 43    &  A0	& 0.78  & 24.6  \\
26   & 212  & 24  &  1.05  & 3.9  &  2.2 & 0.3 & 28    &  A1	& 1.15  & 10.2  \\
27   & 153  & 24  &  3.36  & 4.4  &  7.2 & 0.8 & 108   &  B8	& 1.15  &  9.9  \\
29   & 219  & 22  &  3.57  & 4.4  &  7.6 & 0.9 & 48    &  A0	& 1.15  & 14.5  \\
30   & 127  & 34  & 29.33  & 4.5  & 17.1 & 1.9 & 6945  &  B1.5  & 2.20  &  2.3  \\
31   & 147  & 27  &  1.19  & 4.1  &  3.4 & 0.4 & 96    &  B8	& 1.00  &  5.3  \\
33   & 149  & 24  &  0.45  & 4.1  &  2.2 & 0.3 & 21    &  A2	& 0.75  & 10.0  \\
34   & 188  & 21  &  1.55  & 4.6  &  7.8 & 0.9 & 23    &  A2	& 0.75  & 15.7  \\
37   & 127  & 20  &  3.82  & 5.0  & 19.2 & 2.2 & 63    &  B9	& 0.75  & 12.5  \\
39a  & 133  & 20  &  2.77  & 4.9  & 13.9 & 1.6 & 36    &  A0	& 0.75  & 16.3  \\
39b  & 134  & 21  &  2.09  & 4.7  & 10.5 & 1.2 & 35    &  A0	& 0.75  & 13.1  \\
40   & 224  & 24  &  0.45  & 4.1  &  2.2 & 0.3 & 16    &  A4	& 0.75  &  9.4  \\
41   & 189  & 24  &  0.43  & 4.0  &  2.1 & 0.2 & 15    &  A4	& 0.75  &  8.5  \\
42   & 187  & 22  &  0.89  & 4.4  &  4.5 & 0.5 & 16    &  A4	& 0.75  & 12.3  \\
43   & 143  & 26  & 28.34  & 4.4  & 13.9 & 1.6 & 1385  &  B3	& 2.40  &  6.1  \\
87a  & 204  & 30  &  5.93  & 4.4  &  8.8 & 1.0 & 698   &  B4	& 1.38  &  4.0  \\
87b  & 204  & 31  &  4.45  & 4.3  &  6.6 & 0.7 & 657   &  B4	& 1.38  &  3.4  \\
89   & 194  & 30  &  5.18  & 4.3  &  7.7 & 0.9 & 597   &  B5	& 1.38  &  4.1  \\
\hline
\end{tabular}\\							   	      
\label{tbl:sp}
\scriptsize	
$^{a}$ Spectral types taken from \citet{DeJager1987}\\
$^{b}$ Distances taken from \citet{Sugitani1991} and \citet{Sugitani1994}
\end{center}								        
\end{table*}
\end{center}								        

\begin{center}
\begin{table*}
\begin{center}
\caption{Limiting values of potential core properties determined from upper flux limits in the survey's non-detections.}
\small
\begin{tabular}{ccccc}
\hline
\multicolumn{1}{c}{Source} & {M}	     & {Log$_{\rm{10}}$(n)}  & {A$_{\rm{V}}$} & {A$_{\rm{K}}$}  \\
                           & {(M$_{\odot}$)} & {cm$^{-3}$}	&	    & {}	\\
\hline
06 &  7.80 &  4.1 &  6.1 &  0.7 \\
17 &  0.32 &  4.7 &  5.6 &  0.6 \\
28 &  1.95 &  4.1 &  4.2 &  0.5 \\
77 &  2.48 &  3.8 &  2.6 &  0.3 \\
78 &  2.89 &  3.8 &  3.0 &  0.3 \\
\hline
\end{tabular}
\label{tbl:corepropsnondets}\\
\end{center}
\end{table*}
\end{center}

The assumption has been made in this analysis that all FIR and submillimetre luminosity is due to a single source, enabling an order of magnitude estimate of the stellar content within the protostellar envelope. \citet{Wood1989} showed that, for a realistic mass function, the spectral type of the most massive member in a cluster is only 1.5-2 spectral classes lower than that derived for the single embedded star case. The median luminosity of internal sources in the sample is $\sim$63 L$_{\odot}$, corresponding to a spectral type of B9.

\subsection{Classification of the Internal Sources}
\label{sec:Classification}
\citet{Andre1993} defined Class 0 protostars as those obeying the approximate relationship
L$_{\rm{submm}}$/L$_{\rm{bol}}$ $\gtrsim$ 5 x 10$^{-3}$ (here we adopted the definition
of L$_{\rm{submm}}$ as the luminosity radiated longward of 350 \micron). The clouds that have been modelled using the
modified blackbody analysis all approximately obey this relationship (see Table \ref{tbl:sp}) with 29 of 34 sources having L$_{\rm{submm}}$/L$_{\rm{bol}}$ $\geq$ 5 x 10$^{-3}$ and all sources having L$_{\rm{submm}}$/L$_{\rm{bol}}$ $\geq$ 2.3 x 10$^{-3}$. No correlation was found between the masses, luminosities or temperatures of the sources and the value of their L$_{\rm{submm}}$/L$_{\rm{bol}}$ ratio. \\

The
dust temperatures that have been found for the cores in the sample presented here are higher than that
normally found for starless cores (T$_{\rm{d}}$ $\sim$ 10 K; \citep{Evans1999}). This is expected as internal hot components are observed, these are heating the protostellar cores from within, and this establishes a scenario of ongoing star formation within our sample.\\

The bolometric luminosities derived for the clouds in this survey range from 1 L$_{\odot}$, typical of
class 0 and class I protostellar objects \citep{Andre1993,Chandler2000}, to 6945 L$_{\odot}$, more
indicative of massive star formation \citep{Mueller2002}. The median luminosity found for the sample is
$\sim$63 L$_{\odot}$. This suggests that the sample consists largely of intermediate-mass star forming
regions or stellar clusters, consistent with the findings of SFO91 and \citet{Yamaguchi1999}. This is despite the differences in fluxes found between the IRAS PSC fluxes used by SFO91 and IRAS HiRes fluxes used here.
\subsection{Cloud Classification}
\label{Sec:CC}

\begin{center}
\begin{figure*}
\begin{center}
\includegraphics*[scale=0.45]{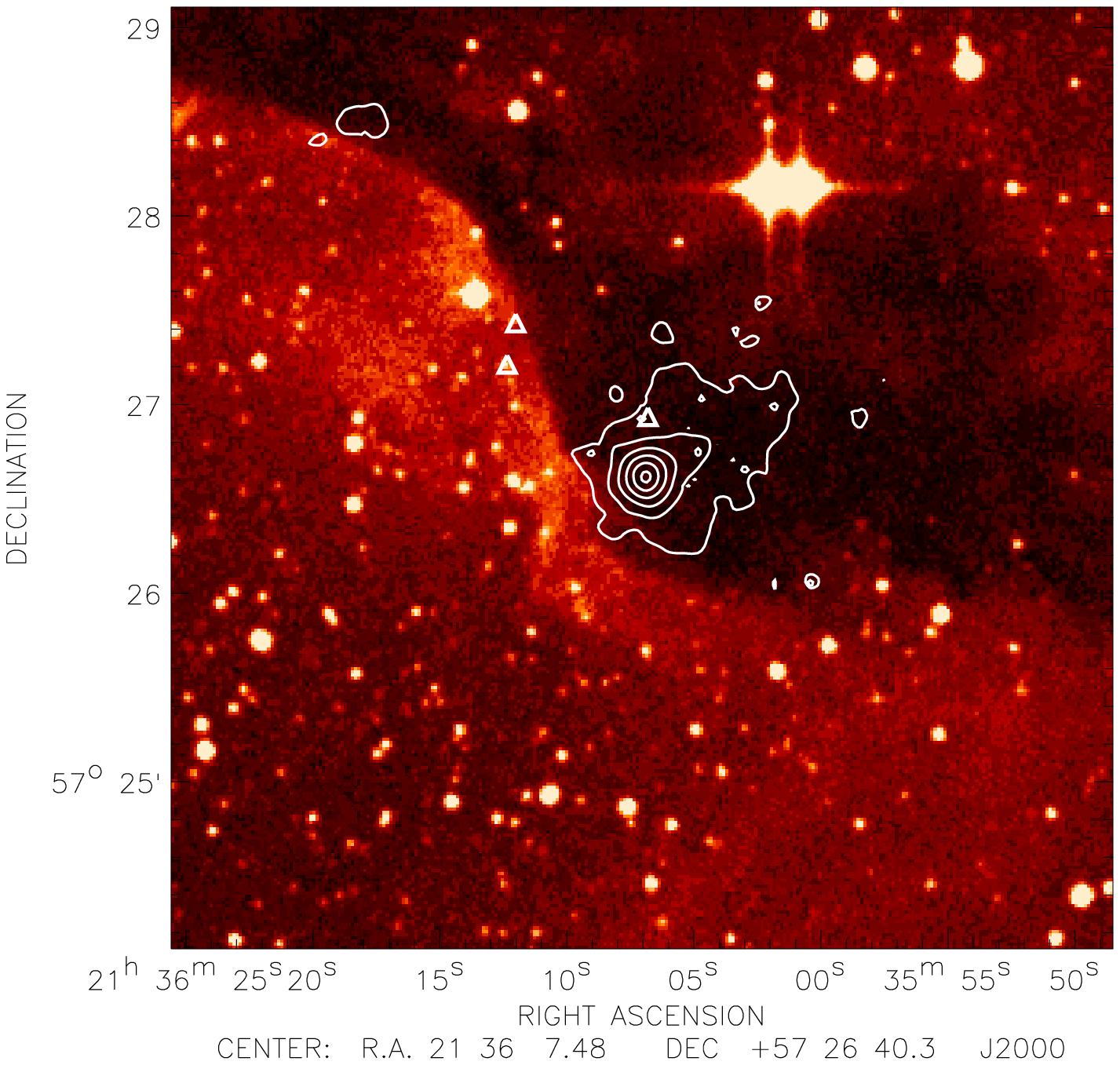}
\caption{SCUBA 850\micron ~contours overlaid on DSS images of SFO 36. Triangles indicate infrared sources from the 2MASS Point Source Catalogue \citep{Cutri2003} that have been identified as YSOs. Contours start at 9$\sigma$ and increase in increments of 20\% of the peak flux. }
\label{fig:scudss36}
\end{center}
\end{figure*}
\end{center}

  SFO91 separated their catalogue into three morphological types based upon the curvature exhibited by the BRC. Their classifications are as follows, (1) Type A, moderately curved rim with a length to width ratio, $l/w$, of less than 0.5; (2) Type B, tightly curved rim with $l/w$ greater than 0.5 and; (3) Type C, cometary rim\footnote{SFO 36 was identified by SFO91 as `A or C'. Based upon the position and morphology of the submillimetre emission presented in these images the identification of SFO 36 as Type `A' morphology is evident (see Fig. \ref{fig:scudss36}).}.\\
  The three rim types are closely related to the development of clouds under the influence of ionising sources. The three general rim types, A, B and C, closely match the \citet{Lefloch1994} RDI model snapshots at 0.036, 0.126 and 1.3 Myr respectively. This strongly suggests that the variation in the large scale morphology represents the duration of each clouds exposure to ionising radiation, rather than a fundamental difference in structure or composition. This is further supported by the fact that \citet{Lefloch1994} found no significant effects upon the morphological evolution of the clouds due to varying initial conditions.

A simplistic interpretation of the results of \citet{Lefloch1994} is that, if their evolutionary scenario is correct, then the fractional number of type `A', `B' and `C' type rims observed should match the fractional lifetime of each phase. This was found not to be the case, however.  The simulations of \citet{Miao2006} have reproduced the morphological results of \citet{Lefloch1994} but incorporate self-gravity and thermal evolution into their models. Whilst \citet{Miao2006} find a similar morphological evolution to \citet{Lefloch1994} they find that the inclusion of self-gravity decreases the timescale for the cometary morphology of their simulated clouds to develop.  The simulations of \citet{Kessel-Deynet2003} include the effects of self-gravity as well as ionisation. These simulations similarly show decreased timescales for morphological evolution in comparison to the results of \citet{Lefloch1994}. They also show fragmentation occuring in BRCs once a type `C' morphology has been achieved (on a timescale of $\sim$0.4 Myr) and subsequently show small `trunk-like' structures emerging from the larger BRC.

\citet{Miao2006} gives representative ages for the `A', `B' and `C' type rims of 0.24, 0.30 and 0.39 Myr respectively.
Using the results of \citet{Miao2006} a close match was found between the number of `A' type rims classified in the SFO catalogue and the lifetime of each of these phases (63\% predicted compared to 66\% observed). If the variation in rim morphology seen across the SFO catalogue may then be attributed to an evolutionary sequence then it is reasonable to search for differences in the physical properties of the different rim types. It is expected that the core properties such as temperature and density would increase with time, while the mass of the cold component would decrease as the protostellar envelope is accreted onto the central object and the cloud is further evaporated by the ionising UV radiation. One would naively expect to find star formation that is more developed in the clouds that are further evolved.

The morphological types of each BRC as identified by SFO91 are tabulated in table \ref{tbl:corepos}. It can be seen that the number of `A' type clouds exceeds the number of `B' types by approximately a factor of 2 (26 as compared to 14) with only 2 `C' type rims. While the timescales reported for `A' type rims versus the total of `B' and `C' type rim timescales in \citet{Miao2006} match those observed, the lack of observed `C' type rims is unexplained. Given the large amount of time attributed to this phase by \citet{Lefloch1994} many of these type rims should be observed. However, without more detailed information on the modelling of the $eventual$ fate of these clouds, it is not possible to draw any conclusions about the lack of observed `C' type rims.

  Grouping the cores in the sample according to the classifications of SFO91 (Table \ref{tbl:rim_types}) does not show any significant differences in the derived core properties (e.g. mass, density and extinction) between the `A' rims and the `B' rims. This is not what might be expected if the rim types are indeed an evolutionary sequence. This suggests two possible scenarios. 
  \begin{enumerate}
  \item The timescale for the development of rim type is short compared to star formation. Thus, assuming that the stars are formed as a result of the interaction of the ionisation front with the BRC, these stars begin to form quickly as a result of this influence. The clouds then rapidly evolve through the suggested development of rim type irrespective of the state of star formation within them.
  \item The formation of stars within the BRCs is unrelated to the interaction of the ionisation front with the BRC. In this case the development of stars within the clouds would be unconnected to the wider cloud evolution.
  \end{enumerate}
  
\begin{center}
\begin{table*}
\caption{Mean core properties grouped as rim type. Values in parentheses are the standard deviation over the sample.}
\begin{center}
\begin{tabular}{cccccc}
\hline
\multicolumn{1}{c}{Rim Type} & {T$_{\rm{Hot}}$} & {T$_{\rm{Cold}}$} & {M} & {Log$_{\rm{10}}$(n)} & {A$_{\rm{k}}$}  \\ 
                             & {(K)}	 & {(K)}     & {(M$_{\odot}$)} & {cm$^{-3}$} & \\
\hline
 A & 165 (45) & 24 (4) & 5.4 (12.5)  & 4.8 (0.5) & 1.1  (1.1) \\
 B & 157 (29) & 24 (4) & 11.7 (11.5) & 5.0 (0.4) & 1.5  (0.9) \\
 C & 127 (1)  & 22 (3) & 2.5 (1.9)   & 5.2 (0.5) & 1.4  (1.2) \\
\hline
\end{tabular}								   	      
\label{tbl:rim_types}\\ 							
\end{center}
\end{table*}									
\end{center}

\subsection{Radio Detections and Ionising Stars}
\label{sec:RDaIS}
\citet{Morgan2004} used archival NVSS \citep{Condon1998} data to look for radio emission from the northern SFO catalogue. Of the 44 clouds observed here, 39 were observed and 31 detected by the NVSS (All of the northern sources that have been observed with SCUBA here were included in \citet{Morgan2004}). All but one of these detections was classified as an ionised boundary layer (IBL) at the edge of the respective bright rim, establishing the interaction of the HII region with the surrounding molecular material. The number of sources with a clearly defined IBL demonstrates that $\sim$70\% of these BRCs are being influenced by the expansion of their respective ionisation fields. Here we find eight clouds in which we have detected submillimetre emission, associated with a prospective protostellar core, which show no detectable radio emission. These clouds are SFO 3, 9, 23, 24, 26, 33, 34 and 39.

\begin{center}
\begin{table*}
\begin{center}
\caption{Comparison of cloud properties between those with and those without associated radio emission, values in parentheses represent the standard deviation within the group.}
\begin{tabular}{ccc}
\hline
		& IBL Detection$^{a}$ & IBL Non-Detection \\
Source Property & Mean Value	  & Mean Value		\\
\hline		
T$_{\rm{Hot}}$(K)			& 154 (38)	    & 157 (34)	         \\
T$_{\rm{Cold}}$(K)			& 24 (4)	    & 23 (3)	         \\
Core Mass(M$\odot$)			& 11 (14)	    & 2 (2)	         \\
Log Density				& 5.0 (0.5)	    & 4.8 (0.4)          \\
A$_{\rm{k}}$				& 1.5 (1.2)	    & 0.7 (0.5)          \\
N$_{\rm{CI}}$$^{b}$			& 3 (2) 	    & 3 (3)	         \\
N$_{\rm{CTTS}}$				& 1 (1) 	    & 0			 \\
L$_{\rm{bol}}$(L$\odot$)		& 651 (1547)	    & 50 (53)	         \\
L$_{\rm{submm}}$/L$_{\rm{bol}}$	& 12e-03 (7.7e-03)  & 14e-03 (6.4e-03)   \\
\hline
\label{tbl:RScomp}
\scriptsize$^{a}$Detection at 21cm
$^{b}$CI = Class I protostellar source
\end{tabular}					      
\end{center}		
\end{table*}								      
\end{center}

\begin{center}
\begin{figure*}
\begin{center}
\includegraphics*[scale=0.50]{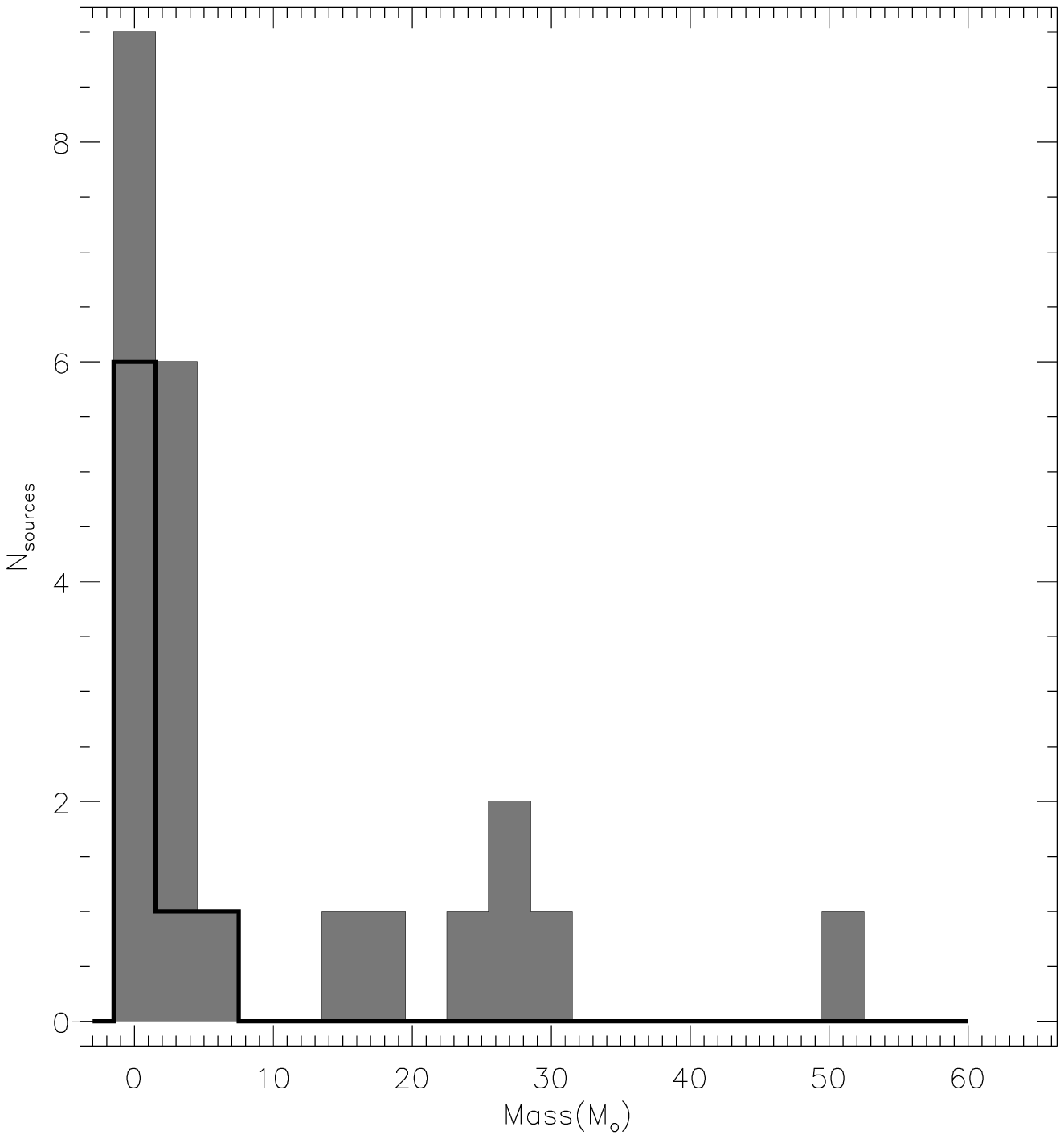}
\end{center}
\caption{Histogram plot of the mass distributions of those sources both detected (filled) and not detected (black line) in the 21cm NVSS}.
\label{fig:Mass_Rdet}
\end{figure*}
\end{center}

  If an interaction between the ionising source and the molecular material exists in the eight clouds not detected at $\sim$ 21 cm, then it is presumably at a much lower level than in the rest of the sample. A comparison between these two sets of sources is therefore warranted to search for differences which may highlight the presence and/or effects of RDI.
  The mean temperatures, masses, densities, extinctions, luminosities, internal stellar contents and the degree of IR source association of the two sets were compared (Table \ref{tbl:RScomp}). Clouds without associated radio emission show a tendency towards lower mass, density, visual extinction and luminosity, though this tendency must be dealt with cautiously due to the size of the standard deviations within the respective groups. It can be seen from Table \ref{tbl:RScomp} that all values agree \emph{within these standard deviations}. A tentative statement is then, the presence of a strong ionisation field in the vicinity of protostars taken from the observed sample may lead to more massive, luminous protostellar sources. Support for this statement may be found by comparing the distribution of source mass for those sources detected and not-detected at radio wavelengths. Figure \ref{fig:Mass_Rdet} shows a histogram of both sets of sources, it can be seen that the distribution of those sources detected at radio wavelengths contains a noticeably higher number of higher mass sources.

  If there exists a correlation between the radio flux emanating from a source rim and the mass/luminosity of the internal protostar embedded within that rim then it is warranted to compare the luminosities and masses of each source to the predicted ionising flux of each rim's ionising source, i.e. the ionising flux we expect to be arriving at each cloud's rim given the spectral type of the star powering the related HII region and the projected distance between that star and the observed bright rim.
  
\begin{center}
\begin{figure*}
\begin{center}
\includegraphics*[scale=0.65]{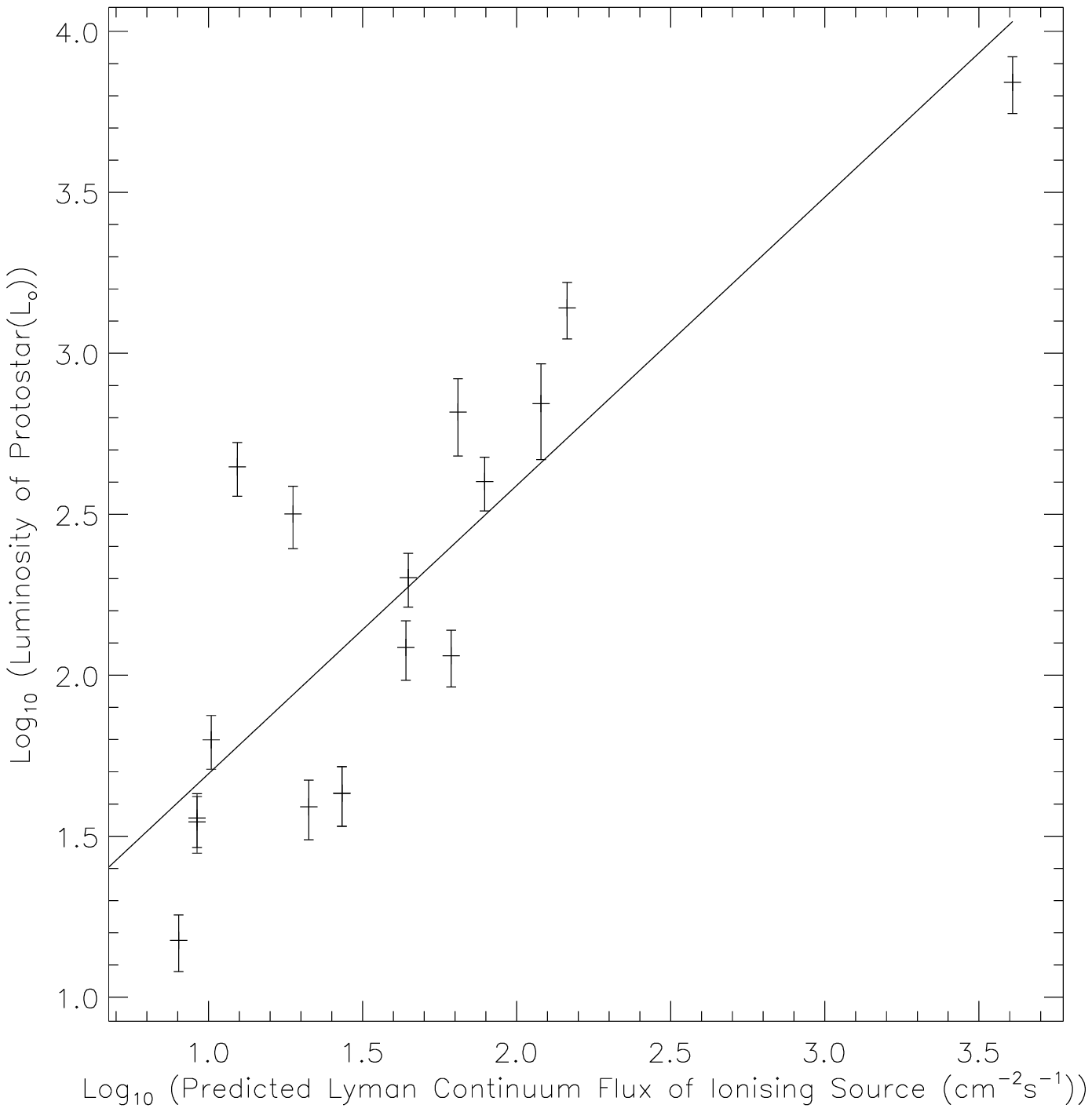}
\end{center}
\caption{Plot of the internal luminosity of protostars vs. the predicted ionising flux arriving at the bright rim of sources with `B' and `C' type morphologies}.
\label{fig:Lbol_vs_PF}
\end{figure*}
\end{center}

  A correlation was found between the predicted ionising flux and the luminosity of each protostellar object. Futhermore, it was found that this correlation appeared to arise solely from the type `B' and type `C' rims within our sample. Type `A' rims showed no such correlation, a plot of the relation between the predicted ionising flux \citep{Morgan2004} and the internal source luminosity for `B' and `C' type rims is presented in Fig.\ref{fig:Lbol_vs_PF}
  
It would then appear that there is a link between the strength of the ionisation field in which the BRCs are embedded and the luminosity of stars that are produced within those clouds, at least for `B' and `C' type rims. This is a very good indication that, at some level, the illuminating source of the HII region plays an important role in the formation of stars within BRCs.

\subsection{The Mass Function of the SFO Catalogue}
\label{sec:mass_fn}  
\begin{center}
\begin{figure*}
\begin{center}
\includegraphics*[scale=0.65]{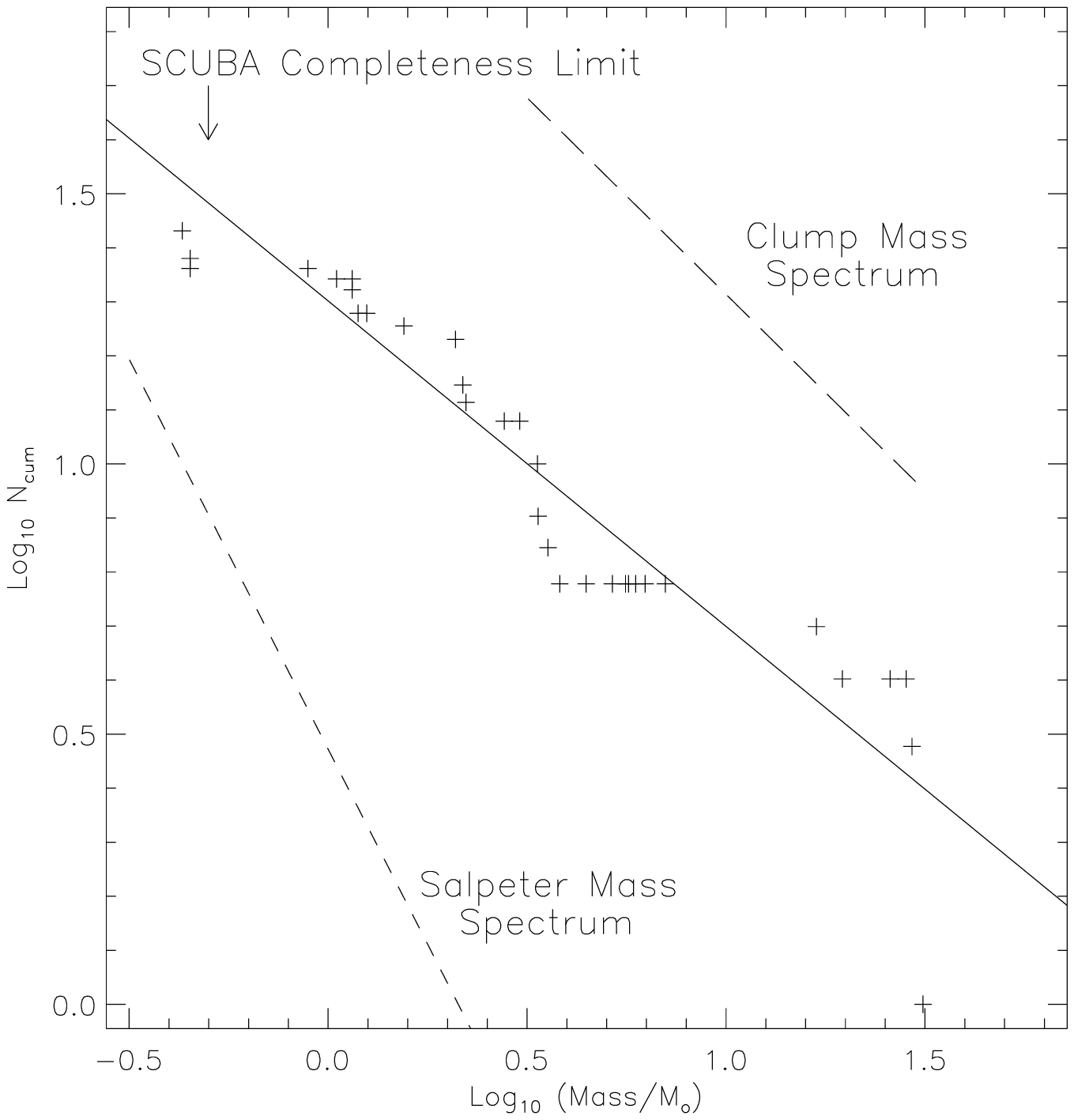}
\caption{The mass spectrum of the sources detected by SCUBA in this survey. The Y-axis represents the cumulative number of sources (N$_{\rm{cum}}$) per mass bin of size 0.05 M$_{\odot}$. The long-dash line represents the giant molecular cloud clump mass function taken from \citet{Blitz1993} and the short-dash line represents the well-known Salpeter slope \citep{Salpeter1955}. The solid line represents a best fit to the data from the presented sample.}
\label{fig:Mass_fn}
\end{center}
\end{figure*}
\end{center}

The `Initial Mass Function' (IMF) is a key diagnostic in distinguishing between various regimes of star formation processes. The Salpeter mass function \citep{Salpeter1955} describing main sequence stars is:
\begin{equation}
\frac{dN(M)}{dM} \propto M^{-2.35}
\end{equation}
 This relation has been refined in different circumstances in an attempt to describe stars that are observed to have number functions differing from that presented in \citet{Salpeter1955}. \citet{Blitz1993} found an observed mass spectrum for `clumps' within Giant Molecular Clouds (GMC) that obeys
\begin{equation}
\frac{dN(M)}{dM} \propto M^{-1.54}
\end{equation}
  The masses that have been found from Eq. \ref{eq:mass} for the detected cores in this survey are presented in Table \ref{tbl:sp}, the corresponding mass spectrum arising from this sample is presented in Figure \ref{fig:Mass_fn}. The Salpeter and Blitz slopes are presented for comparison and have values of -2.35 and -1.54 respectively, while a least-squares fit to our data gives a slope of -1.67.
  
We estimate that our sample is complete to 3$\sigma$ at $\sim$0.5 M$_{\odot}$. Therefore, when fitting a line to the mass spectrum presented here we have discarded sources with derived masses below this value. The usefulness of a comparison with the Salpeter slope may not be completely valid as the
Salpeter slope describes stars upon the main sequence, whereas this sample is comprised of YSOs (or possibly multiple YSOs) which are
fundamentally different astronomical objects to main sequence stars. However, an important question is addressed by a
comparison of the presented sample's mass spectrum with the Salpeter slope. \citet{Alves2007} found that the mass function of dense molecular cores is similar in shape to the stellar IMF but may be scaled to higher masses. Their observations of 159 dense cores revealed a `Salpeter-like power law'. It can be seen that
the slope of our mass spectrum is in fact significantly shallower than that of the Salpeter function and somewhat shallower than the Blitz `clump' function. From this, we can determine that these sources are more efficiently producing higher mass stars than may be expected or, as is more likely, these sources contain embedded $clusters$ of stellar sources. This increased efficiency is apparent \textit{at a given mass}, therefore, the observed difference in slope may well be interpreted as a lack of low-mass sources rather than a preponderance of high-mass sources.

\begin{center}
\begin{figure*}
\begin{center}
\includegraphics*[scale=0.65]{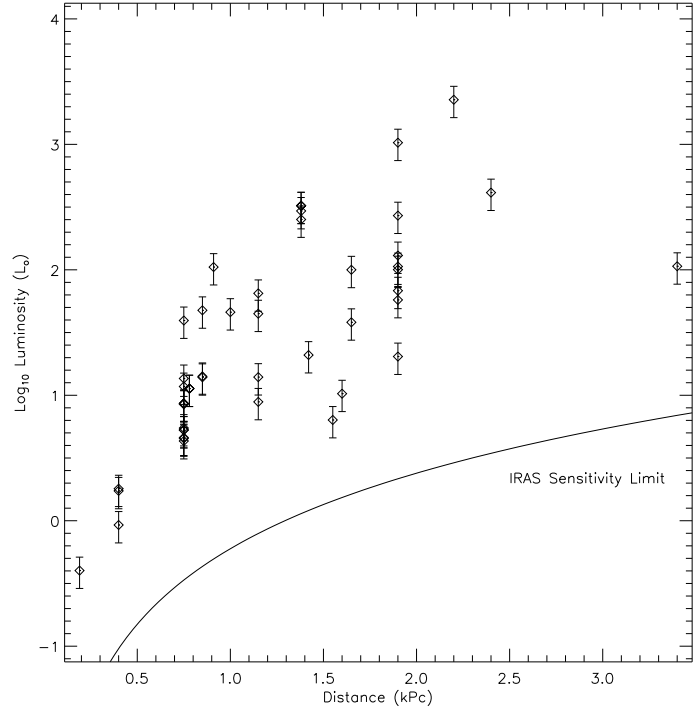}
\caption{The luminosities of the SCUBA sample plotted as a function of distance (diamonds), the solid line represents the sensitivity limit of the IRAS.}
\label{fig:IRAS_limit}
\end{center}
\end{figure*}
\end{center}

The sensitivity of IRAS limits the sample to sources above $\sim$ 3 L$_{\odot}$ at a mean distance of 1.29
kpc. By plotting the luminosity/distance relationship of these sources\footnote{In creating this plot only 60 and 100 \micron ~fluxes were used and compared to the flux limits of IRAS at these wavelengths.} (Fig.
\ref{fig:IRAS_limit}) it can be seen that the sample lies well above this sensitivity limit

\subsubsection{Separating the Sample}
  If, as suggested in Section \ref{sec:RDaIS}, the protostars found within the different cloud morphologies have some fundamental difference in their formation mechanism then it may become more evident by separating the sample on a morphological basis and plotting mass spectra separately.
  
  Here we compare our source sample with data taken from \citet{Mitchell2001}, in which 850 \micron
~SCUBA maps are presented of 67 discrete continuum sources in the Orion B molecular cloud, some of which
are shown to be star-forming. \citet{Mitchell2001} define a mass for each `clump' through assuming a
dust temperature, T$_{\rm{d}}$, via:

\begin{equation}
M_{\rm{clump} = }1.50 \times S_{\rm{850}} [exp \left ( \frac{17 K}{T_{\rm{d}}}\right )-1] \times \left ( \frac{\kappa_{\rm{850}}}{0.01~cm^{2}~g^{-1}}\right )^{-1} M_{\odot}
\end{equation}

where S$_{\rm{850}}$ is the 850 \micron ~flux enclosed within the clump boundary measured in Janskys and
$\kappa_{\rm{850}}$ is the mass absorption coefficient at 850 \micron, functionally equivalent to the
reciprocal of the mass conversion factor $C_{\rm{\nu}}$ (see Section \ref{sec:DC}). \citet{Mitchell2001} use
fluxes for each clump taken from a 21\arcsec ~aperture and assume a value for $\kappa_{\rm{850}}$ of 0.01 cm
g$^{-1}$ to find masses for a small sample of the clumps in their survey. In order to find masses for
the complete sample of their observations we have recalculated their dust masses assuming a dust temperature of 20 K. As the 850
\micron ~fluxes within a 21\arcsec ~aperture are not available, and to maintain consistency with the
method presented here, we have used the total integrated flux along with a different value of
$\kappa_{\rm{850}}$=0.02 cm$^{2}$g$^{-1}$. The masses found here agree well with those presented for the
smaller sample within \citet{Mitchell2001}.\\

Plotting the mass spectrum of the clumps observed within \citet{Mitchell2001} may provide a more illuminating comparison to our sample than the Salpeter function, or the Blitz observed
clump mass spectrum. As well as the protostellar clumps being similar objects to this sample (the
important difference being the illumination of the clouds by nearby massive stars) the methods used to
derive their masses are identical, whereas the Salpeter and Blitz spectra are based upon various
observations and assumptions. It should be noted that Orion B is a single distance sample, while this
sample represents BRCs at a range of distances which may not be well known. This could possibly
introduce errors into the sample.  As with our own mass spectrum lower mass sources have not been used in a
best-fit slope to the mass spectrum in order to reduce the effects that an incomplete sample may have on
the determination.

\begin{center}
\begin{figure*}[!ht]
\begin{center}
\includegraphics*[scale=0.50]{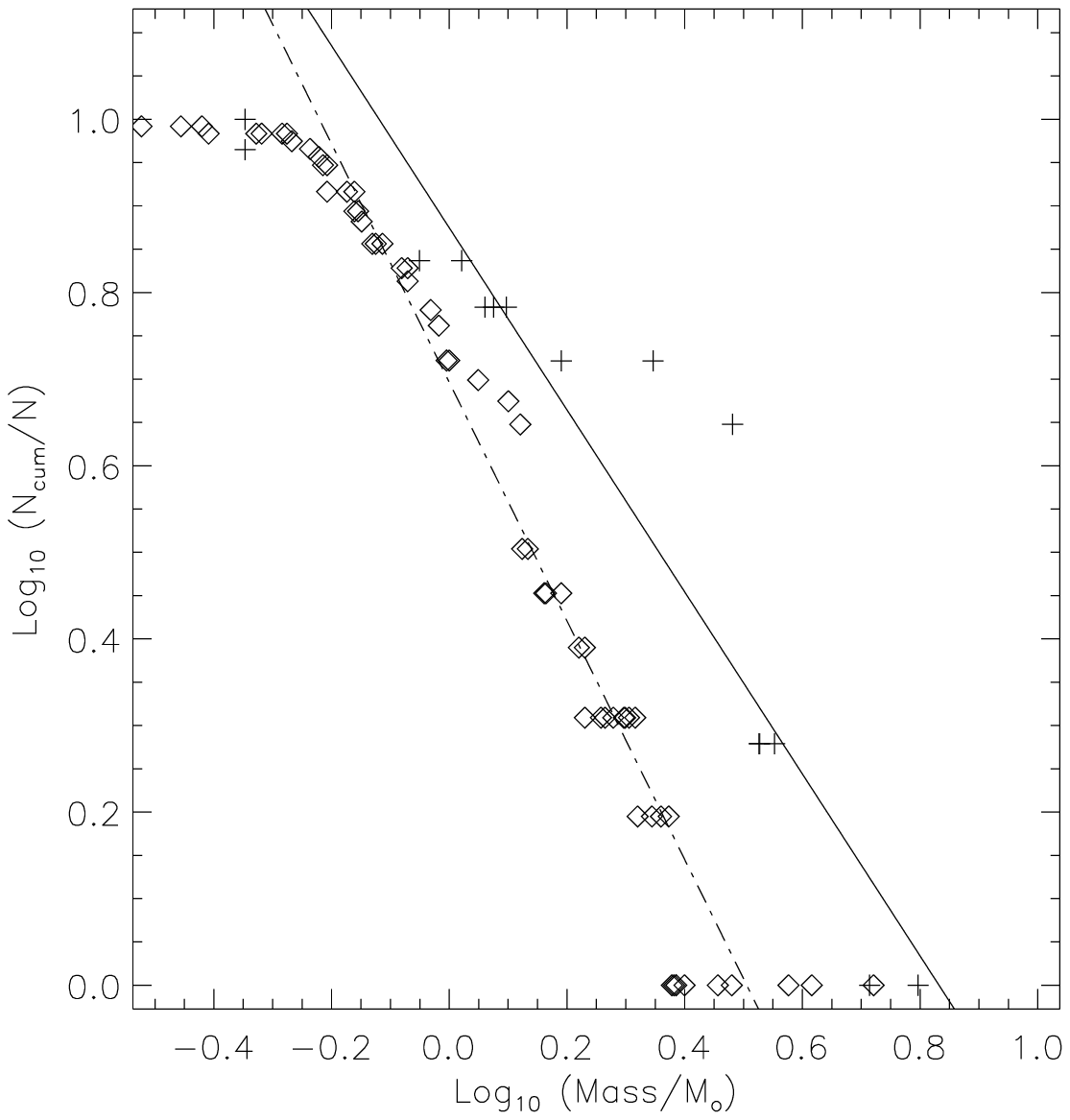}
\includegraphics*[scale=0.50]{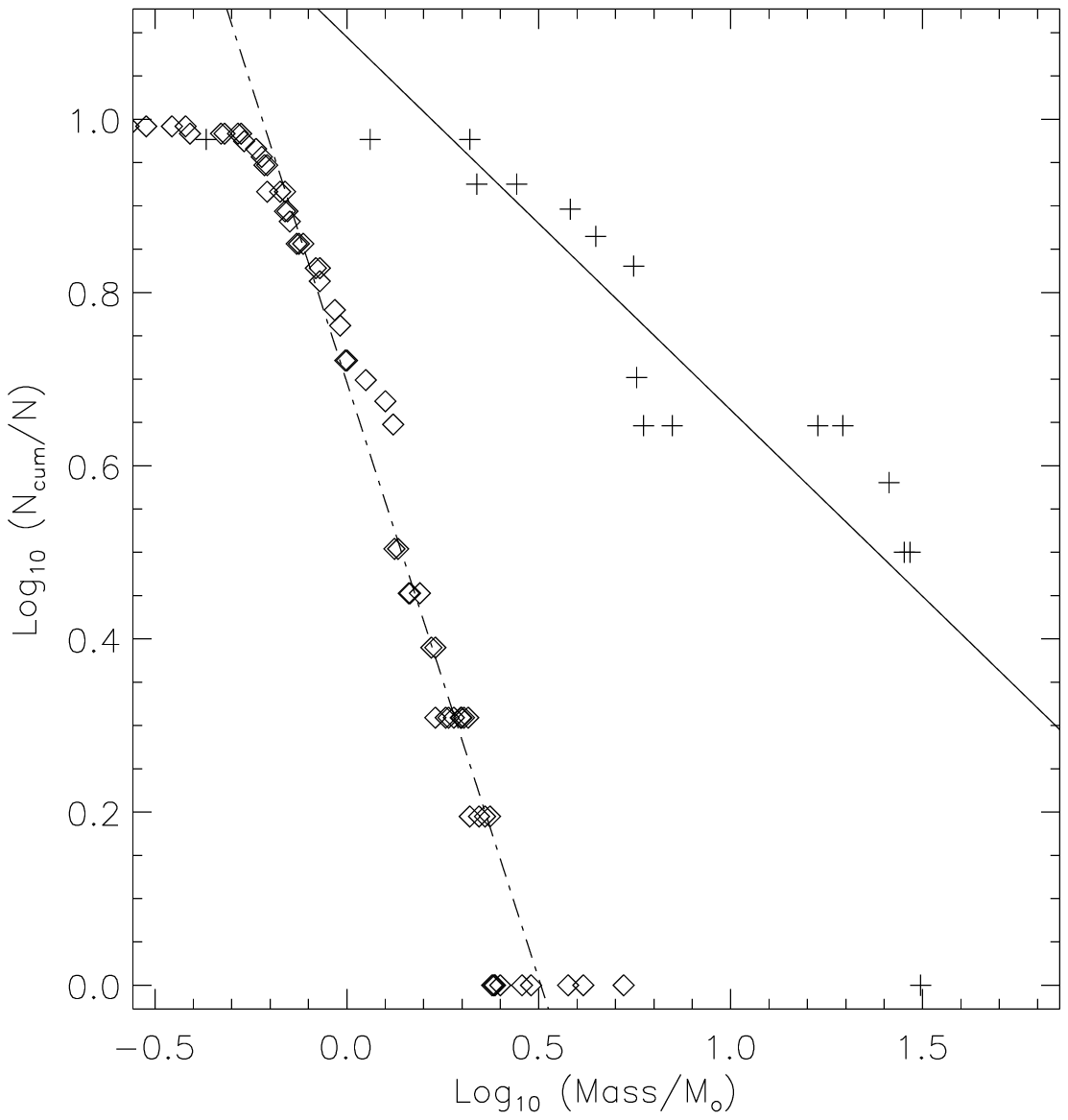}
\caption{The mass spectrum of the `A' type (left) and `B' and `C' type (right) sources detected by SCUBA in this survey (pluses) and in the survey of \citet{Mitchell2001} (diamonds). The Y-axis represents the log of the cumulative number of sources (N$_{\rm{cum}}$) per mass bin of size 0.05 M$_{\odot}$ as a fraction of N, where N is the total number of sources. The solid line is a best fit to our data above the mean sensitivity limit, the dash-dot line is a best fit to the data of \citet{Mitchell2001}. The outlier of SFO 14 with mass 51.79 M$_{\odot}$ has been neglected from the fit for the `A' type rims as it appears anomalous, see Section. \ref{Sec:Discussion}}
\label{fig:Mitchell_Mass_fn}
\end{center}
\end{figure*}
\end{center}

The mass spectra presented in Fig.\ref{fig:Mitchell_Mass_fn} show that clouds with type `A' morphologies appear to follow a mass function close to that of the survey of \citet{Mitchell2001}, while clouds with type `B' and `C' morphologies have a significantly shallower mass function. It is noticeable that the best fit line to our `A' type rim mass function may be skewed by a few outliers at around 3-4 M$_{\odot}$. The number of data points in this plot is low and thus we must try to fill in the graph around these points in order to explore this function further. However, the suggestion from these spectra is that `A' type rims seem to follow a mass law similar to that found in other, non-triggered regions. Whereas type `B' and type `C' rims have a significantly shallower mass spectrum, indicating a tendency toward greater numbers of higher mass sources.

\section{Discussion}
\label{Sec:Discussion}
  Our survey of the submillimetre emission found within BRCs has revealed the presence of embedded cores. The positions of peak submillimetre flux are consistent with IRAS PSC sources, previously discussed by SFO91, SO94 and \citet{Morgan2004} and situated near the edge of known HII regions. \citet{Morgan2004} established the presence of IBLs and photon dominated regions (PDRs) at the interface between the HII regions and the molecular clouds composing the BRCs. Previous studies \citep{Morgan2004,Thompson2004a,Urquhart2006,Urquhart2007} also show that many of these clouds are likely to be in a \textit{post-shocked} state and that any observed star formation is therefore a possible result of the cloud's interaction with the external ionising source.
  
  The embedded cores are likely to contain class 0 protostars with few exceptions (29 of 34 sources obey the \citet{Andre1993} classification for class 0 protostars of L$_{\rm{submm}}$/L$_{\rm{bol}}$ $\gtrsim$ 5 x 10$^{-3}$). Dust temperatures found towards the cores exceed those found towards starless cores (\mbox{$\sim$ 10 K} \citep{Evans1999}) with an average of 24 K. This combination of observed submillimetre flux excess and high dust temperatures clearly establishes a scenario of ongoing early star formation within the sample of BRCs. Bolometric luminosities of the sources have a large range, from 1 L$_{\odot}$ for SFO 4 to 6945 L$_{\odot}$ for SFO 30. However, the majority of sources have L$_{bol}$ $>$ 10 L$_{\odot}$ suggesting that the sample consists largely of intermediate to high-mass star-forming regions. Though some, if not all, of the higher luminosity sources are likely host to protostellar $clusters$ rather than individual protostars.
  
  Bright-rimmed clouds are the potential results of RDI and it has been suggested by theoretical investigations that they may follow a morphological evolutionary sequence \citep{Lefloch1994,Kessel-Deynet2003,Miao2006}. In this scenario different morphological classifications may host different classes, or epochs, of star formation. A preliminary analysis revealed that this was not the case. The sample was separated into each morphological grouping and average masses, luminosities, dust temperatures, extinctions and densities were found to be similar for all sources with no significant deviations; no link between rim morphology and embedded core properties could be made. This strongly suggests that the development of rim type and the ongoing star formation within the BRCs are not linked, or that they occur upon differential timescales such that, once initiated, star formation proceeds independently of any morphological changes occuring to the BRC as a whole.
  
   The detection of radio free-free emission associated with a large number of the clouds in our sample \citep{Morgan2004} establishes the presence of IBLs at the edges of many of these clouds. This indicates that there is some interaction occuring between the clouds and their environment, even if this interaction is unconnected to the internal protostellar activity. It was found that clouds associated with radio-detected IBLs contain protostars, or protostellar clusters, with higher masses, densities and extinctions (column densities) than those not detected at radio wavelengths. This tendency was not found to be significantly large to definitively separate our sample and may simply be due to skewing by a few, high-mass, sources. However, when the predicted strength of the ionisation field within which the clouds are situated was investigated it was found that a significant correlation exists between the strength of this field and the luminosity of the embedded protostars or protostellar clusters. Furthermore, it was found that this correlation arises solely from the `B' and `C' morphological type rims. This clearly indicates the interaction of the external ionisation field with star formation occuring within the clouds, at least in `B' and `C' morphological type rims. The absence of any similar correlation in `A' type rims provides reasonable evidence that these clouds are not subject to RDI triggering processes. 
  
  The evidence of interaction between ambient ionisation field strength and star formation does not rule out the possibility of formation via the `collect and collapse' method proposed by \citet{Elmegreen1977}. In this scenario the expanding nebula around the star ionising the HII region generates gravitationally unstable, shocked layers of gas which are fragmented and have high masses \citep{Whitworth1994}. This mechanism is generally long-lived and acts upon large scales \citep{Zavagno2006} and is capable of producing the flat edged structures seen in `A' type rims such as SFO 41 and SFO 88 (Figs.\ref{fig:images41} \& \ref{fig:images88}).
  
  It appears that star formation occuring within different morphological types of BRC follows different formation processes. The star formation found within `B' and `C' morphological rim types resulting in some part from the influence of nearby massive stars, while the formation found within `A' type rims evolves presumably independently of the environment within which the BRCs are located. This scenario contradicts the evolutionary scenario of rim type predicted by \citet{Lefloch1994}, \citet{Kessel-Deynet2003} and \citet{Miao2006}, at least as connected to protostellar development. This inherently implies that any star formation occuring within `A' type rims must either have existed before the ionisation of the rim or that any rim ionisation occuring is not sufficiently dramatic to influence any subsequent star formation.
  
  In the light of the differences between `A' type rims and `B' and `C' type rims, we sought to further solidify any evidence of this exterior influence upon embedded star formation. An analysis of the mass functions of the two sub-classes of BRC revealed that the star formation within `A' type rims follows a similar function to protostars found in an environment not subjected to exterior ionisation influence. However, `B' and `C' type rims appear to follow a mass function more pre-disposed toward a higher efficiency at higher masses (see Fig.\ref{fig:Mitchell_Mass_fn}).
This indicates that, in `B' and `C' type rims, relatively more massive objects are formed than low-mass stars in comparison to the Salpeter IMF and thus may make an important contribution to the overall intermediate to high-mass stellar population.

\section{Conclusions}  
Our main findings are as follows:
\begin{enumerate}
\item
The SCUBA observations presented here generally show centrally condensed cores interior to the rims of BRCs. The morphology of the cores themselves suggests the presence of bound condensations, indicative of star formation. The morphology of the larger area, incorporating the entire bright rim, is, in general, supportive of the scenario seen in RDI models; a dense core at the head of an elongated column. Support for the interaction of multiple layers is given by observations at multiple wavelengths i.e. radio observations tracing the hot ionised layer bordering the PDR observed in the near infrared which borders the molecular material, the dense regions of which are seen at submillimetre wavelengths.

\item
The physical properties of the submillimetre cores were determined using modified blackbody analyses. The dust temperatures found for the sources are high compared to starless cores (average T$_{\rm{d}}$ = 24 K), indicating the presence of internal stellar sources in these clouds. Based on their L$_{\rm{submm}}$/L$_{\rm{bol}}$ ratio these sources are consistent with being Class 0 protostars. In addition, the indicated stellar sources embedded within the clouds may be classified via the hypothesis that all illumination is due to a single illuminating source. This leads to spectral classifications of intermediate to high mass stellar sources within the cores, whether these sources are isolated singular sources or weaker sources, distributed in clusters.

\item
Observations of mid-infrared sources taken from the 2MASS further support the proposed scenario of early star formation across our sample by revealing the
presence of young stars in the vicinity of the majority (89\%) of our sample. The spectral type of the associated sources is not determinable from the current data, however the
observed stars are known to be in the relatively early stages of their formation.

\item
Both the mass functions and correlations of source luminosity with ionisation field strength suggest some fundamental difference between the subsets of `A' type rims and `B' and `C' type rims. It is suggested that `B' and `C' type rims represent true `triggered' star formation as enabled by the RDI process. There is no reason to rule out the possibility that `A' type rims are also the result of triggering, though perhaps via a different mechanism i.e. the `collect and collapse' model \citep{Elmegreen1977}.
  The correlation between ionisation field strength and source luminosity is highly indicative of `B' and `C' type rims forming with at least some degree of interaction with their UV surroundings. By increasing the size of our sample and incorporating other BRCs that have yet to begin their star formation this hypothesis may be investigated further.

\end{enumerate}

$Acknowledgements$
The authors would like to thank both an anonymous referee and Malcolm Walmsley for their useful and illuminating comments. They have improved the quality of this work.

The authors would like to thank the JCMT support staff for a pleasant and productive observing run; Helen Kirk for assistance with data reduction issues and D.J. Pisano for his insight and helpful discussions.

The Digitized Sky Surveys were produced at the Space Telescope Science Institute under U.S. Government grant NAG W-2166. The images of these surveys are based on photographic data obtained using the Oschin Schmidt Telescope on Palomar Mountain and the UK Schmidt Telescope. The plates were processed into the present compressed digital form with the permission of these institutions.

This publication makes use of data products from the Two Micron All Sky Survey, which is a joint project of the University of Massachusetts and the Infrared Processing and Analysis Center/California Institute of Technology, funded by the National Aeronautics and Space Administration and the National Science Foundation.

This research made use of data products from the Midcourse Space Experiment. Processing of the data was funded by the Ballistic Missile Defense Organization with additional support from NASA Office of Space Science.

This research has made use of the NASA/ IPAC Infrared Science Archive, which is operated by the Jet Propulsion Laboratory, California Institute of Technology, under contract with the National Aeronautics and Space Administration.

This research has made use of the SIMBAD database,
operated at CDS, Strasbourg, France

This research has made use of NASA's Astrophysics Data System.

\bibliography{8104}

\begin{thebibliography}{61}
\expandafter\ifx\csname natexlab\endcsname\relax\def\natexlab#1{#1}\fi

\bibitem[{{Alves} {et~al.}(2007){Alves}, {Lombardi}, \& {Lada}}]{Alves2007}
{Alves}, J., {Lombardi}, M., \& {Lada}, C.~J. 2007, \aap, 462, L17

\bibitem[{{Andre} {et~al.}(1993){Andre}, {Ward-Thompson}, \&
  {Barsony}}]{Andre1993}
{Andre}, P., {Ward-Thompson}, D., \& {Barsony}, M. 1993, \apj, 406, 122

\bibitem[{{Aumann} {et~al.}(1990){Aumann}, {Fowler}, \& {Melnyk}}]{Aumann1990}
{Aumann}, H.~H., {Fowler}, J.~W., \& {Melnyk}, M. 1990, \aj, 99, 1674

\bibitem[{{Bertoldi}(1989)}]{Bertoldi1989}
{Bertoldi}, F. 1989, \apj, 346, 735

\bibitem[{{Beuther} {et~al.}(2002){Beuther}, {Walsh}, {Schilke}, {Sridharan},
  {Menten}, \& {Wyrowski}}]{Beuther2002}
{Beuther}, H., {Walsh}, A., {Schilke}, P., {et~al.} 2002, \aap, 390, 289

\bibitem[{{Blitz}(1993)}]{Blitz1993}
{Blitz}, L. 1993, in Protostars and Planets III, 125--161

\bibitem[{{Campeggio} {et~al.}(2007){Campeggio}, {Strafella}, {Maiolo}, {Elia},
  \& {Aiello}}]{Campeggio2007}
{Campeggio}, L., {Strafella}, F., {Maiolo}, B., {Elia}, D., \& {Aiello}, S.
  2007, \apj, 668, 316

\bibitem[{{Chandler} \& {Richer}(2000)}]{Chandler2000}
{Chandler}, C.~J. \& {Richer}, J.~S. 2000, \apj, 530, 851

\bibitem[{{Codella} {et~al.}(2001){Codella}, {Bachiller}, {Nisini}, {Saraceno},
  \& {Testi}}]{Codella2001}
{Codella}, C., {Bachiller}, R., {Nisini}, B., {Saraceno}, P., \& {Testi}, L.
  2001, \aap, 376, 271

\bibitem[{{Condon} {et~al.}(1998){Condon}, {Cotton}, {Greisen}, {Yin},
  {Perley}, {Taylor}, \& {Broderick}}]{Condon1998}
{Condon}, J.~J., {Cotton}, W.~D., {Greisen}, E.~W., {et~al.} 1998, \aj, 115,
  1693

\bibitem[{{Currie} \& {Berry}(2002)}]{Kappa2002}
{Currie}, M.~J. \& {Berry}, D.~S. 2002, Starlink Project, CCLRC

\bibitem[{{Cutri} {et~al.}(2003){Cutri}, {Skrutskie}, {van Dyk}, {Beichman},
  {Carpenter}, {Chester}, {Cambresy}, {Evans}, {Fowler}, {Gizis}, \& {and 15
  co-authors.}}]{Cutri2003}
{Cutri}, R.~M., {Skrutskie}, M.~F., {van Dyk}, S., {et~al.} 2003, VizieR Online
  Data Catalog, 2246, 0

\bibitem[{{De Jager} \& {Nieuwenhuijzen}(1987)}]{DeJager1987}
{De Jager}, C. \& {Nieuwenhuijzen}, H. 1987, \aap, 177, 217

\bibitem[{{Deharveng} {et~al.}(2005){Deharveng}, {Zavagno}, \&
  {Caplan}}]{Deharveng2005}
{Deharveng}, L., {Zavagno}, A., \& {Caplan}, J. 2005, \aap, 433, 565

\bibitem[{{Dent} {et~al.}(1998){Dent}, {Matthews}, \&
  {Ward-Thompson}}]{Dent1998}
{Dent}, W.~R.~F., {Matthews}, H.~E., \& {Ward-Thompson}, D. 1998, \mnras, 301,
  1049

\bibitem[{{Dobashi} {et~al.}(2001){Dobashi}, {Yonekura}, {Matsumoto}, {Momose},
  {Sato}, {Bernard}, \& {Ogawa}}]{Dobashi2001}
{Dobashi}, K., {Yonekura}, Y., {Matsumoto}, T., {et~al.} 2001, \pasj, 53, 85

\bibitem[{{Draine} \& {Lee}(1984)}]{Draine1984}
{Draine}, B.~T. \& {Lee}, H.~M. 1984, \apj, 285, 89

\bibitem[{{Draper} {et~al.}(2004){Draper}, {Gray}, \& {Berry}}]{Gaia2004}
{Draper}, P.~W., {Gray}, N., \& {Berry}, D.~S. 2004, Starlink Project, CCLRC

\bibitem[{{Elmegreen}(1998)}]{Elmegreen1998}
{Elmegreen}, B.~G. 1998, in ASP Conf. Ser. 148: Origins, ed. C.~E. {Woodward},
  J.~M. {Shull}, \& H.~A. {Thronson}, Jr., 150--+

\bibitem[{{Elmegreen} \& {Lada}(1977)}]{Elmegreen1977}
{Elmegreen}, B.~G. \& {Lada}, C.~J. 1977, \apj, 214, 725

\bibitem[{{Evans}(1999)}]{Evans1999}
{Evans}, N.~J. 1999, \araa, 37, 311

\bibitem[{{Fa{\'u}ndez} {et~al.}(2004){Fa{\'u}ndez}, {Bronfman}, {Garay},
  {Chini}, {Nyman}, \& {May}}]{Faundez2004}
{Fa{\'u}ndez}, S., {Bronfman}, L., {Garay}, G., {et~al.} 2004, \aap, 426, 97

\bibitem[{{Hildebrand}(1983)}]{Hildebrand1983}
{Hildebrand}, R.~H. 1983, \qjras, 24, 267

\bibitem[{{Hogerheijde} \& {Sandell}(2000)}]{Hogerheijde2000}
{Hogerheijde}, M.~R. \& {Sandell}, G. 2000, \apj, 534, 880

\bibitem[{{Holland} {et~al.}(1999){Holland}, {Robson}, {Gear}, {Cunningham},
  {Lightfoot}, {Jenness}, {Ivison}, {Stevens}, {Ade}, {Griffin}, {Duncan},
  {Murphy}, \& {Naylor}}]{Holland1999}
{Holland}, W.~S., {Robson}, E.~I., {Gear}, W.~K., {et~al.} 1999, \mnras, 303,
  659

\bibitem[{{Hurt} \& {Barsony}(1996)}]{Hurt1996}
{Hurt}, R.~L. \& {Barsony}, M. 1996, \apjl, 460, L45+

\bibitem[{{Jenness} \& {Lightfoot}(2000)}]{Surf2000}
{Jenness}, T. \& {Lightfoot}, J.~F. 2000, Starlink Project, CCLRC

\bibitem[{{Kessel-Deynet} \& {Burkert}(2003)}]{Kessel-Deynet2003}
{Kessel-Deynet}, O. \& {Burkert}, A. 2003, \mnras, 338, 545

\bibitem[{{Koornneef}(1983)}]{Koornneef1983}
{Koornneef}, J. 1983, \aap, 128, 84

\bibitem[{{Kruegel} \& {Siebenmorgen}(1994)}]{Kruegel1994}
{Kruegel}, E. \& {Siebenmorgen}, R. 1994, \aap, 288, 929

\bibitem[{{Lada} \& {Adams}(1992)}]{Lada1992}
{Lada}, C.~J. \& {Adams}, F.~C. 1992, \apj, 393, 278

\bibitem[{{Larson}(2003)}]{Larson2003}
{Larson}, R.~B. 2003, Reports on Progress in Physics, 66, 1652

\bibitem[{{Lefloch} \& {Lazareff}(1994)}]{Lefloch1994}
{Lefloch}, B. \& {Lazareff}, B. 1994, \aap, 289, 559

\bibitem[{{Lefloch} \& {Lazareff}(1995)}]{Lefloch1995}
{Lefloch}, B. \& {Lazareff}, B. 1995, \aap, 301, 522

\bibitem[{{Lefloch} {et~al.}(1997){Lefloch}, {Lazareff}, \&
  {Castets}}]{Lefloch1997}
{Lefloch}, B., {Lazareff}, B., \& {Castets}, A. 1997, \aap, 324, 249

\bibitem[{{Levine} \& {Surace}(1993)}]{Levine1993}
{Levine}, D.~M. \& {Surace}, J. 1993, IPAC Users' Guide (5th ed.;
  Pasadena:IPAC)

\bibitem[{{Megeath} \& {Wilson}(1997)}]{Megeath1997}
{Megeath}, S.~T. \& {Wilson}, T.~L. 1997, \aj, 114, 1106

\bibitem[{{Meyer} {et~al.}(1997){Meyer}, {Calvet}, \&
  {Hillenbrand}}]{Meyer1997}
{Meyer}, M.~R., {Calvet}, N., \& {Hillenbrand}, L.~A. 1997, \aj, 114, 288

\bibitem[{{Miao} {et~al.}(2006){Miao}, {White}, {Nelson}, {Thompson}, \&
  {Morgan}}]{Miao2006}
{Miao}, J., {White}, G.~J., {Nelson}, R., {Thompson}, M., \& {Morgan}, L. 2006,
  \mnras, 369, 143

\bibitem[{{Mitchell} {et~al.}(2001){Mitchell}, {Johnstone},
  {Moriarty-Schieven}, {Fich}, \& {Tothill}}]{Mitchell2001}
{Mitchell}, G.~F., {Johnstone}, D., {Moriarty-Schieven}, G., {Fich}, M., \&
  {Tothill}, N.~F.~H. 2001, \apj, 556, 215

\bibitem[{{Morgan} {et~al.}(2004){Morgan}, {Thompson}, {Urquhart}, {White}, \&
  {Miao}}]{Morgan2004}
{Morgan}, L.~K., {Thompson}, M.~A., {Urquhart}, J.~S., {White}, G.~J., \&
  {Miao}, J. 2004, \aap, 426, 535

\bibitem[{{Mueller} {et~al.}(2002){Mueller}, {Shirley}, {Evans}, \&
  {Jacobson}}]{Mueller2002}
{Mueller}, K.~E., {Shirley}, Y.~L., {Evans}, N.~J., \& {Jacobson}, H.~R. 2002,
  \apjs, 143, 469

\bibitem[{{Ogura} {et~al.}(2002){Ogura}, {Sugitani}, \& {Pickles}}]{Ogura2002}
{Ogura}, K., {Sugitani}, K., \& {Pickles}, A. 2002, \aj, 123, 2597

\bibitem[{{Privett} {et~al.}(1998){Privett}, {Jenness}, \&
  {Lightfoot}}]{Fluxes1998}
{Privett}, G., {Jenness}, T., \& {Lightfoot}, J.~F. 1998, Starlink Project,
  CCLRC

\bibitem[{{Rieke} \& {Lebofsky}(1985)}]{Rieke1985}
{Rieke}, G.~H. \& {Lebofsky}, M.~J. 1985, \apj, 288, 618

\bibitem[{{Salpeter}(1955)}]{Salpeter1955}
{Salpeter}, E.~E. 1955, \apj, 121, 161

\bibitem[{{Sharpless}(1959)}]{Sharpless1959}
{Sharpless}, S. 1959, \apjs, 4, 257

\bibitem[{{Sridharan} {et~al.}(2002){Sridharan}, {Beuther}, {Schilke},
  {Menten}, \& {Wyrowski}}]{Sridharan2002}
{Sridharan}, T.~K., {Beuther}, H., {Schilke}, P., {Menten}, K.~M., \&
  {Wyrowski}, F. 2002, \apj, 566, 931

\bibitem[{{Sugitani} {et~al.}(1991){Sugitani}, {Fukui}, \&
  {Ogura}}]{Sugitani1991}
{Sugitani}, K., {Fukui}, Y., \& {Ogura}, K. 1991, \apjs, 77, 59

\bibitem[{{Sugitani} \& {Ogura}(1994)}]{Sugitani1994}
{Sugitani}, K. \& {Ogura}, K. 1994, \apjs, 92, 163

\bibitem[{{Thompson} {et~al.}(2004{\natexlab{a}}){Thompson}, {Urquhart}, \&
  {White}}]{Thompson2004}
{Thompson}, M.~A., {Urquhart}, J.~S., \& {White}, G.~J. 2004{\natexlab{a}},
  \aap, 415, 627

\bibitem[{{Thompson} \& {White}(2004)}]{Thompson2004b}
{Thompson}, M.~A. \& {White}, G.~J. 2004, \aap, 419, 599

\bibitem[{{Thompson} {et~al.}(2004{\natexlab{b}}){Thompson}, {White}, {Morgan},
  {Miao}, {Fridlund}, \& {Huldtgren-White}}]{Thompson2004a}
{Thompson}, M.~A., {White}, G.~J., {Morgan}, L.~K., {et~al.}
  2004{\natexlab{b}}, \aap, 414, 1017

\bibitem[{{Urquhart} {et~al.}(2007){Urquhart}, {Thompson}, {Morgan},
  {Pestalozzi}, {White}, \& {Muna}}]{Urquhart2007}
{Urquhart}, J.~S., {Thompson}, M.~A., {Morgan}, L.~K., {et~al.} 2007, \aap,
  467, 1125

\bibitem[{{Urquhart} {et~al.}(2004){Urquhart}, {Thompson}, {Morgan}, \&
  {White}}]{Urquhart2004}
{Urquhart}, J.~S., {Thompson}, M.~A., {Morgan}, L.~K., \& {White}, G.~J. 2004,
  \aap, 428, 723

\bibitem[{{Urquhart} {et~al.}(2006){Urquhart}, {Thompson}, {Morgan}, \&
  {White}}]{Urquhart2006}
{Urquhart}, J.~S., {Thompson}, M.~A., {Morgan}, L.~K., \& {White}, G.~J. 2006,
  \aap, 450, 625

\bibitem[{{Whitworth} {et~al.}(1994){Whitworth}, {Bhattal}, {Chapman},
  {Disney}, \& {Turner}}]{Whitworth1994}
{Whitworth}, A.~P., {Bhattal}, A.~S., {Chapman}, S.~J., {Disney}, M.~J., \&
  {Turner}, J.~A. 1994, \mnras, 268, 291

\bibitem[{{Wood} \& {Churchwell}(1989)}]{Wood1989}
{Wood}, D.~O.~S. \& {Churchwell}, E. 1989, \apj, 340, 265

\bibitem[{{Yamaguchi} {et~al.}(1999){Yamaguchi}, {Saito}, {Mizuno}, {Mine},
  {Mizuno}, {Ogawa}, \& {Fukui}}]{Yamaguchi1999}
{Yamaguchi}, R., {Saito}, H., {Mizuno}, N., {et~al.} 1999, \pasj, 51, 791

\bibitem[{{Zavagno} {et~al.}(2006){Zavagno}, {Deharveng}, {Comer{\'o}n},
  {Brand}, {Massi}, {Caplan}, \& {Russeil}}]{Zavagno2006}
{Zavagno}, A., {Deharveng}, L., {Comer{\'o}n}, F., {et~al.} 2006, \aap, 446,
  171

\bibitem[{{Zinnecker} {et~al.}(1993){Zinnecker}, {McCaughrean}, \&
  {Wilking}}]{Zinnecker1993}
{Zinnecker}, H., {McCaughrean}, M.~J., \& {Wilking}, B.~A. 1993, in Protostars
  and Planets III, ed. E.~H. {Levy} \& J.~I. {Lunine}, 429--495

\end{thebibliography}
\Online
\appendix
\section{Description Of Plots and Images}
Data for all of the SFO clouds are presented in the following pages. SCUBA 850 \micron ~and 450 \micron ~contours are overlaid on DSS images. Infrared sources from the 2MASS Point Source Catalogue \citep{Cutri2003} that have been identified as YSOs are shown as triangles.

J-H versus H-K$_{\rm{s}}$ colours of the 2MASS sources associated with each cloud are plotted along with a figure of the SED plot of the object composed from a best fit to various observed fluxes where possible. J-H versus H-K$_{\rm{s}}$ colour diagrams have solid lines representing the unreddened loci of main sequence and giant stars from \citet{Koornneef1983}. Dotted lines represent the classical T-Tauri locus of \citet{Meyer1997}. Reddening tracks are shown by dashed lines and the dash-dot line represents the reddening track of the earliest main-sequence stars.
\newpage

\begin{center}
\begin{figure*}
\begin{center}
\includegraphics*[height=6cm]{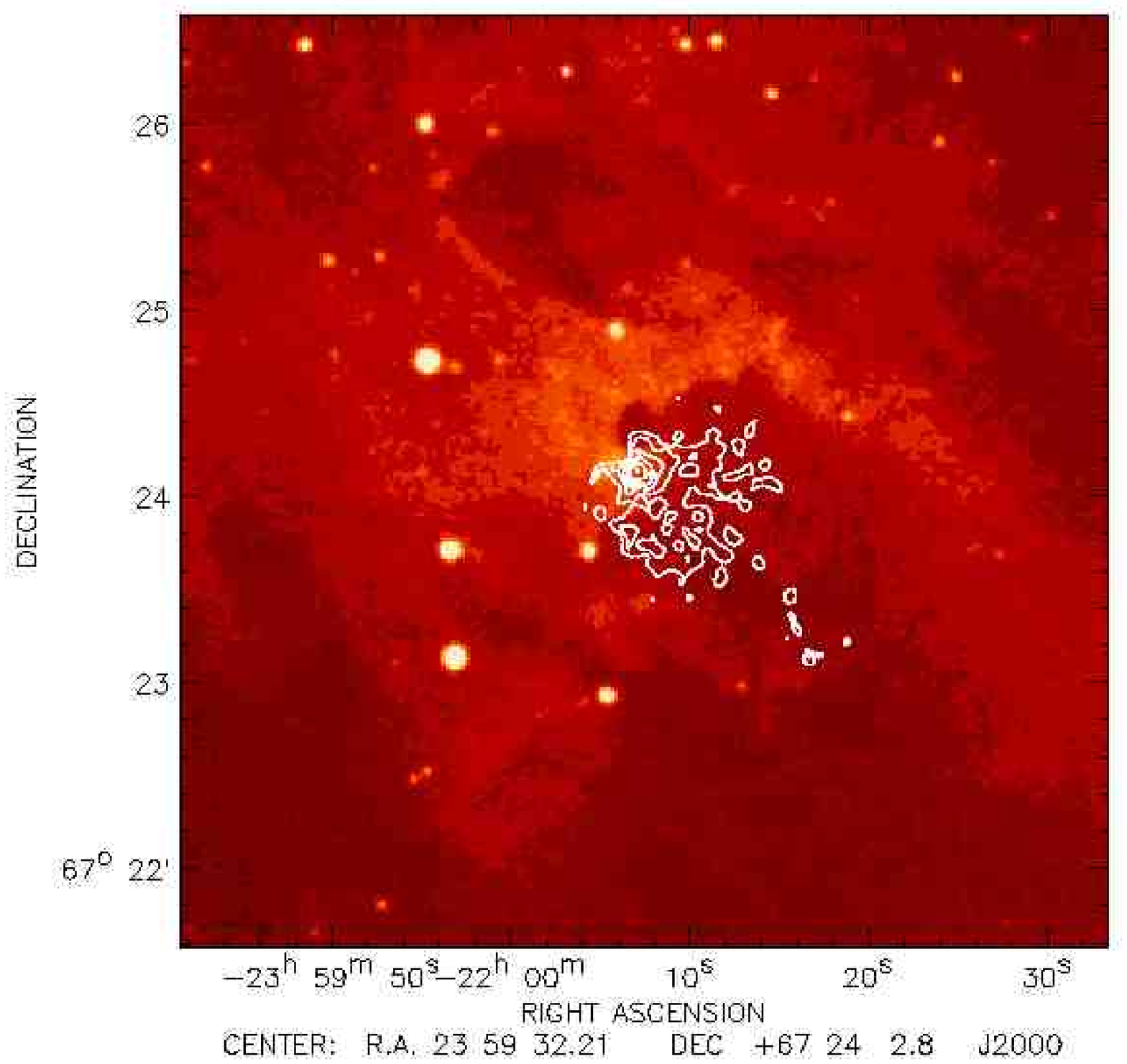}
\includegraphics*[height=6cm]{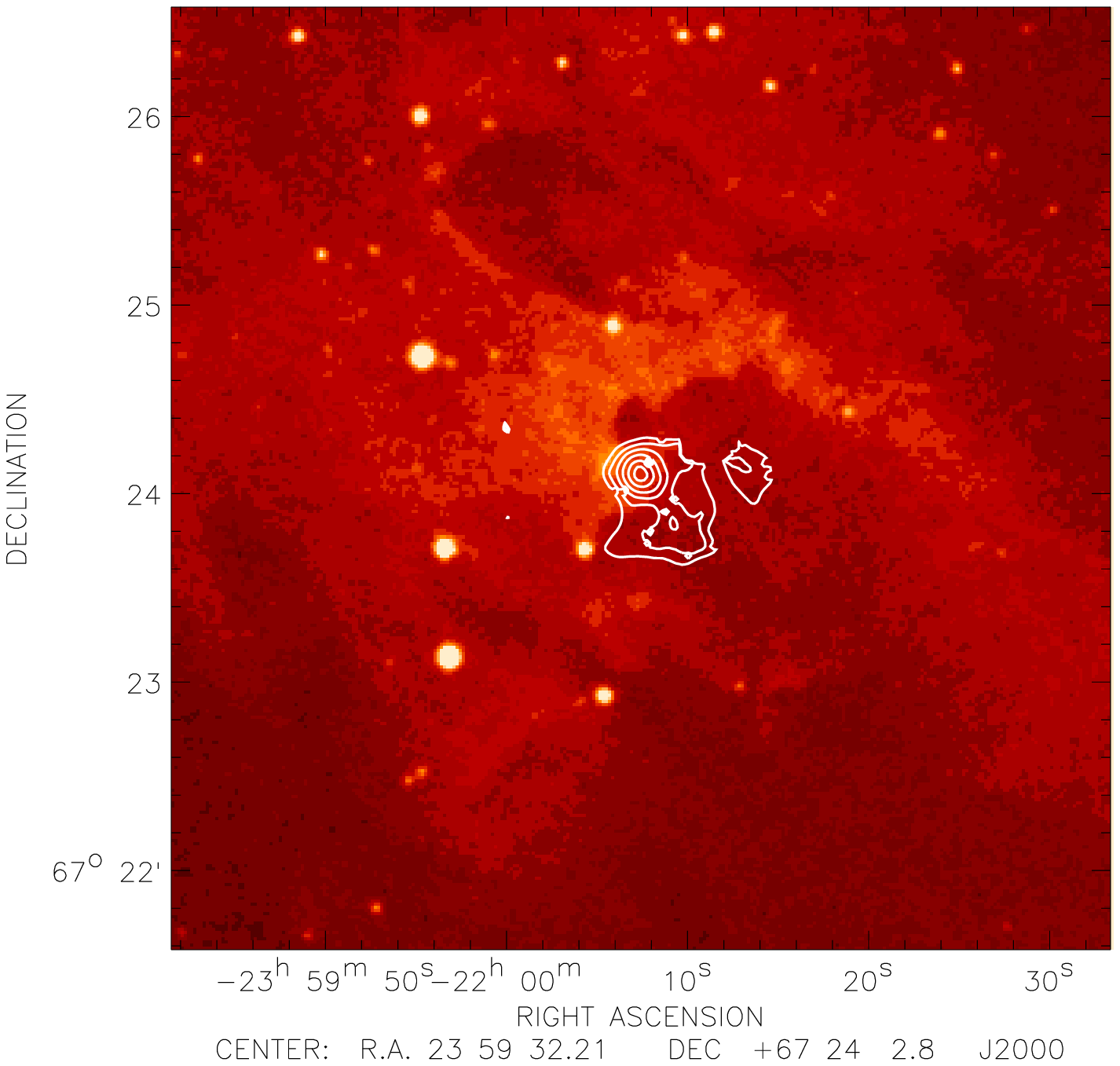}\\
\includegraphics*[height=6cm]{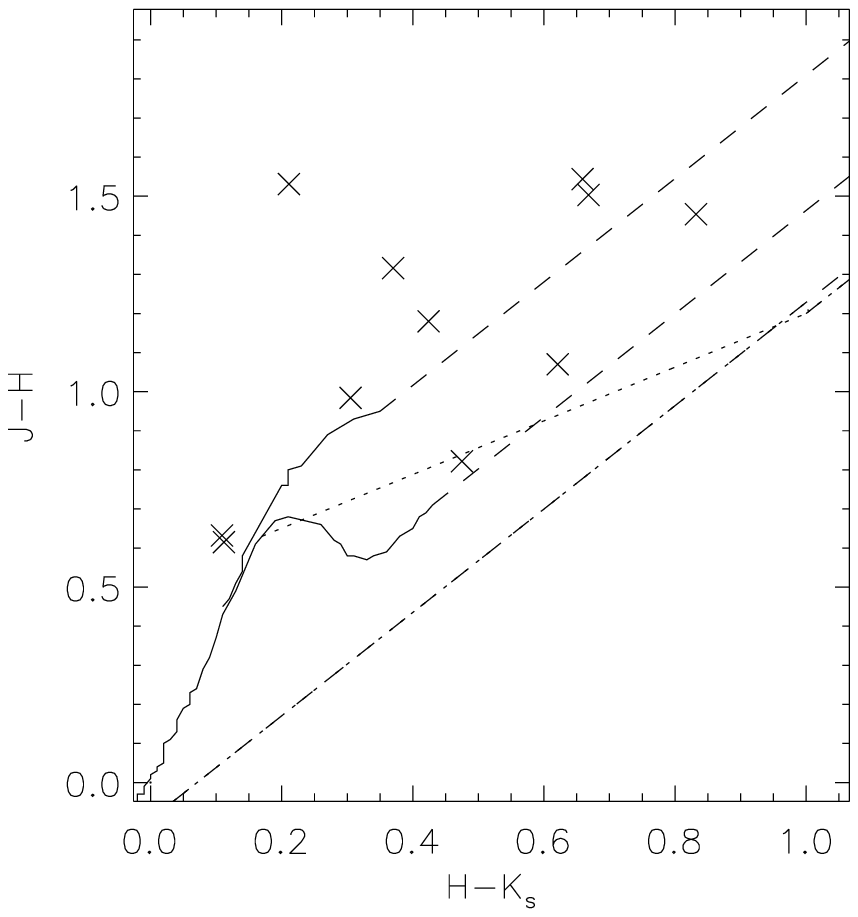}
\includegraphics*[height=6cm]{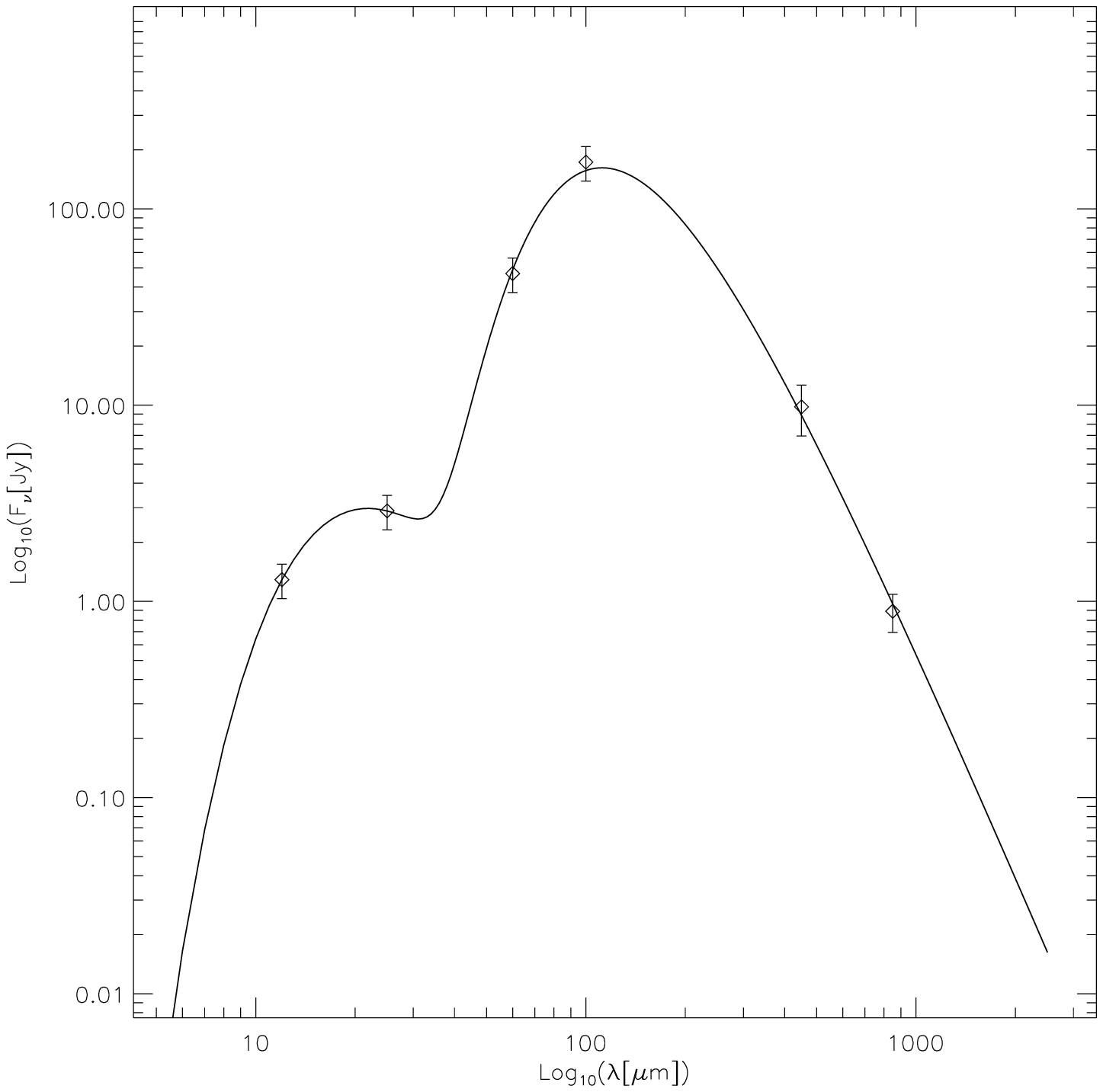}\\
\end{center}
\caption{Plots and images associated with the object SFO 1. The top images show SCUBA 450 \micron ~(left) and 850 \micron ~(right) contours overlaid on a DSS image, infrared sources from the 2MASS Point Source Catalogue \citep{Cutri2003} that have been identified as YSOs are shown as triangles. 850 \micron ~contours start at 4$\sigma$ and increase in increments of 20\% of the peak flux value, 450 \micron ~contours start at 3$\sigma$ and increase in increments of 20\% of the peak flux value. 
\indent The bottom left plot shows the J-H versus H-K$_{\rm{s}}$ colours of the 2MASS sources associated with the cloud while the bottom right image shows the SED plot of the object composed from a best fit to various observed fluxes.}
\label{fig:images}
\end{figure*}
\end{center}

\newpage

\begin{center}
\begin{figure*}
\begin{center}
\includegraphics*[height=6cm]{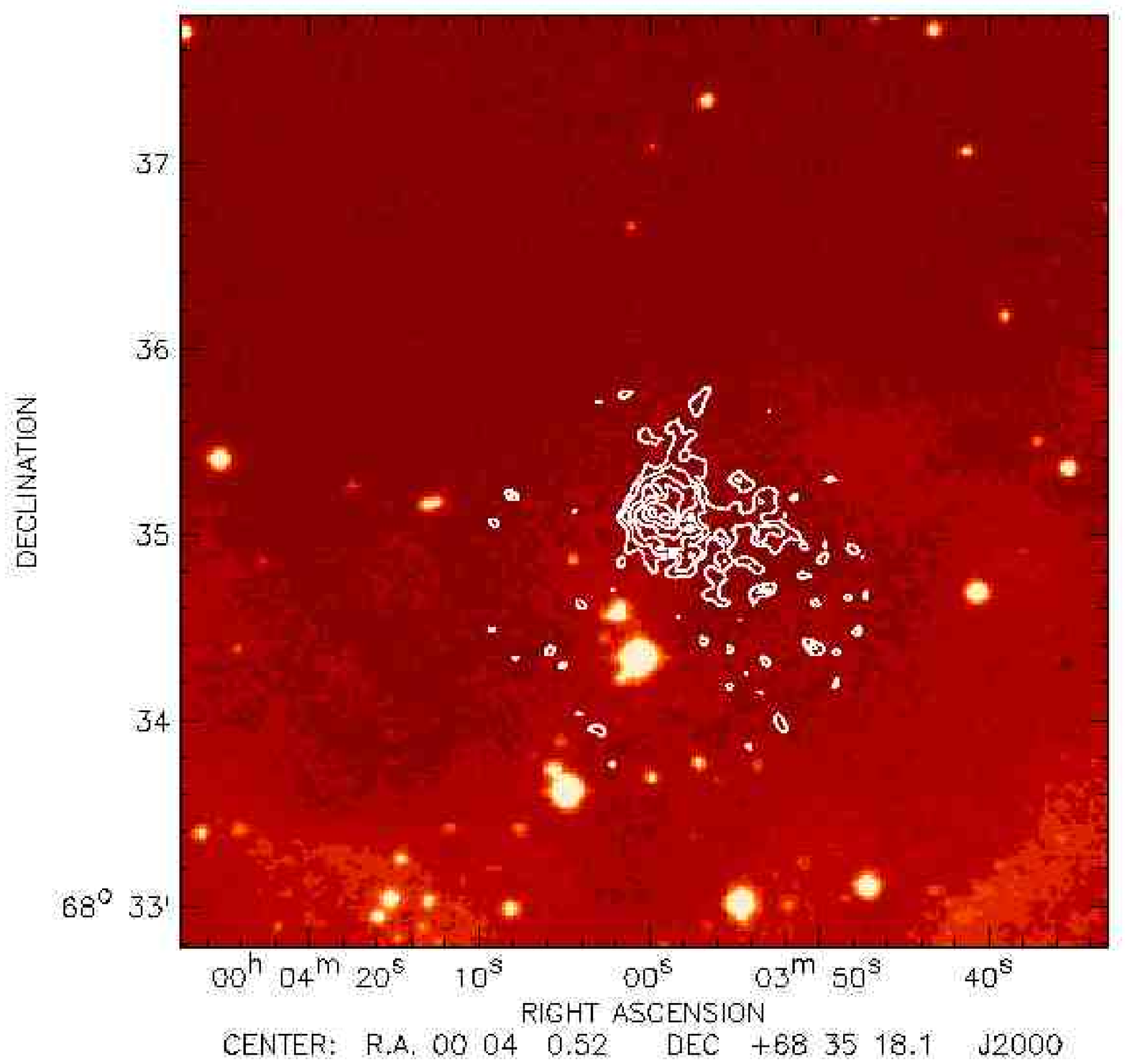}
\includegraphics*[height=6cm]{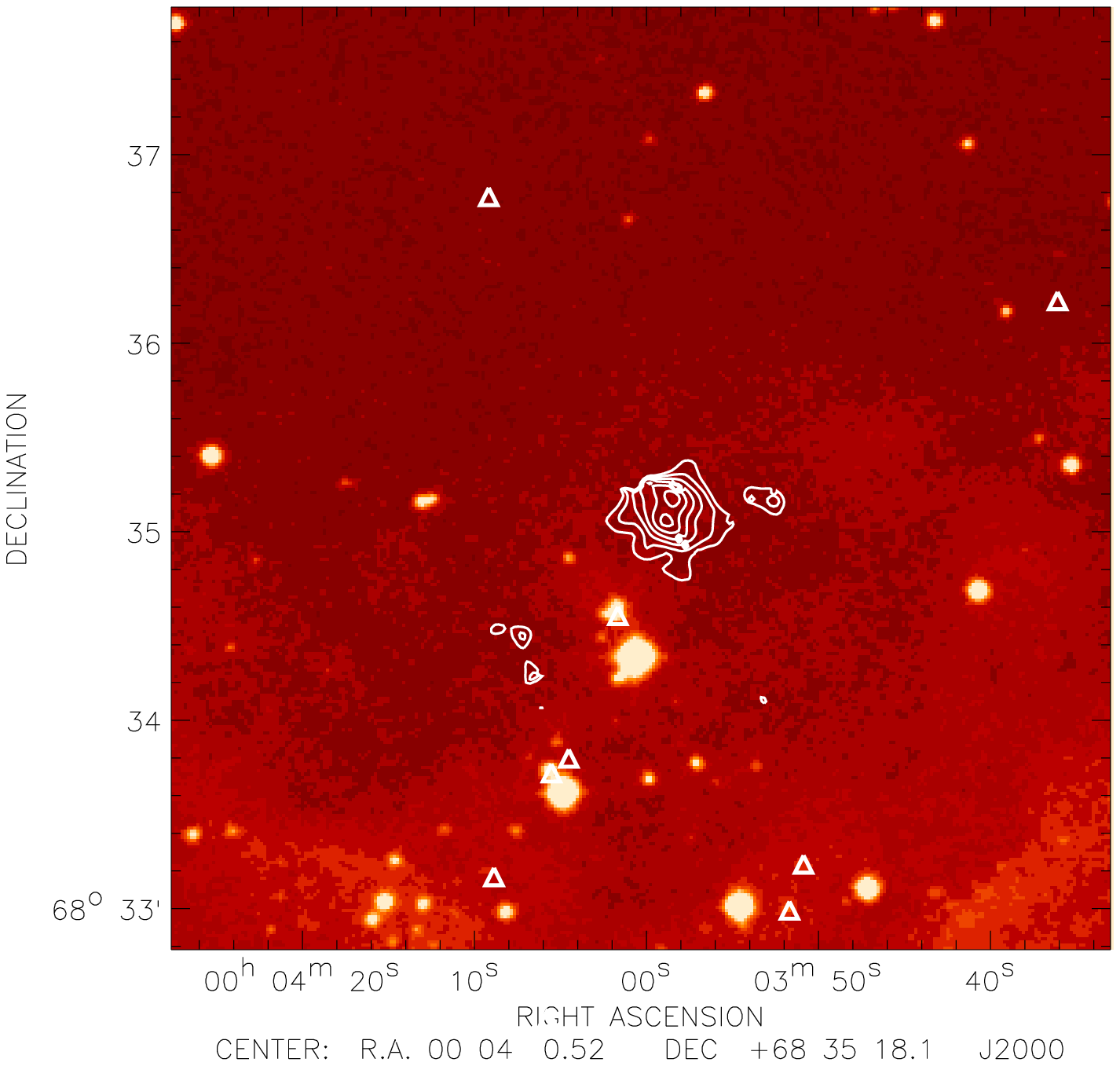}\\
\includegraphics*[height=6cm]{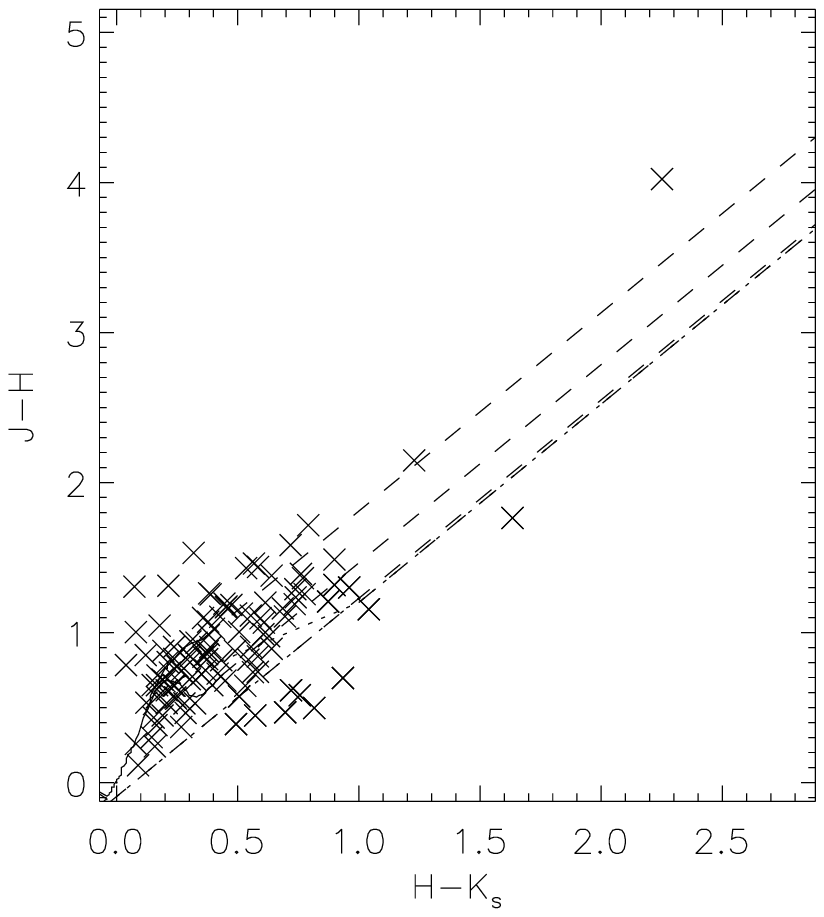}
\includegraphics*[height=6cm]{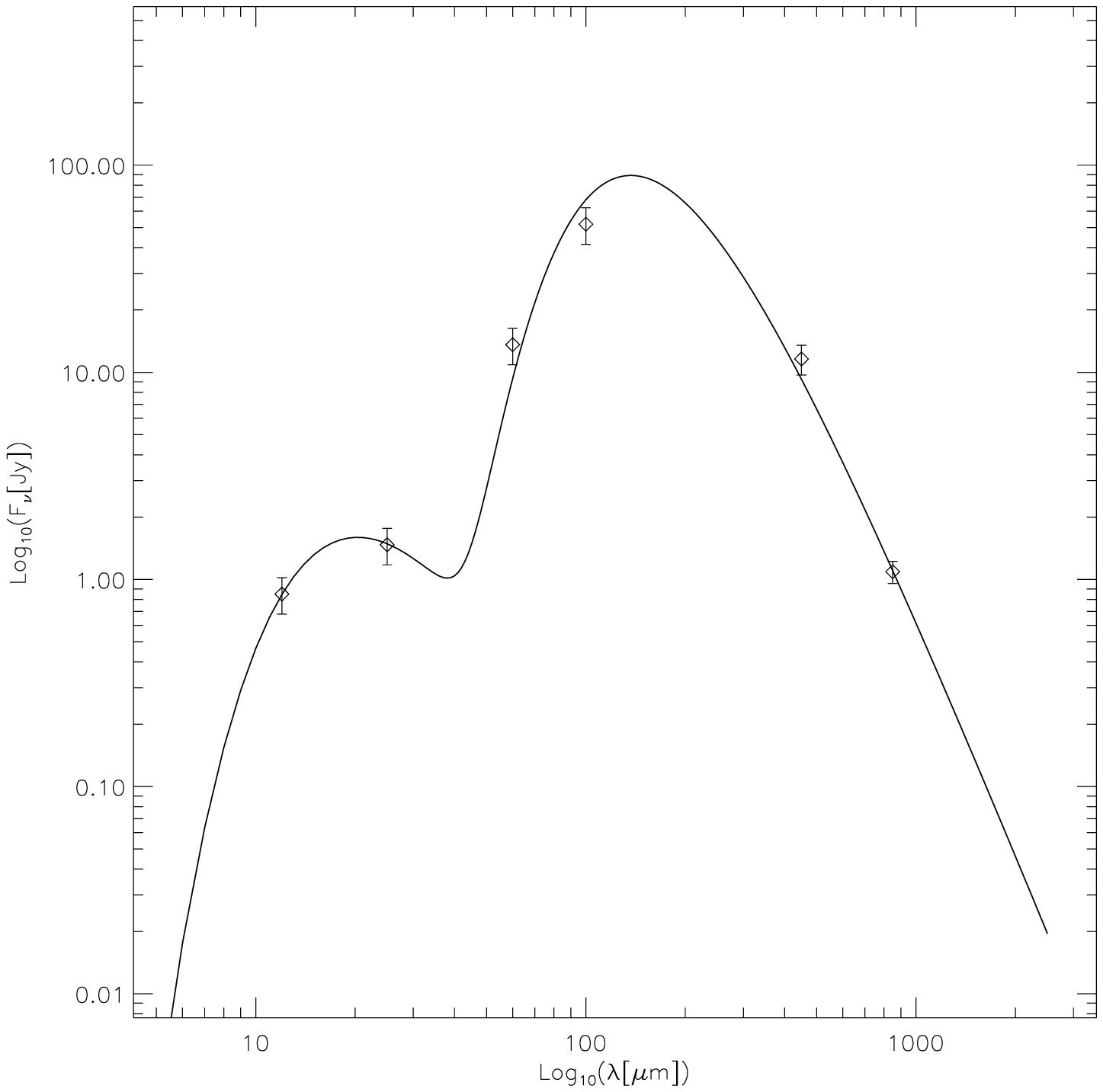}\\
\end{center}
\caption{Plots and images associated with the object SFO 2. The top images show SCUBA 450 \micron ~(left) and 850 \micron ~(right) contours overlaid on a DSS image, infrared sources from the 2MASS Point Source Catalogue \citep{Cutri2003} that have been identified as YSOs are shown as triangles.  850 \micron ~contours start at 9$\sigma$ and increase in increments of 20\% of the peak flux value, 450 \micron ~contours start at 5$\sigma$ and increase in increments of 20\% of the peak flux value.
\indent The bottom left plot shows the J-H versus H-K$_{\rm{s}}$ colours of the 2MASS sources associated with the cloud while the bottom right image shows the SED plot of the object composed from a best fit to various observed fluxes.}
\end{figure*}
\end{center}

\newpage

\begin{center}
\begin{figure*}
\begin{center}
\includegraphics*[height=6cm]{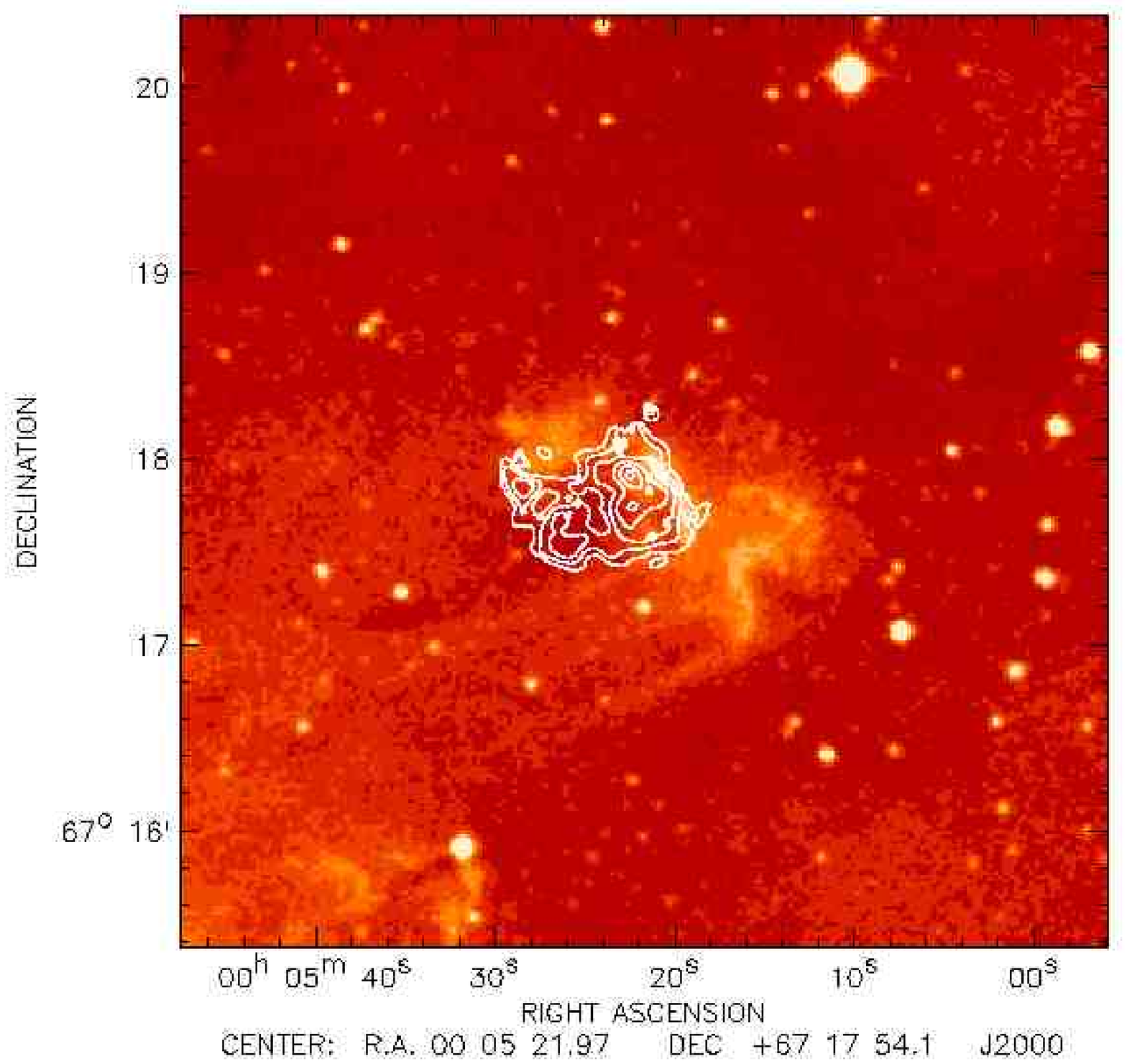}\\
\includegraphics*[height=6cm]{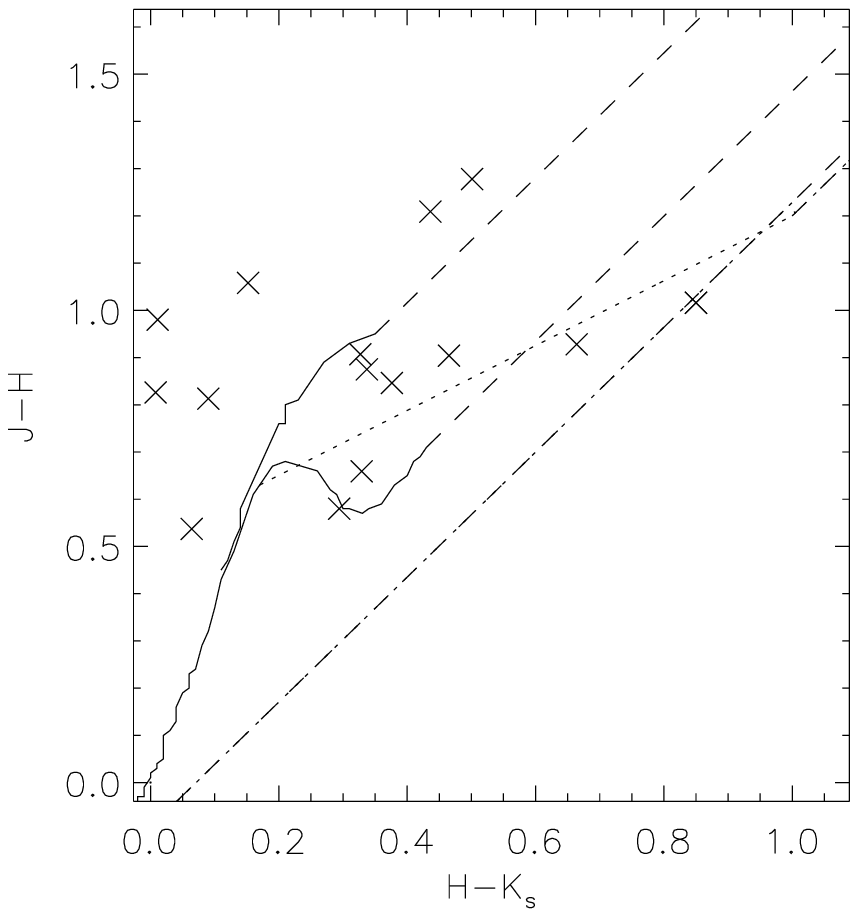}
\includegraphics*[height=6cm]{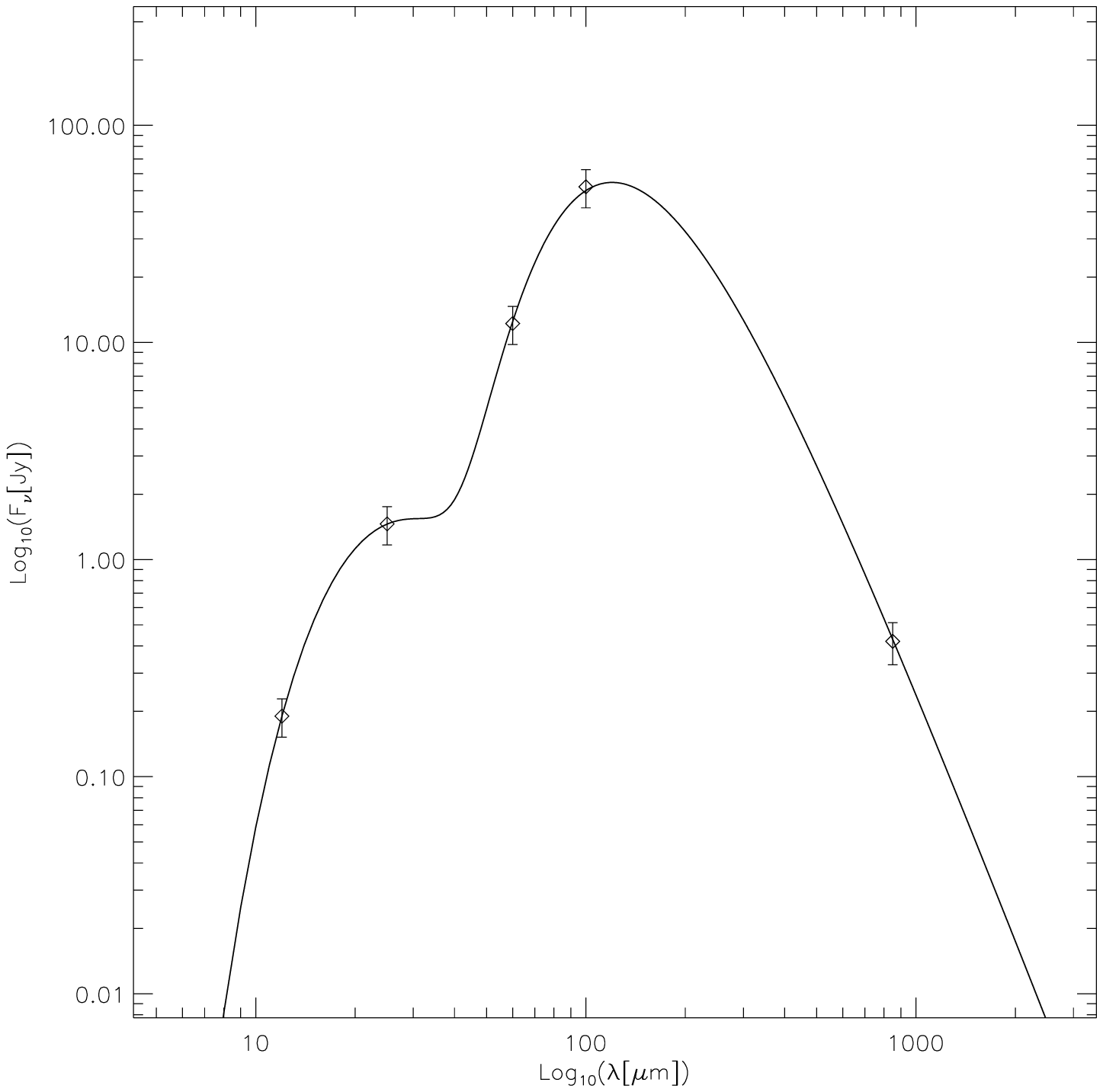}\\
\caption{Plots and images associated with the object SFO 3. The top image shows SCUBA 850 \micron ~contours overlaid on a DSS image, infrared sources from the 2MASS Point Source Catalogue \citep{Cutri2003} are shown as triangles.  850 \micron ~contours start at 3$\sigma$ and increase in increments of 20\% of the peak flux value.
\indent The bottom left plot shows the J-H versus H-K$_{\rm{s}}$ colours of the 2MASS sources associated with the cloud while the bottom right image shows the SED plot of the object composed from a best fit to various observed fluxes.}
\end{center}
\end{figure*}
\end{center}

\newpage

\begin{center}
\begin{figure*}
\begin{center}
\includegraphics*[height=6cm]{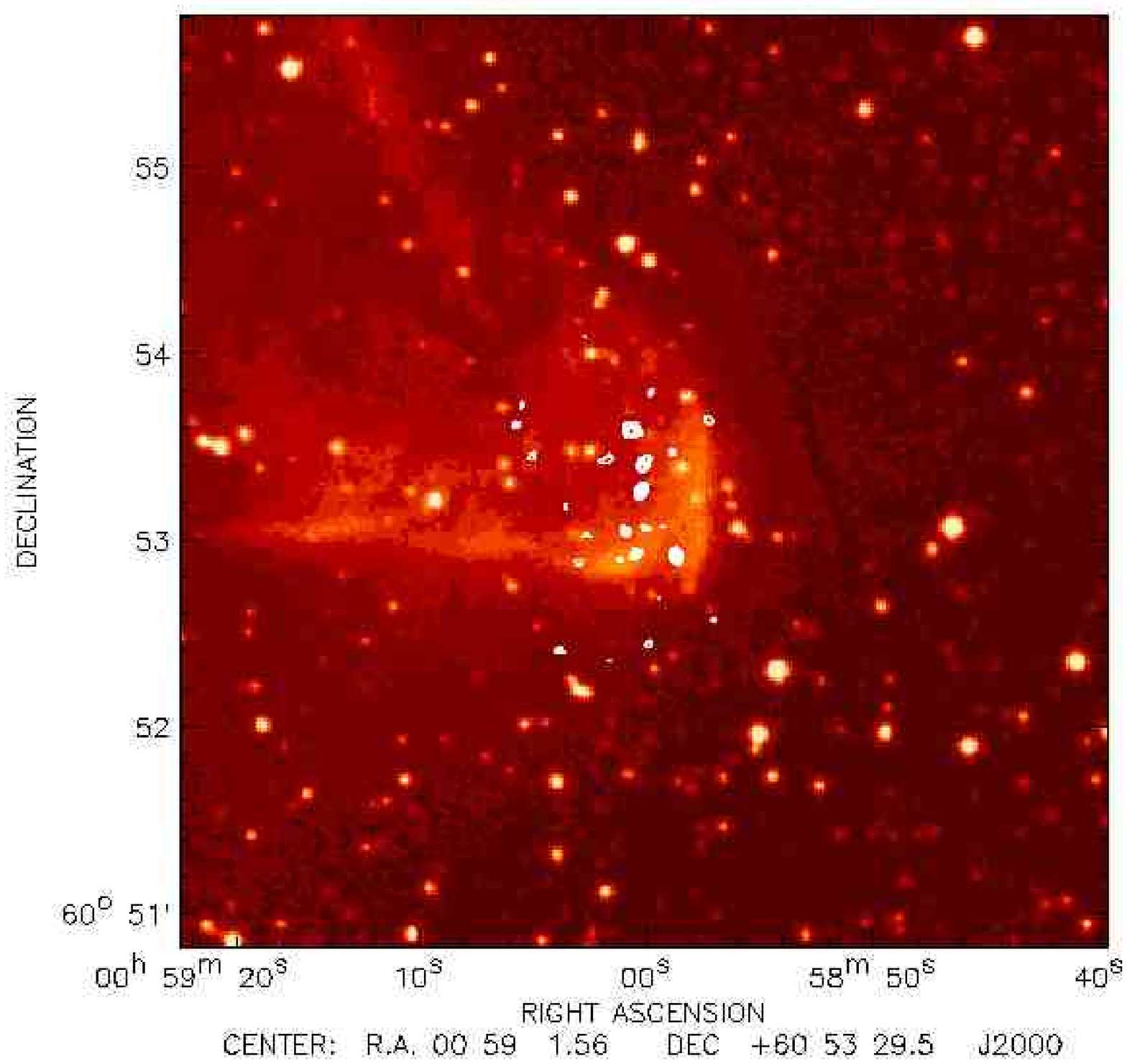}
\includegraphics*[height=6cm]{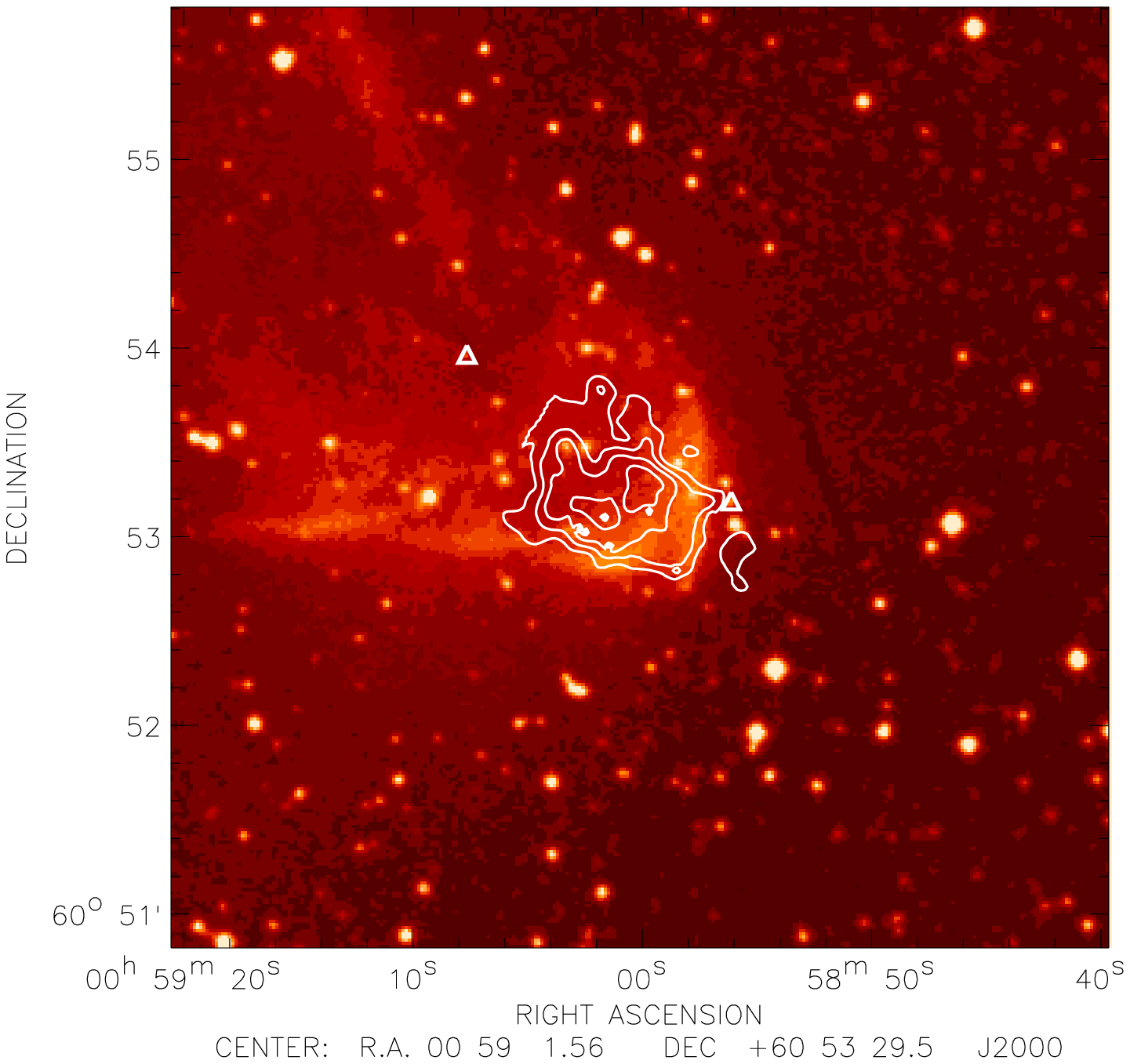}\\
\includegraphics*[height=6cm]{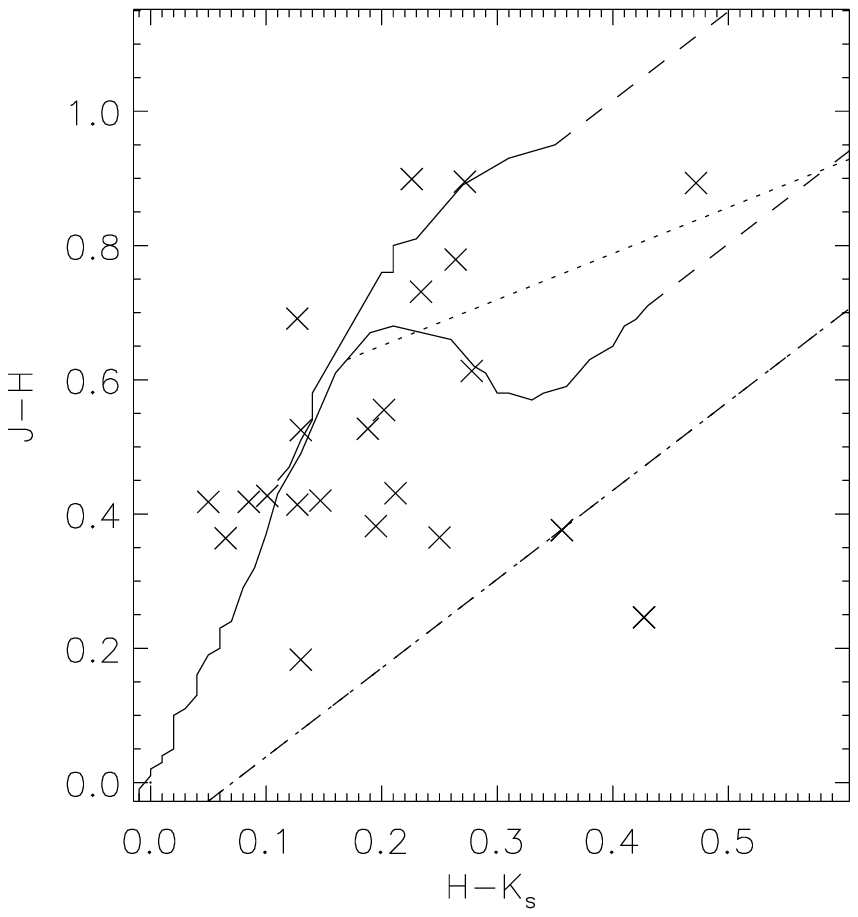}
\includegraphics*[height=6cm]{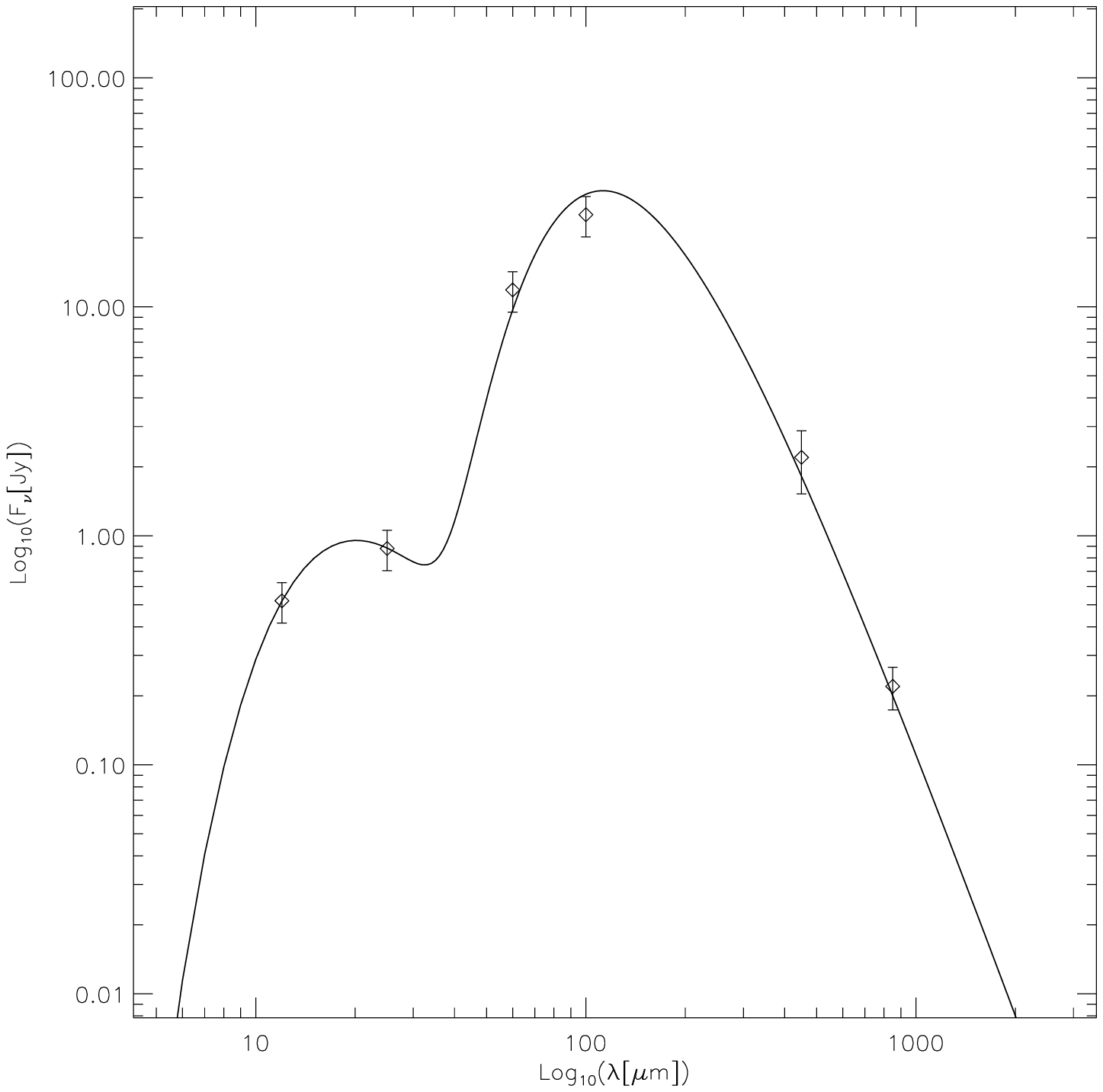}\\
\end{center}
\caption{Plots and images associated with the object SFO 4. The top images show SCUBA 450 \micron ~(left) and 850 \micron ~(right) contours overlaid on a DSS image, infrared sources from the 2MASS Point Source Catalogue \citep{Cutri2003} that have been identified as YSOs are shown as triangles.  850 \micron ~contours start at 4$\sigma$ and increase in increments of 20\% of the peak flux value, 450 \micron ~contours start at 5$\sigma$ and increase in increments of 20\% of the peak flux value.
\indent The bottom left plot shows the J-H versus H-K$_{\rm{s}}$ colours of the 2MASS sources associated with the cloudwhile the bottom right image shows the SED plot of the object composed from a best fit to various observed fluxes.}
\end{figure*}
\end{center}

\newpage

\begin{center}
\begin{figure*}
\begin{center}
\includegraphics*[height=6cm]{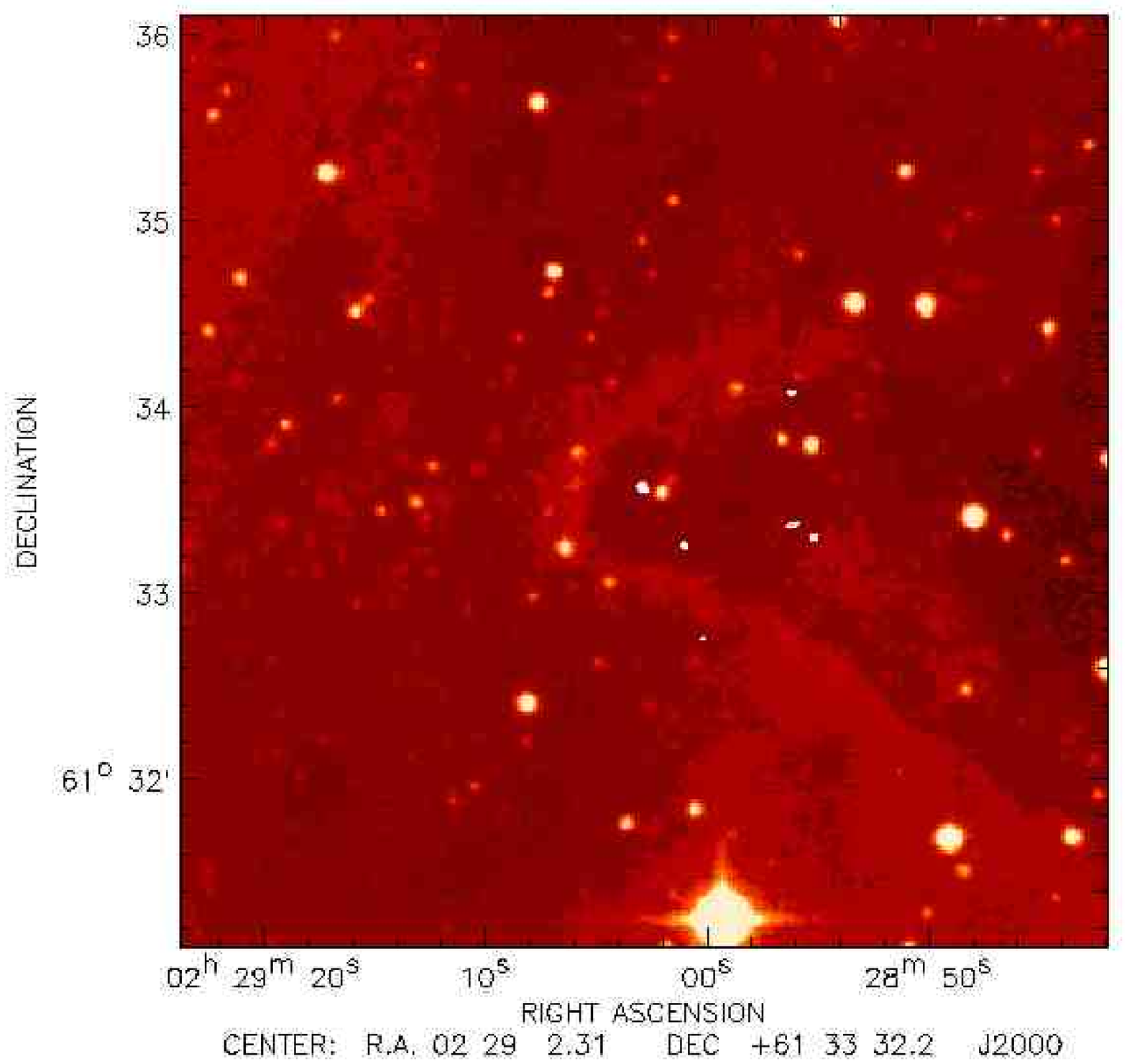}
\includegraphics*[height=6cm]{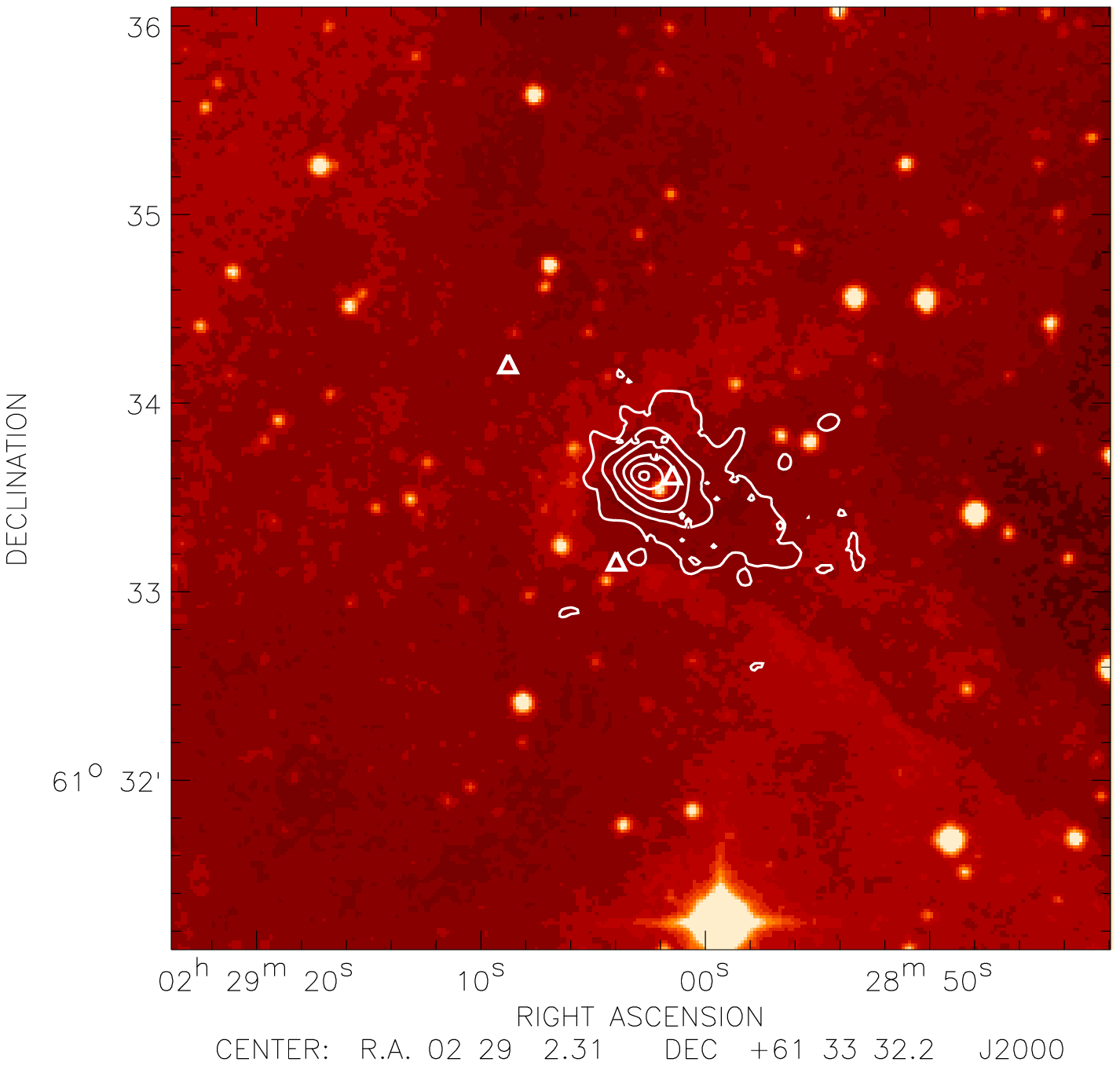}\\
\includegraphics*[height=6cm]{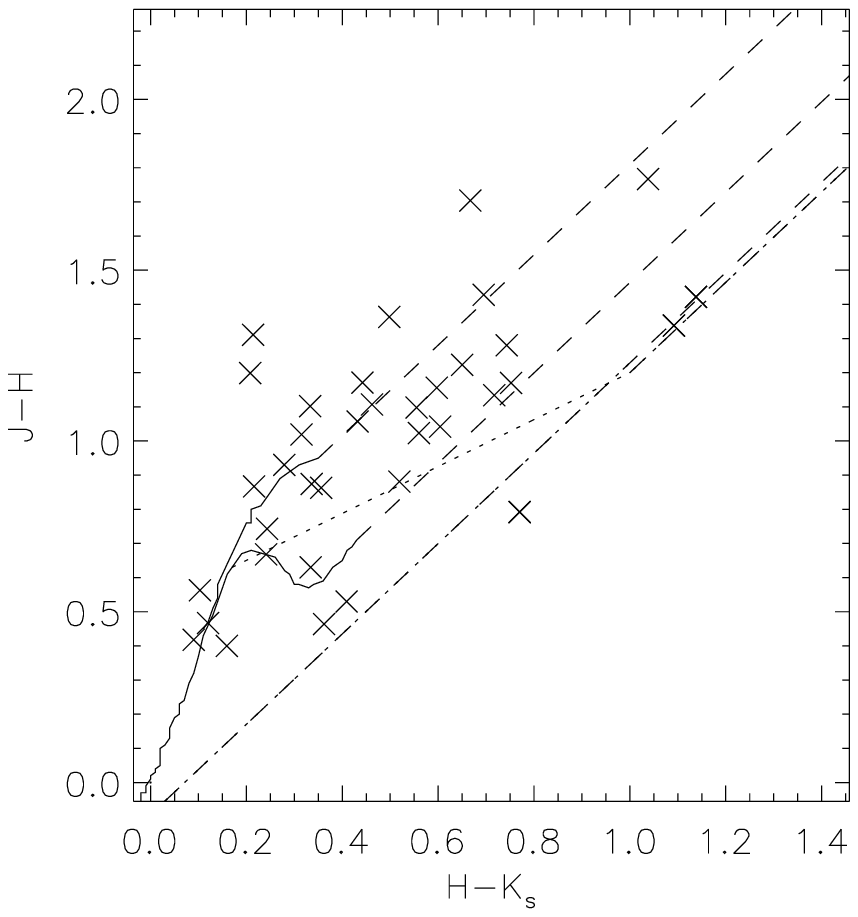}
\includegraphics*[height=6cm]{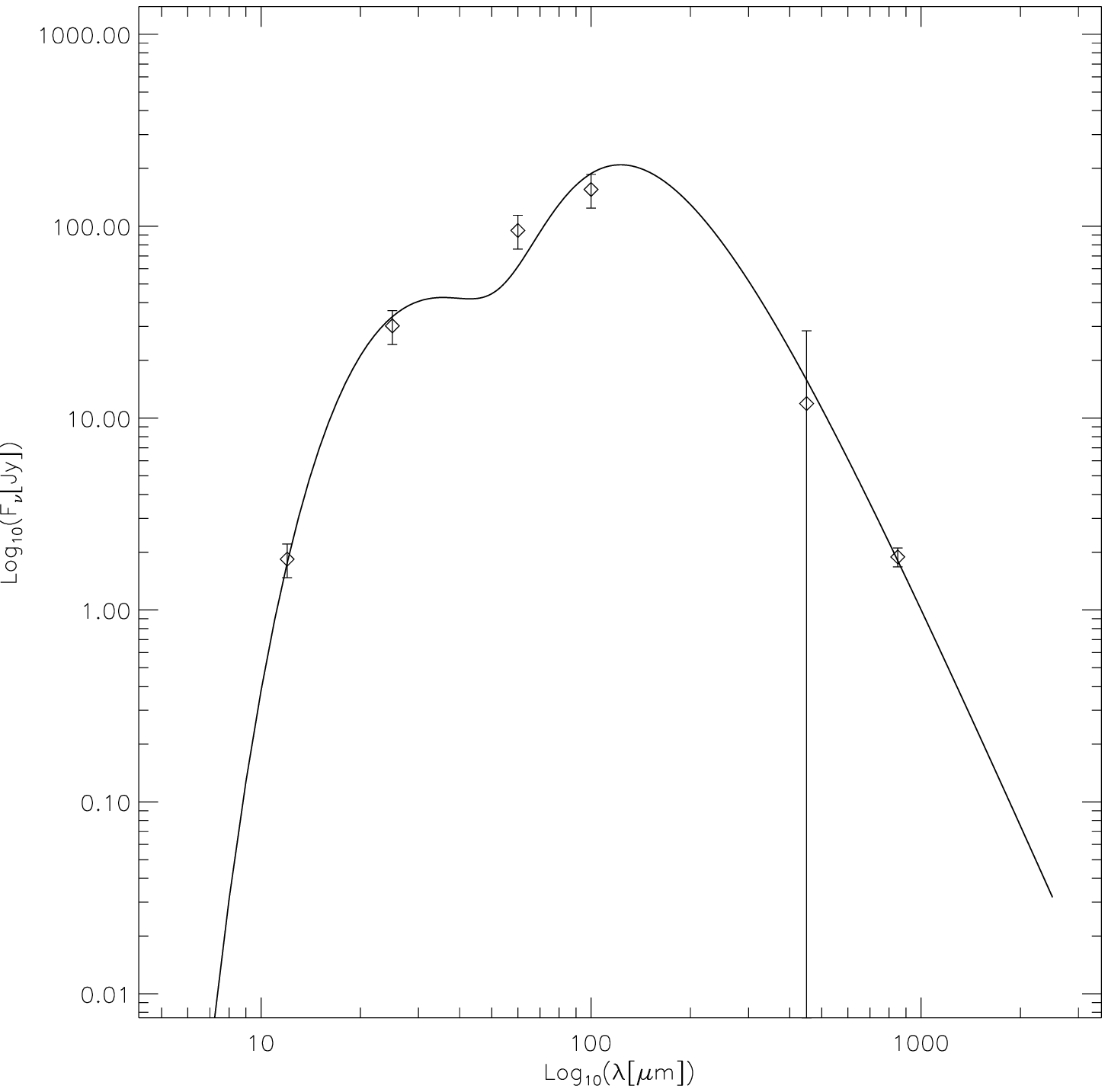}\\
\end{center}
\caption{Plots and images associated with the object SFO 5.  The top images show SCUBA 450 \micron ~(left) and 850 \micron ~(right) contours overlaid on a DSS image, infrared sources from the 2MASS Point Source Catalogue \citep{Cutri2003} that have been identified as YSOs are shown as triangles.  850 \micron ~contours start at 6$\sigma$ and increase in increments of 20\% of the peak flux value, 450 \micron ~contours start at 3$\sigma$ and increase in increments of 20\% of the peak flux value.
\indent The bottom left plot shows the J-H versus H-K$_{\rm{s}}$ colours of the 2MASS sources associated with the cloud while the bottom right image shows the SED plot of the object composed from a best fit to various observed fluxes.}
\end{figure*}
\end{center}

\newpage

\begin{center}
\begin{figure*}
\begin{center}
\includegraphics*[height=6cm]{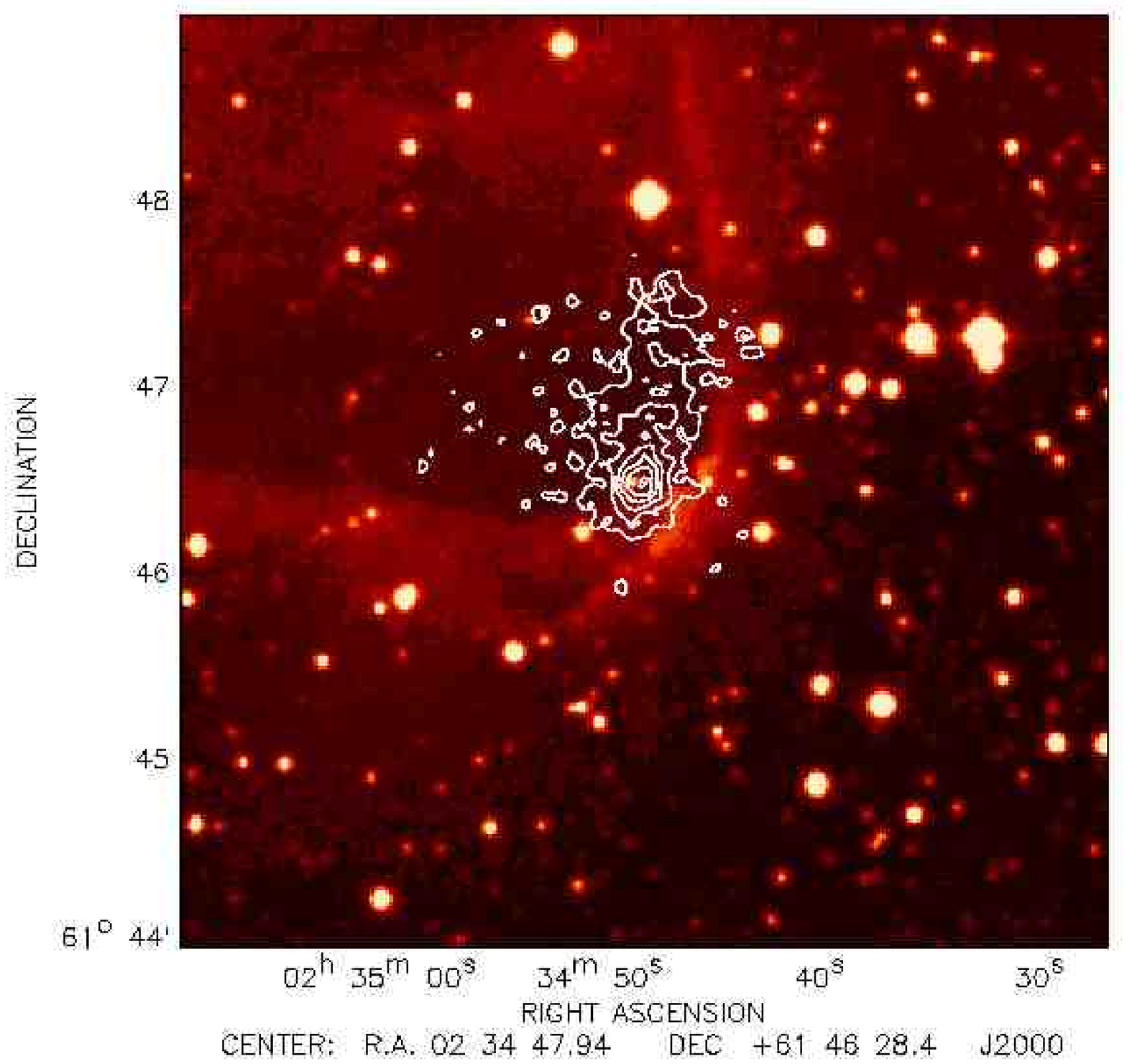}
\includegraphics*[height=6cm]{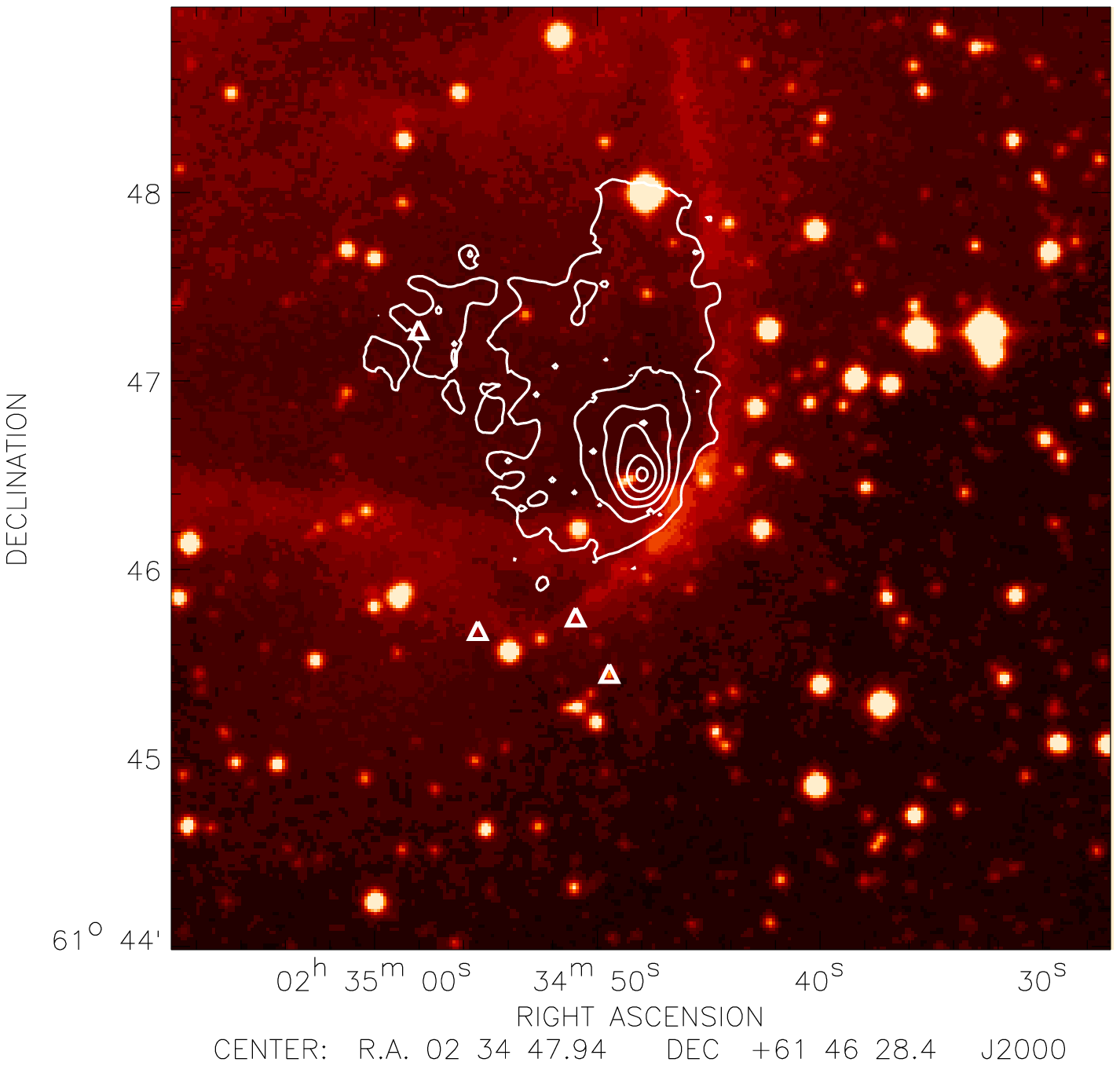}\\
\includegraphics*[height=6cm]{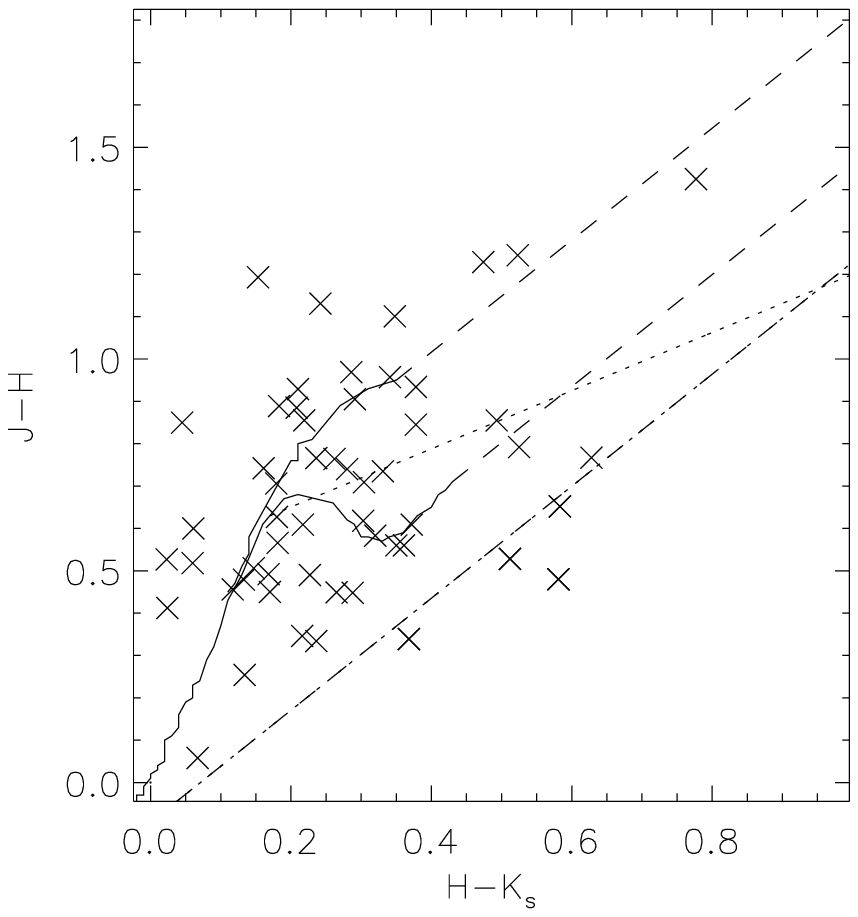}
\includegraphics*[height=6cm]{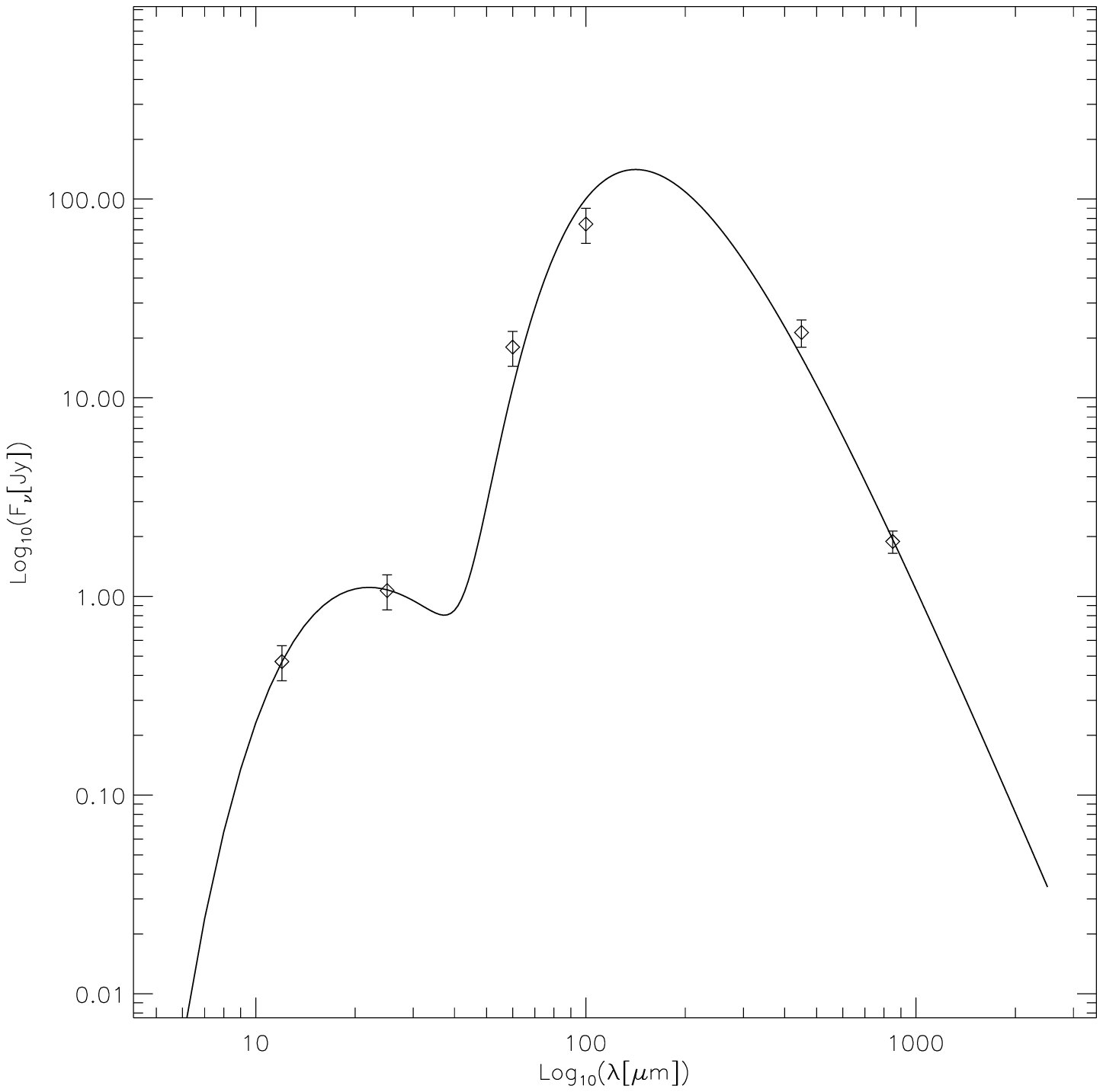}\\
\end{center}
\caption{Plots and images associated with the object SFO 7. The top images show SCUBA 450 \micron ~(left) and 850 \micron ~(right) contours overlaid on a DSS image, infrared sources from the 2MASS Point Source Catalogue \citep{Cutri2003} that have been identified as YSOs are shown as triangles.  850 \micron ~contours start at 3$\sigma$ and increase in increments of 20\% of the peak flux value, 450 \micron ~contours start at 3$\sigma$ and increase in increments of 20\% of the peak flux value.
\indent The bottom left plot shows the J-H versus H-K$_{\rm{s}}$ colours of the 2MASS sources associated with the cloud while the bottom right image shows the SED plot of the object composed from a best fit to various observed fluxes.}
\end{figure*}
\end{center}

\newpage

\begin{center}
\begin{figure*}
\begin{center}
\includegraphics*[height=6cm]{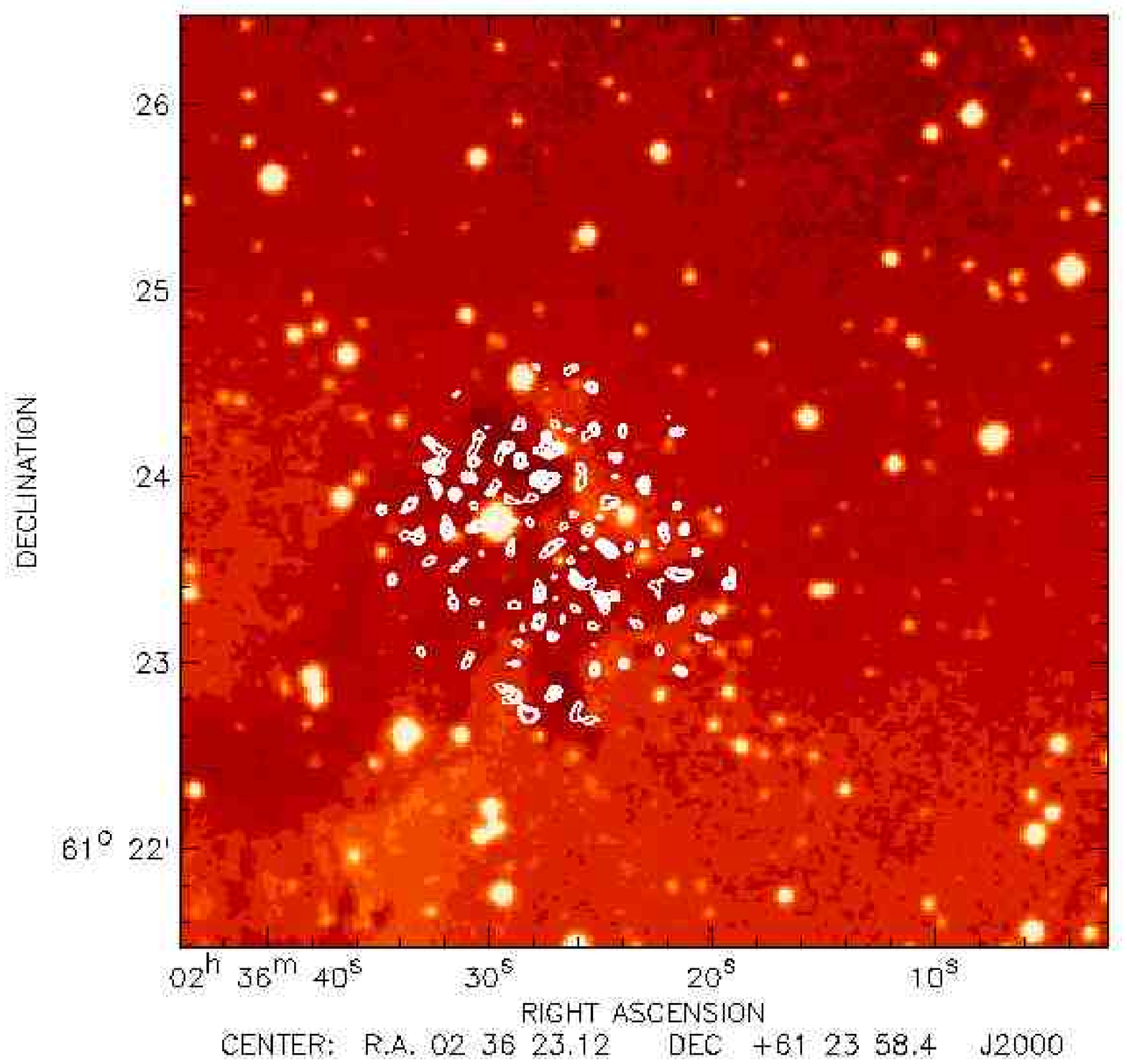}
\includegraphics*[height=6cm]{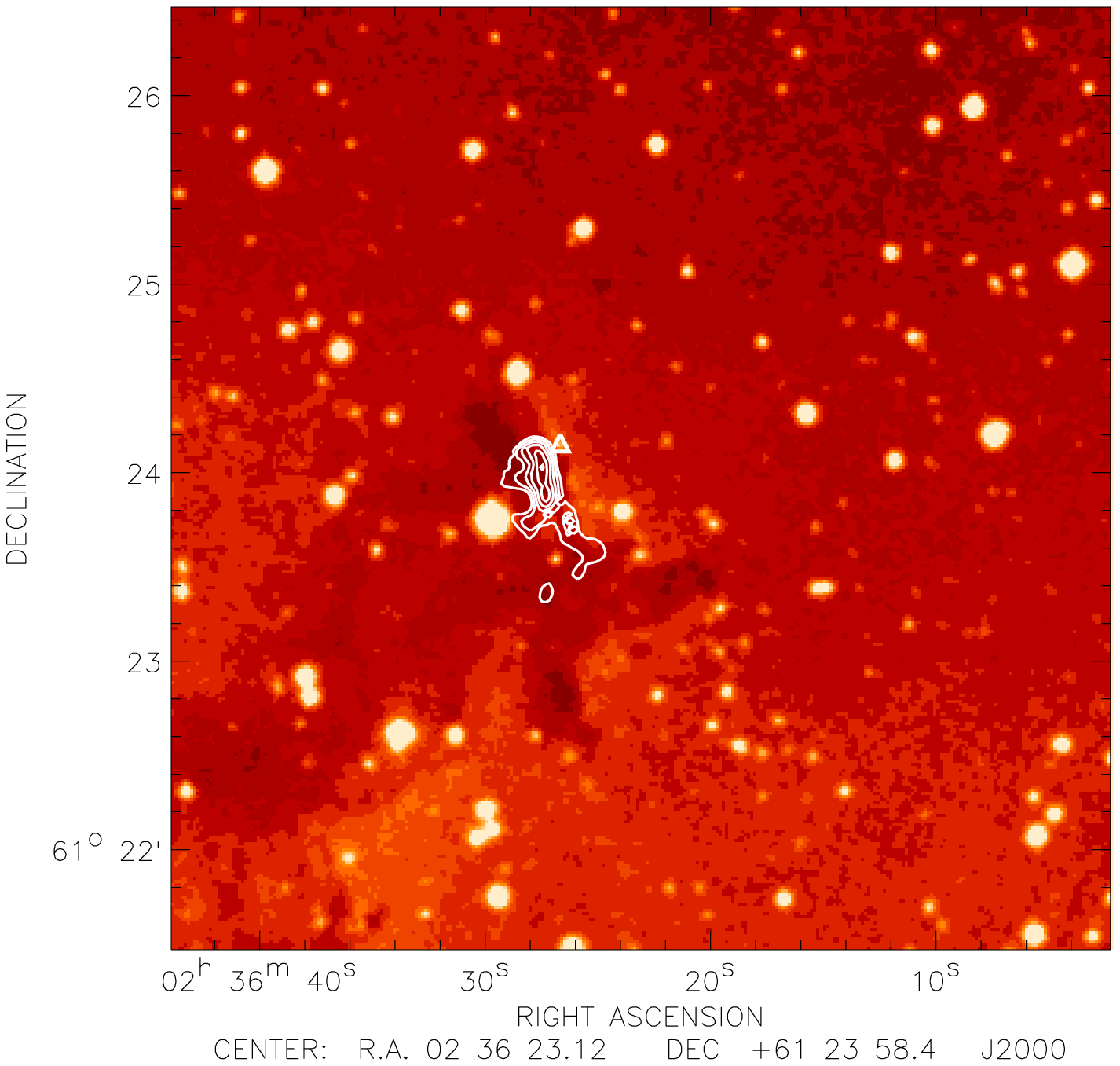}\\
\includegraphics*[height=6cm]{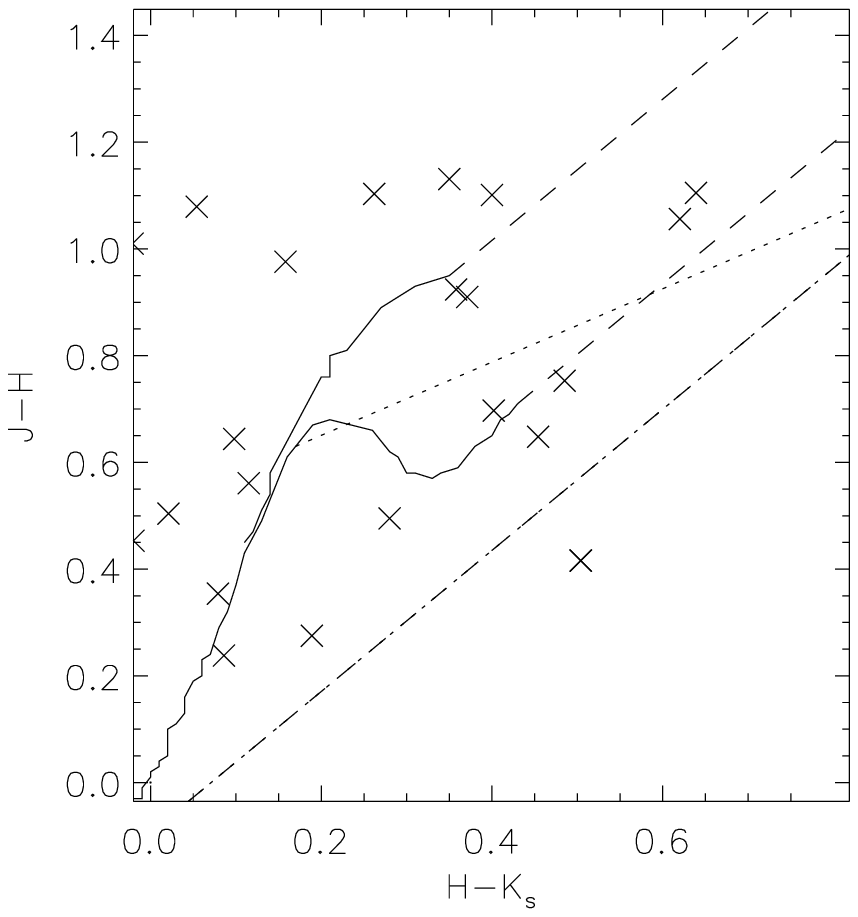}
\includegraphics*[height=6cm]{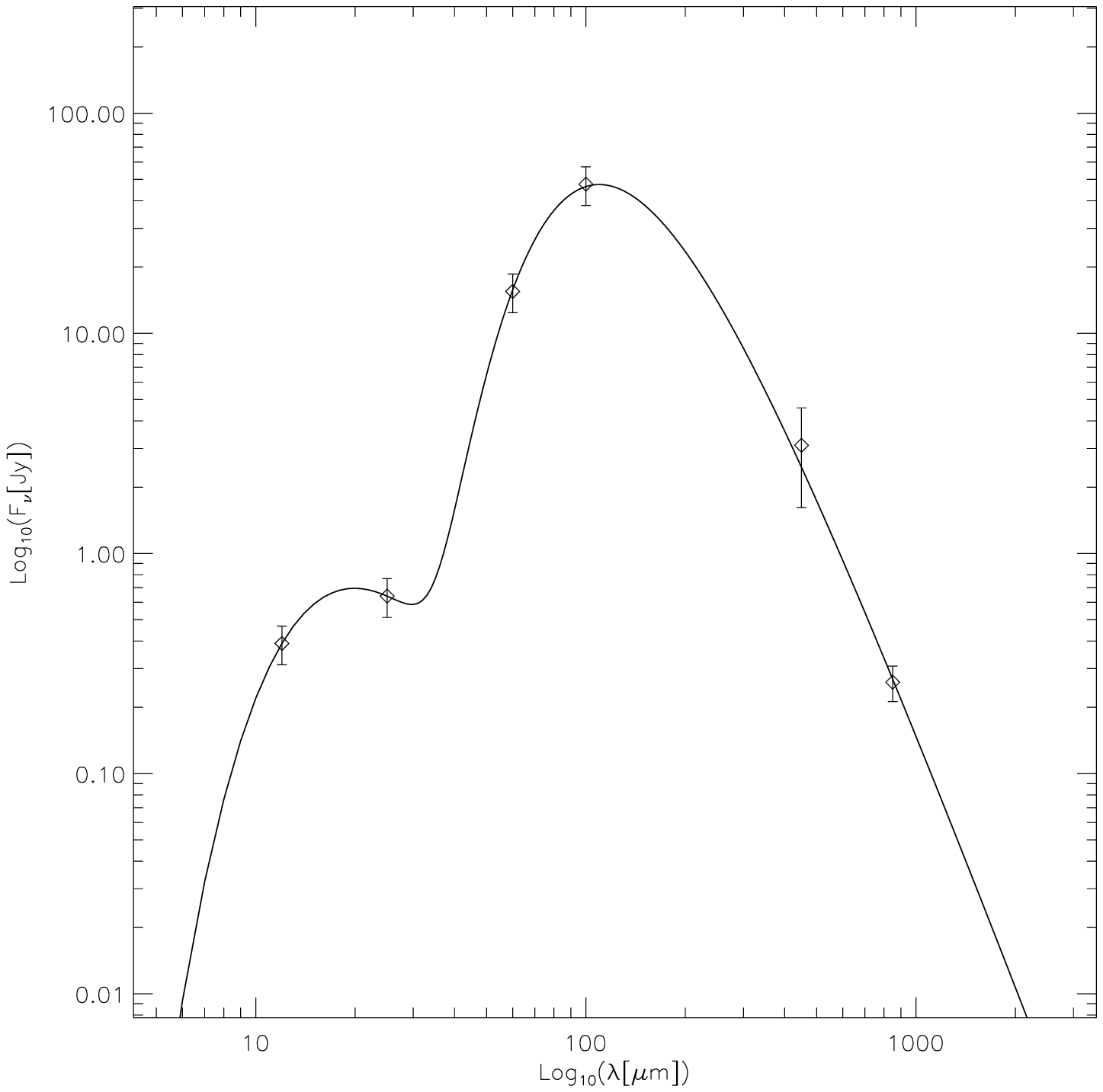}\\
\end{center}
\caption{Plots and images associated with the object SFO 9. The top images show SCUBA 450 \micron ~(left) and 850 \micron ~(right) contours overlaid on a DSS image, infrared sources from the 2MASS Point Source Catalogue \citep{Cutri2003} that have been identified as YSOs are shown as triangles.  850 \micron ~contours start at 6$\sigma$ and increase in increments of 20\% of the peak flux value, 450 \micron ~contours start at 3$\sigma$ and increase in increments of 20\% of the peak flux value.
\indent The bottom left plot shows the J-H versus H-K$_{\rm{s}}$ colours of the 2MASS sources associated with the cloud while the bottom right image shows the SED plot of the object composed from a best fit to various observed fluxes.}
\end{figure*}
\end{center}

\newpage

\begin{center}
\begin{figure*}
\begin{center}
\includegraphics*[height=6cm]{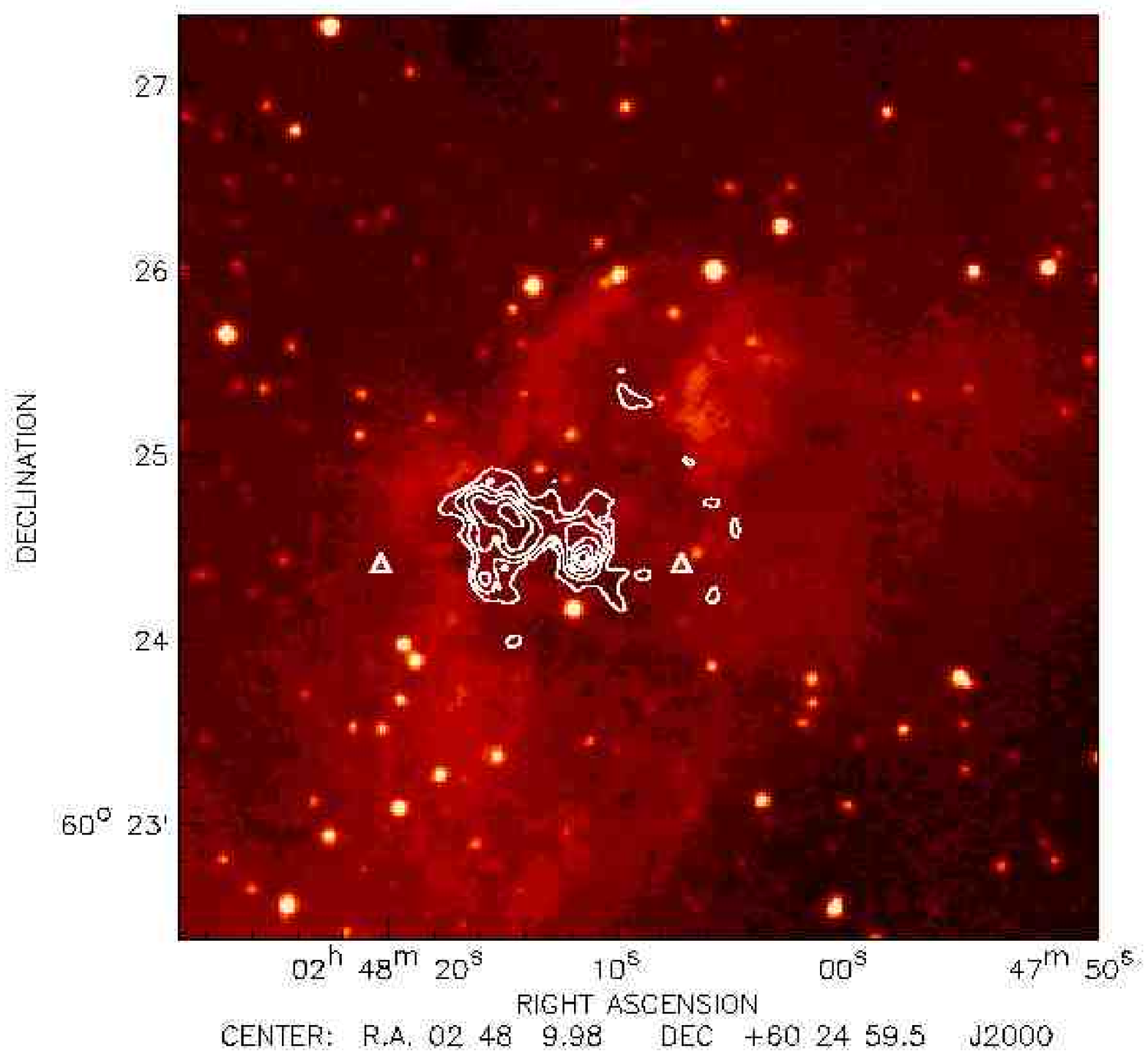}\\
\includegraphics*[height=6cm]{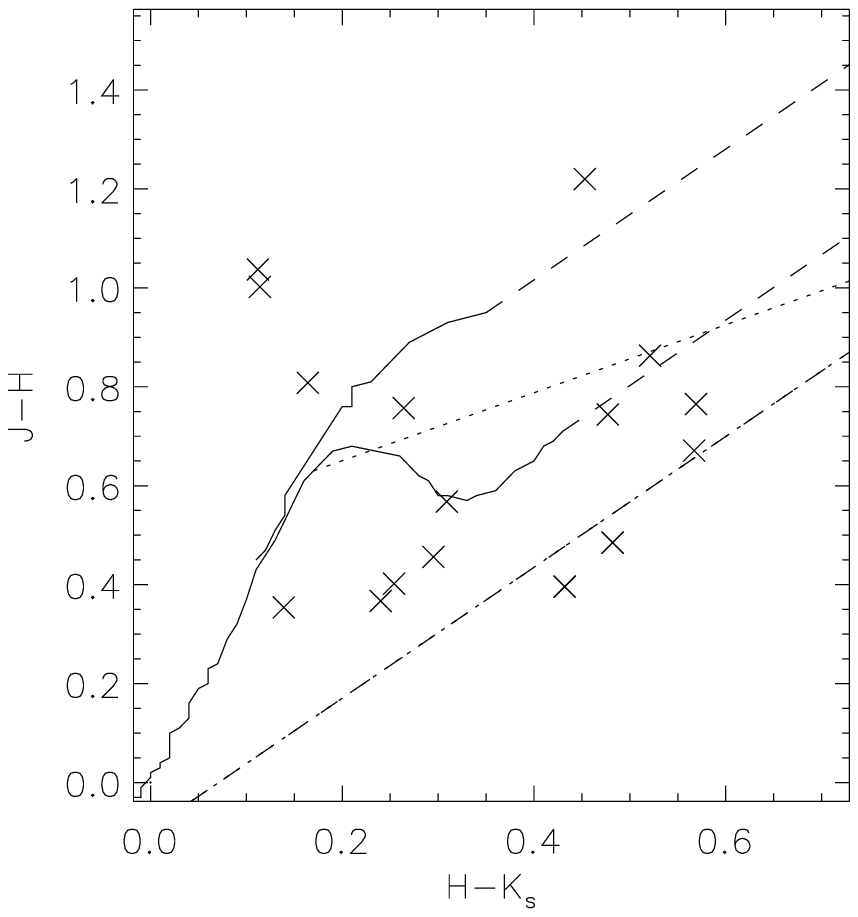}
\includegraphics*[height=6cm]{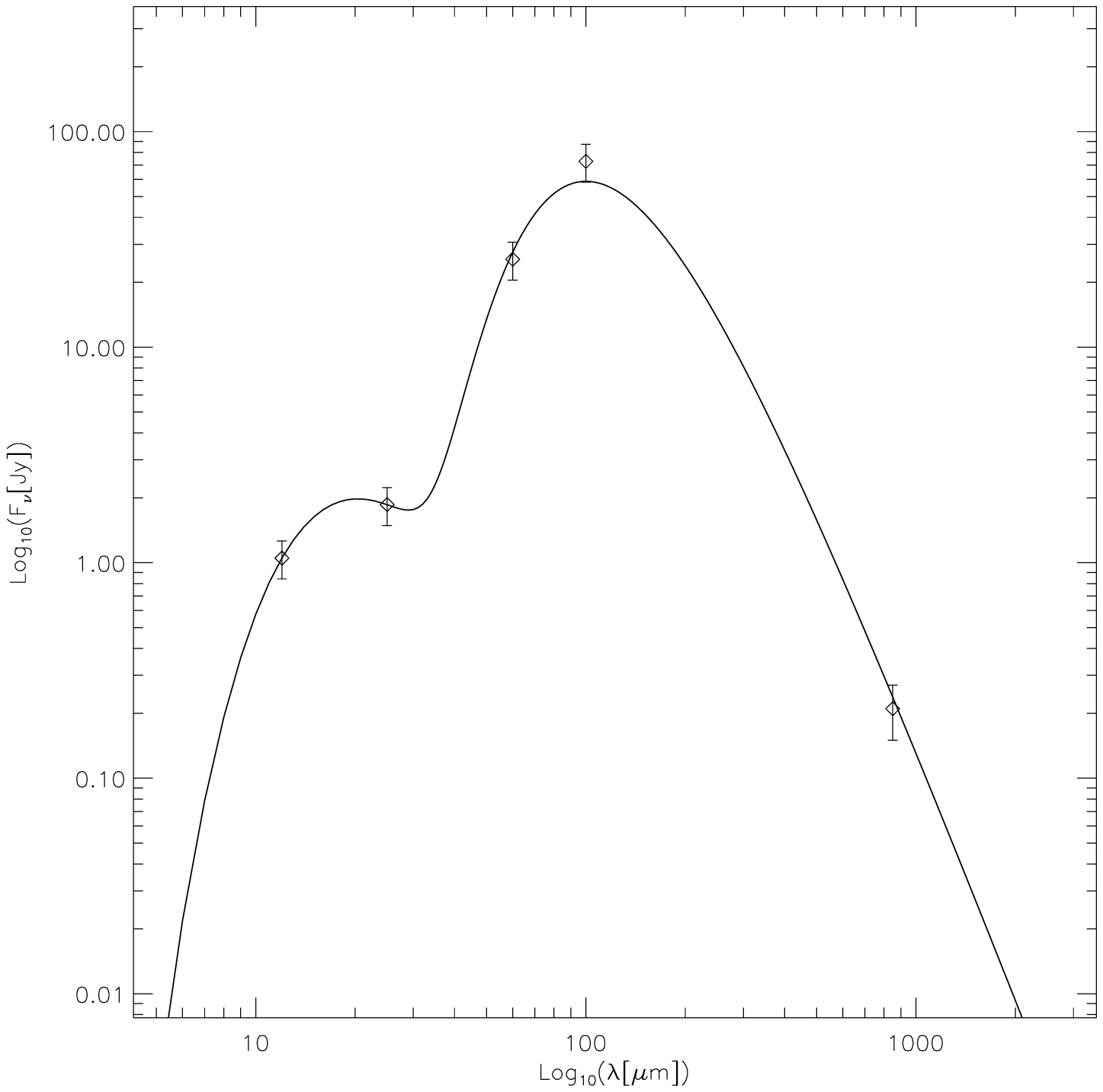}\\
\end{center}
\caption{Plots and images associated with the object SFO 10. The top image shows 850 \micron ~contours overlaid on a DSS image, infrared sources from the 2MASS Point Source Catalogue \citep{Cutri2003} are shown as triangles.  850 \micron ~contours start at 3$\sigma$ and increase in increments of 20\% of the peak flux value.
\indent The bottom left plot shows the J-H versus H-K$_{\rm{s}}$ colours of the 2MASS sources associated with the cloud while the bottom right image shows the SED plot of the object composed from a best fit to various observed fluxes.}
\end{figure*}
\end{center}

\newpage

\begin{center}
\begin{figure*}
\begin{center}
\includegraphics*[height=6cm]{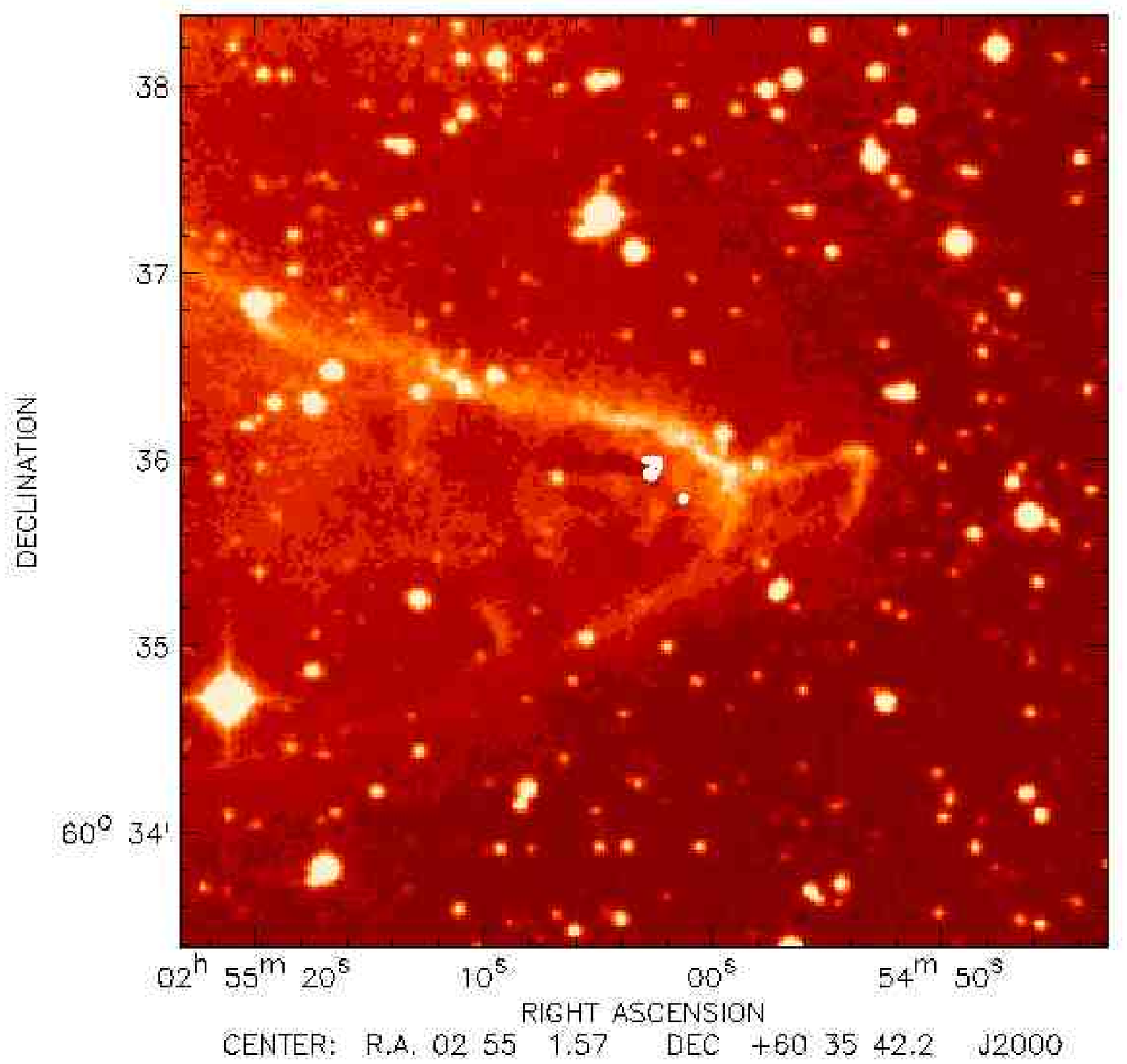}
\includegraphics*[height=6cm]{8104A9b.eps}\\
\includegraphics*[height=6cm]{8104A9c.eps}
\includegraphics*[height=6cm]{8104A9d.eps}\\
\end{center}
\caption{Plots and images associated with the object SFO 12. The top images show SCUBA 450 \micron ~(left) and 850 \micron ~(right) contours overlaid on a DSS image, infrared sources from the 2MASS Point Source Catalogue \citep{Cutri2003} that have been identified as YSOs are shown as triangles.  850 \micron ~contours start at 4$\sigma$ and increase in increments of 20\% of the peak flux value, 450 \micron ~contours start at 3$\sigma$ and increase in increments of 20\% of the peak flux value.
\indent The bottom left plot shows the J-H versus H-K$_{\rm{s}}$ colours of the 2MASS sources associated with the cloud while the bottom right image shows the SED plot of the object composed from a best fit to various observed fluxes.}
\end{figure*}
\end{center}

\newpage

\begin{center}
\begin{figure*}
\begin{center}
\includegraphics*[height=6cm]{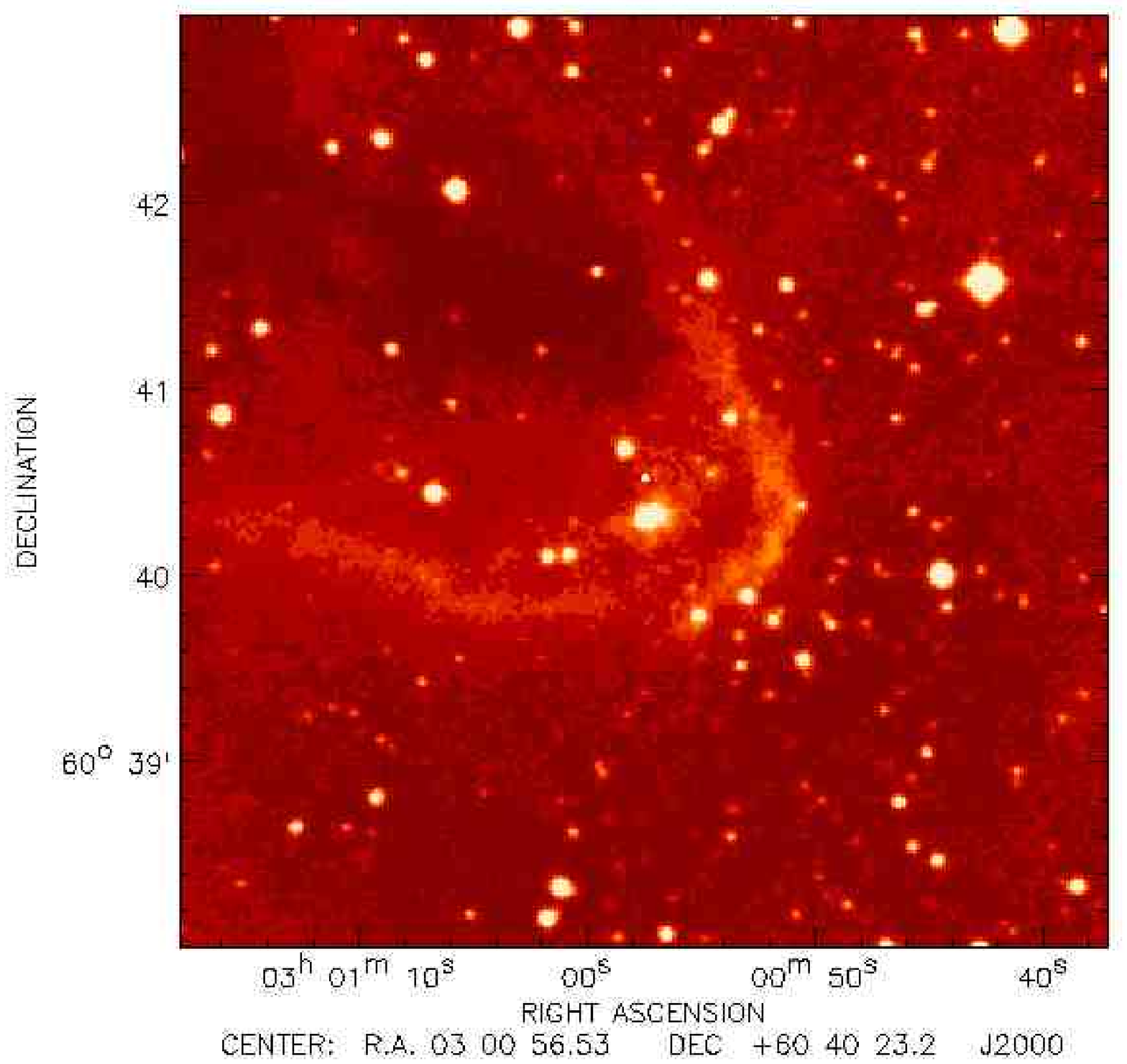}
\includegraphics*[height=6cm]{8104A10b.eps}\\
\includegraphics*[height=6cm]{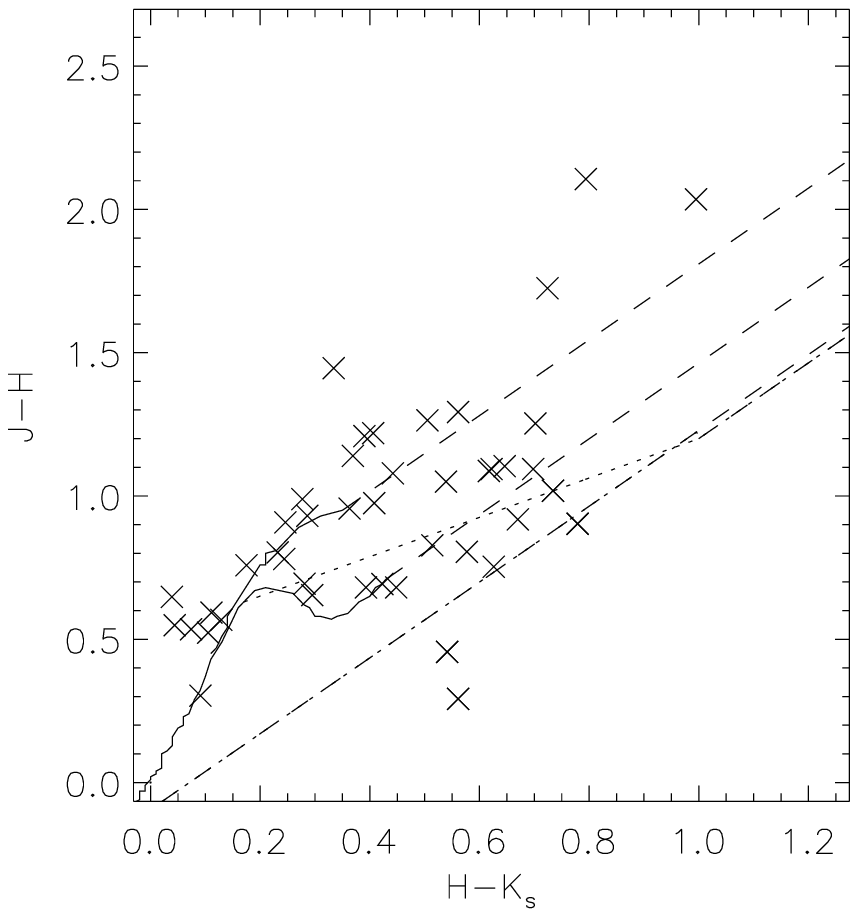}
\includegraphics*[height=6cm]{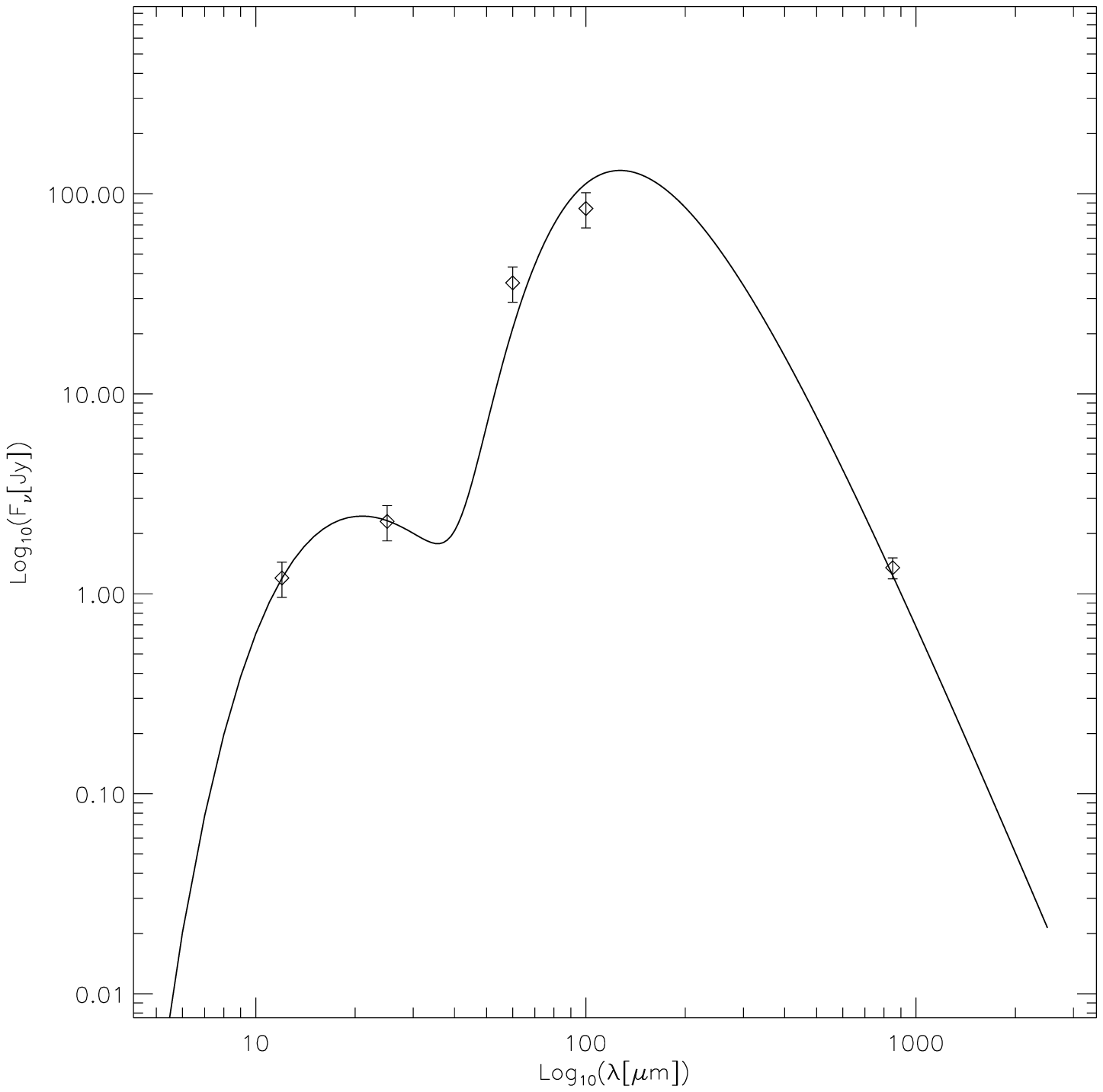}\\
\end{center}
\caption{Plots and images associated with the object SFO 13. The top images show SCUBA 450 \micron ~(left) and 850 \micron ~(right) contours overlaid on a DSS image, infrared sources from the 2MASS Point Source Catalogue \citep{Cutri2003} that have been identified as YSOs are shown as triangles.  850 \micron ~contours start at 6$\sigma$ and increase in increments of 20\% of the peak flux value, 450 \micron ~contours start at 3$\sigma$ and increase in increments of 20\% of the peak flux value.
\indent The bottom left plot shows the J-H versus H-K$_{\rm{s}}$ colours of the 2MASS sources associated with the cloud while the bottom right image shows the SED plot of the object composed from a best fit to various observed fluxes.}
\end{figure*}
\end{center}

\newpage

\begin{center}
\begin{figure*}
\begin{center}
\includegraphics*[height=6cm]{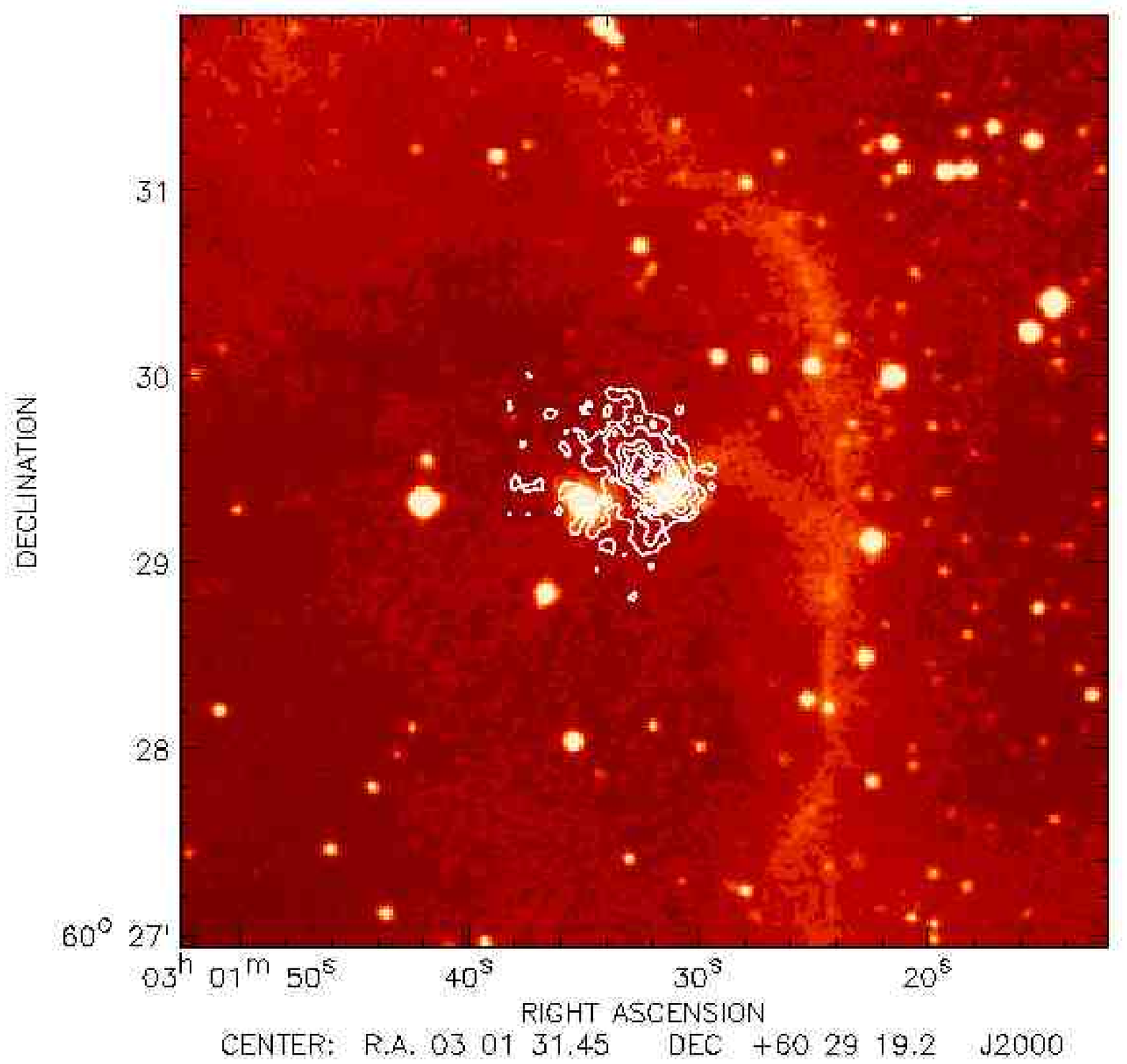}
\includegraphics*[height=6cm]{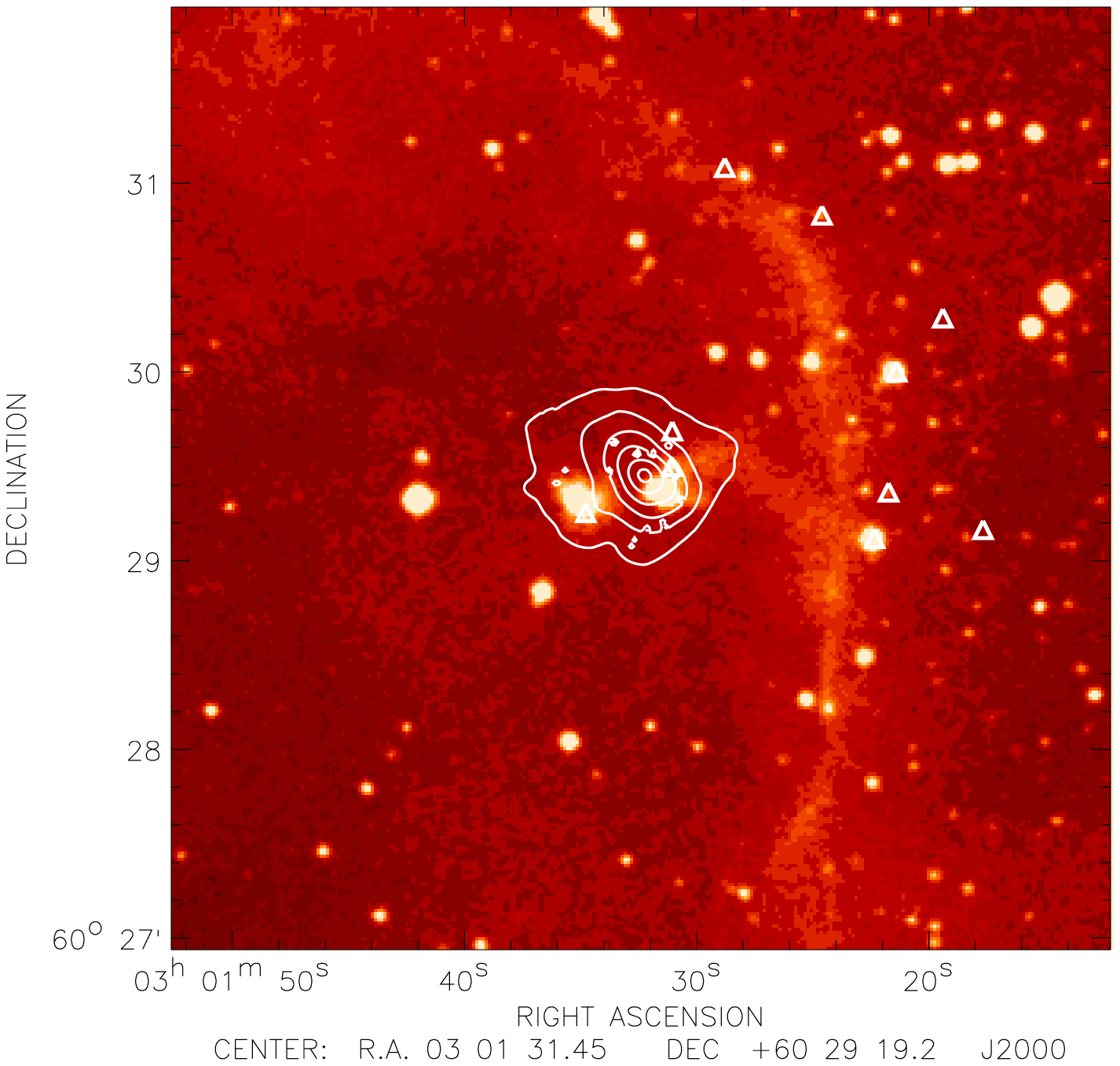}\\
\includegraphics*[height=6cm]{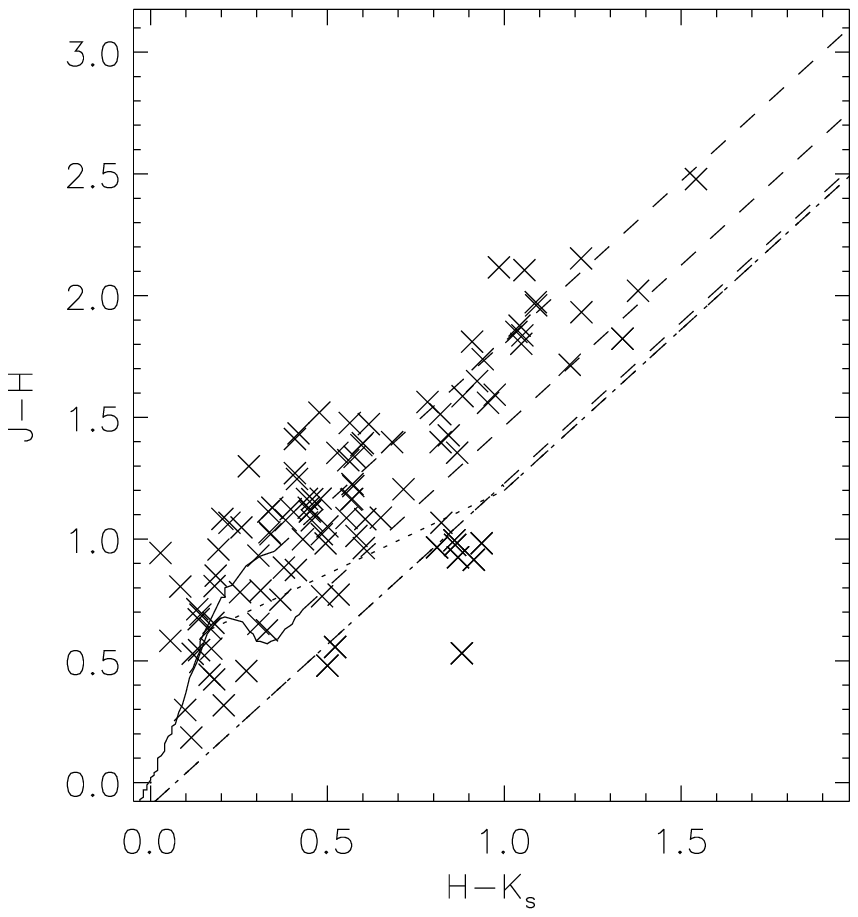}
\includegraphics*[height=6cm]{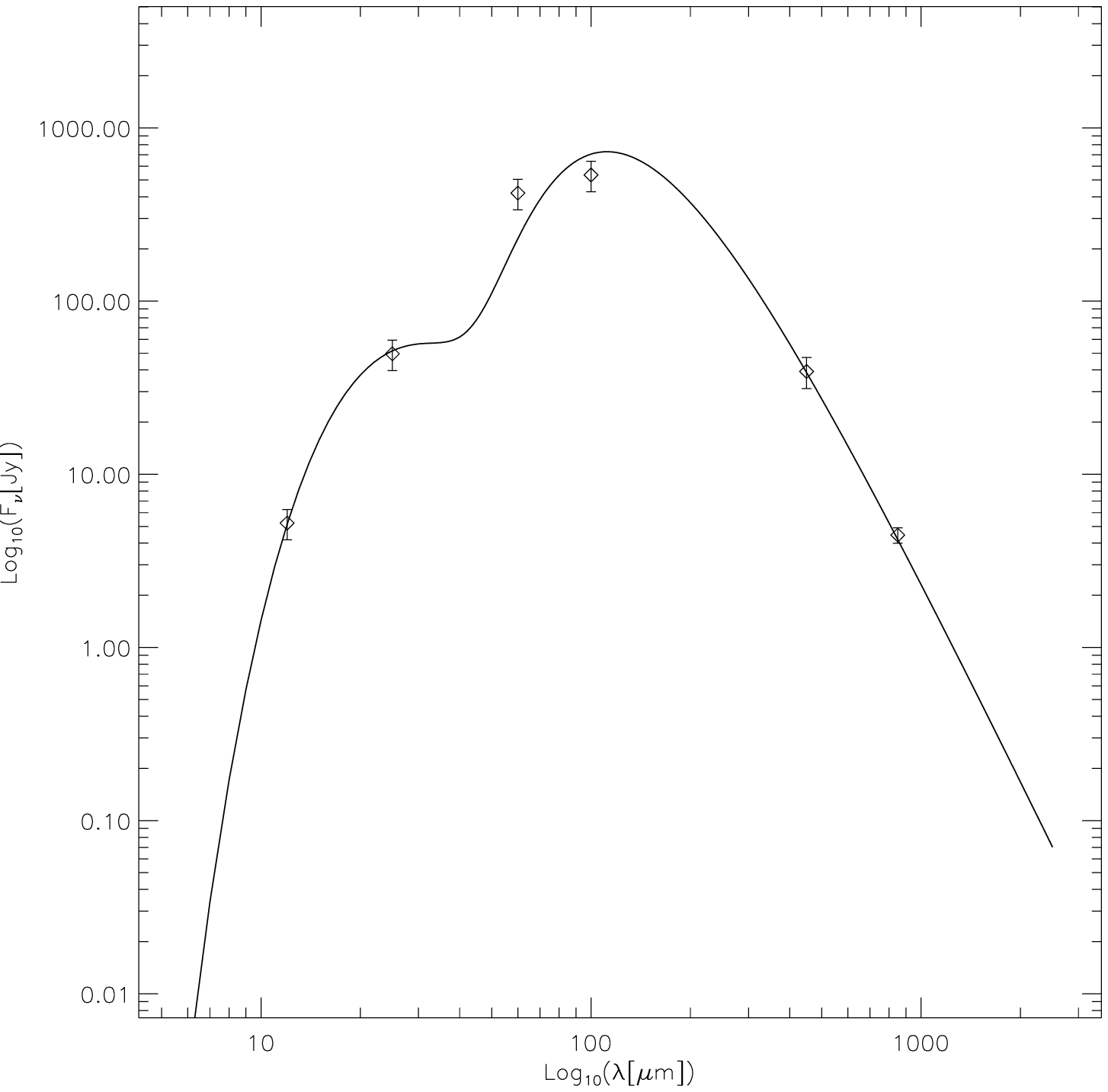}\\
\end{center}
\caption{Plots and images associated with the object SFO 14. The top images show SCUBA 450 \micron ~(left) and 850 \micron ~(right) contours overlaid on a DSS image, infrared sources from the 2MASS Point Source Catalogue \citep{Cutri2003} that have been identified as YSOs are shown as triangles.  850 \micron ~contours start at 15$\sigma$ and increase in increments of 20\% of the peak flux value, 450 \micron ~contours start at 3$\sigma$ and increase in increments of 20\% of the peak flux value.
\indent The bottom left plot shows the J-H versus H-K$_{\rm{s}}$ colours of the 2MASS sources associated with the cloud while the bottom right image shows the SED plot of the object composed from a best fit to various observed fluxes.}
\end{figure*}
\end{center}

\newpage

\begin{center}
\begin{figure*}
\begin{center}
\includegraphics*[height=6cm]{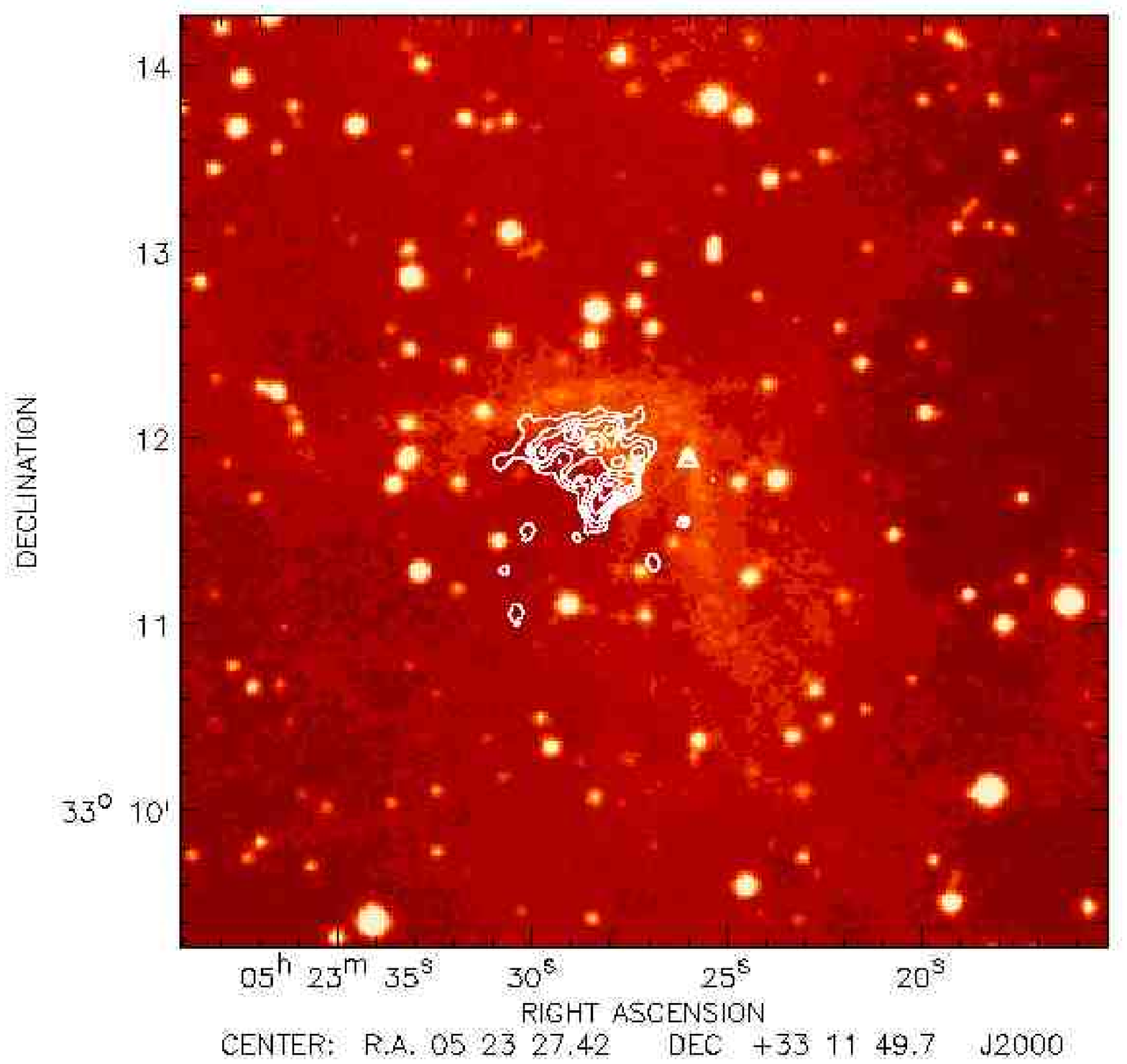}\\
\includegraphics*[height=6cm]{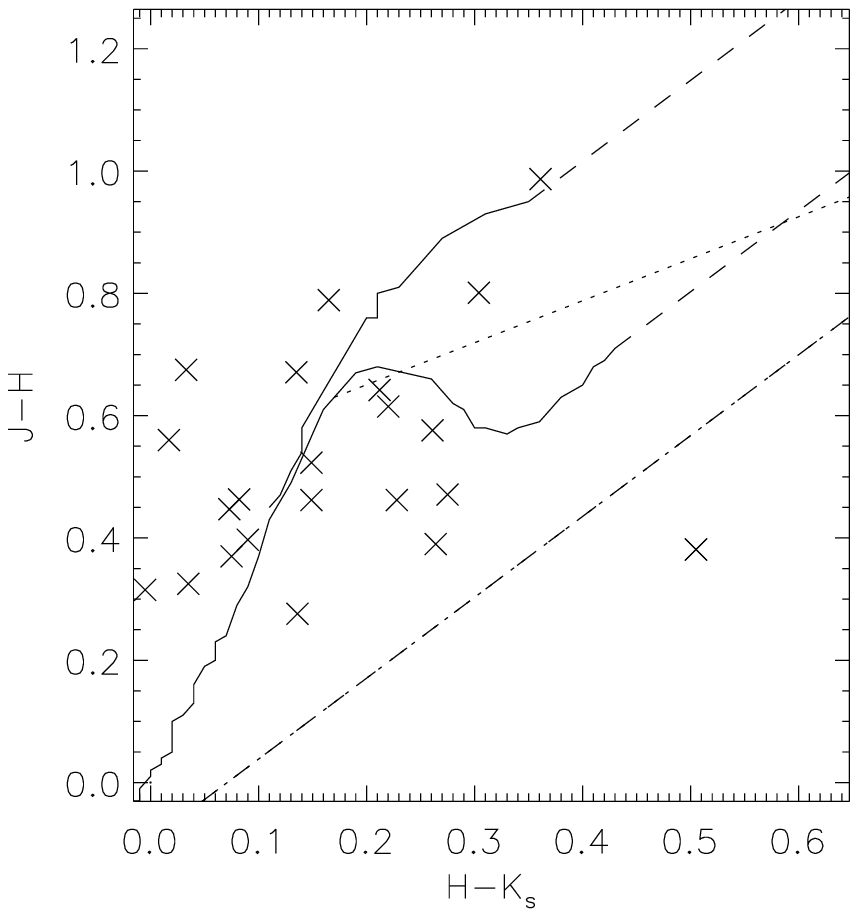}
\includegraphics*[height=6cm]{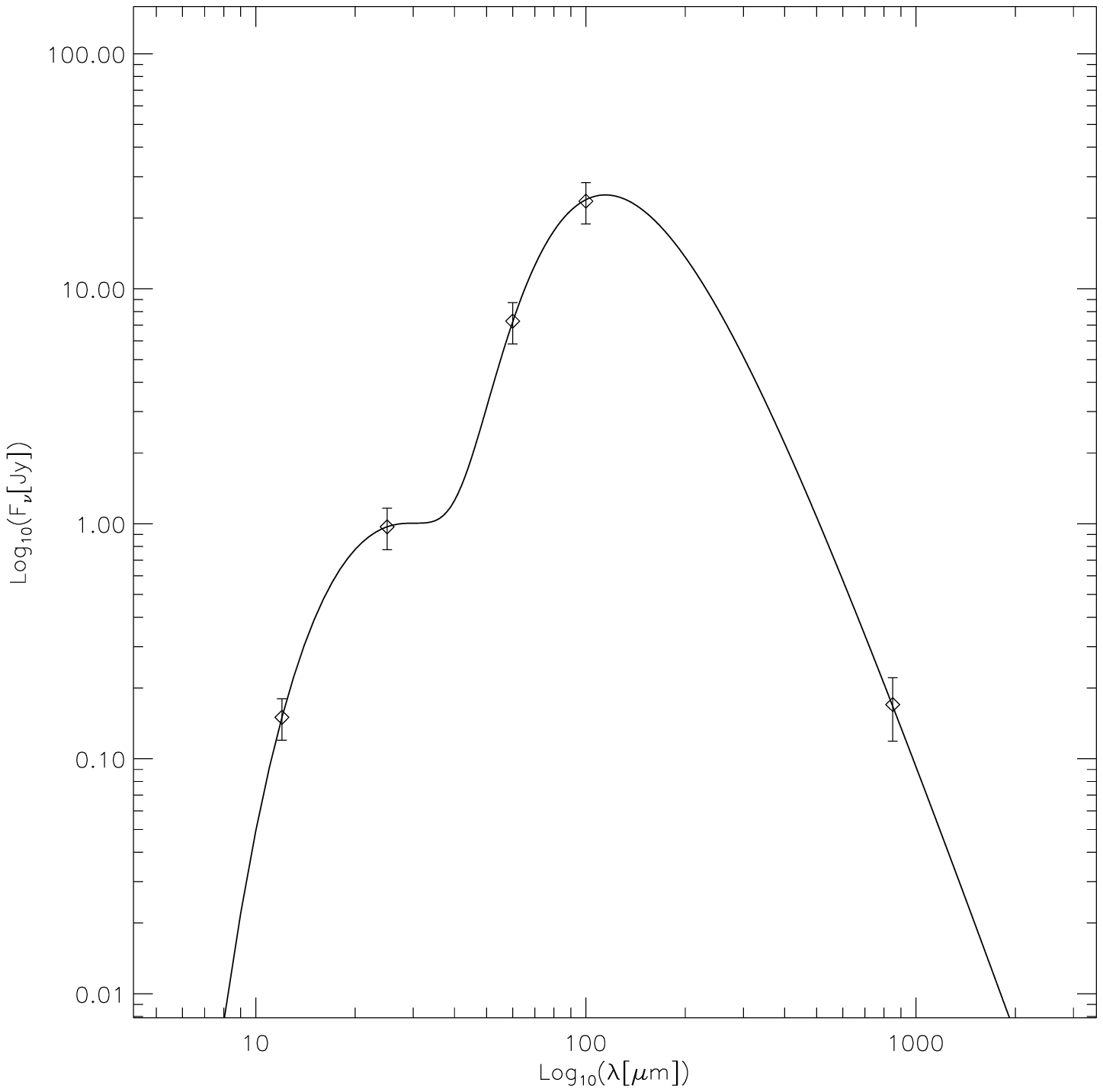}\\
\end{center}
\caption{Plots and images associated with the object SFO 15. The top image shows SCUBA 850 \micron ~contours overlaid on a DSS image, infrared sources from the 2MASS Point Source Catalogue \citep{Cutri2003} are shown as triangles.  850 \micron ~contours start at 3$\sigma$ and increase in increments of 20\% of the peak flux value.
\indent The bottom left plot shows the J-H versus H-K$_{\rm{s}}$ colours of the 2MASS sources associated with the cloud while the bottom right image shows the SED plot of the object composed from a best fit to various observed fluxes.}
\end{figure*}
\end{center}

\newpage

\begin{center}
\begin{figure*}
\begin{center}
\includegraphics*[height=6cm]{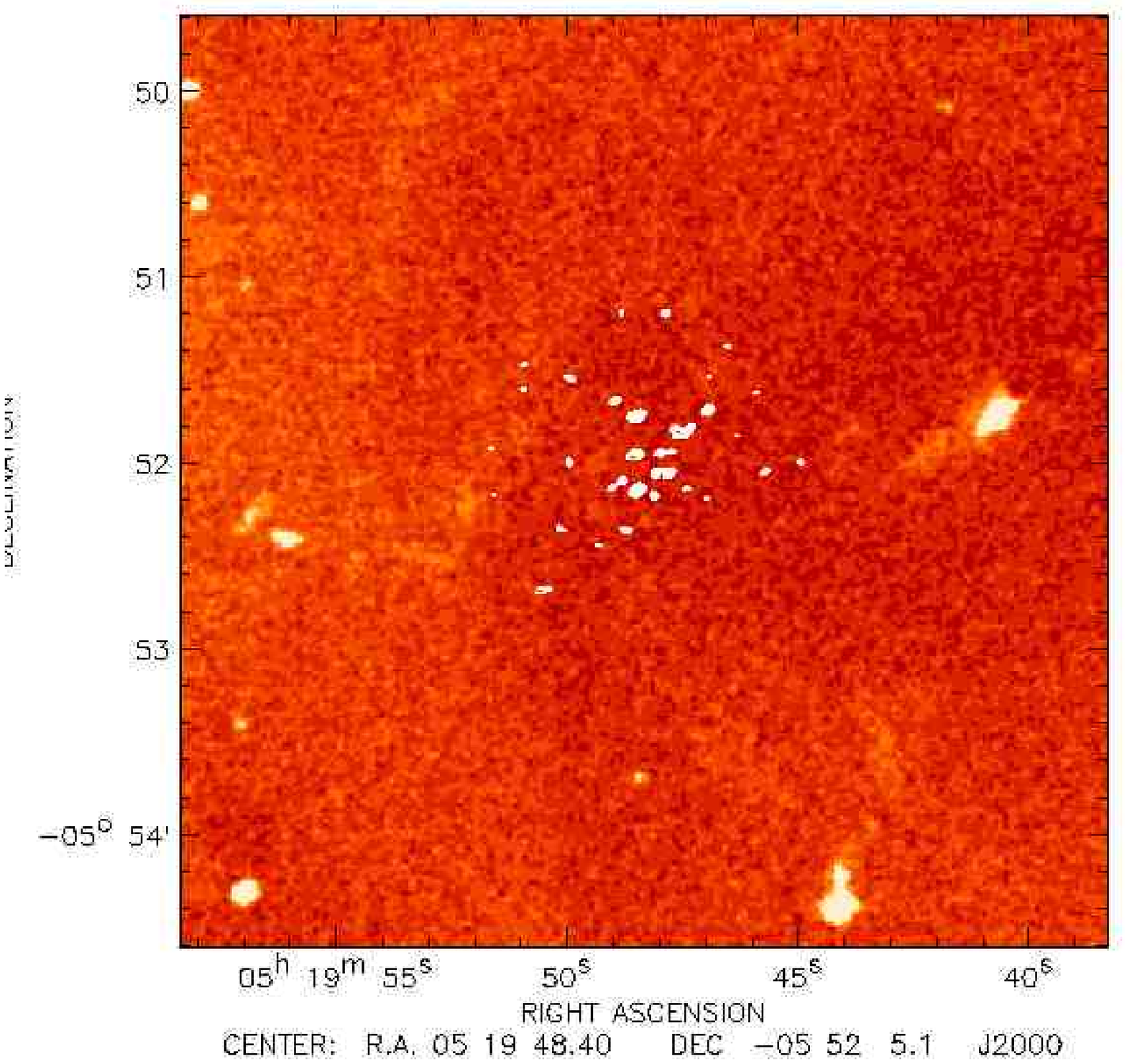}
\includegraphics*[height=6cm]{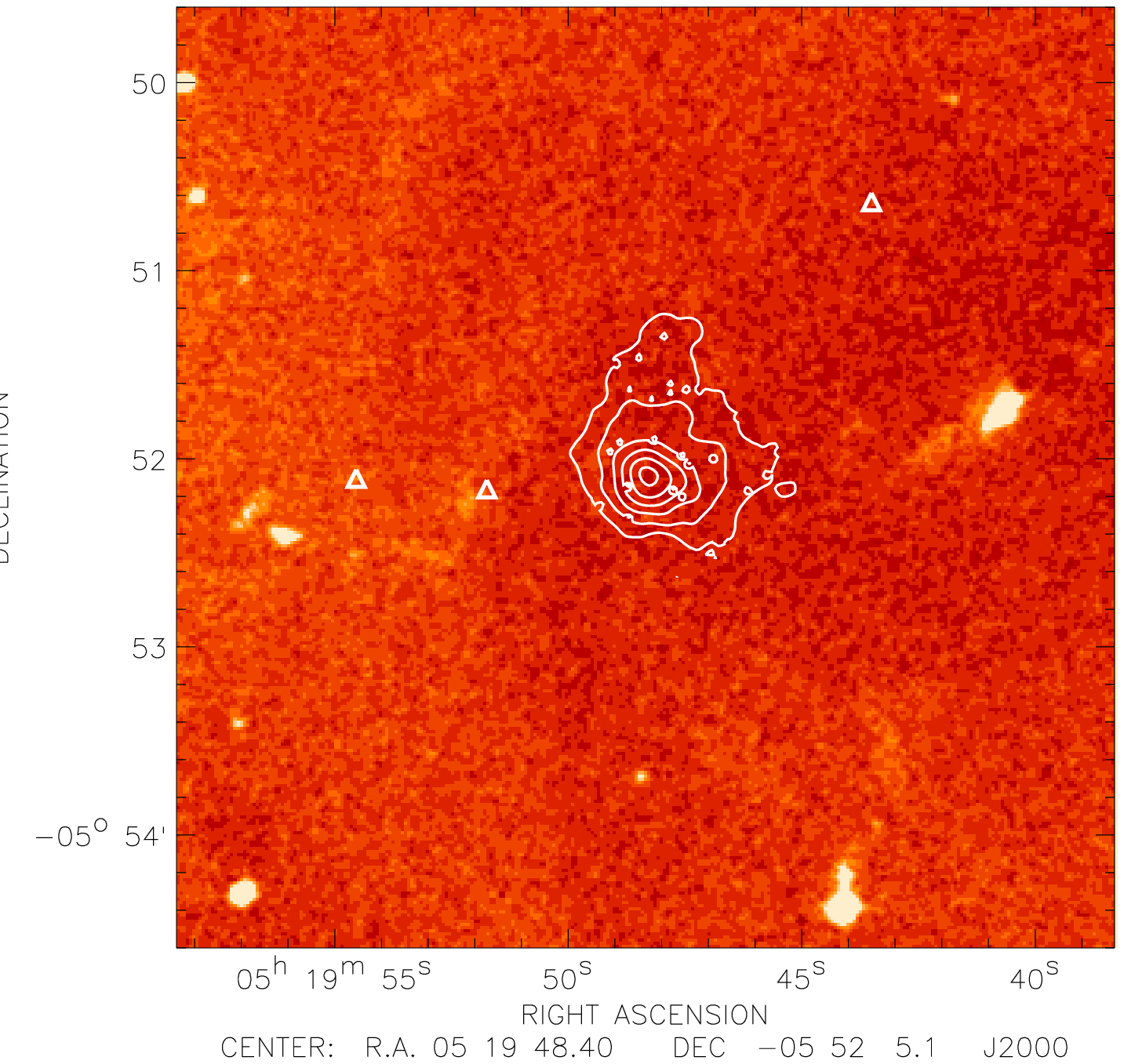}\\
\includegraphics*[height=6cm]{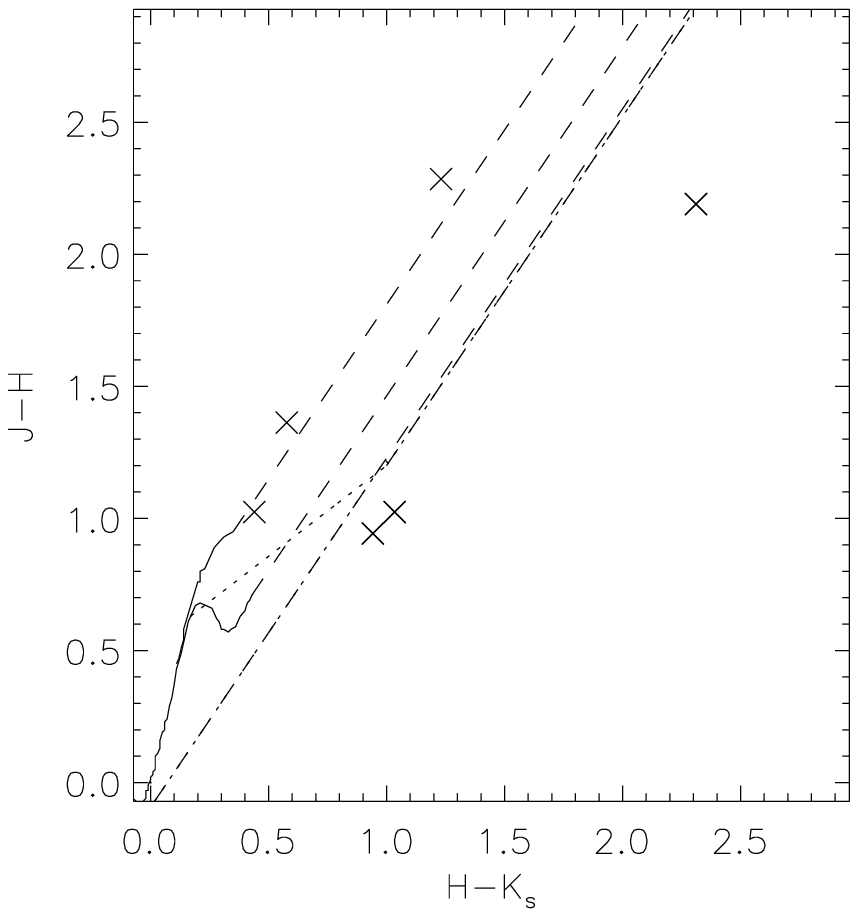}
\includegraphics*[height=6cm]{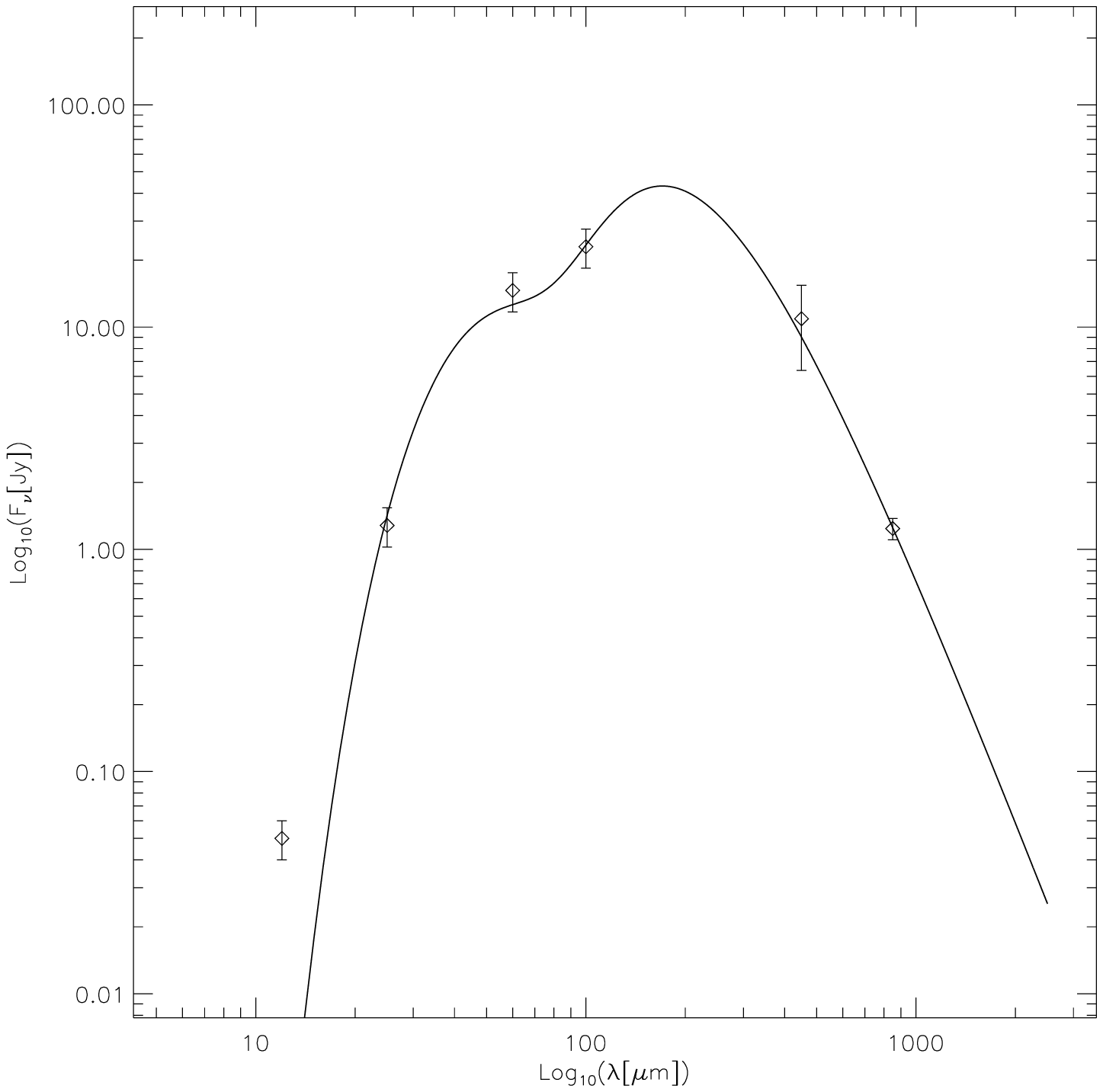}\\
\end{center}
\caption{Plots and images associated with the object SFO 16. The top images show SCUBA 450 \micron ~(left) and 850 \micron ~(right) contours overlaid on a DSS image, infrared sources from the 2MASS Point Source Catalogue \citep{Cutri2003} that have been identified as YSOs are shown as triangles.  850 \micron ~contours start at 9$\sigma$ and increase in increments of 20\% of the peak flux value, 450 \micron ~contours start at 4$\sigma$ and increase in increments of 20\% of the peak flux value.
\indent The bottom left plot shows the J-H versus H-K$_{\rm{s}}$ colours of the 2MASS sources associated with the cloud while the bottom right image shows the SED plot of the object composed from a best fit to various observed fluxes.}
\end{figure*}
\end{center}

\newpage
\clearpage

\begin{center}
\begin{figure*}
\begin{center}
\includegraphics*[height=6cm]{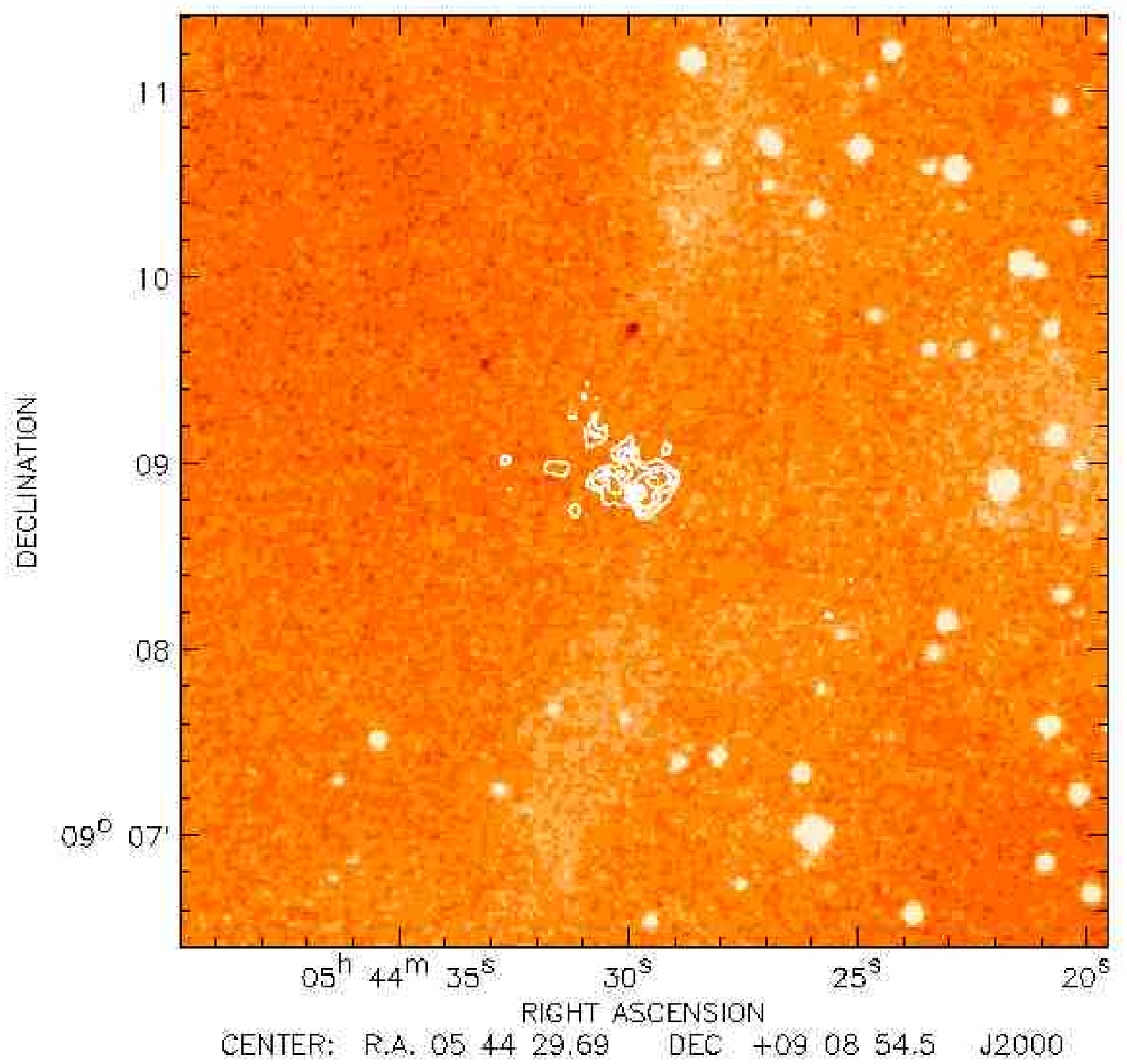}
\includegraphics*[height=6cm]{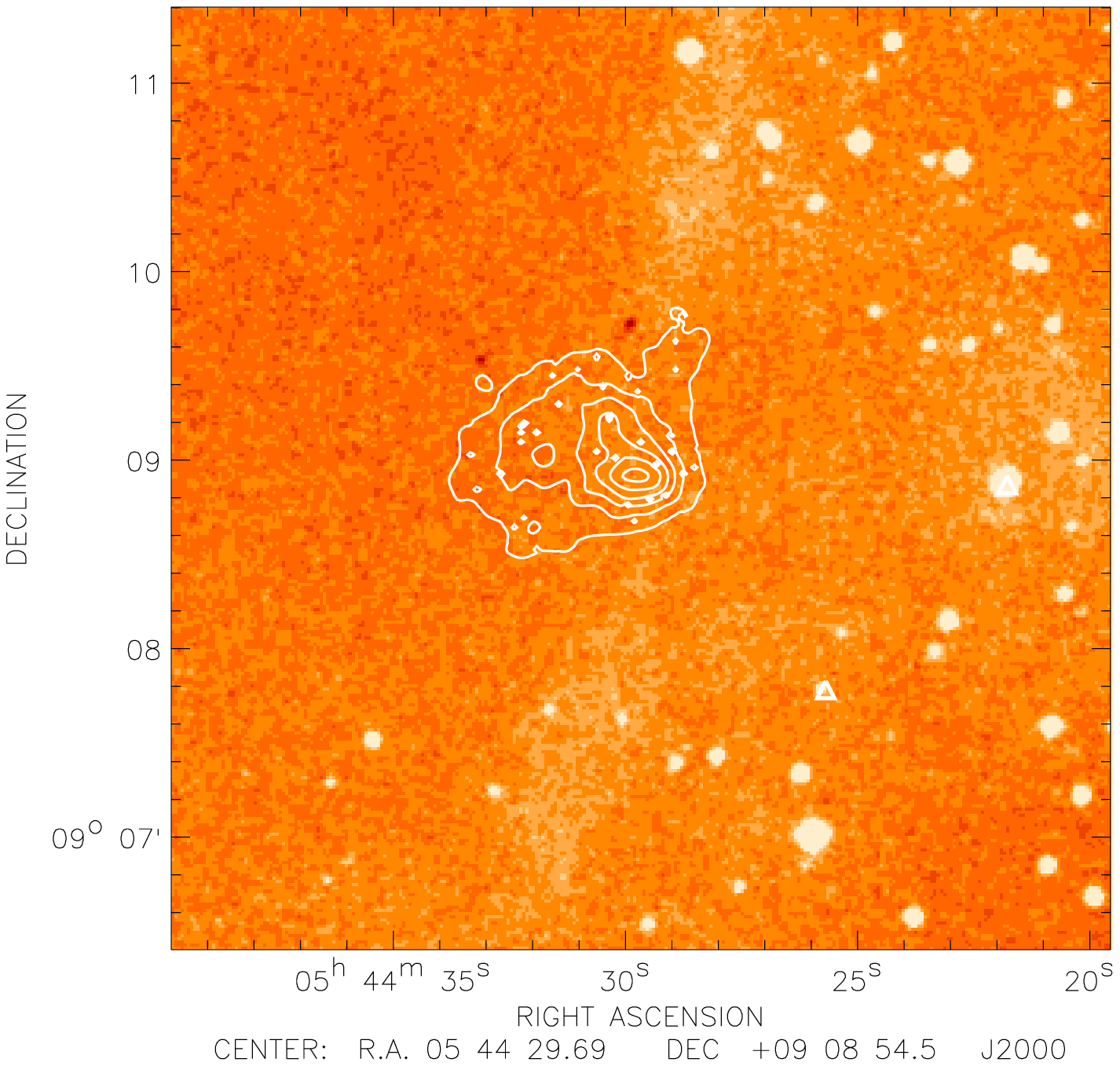}\\
\includegraphics*[height=6cm]{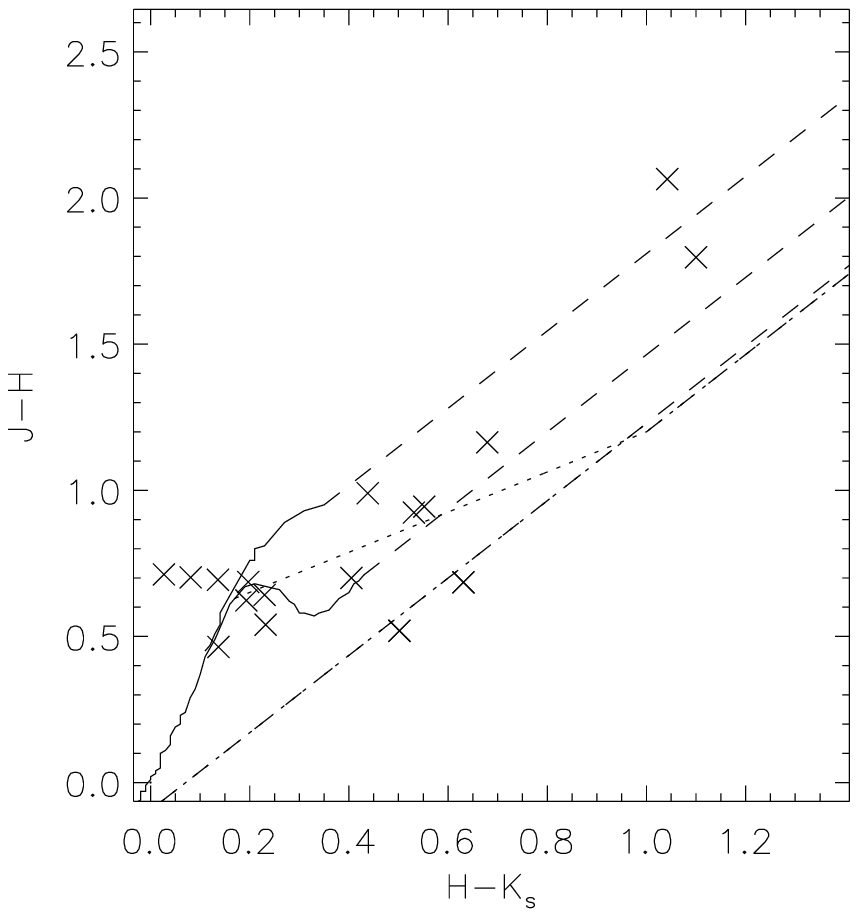}
\includegraphics*[height=6cm]{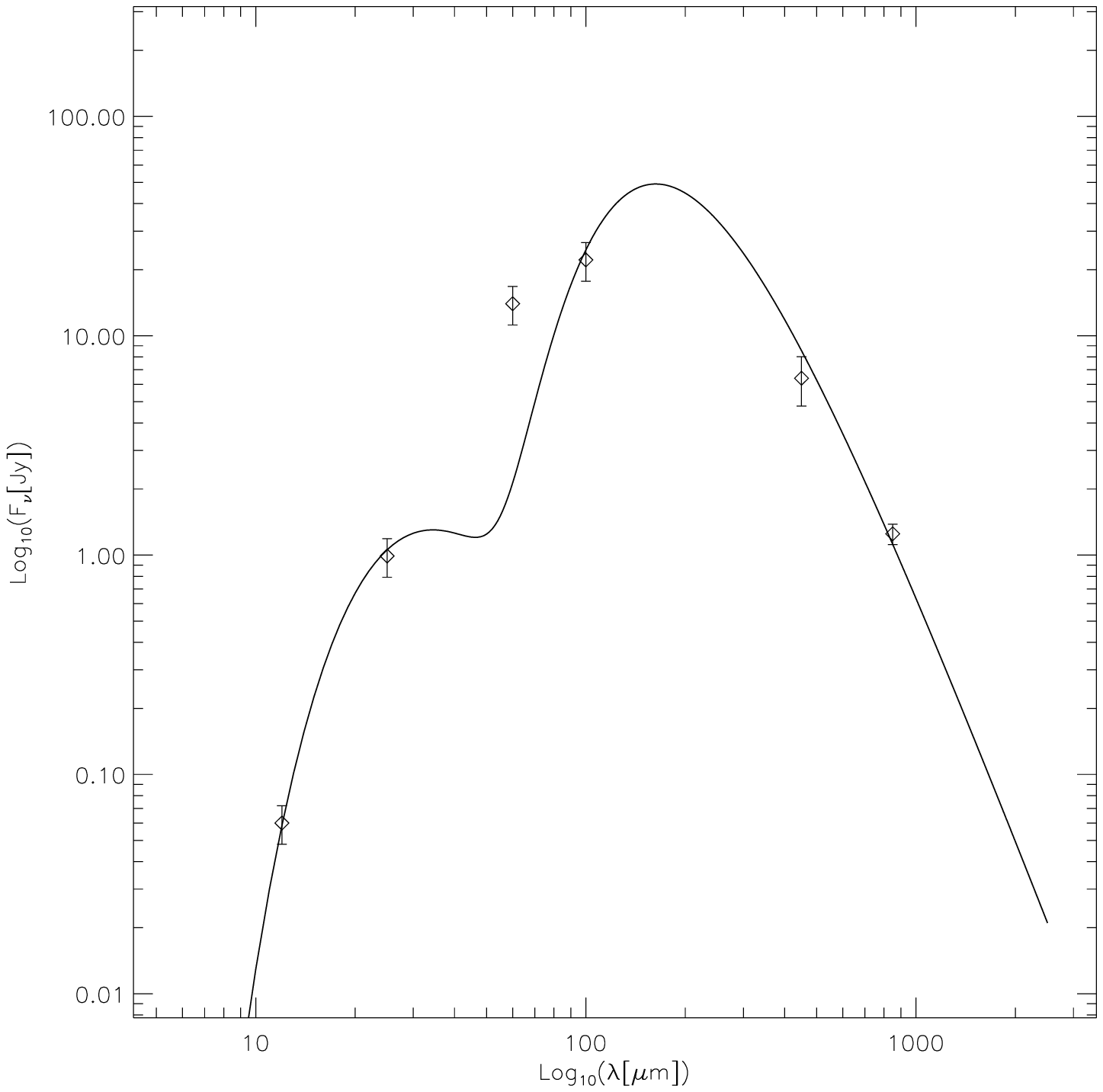}\\
\end{center}
\caption{Plots and images associated with the object SFO 18. The top images show SCUBA 450 \micron ~(left) and 850 \micron ~(right) contours overlaid on a DSS image, infrared sources from the 2MASS Point Source Catalogue \citep{Cutri2003} that have been identified as YSOs are shown as triangles.  850 \micron ~contours start at 9$\sigma$ and increase in increments of 20\% of the peak flux value, 450 \micron ~contours start at 4$\sigma$ and increase in increments of 20\% of the peak flux value.
\indent The bottom left plot shows the J-H versus H-K$_{\rm{s}}$ colours of the 2MASS sources associated with the cloud while the bottom right image shows the SED plot of the object composed from a best fit to various observed fluxes.}
\end{figure*}
\end{center}

\newpage

\begin{center}
\begin{figure*}
\begin{center}
\includegraphics*[height=6cm]{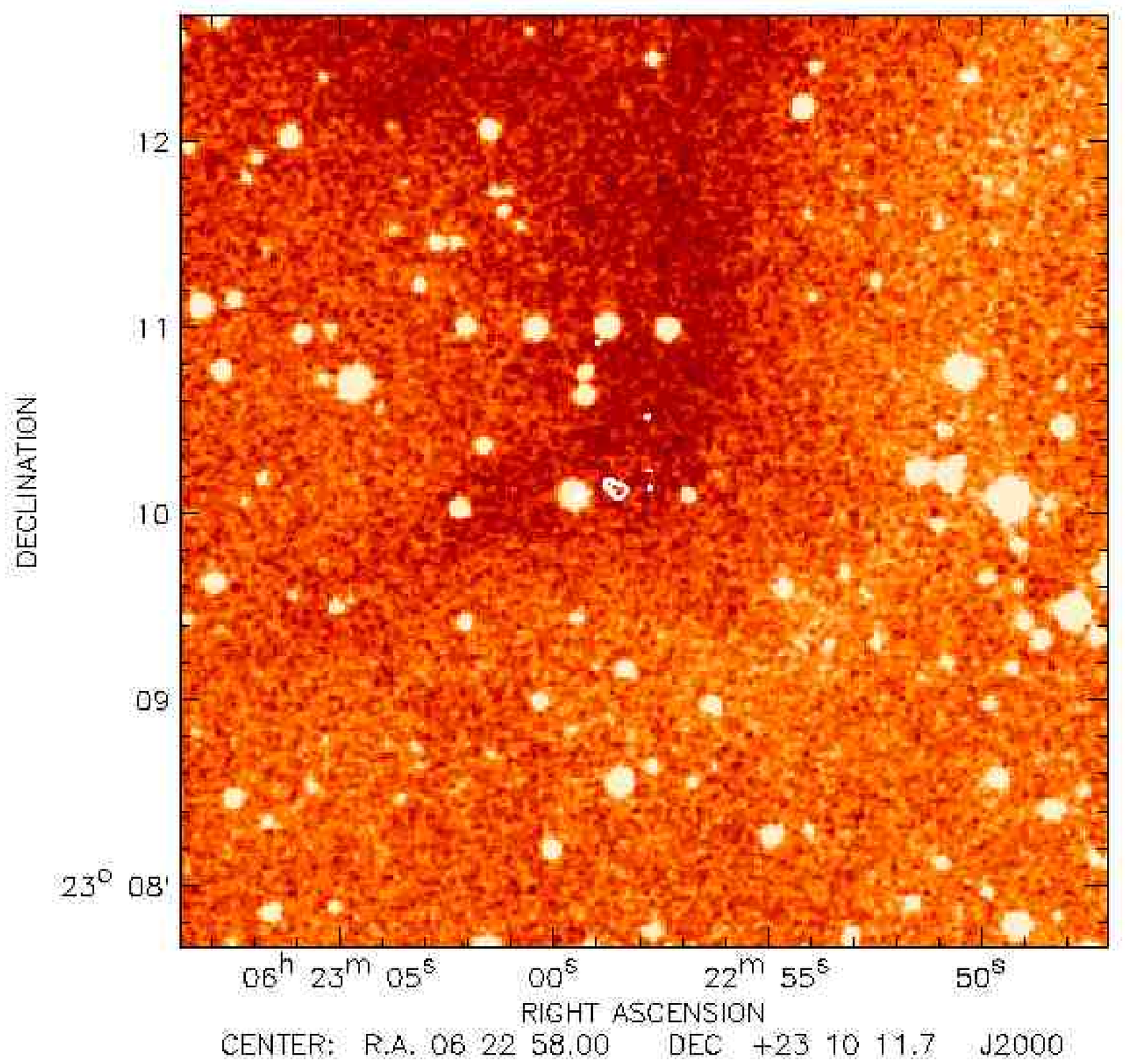}
\includegraphics*[height=6cm]{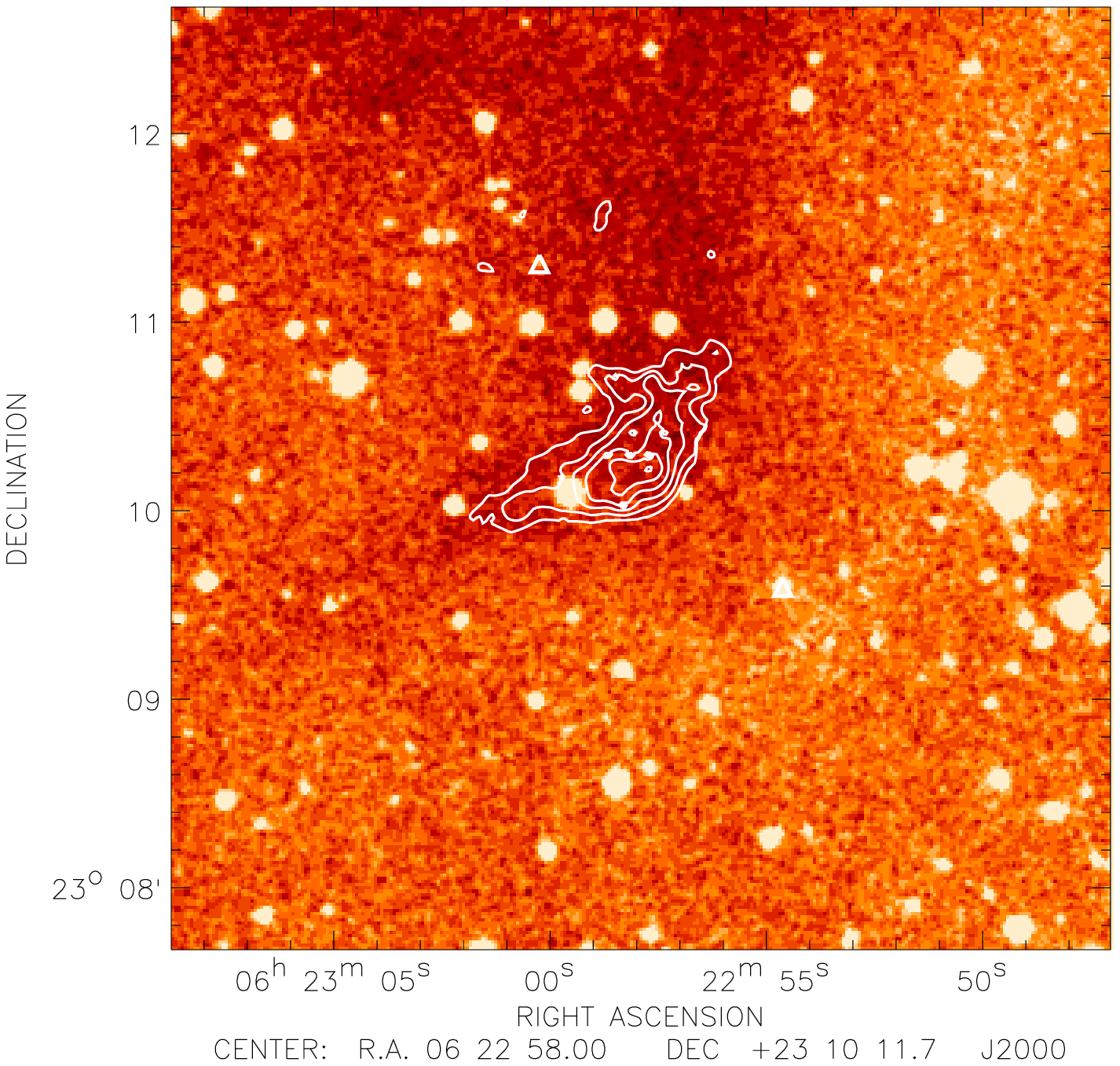}\\
\includegraphics*[height=6cm]{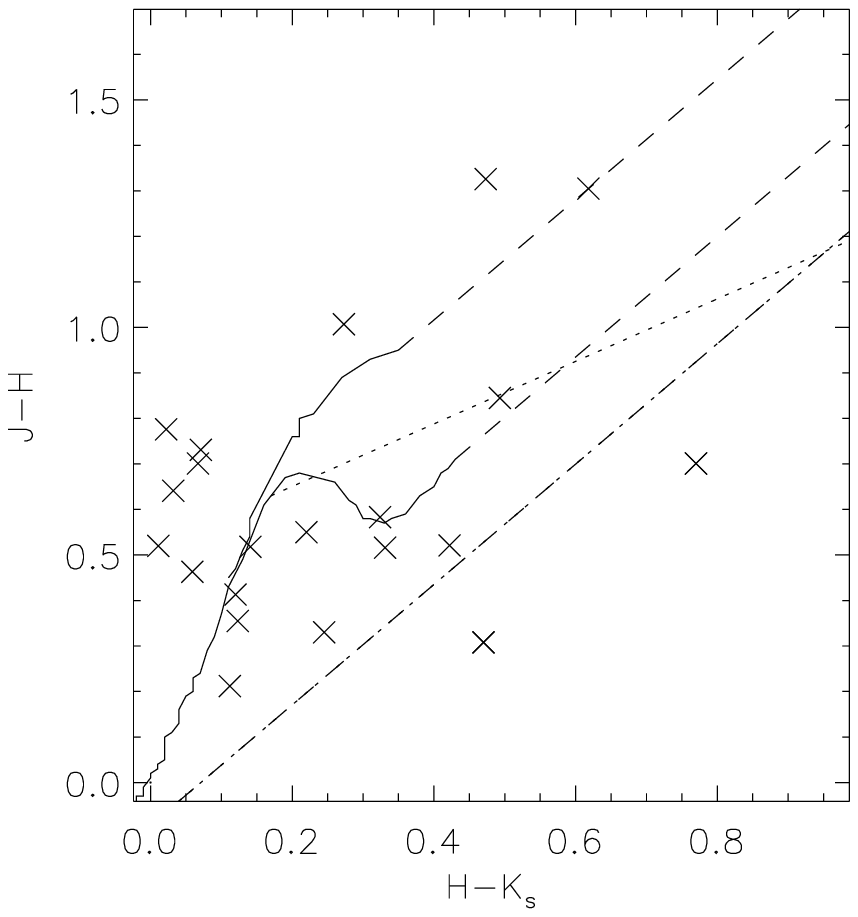}
\includegraphics*[height=6cm]{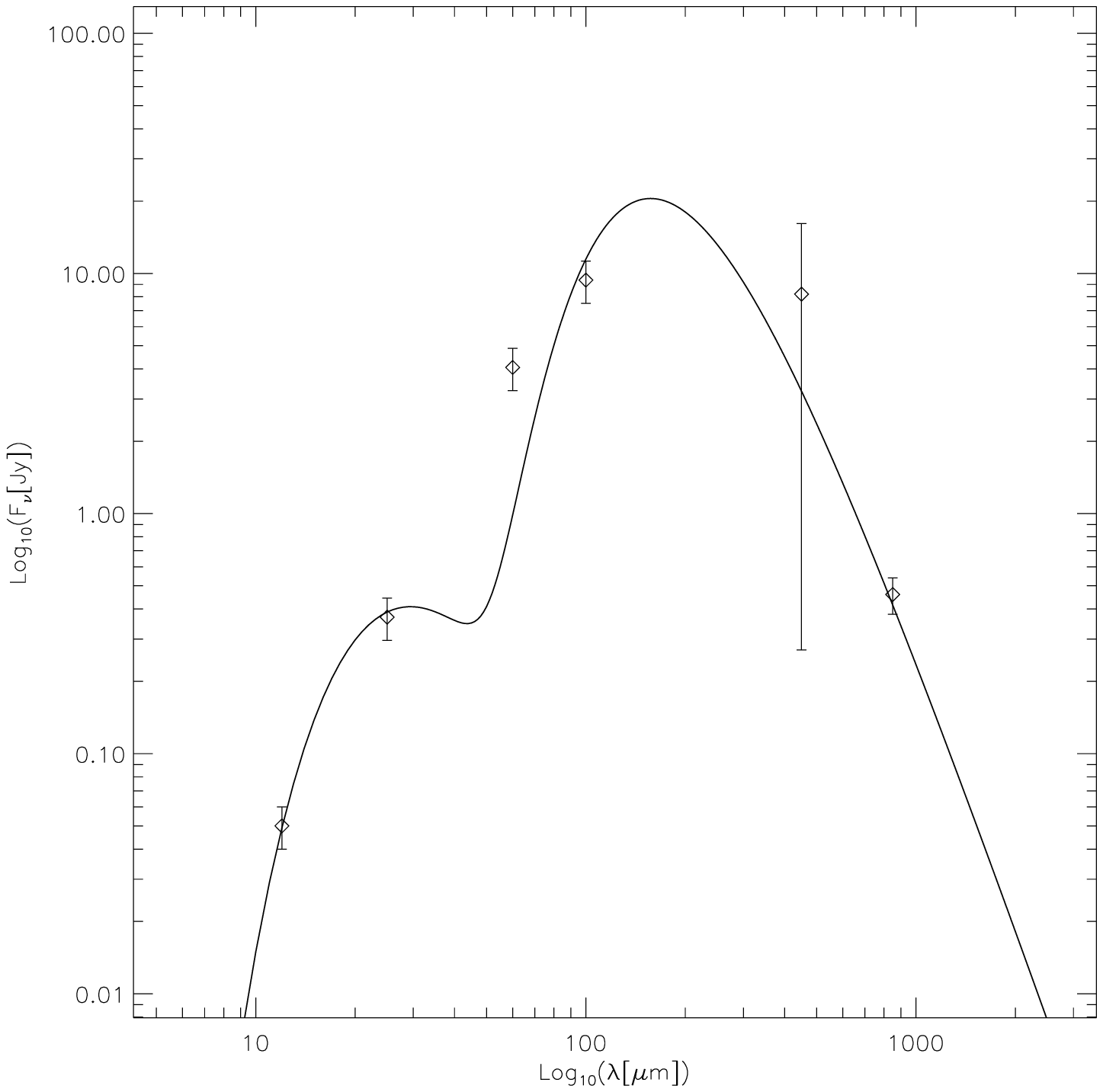}\\
\end{center}
\caption{Plots and images associated with the object SFO 23. The top images show SCUBA 450 \micron ~(left) and 850 \micron ~(right) contours overlaid on a DSS image, infrared sources from the 2MASS Point Source Catalogue \citep{Cutri2003} that have been identified as YSOs are shown as triangles.  850 \micron ~contours start at 3$\sigma$ and increase in increments of 20\% of the peak flux value, 450 \micron ~contours start at 3$\sigma$ and increase in increments of 20\% of the peak flux value.
\indent The bottom left plot shows the J-H versus H-K$_{\rm{s}}$ colours of the 2MASS sources associated with the cloud while the bottom right image shows the SED plot of the object composed from a best fit to various observed fluxes.}
\end{figure*}
\end{center}

\newpage

\begin{center}
\begin{figure*}
\begin{center}
\includegraphics*[height=6cm]{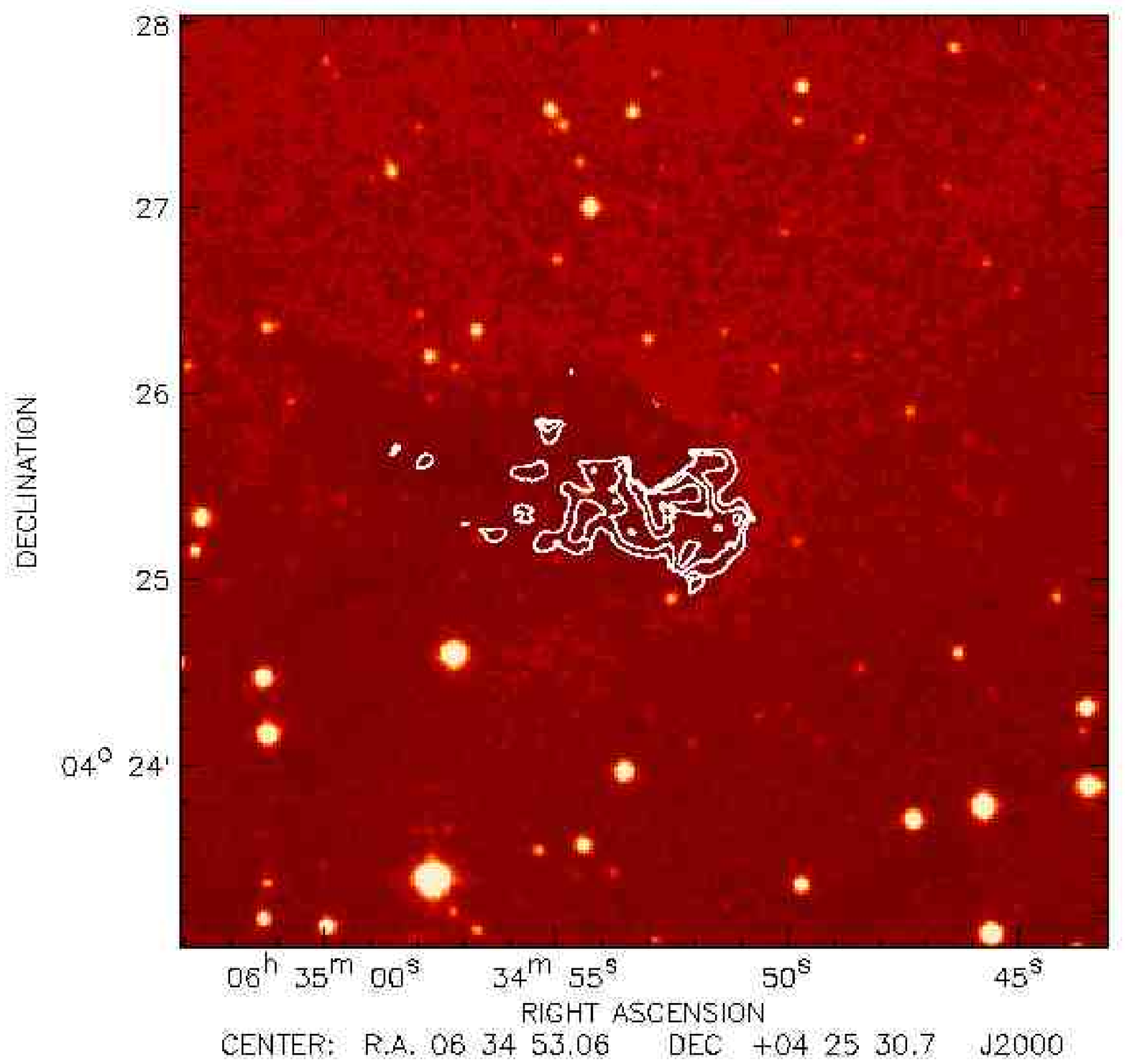}
\includegraphics*[height=6cm]{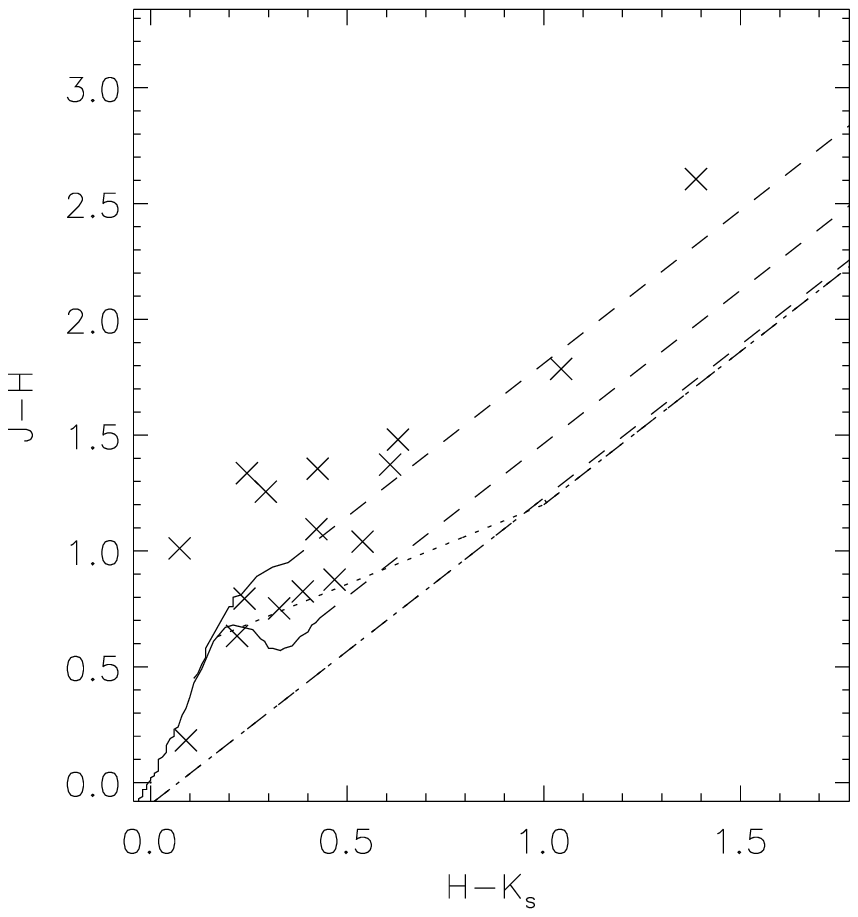}
\end{center}
\caption{Plots and images associated with the object SFO 24. The left image shows SCUBA 850 \micron ~contours overlaid on a DSS image, infrared sources from the 2MASS Point Source Catalogue \citep{Cutri2003} are shown as triangles.  850 \micron ~contours start at 3$\sigma$ and increase in increments of 33\ of the peak flux value
\indent The right plot shows the J-H versus H-K$_{\rm{s}}$ colours of the 2MASS sources associated with the cloud.}
\end{figure*}
\end{center}

\newpage

\begin{center}
\begin{figure*}
\begin{center}
\includegraphics*[height=6cm]{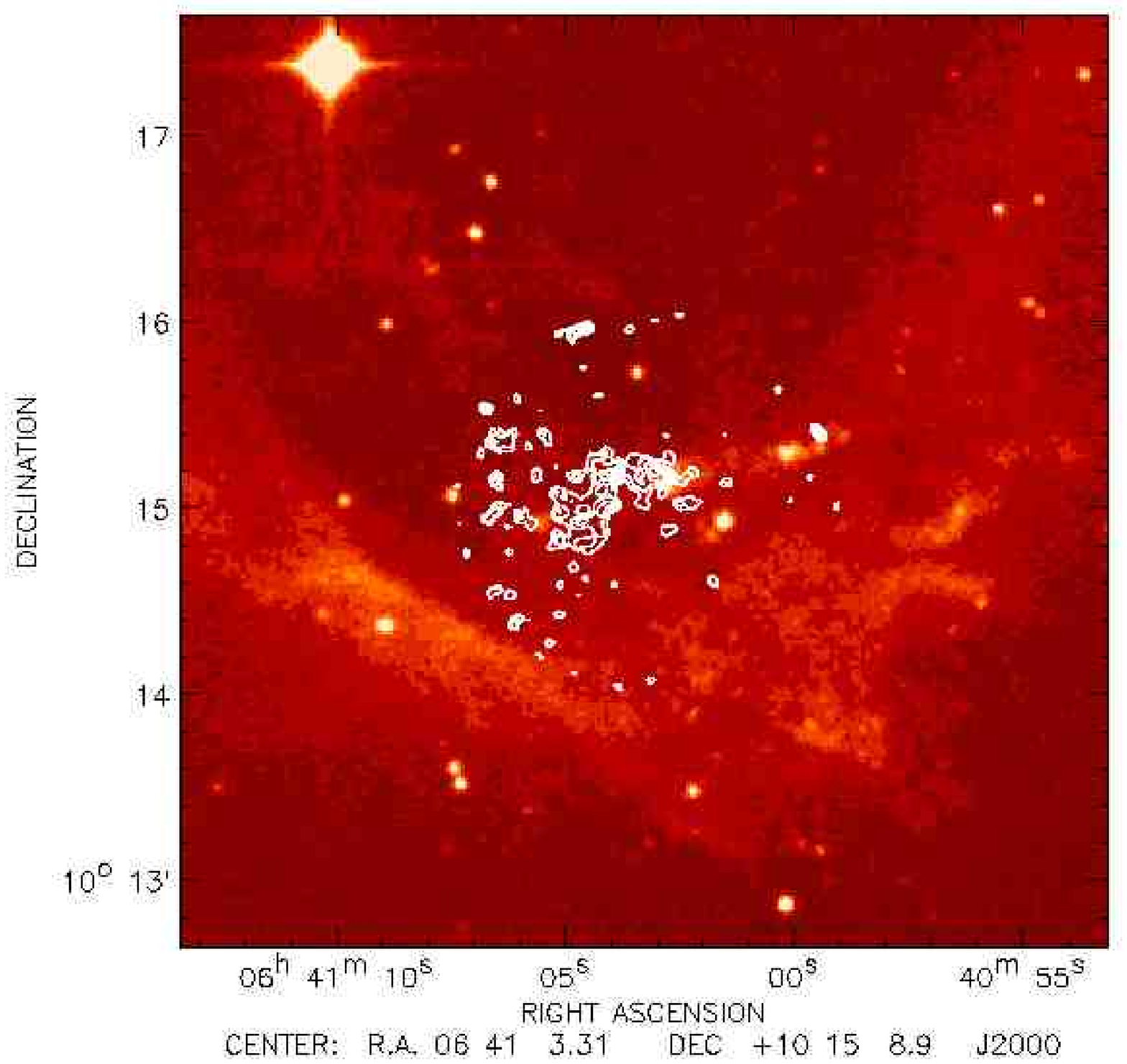}
\includegraphics*[height=6cm]{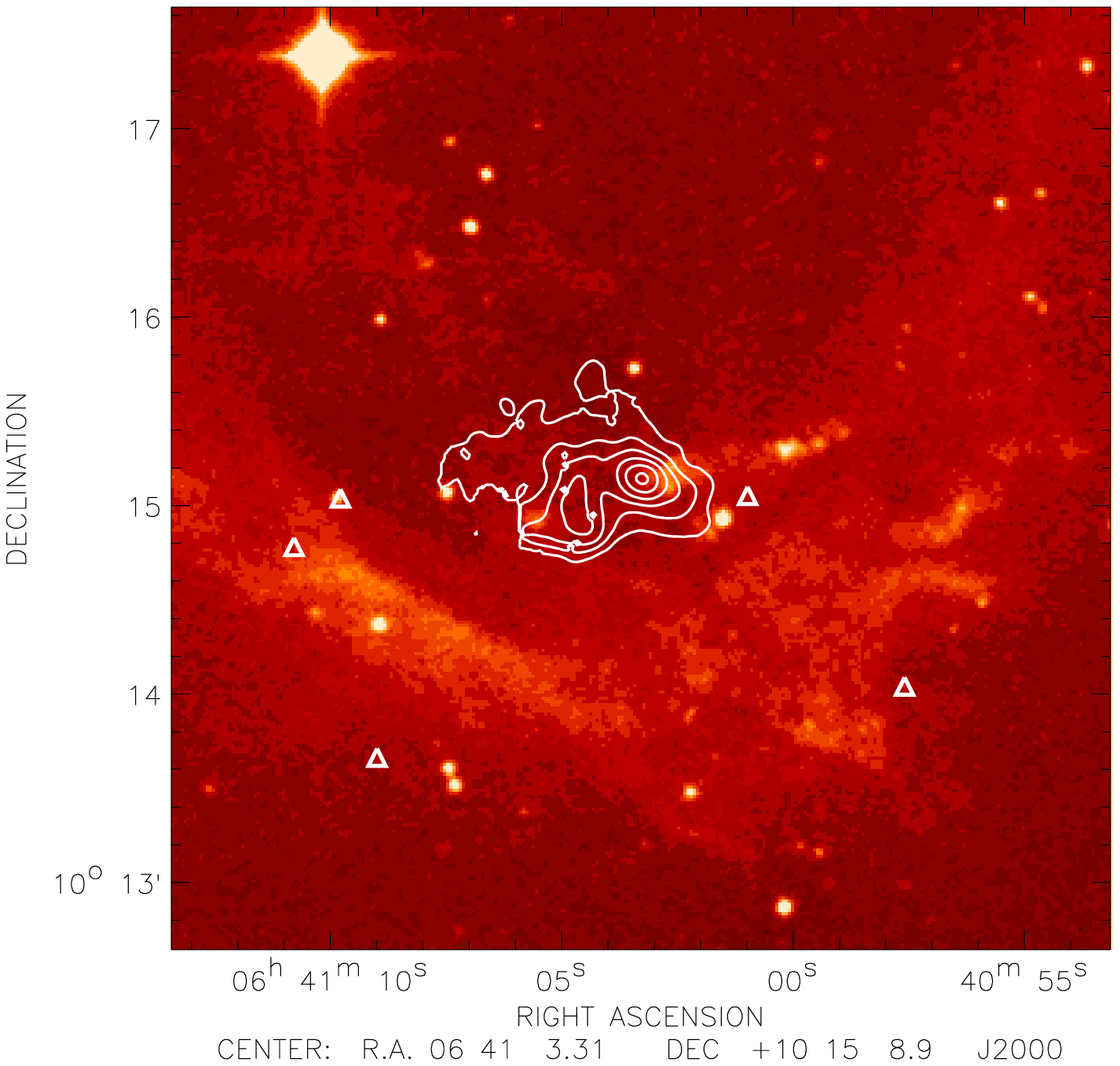}\\
\includegraphics*[height=6cm]{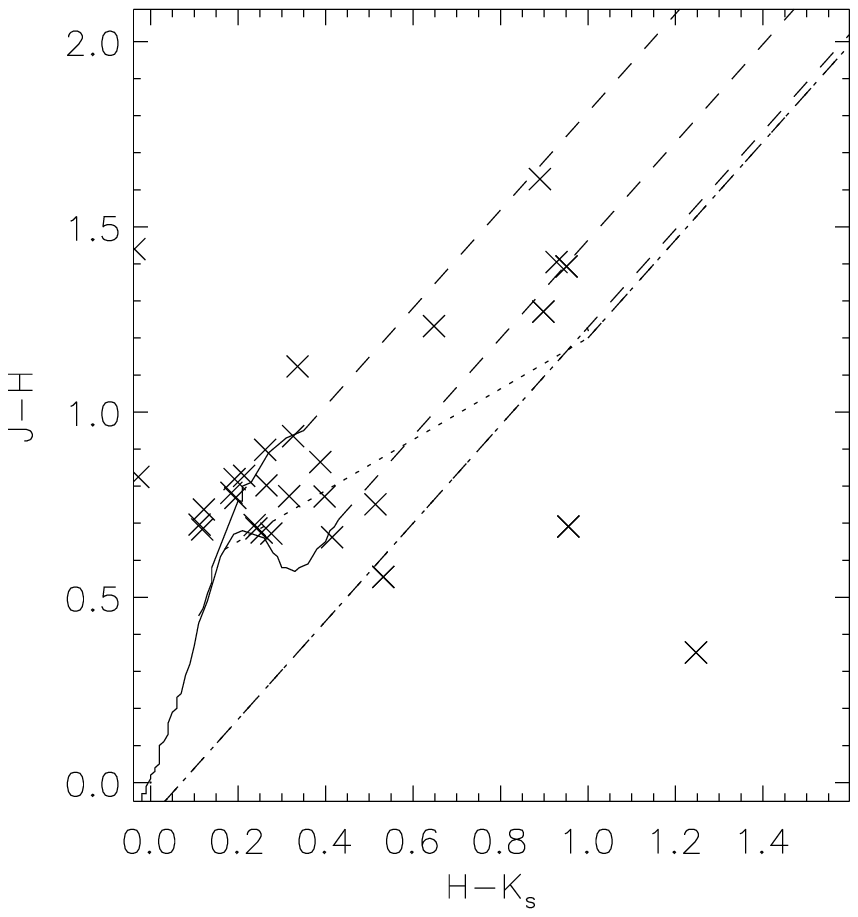}\\
\includegraphics*[height=6cm]{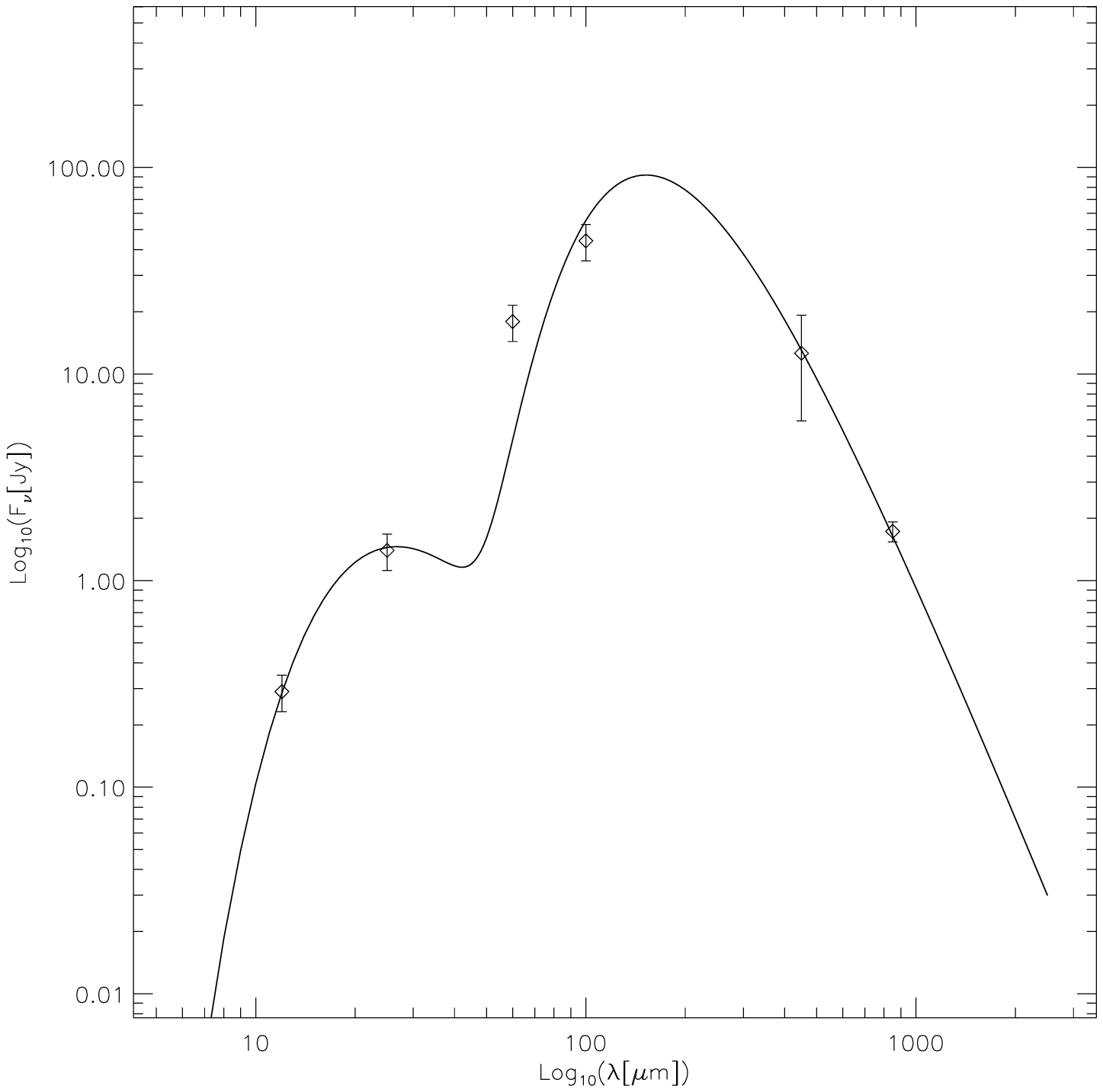}
\includegraphics*[height=6cm]{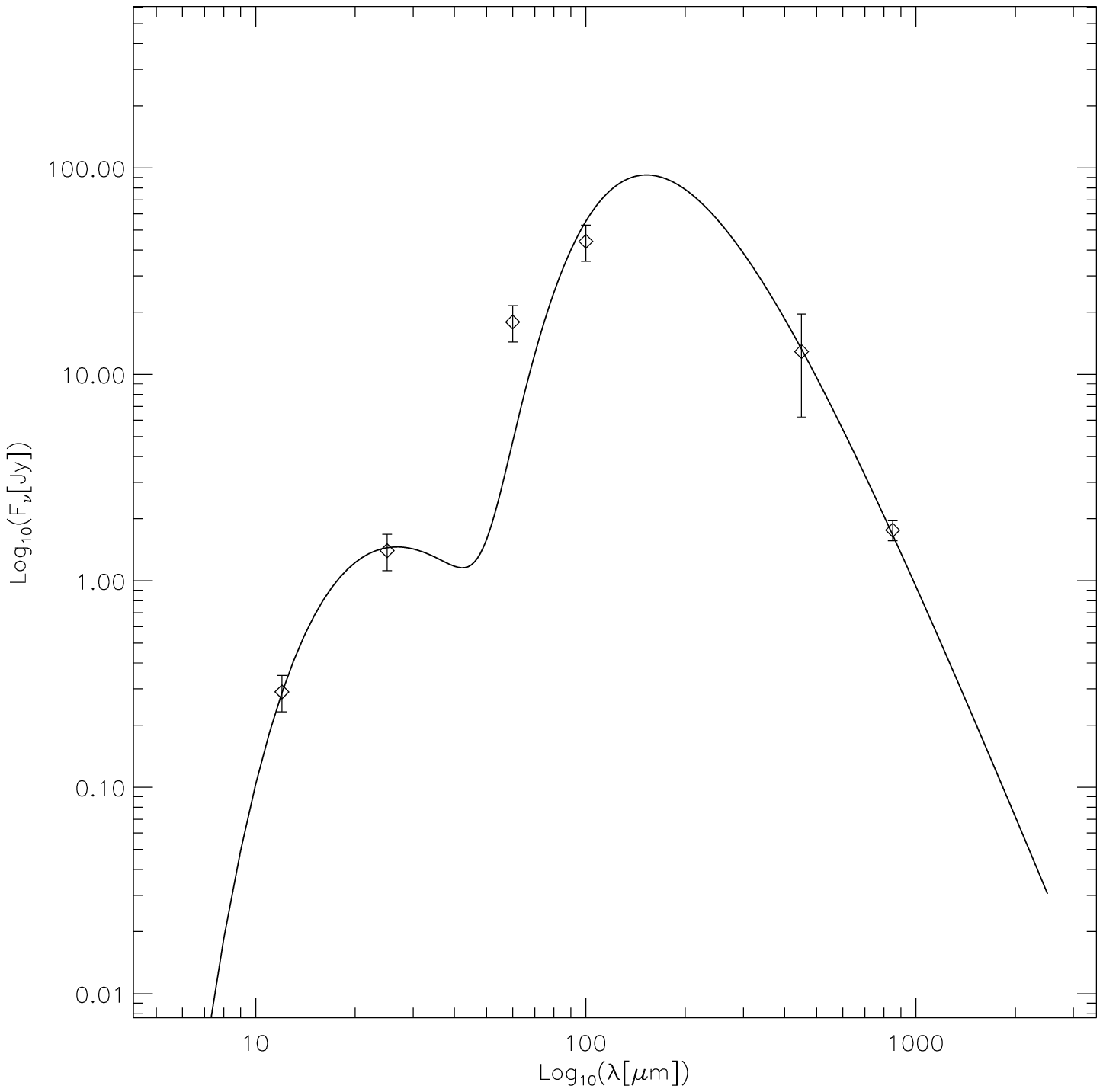}\\
\end{center}
\caption{Plots and images associated with the object SFO 25. The top images show SCUBA 450 \micron ~(left) and 850 \micron ~(right) contours overlaid on a DSS image, infrared sources from the 2MASS Point Source Catalogue \citep{Cutri2003} that have been identified as YSOs are shown as triangles.  850 \micron ~contours start at 9$\sigma$ and increase in increments of 20\% of the peak flux value, 450 \micron ~contours start at 2$\sigma$ and increase in increments of 20\% of the peak flux value.
\indent The central plot shows the J-H versus H-K$_{\rm{s}}$ colours of the 2MASS sources associated with the cloud while the bottom images show the SED plots of the separate core objects composed from a best fit to various observed fluxes.}
\end{figure*}
\end{center}

\newpage

\begin{center}
\begin{figure*}
\begin{center}
\includegraphics*[height=6cm]{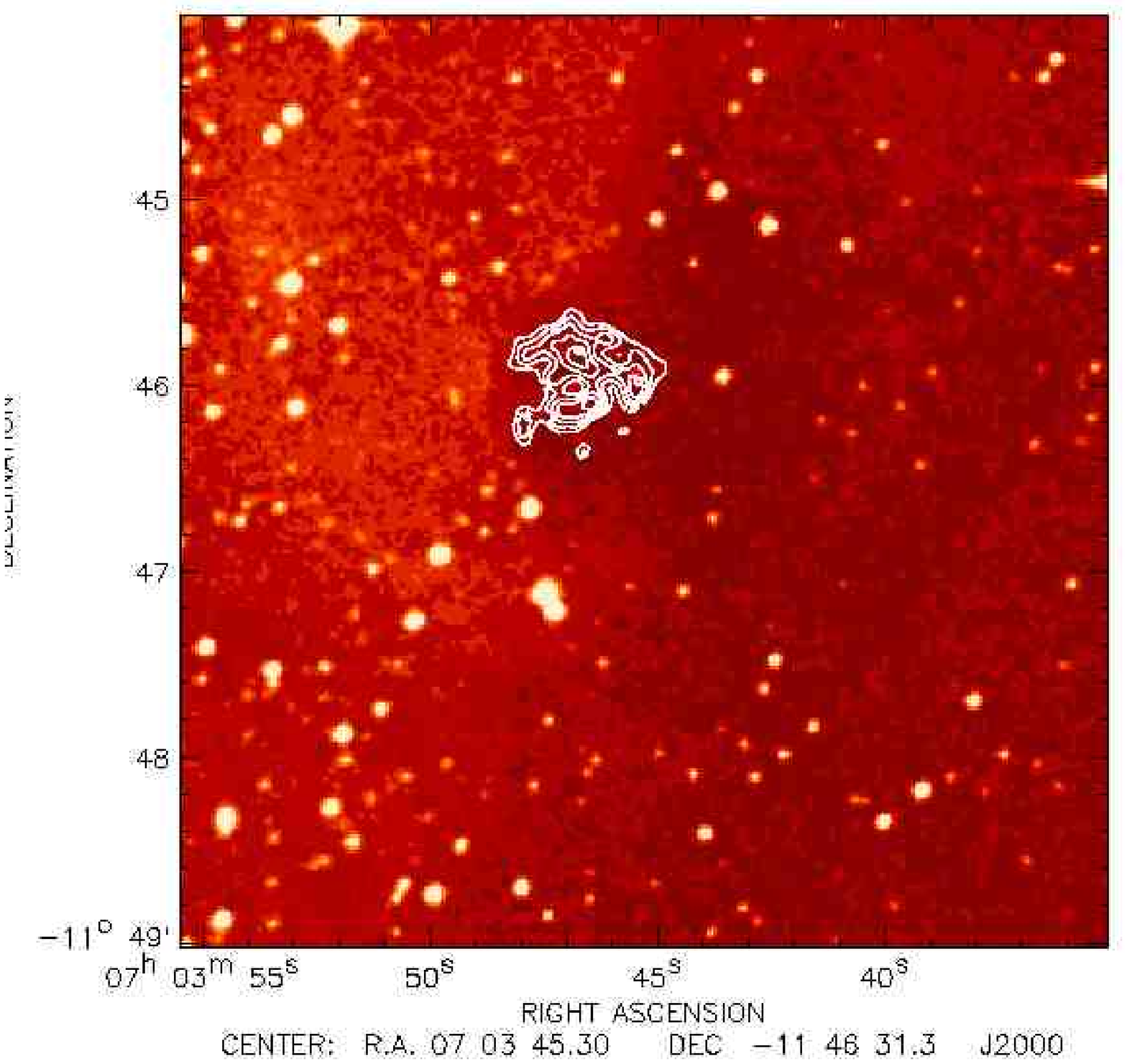}\\
\includegraphics*[height=6cm]{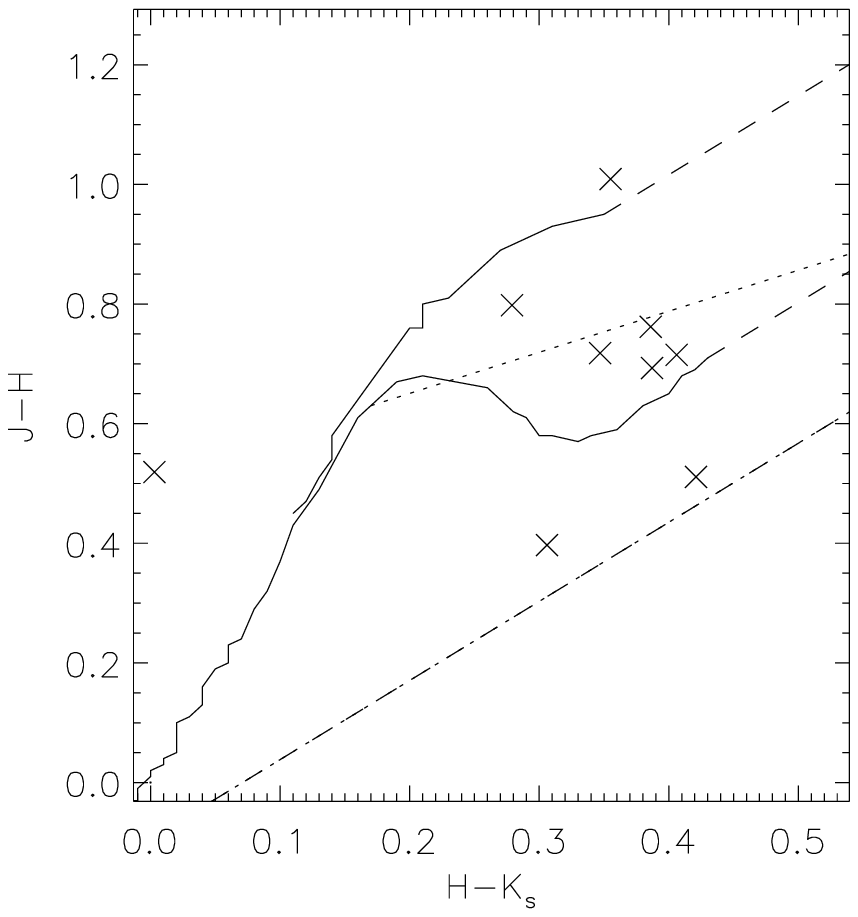}
\includegraphics*[height=6cm]{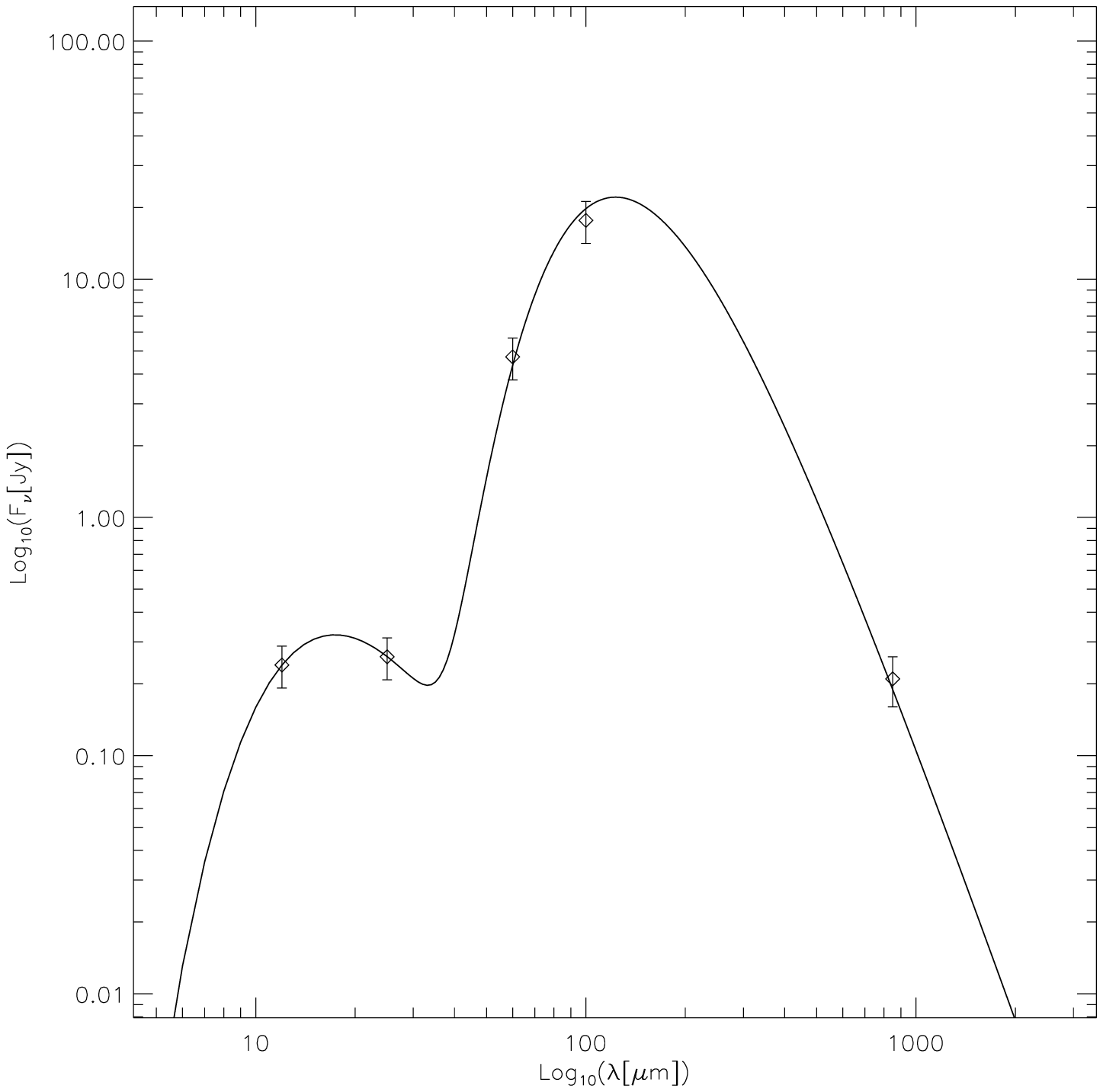}\\
\end{center}
\caption{Plots and images associated with the object SFO 26. The top image shows SCUBA 850 \micron ~contours overlaid on a DSS image, infrared sources from the 2MASS Point Source Catalogue \citep{Cutri2003} are shown as triangles.  850 \micron ~contours start at 3$\sigma$ and increase in increments of 20\% of the peak flux value.
\indent The bottom left plot shows the J-H versus H-K$_{\rm{s}}$ colours of the 2MASS sources associated with the cloud while the bottom right image shows the SED plot of the object composed from a best fit to various observed fluxes.}
\end{figure*}
\end{center}

\newpage
\clearpage

\begin{center}
\begin{figure*}
\begin{center}
\includegraphics*[height=6cm]{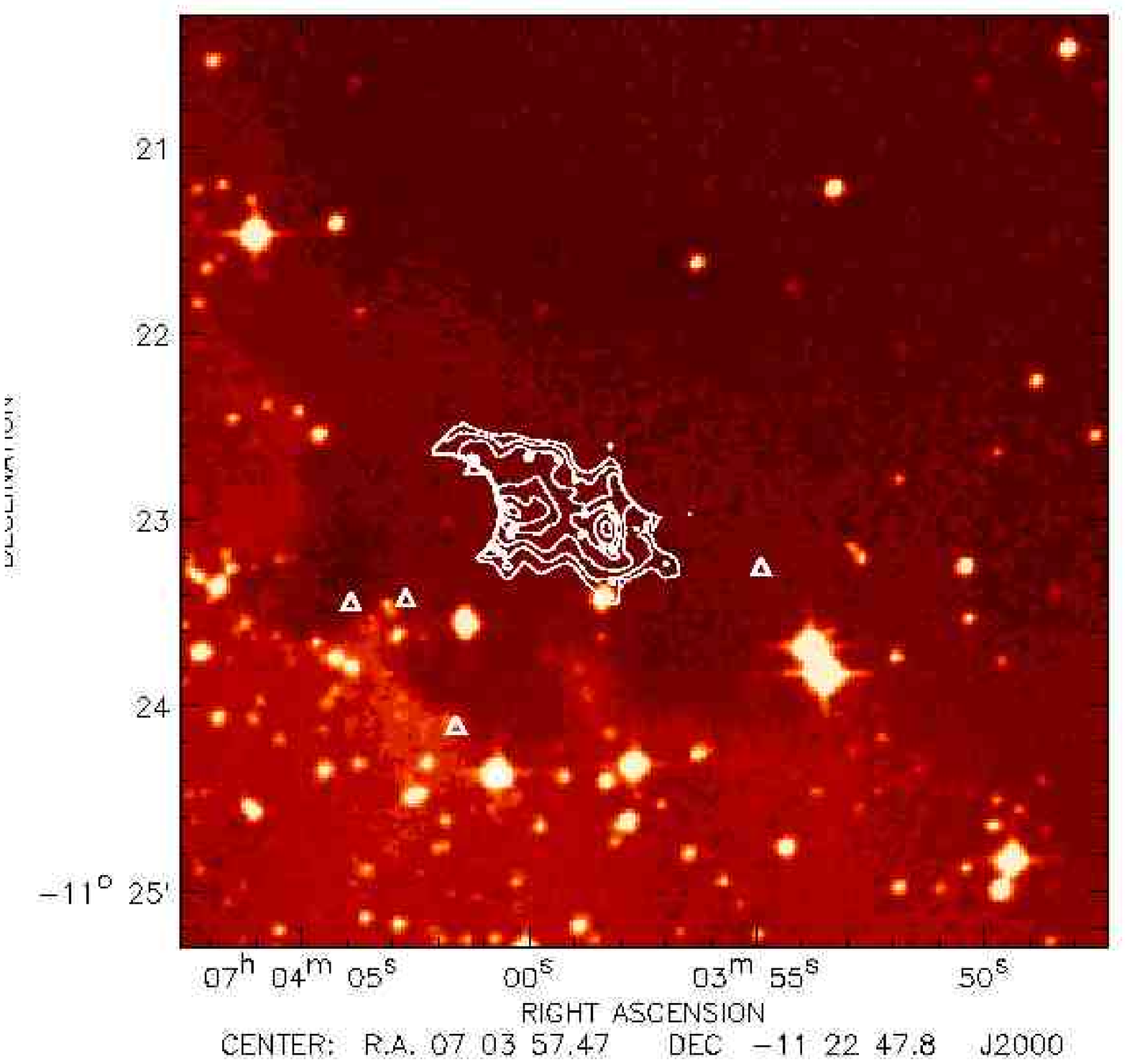}\\
\includegraphics*[height=6cm]{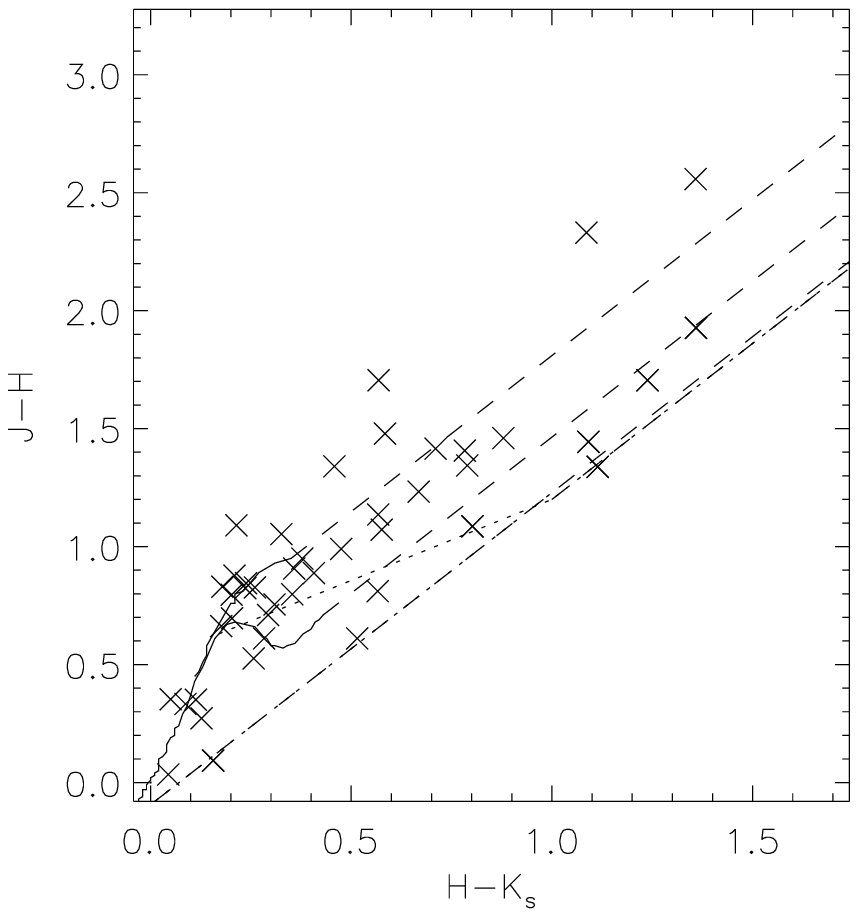}
\includegraphics*[height=6cm]{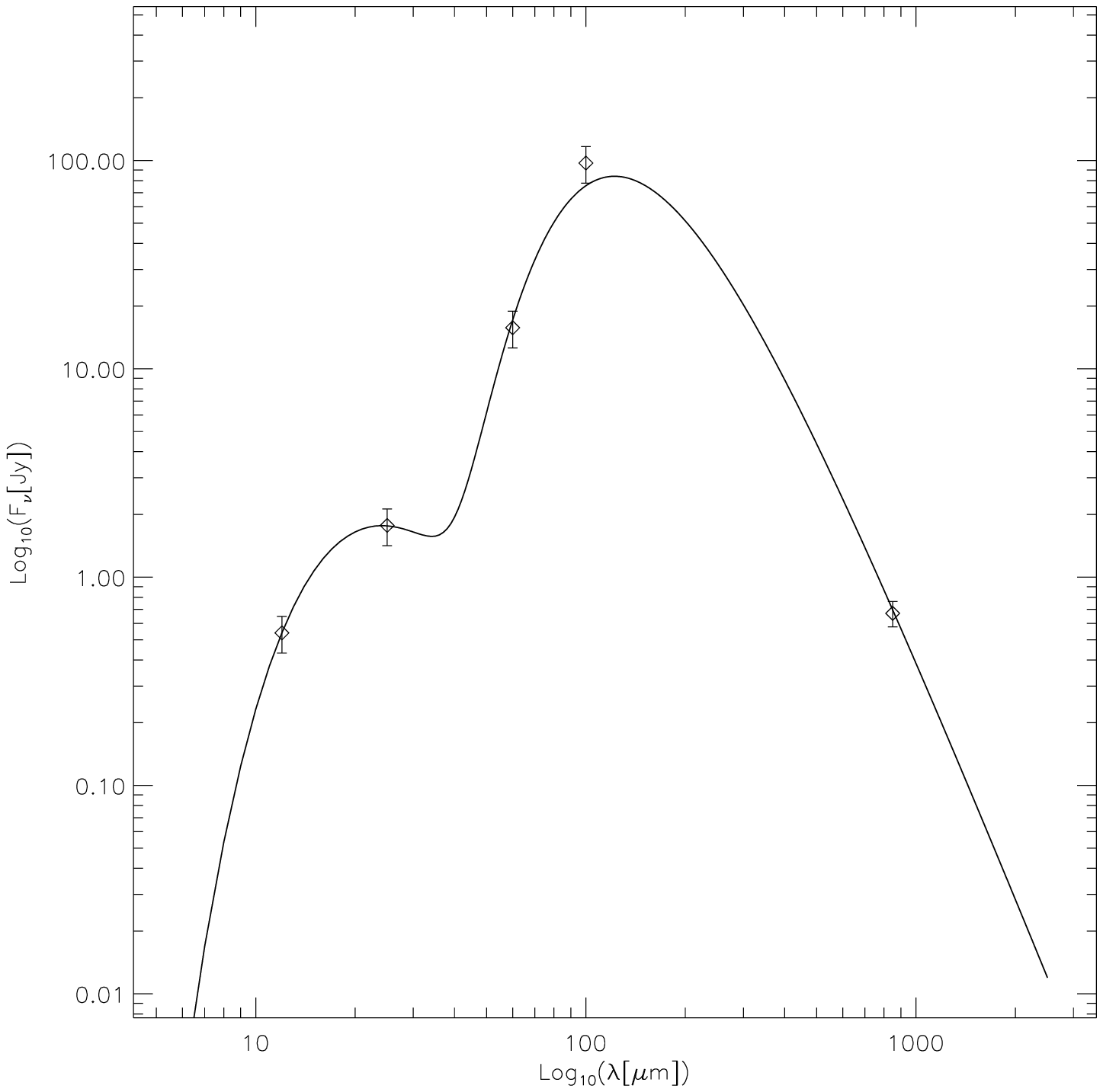}\\
\end{center}
\caption{Plots and images associated with the object SFO 27. The top image shows SCUBA 850 \micron ~contours overlaid on a DSS image, infrared sources from the 2MASS Point Source Catalogue \citep{Cutri2003} are shown as triangles.  850 \micron ~contours start at 8$\sigma$ and increase in increments of 20\% of the peak flux value.
\indent The bottom left plot shows the J-H versus H-K$_{\rm{s}}$ colours of the 2MASS sources associated with the cloud while the bottom right image shows the SED plot of the object composed from a best fit to various observed fluxes.}
\end{figure*}
\end{center}

\newpage

\begin{center}
\begin{figure*}
\begin{center}
\includegraphics*[height=6cm]{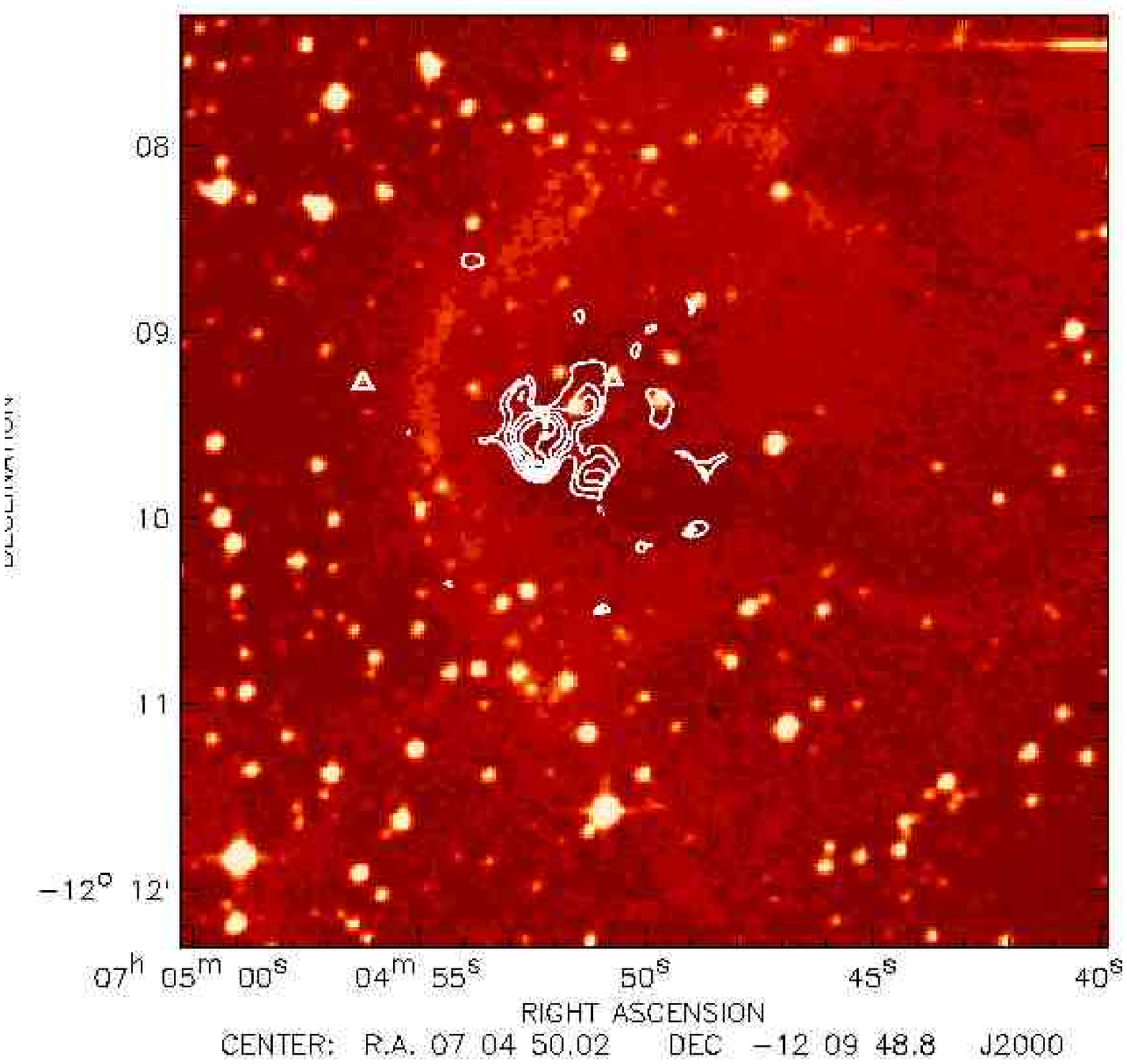}\\
\includegraphics*[height=6cm]{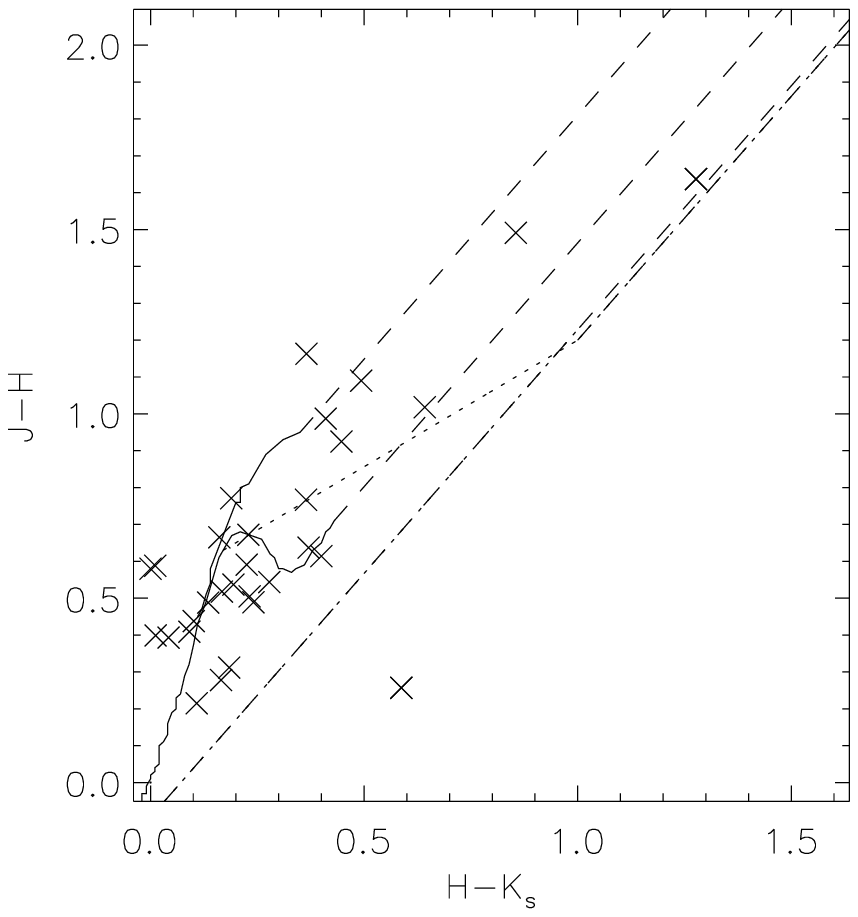}
\includegraphics*[height=6cm]{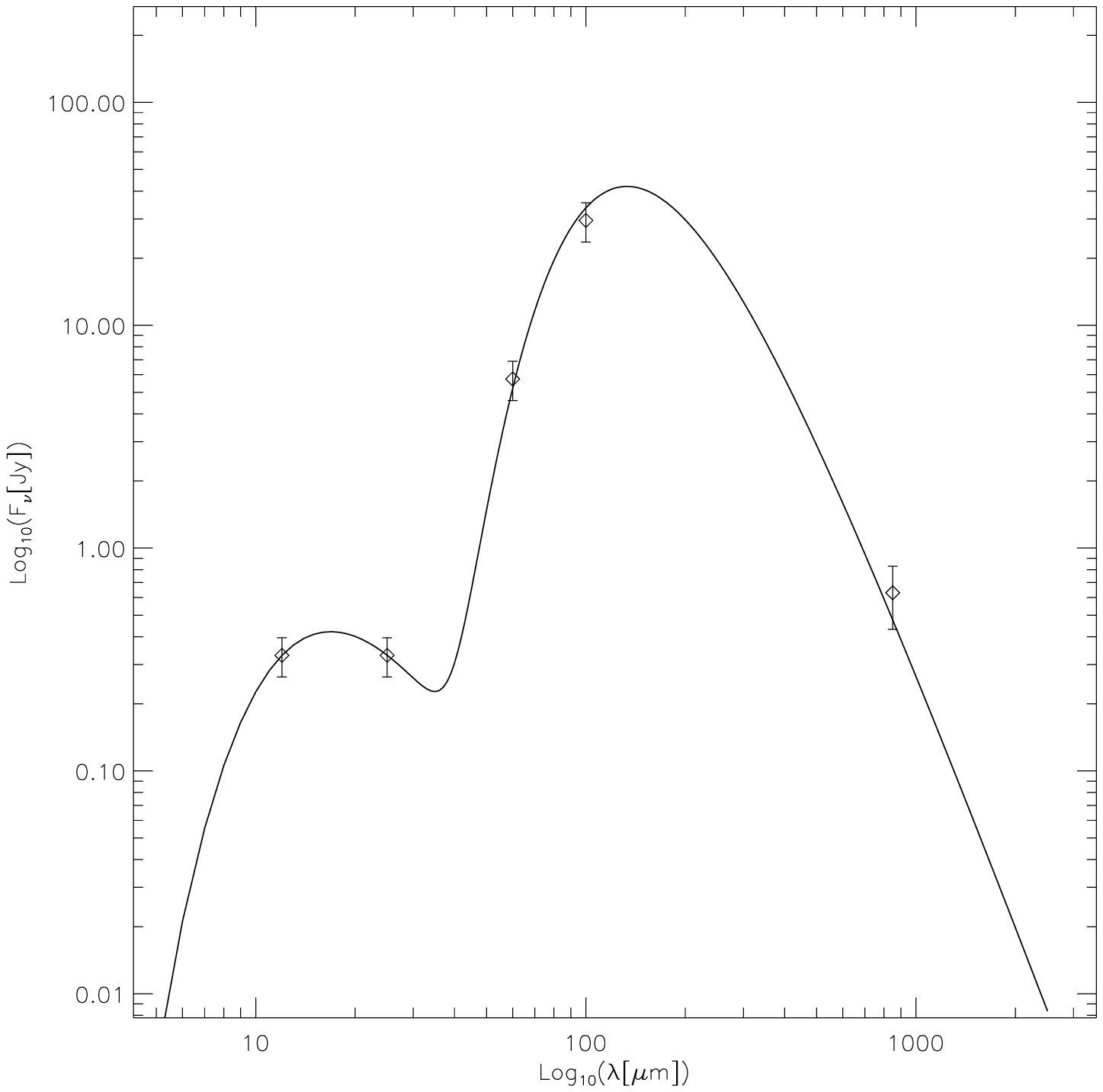}\\
\end{center}
\caption{Plots and images associated with the object SFO 29. The top image shows SCUBA 850 \micron ~contours overlaid on a DSS image, infrared sources from the 2MASS Point Source Catalogue \citep{Cutri2003} are shown as triangles.  850 \micron ~contours start at 3$\sigma$ and increase in increments of 20\% of the peak flux value.
\indent The bottom left plot shows the J-H versus H-K$_{\rm{s}}$ colours of the 2MASS sources associated with the cloud while the bottom right image shows the SED plot of the object composed from a best fit to various observed fluxes.}
\end{figure*}
\end{center}

\newpage

\begin{center}
\begin{figure*}
\begin{center}
\includegraphics*[height=6cm]{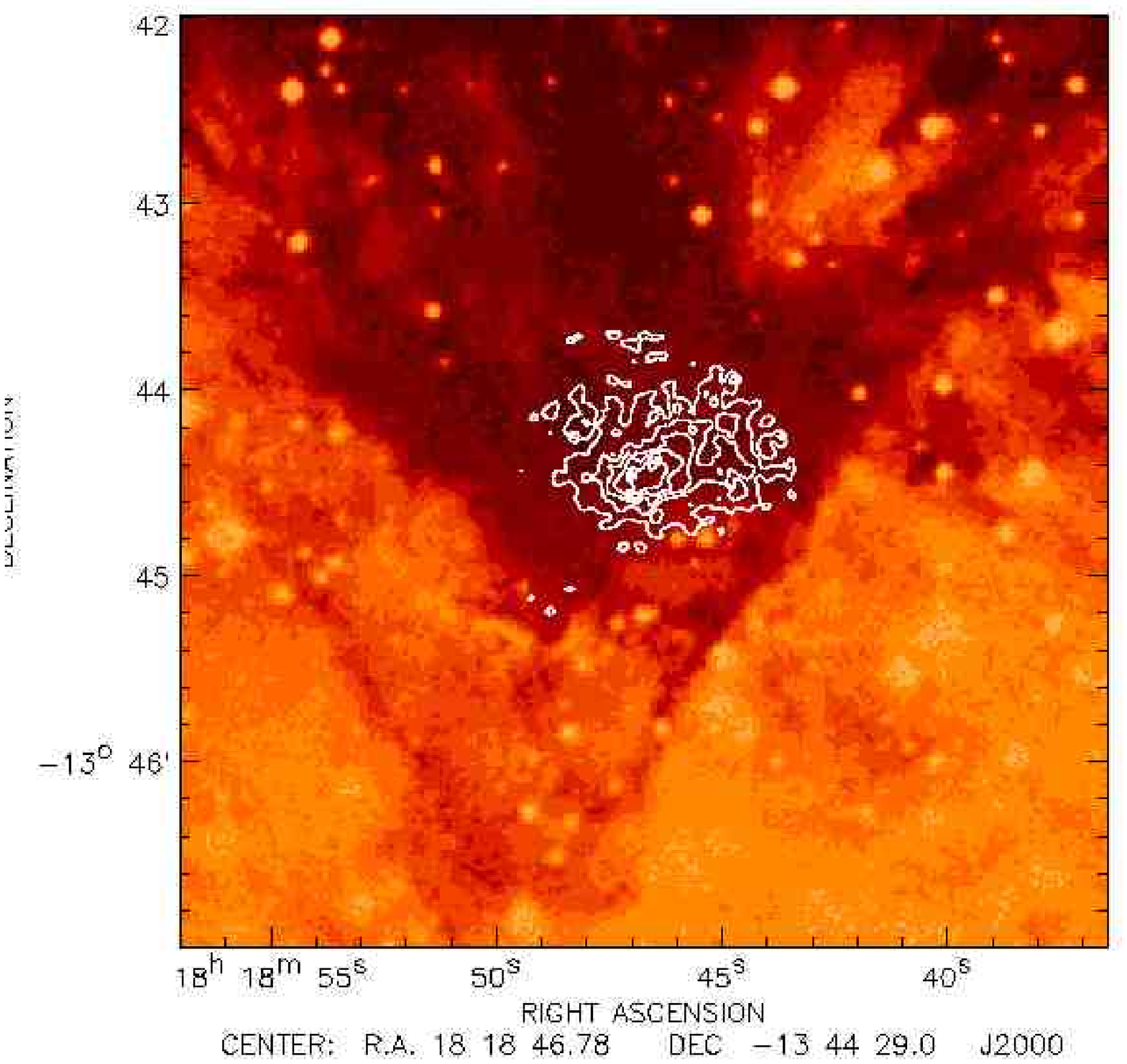}
\includegraphics*[height=6cm]{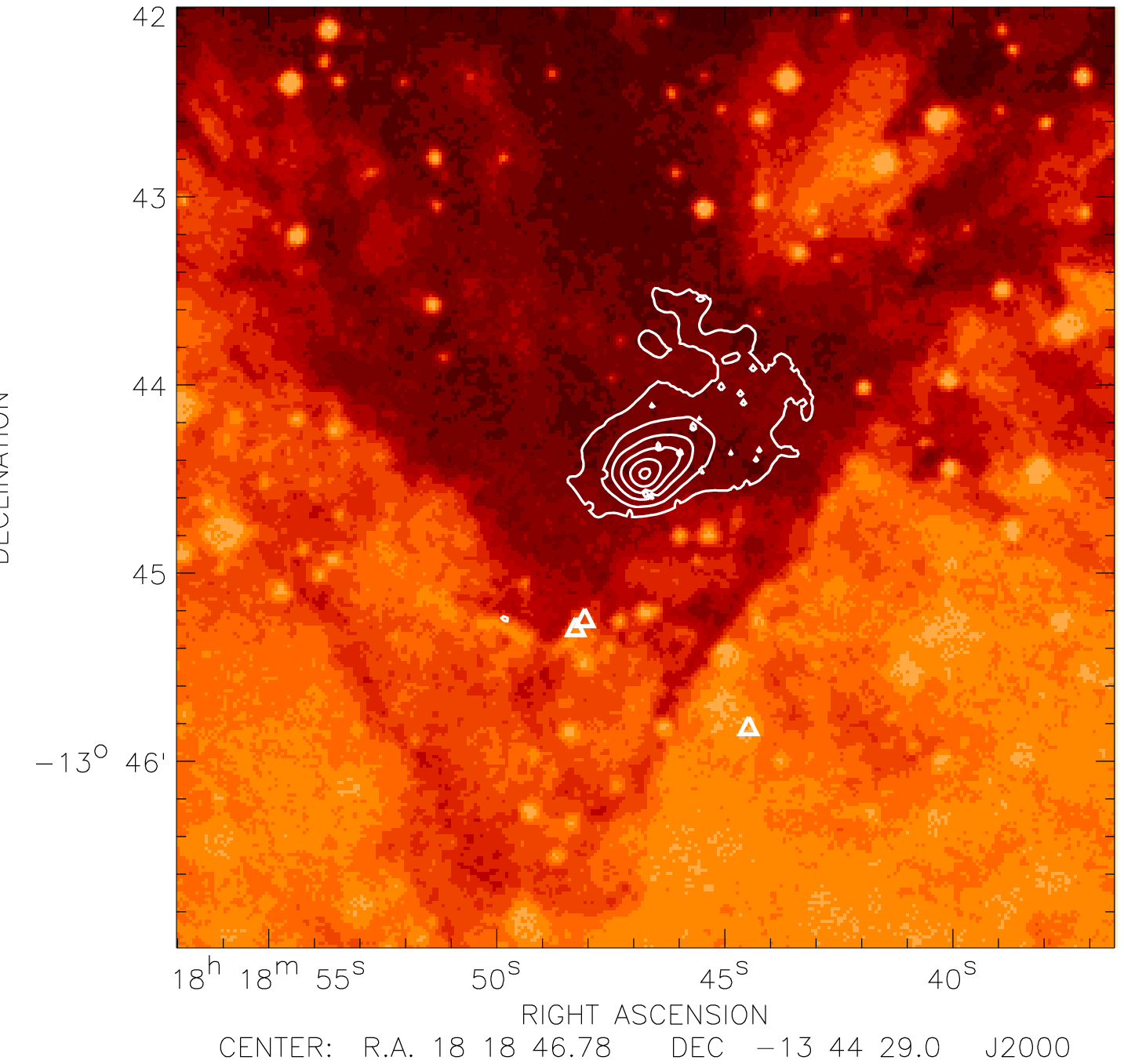}\\
\includegraphics*[height=6cm]{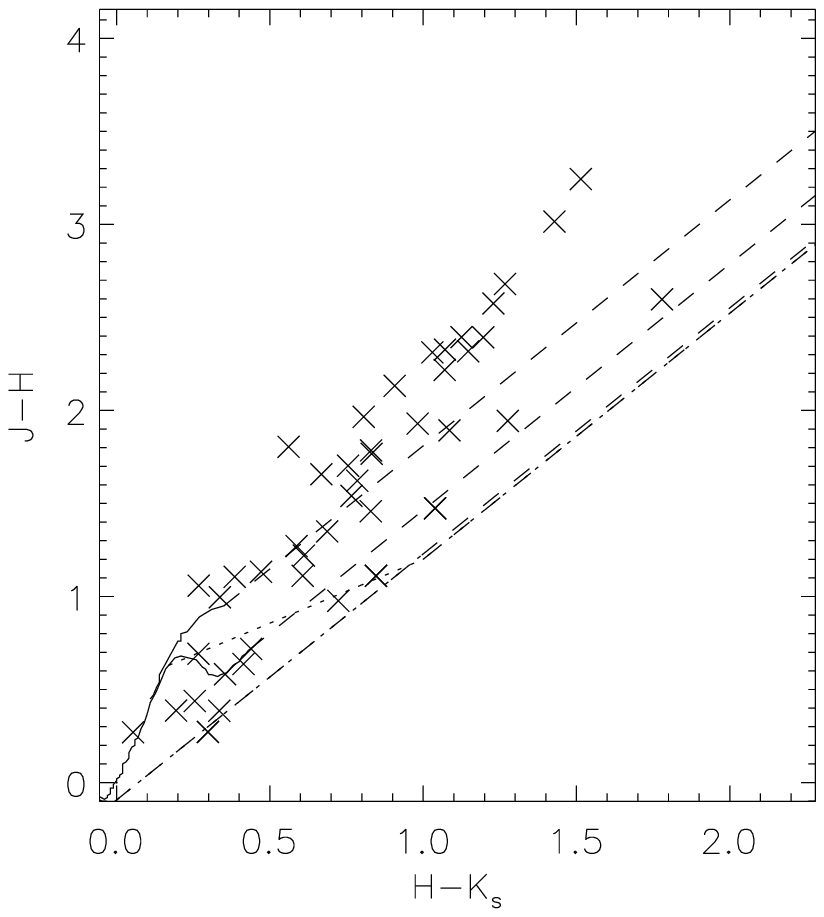}
\includegraphics*[height=6cm]{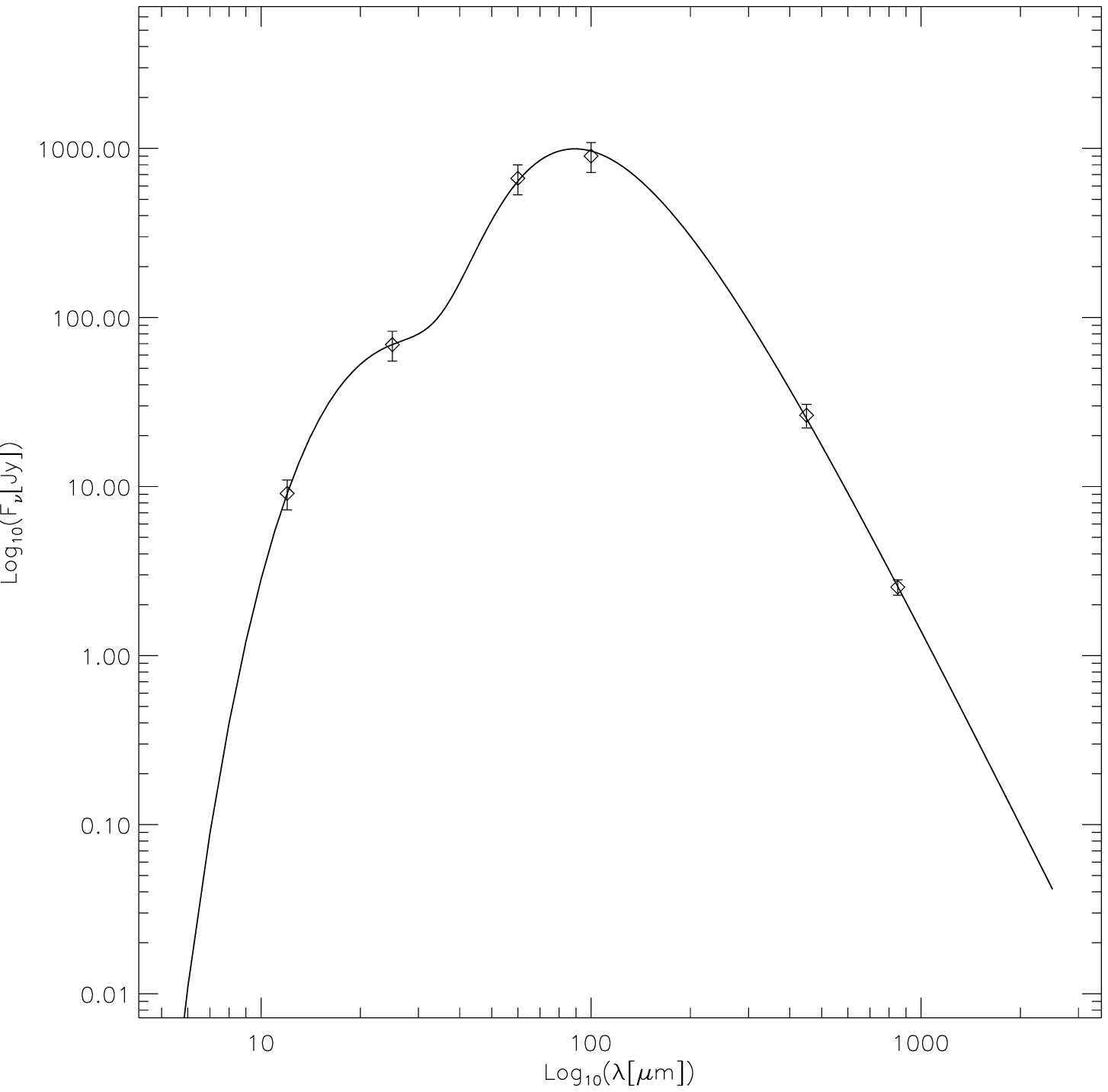}\\
\end{center}
\caption{Plots and images associated with the object SFO 30. The top images show SCUBA 450 \micron ~(left) and 850 \micron ~(right) contours overlaid on a DSS image, infrared sources from the 2MASS Point Source Catalogue \citep{Cutri2003} that have been identified as YSOs are shown as triangles.  850 \micron ~contours start at 12$\sigma$ and increase in increments of 20\% of the peak flux value, 450 \micron ~contours start at 3$\sigma$ and increase in increments of 20\% of the peak flux value.
\indent The bottom left plot shows the J-H versus H-K$_{\rm{s}}$ colours of the 2MASS sources associated with the cloud while the bottom right image shows the SED plot of the object composed from a best fit to various observed fluxes.}
\end{figure*}
\end{center}

\newpage

\begin{center}
\begin{figure*}
\begin{center}
\includegraphics*[height=6cm]{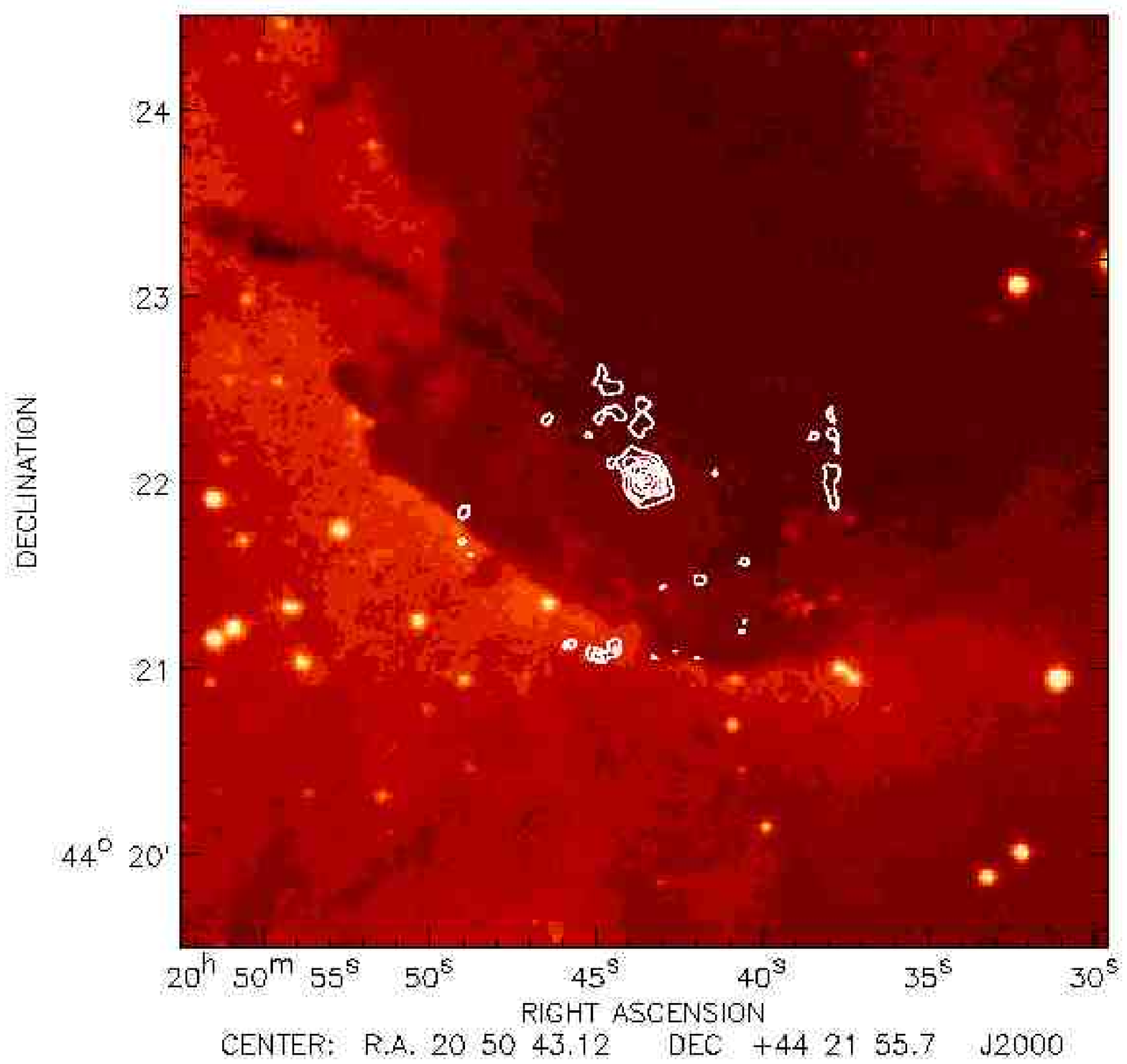}
\includegraphics*[height=6cm]{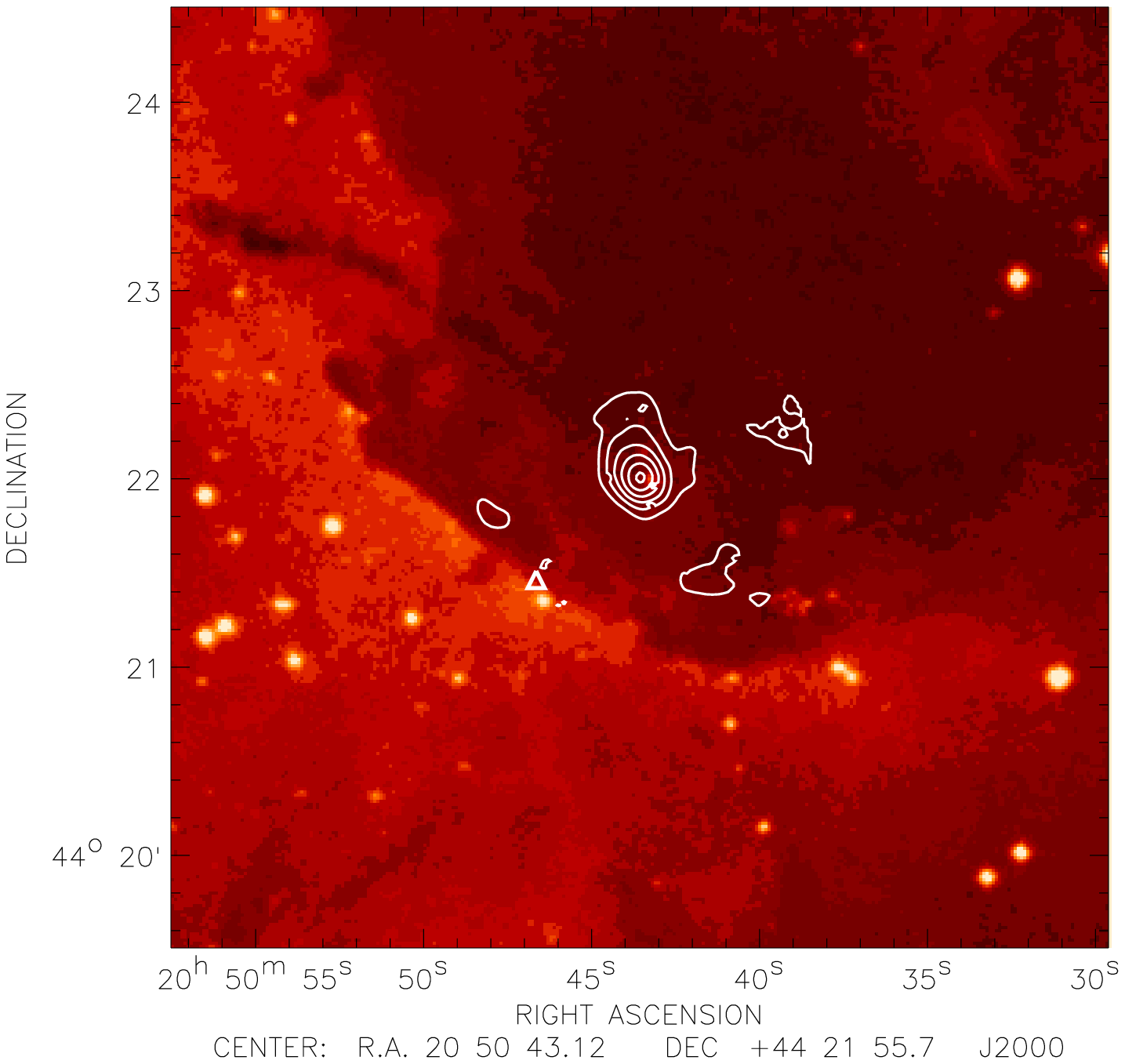}\\
\includegraphics*[height=6cm]{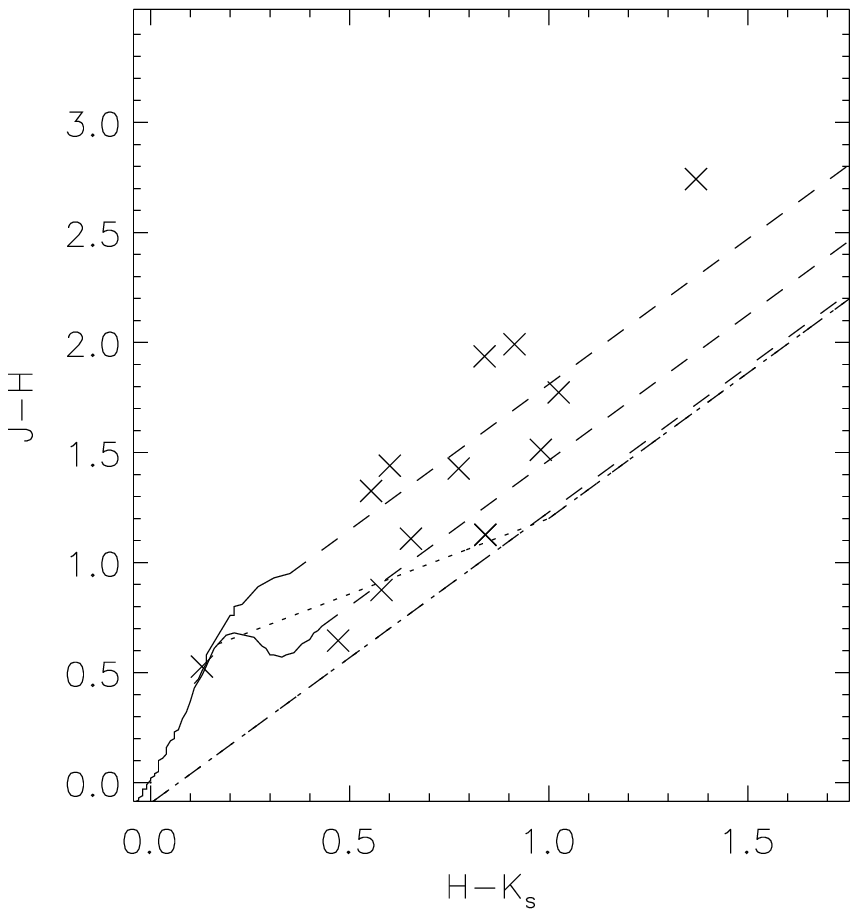}
\includegraphics*[height=6cm]{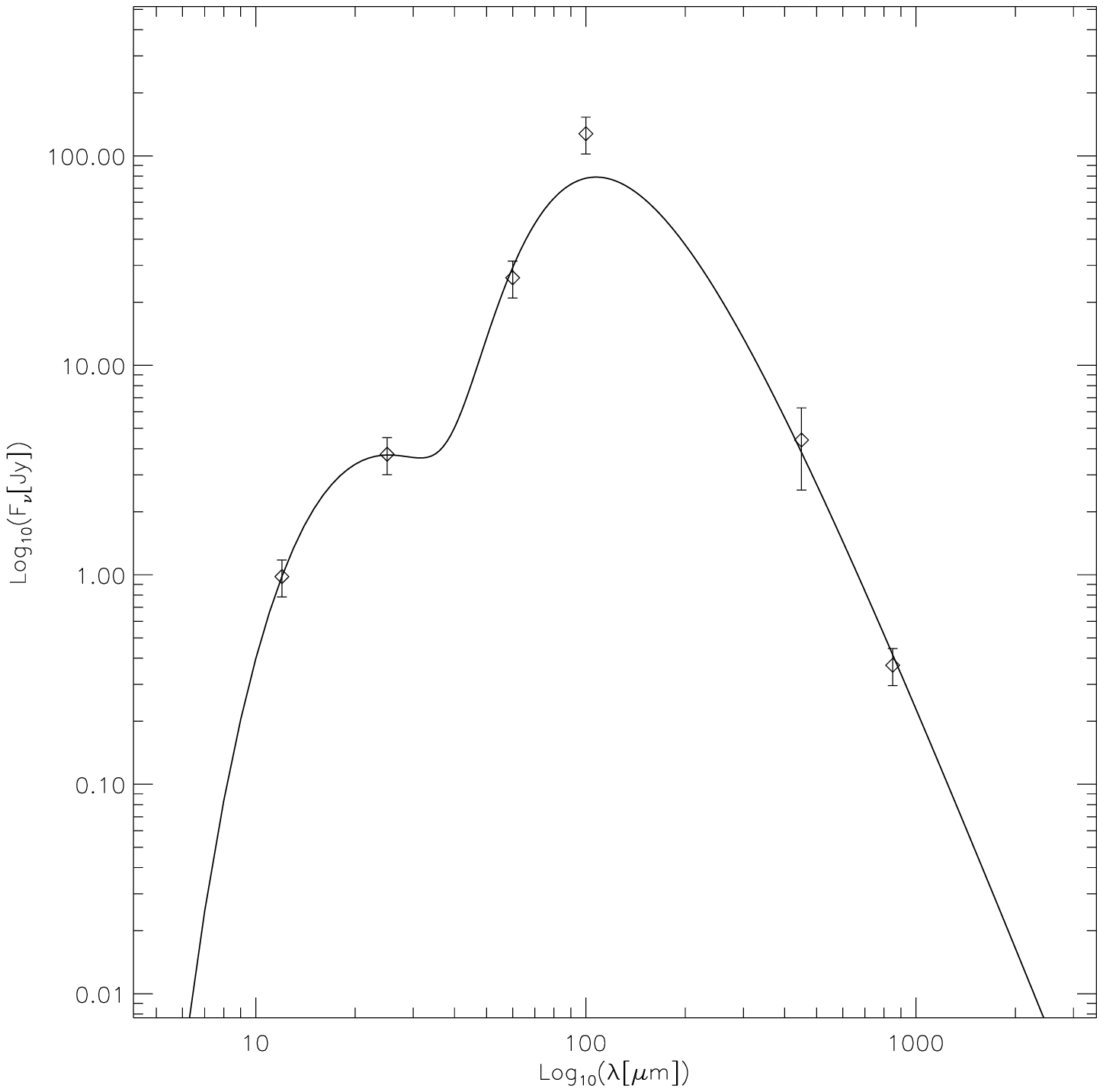}\\
\end{center}
\caption{Plots and images associated with the object SFO 31. The top images show SCUBA 450 \micron ~(left) and 850 \micron ~(right) contours overlaid on a DSS image, infrared sources from the 2MASS Point Source Catalogue \citep{Cutri2003} that have been identified as YSOs are shown as triangles.  850 \micron ~contours start at 3$\sigma$ and increase in increments of 20\% of the peak flux value, 450 \micron ~contours start at 3$\sigma$ and increase in increments of 20\% of the peak flux value.
\indent The bottom left plot shows the J-H versus H-K$_{\rm{s}}$ colours of the 2MASS sources associated with the cloud while the bottom right image shows the SED plot of the object composed from a best fit to various observed fluxes.}
\end{figure*}
\end{center}

\newpage

\begin{center}
\begin{figure*}
\begin{center}
\includegraphics*[height=6cm]{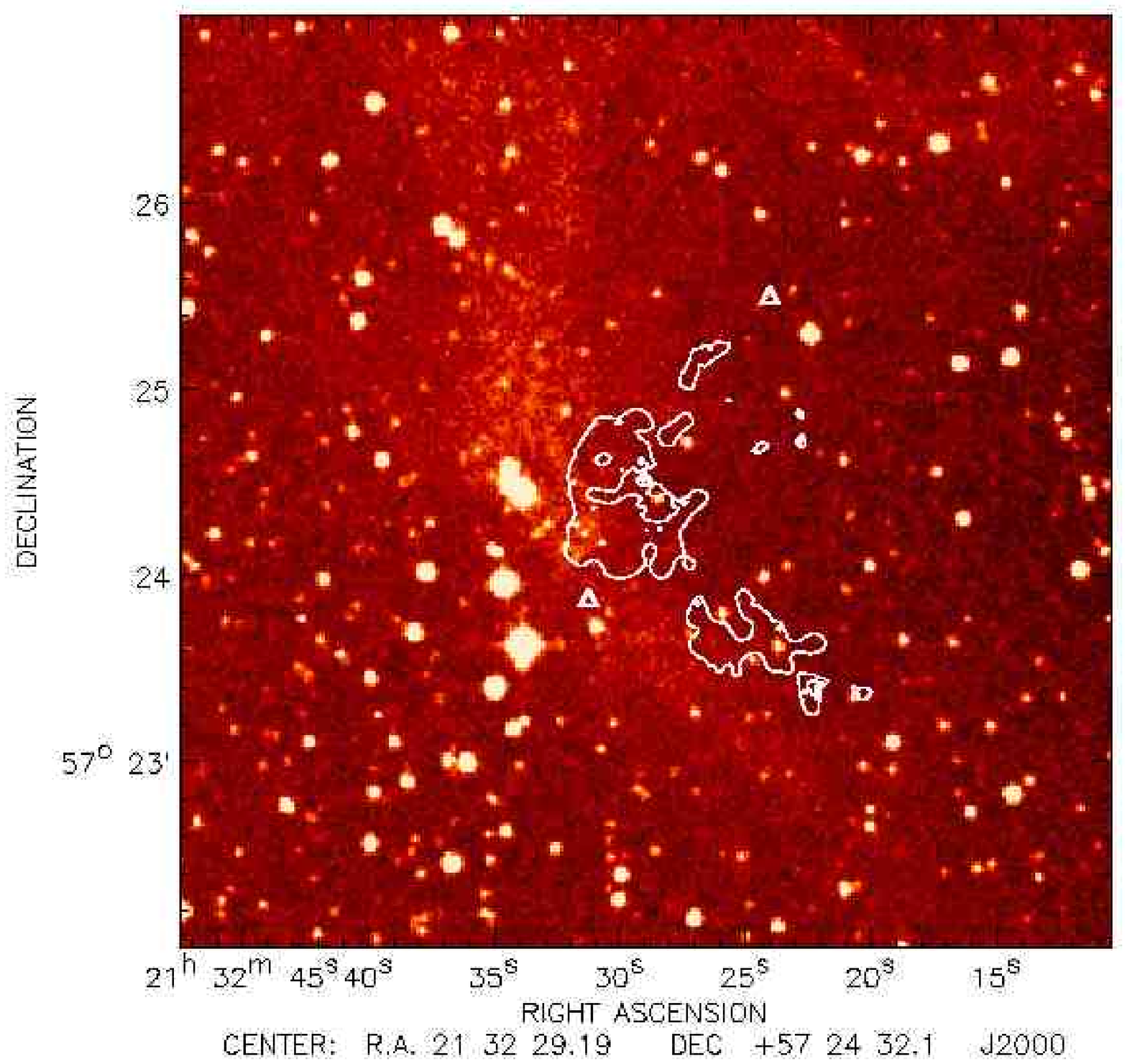}
\includegraphics*[height=6cm]{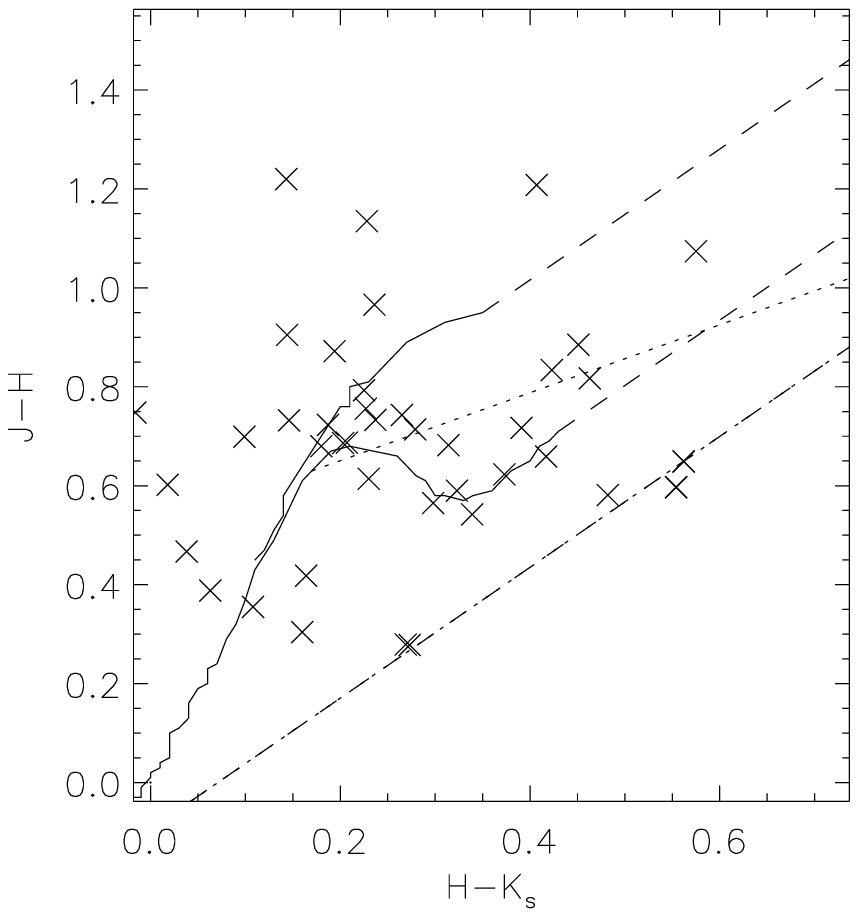}
\end{center}
\caption{Plots and images associated with the object SFO 32. The left image shows SCUBA 850 \micron ~contours overlaid on a DSS image, infrared sources from the 2MASS Point Source Catalogue \citep{Cutri2003} are shown as triangles.  850 \micron ~contours start at 3$\sigma$ and increase in increments of 20\% of the peak flux value.
\indent The right plot shows the J-H versus H-K$_{\rm{s}}$ colours of the 2MASS sources associated with the cloud.}
\end{figure*}
\end{center}

\newpage
\clearpage

\begin{center}
\begin{figure*}
\begin{center}
\includegraphics*[height=6cm]{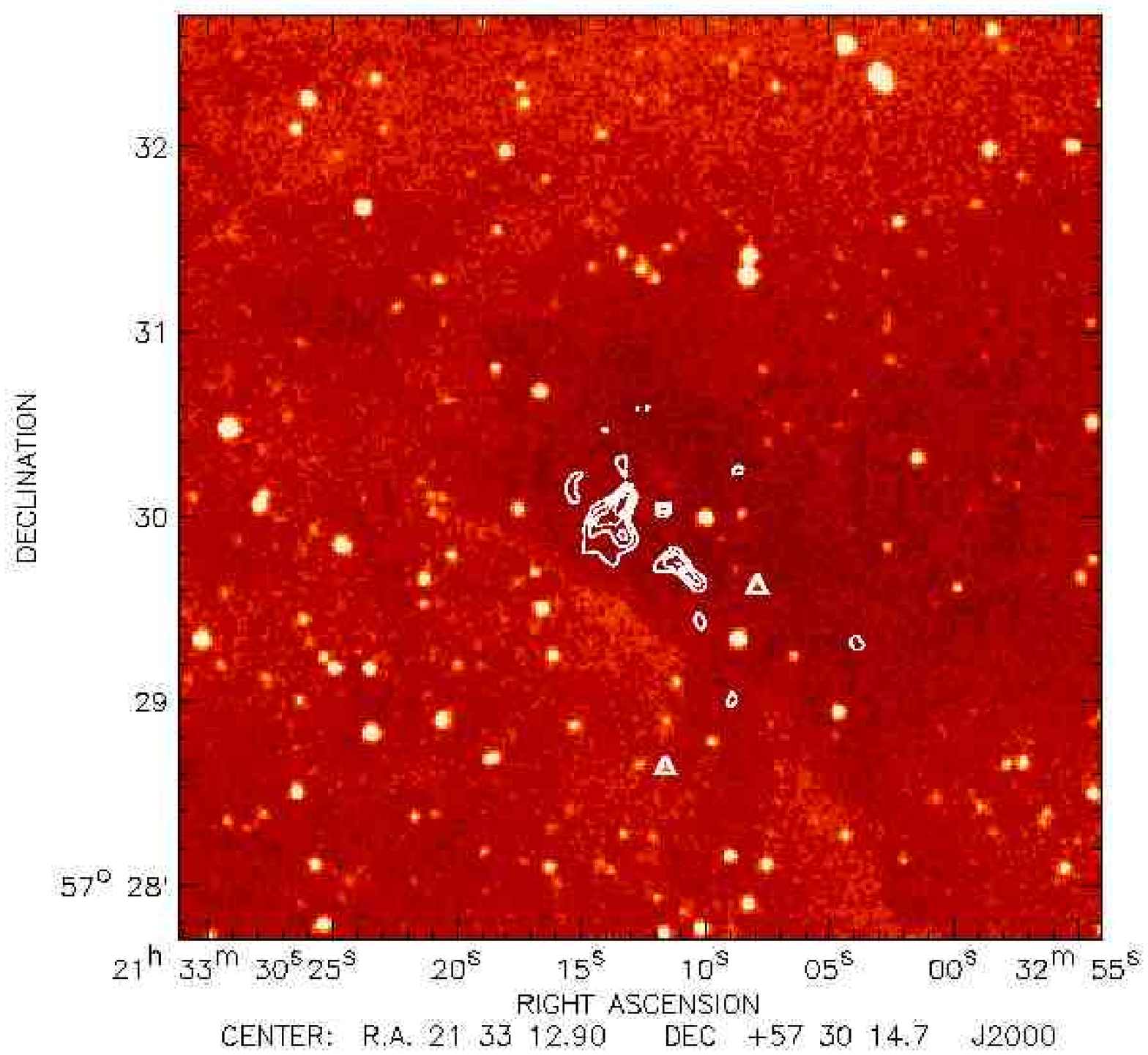}\\
\includegraphics*[height=6cm]{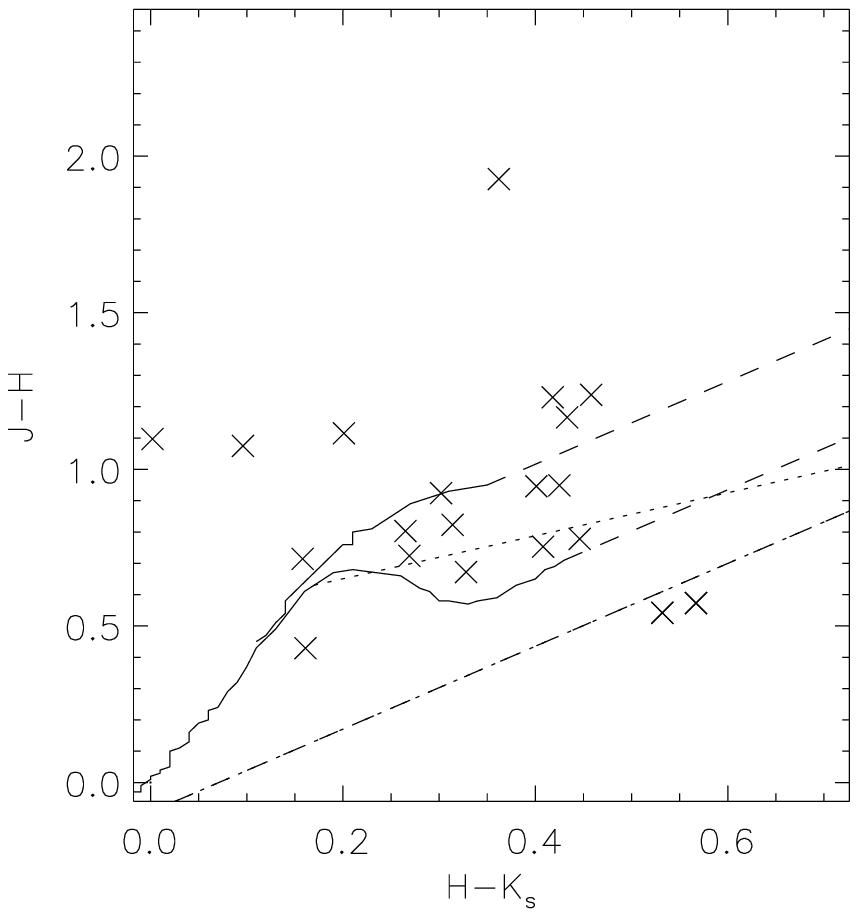}
\includegraphics*[height=6cm]{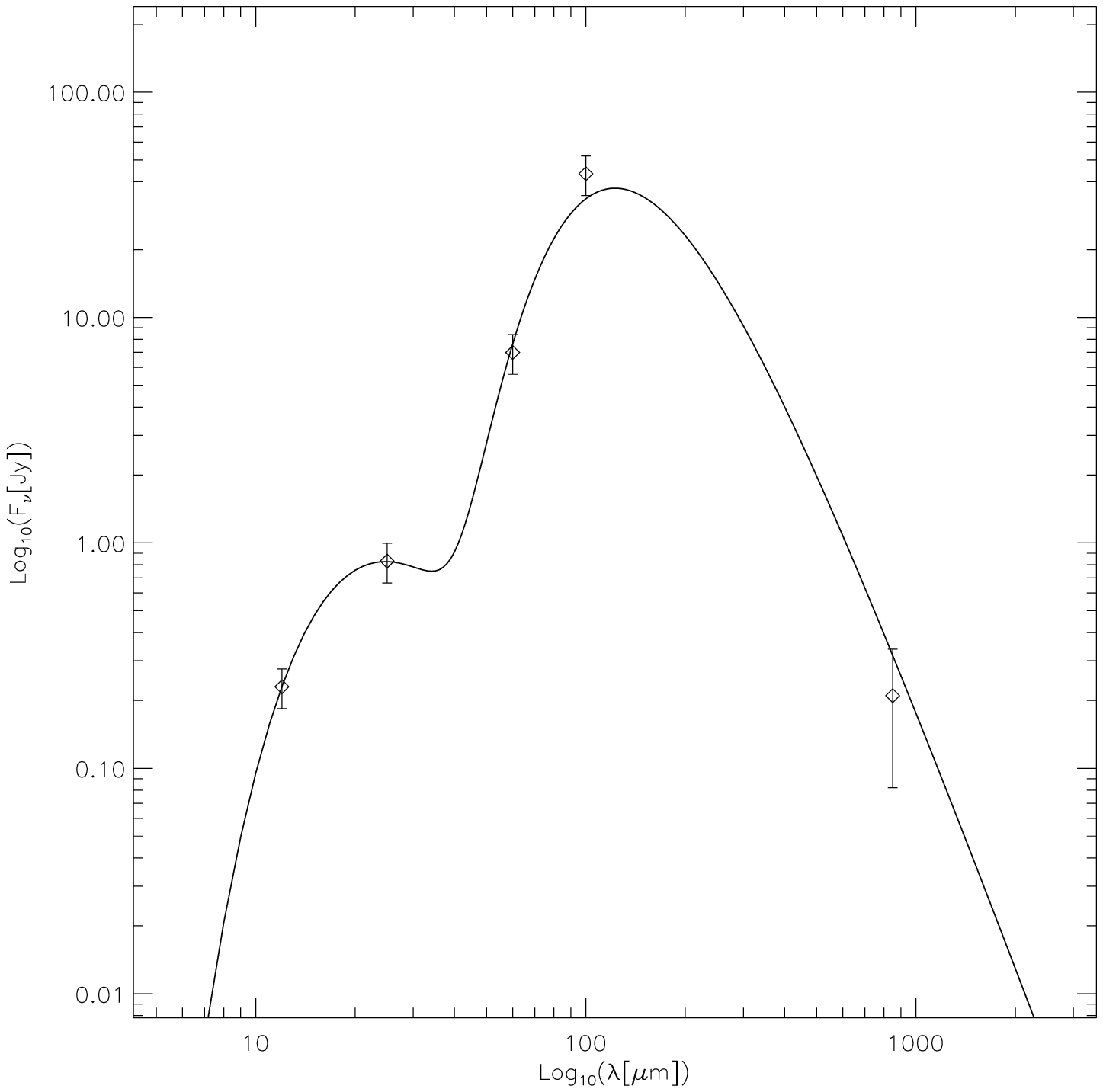}\\
\end{center}
\caption{Plots and images associated with the object SFO 33. The top image shows SCUBA 850 \micron ~contours overlaid on a DSS image, infrared sources from the 2MASS Point Source Catalogue \citep{Cutri2003} are shown as triangles.  850 \micron ~contours start at 4$\sigma$ and increase in increments of 33\ of the peak flux value.
\indent The bottom left plot shows the J-H versus H-K$_{\rm{s}}$ colours of the 2MASS sources associated with the cloud while the bottom right image shows the SED plot of the object composed from a best fit to various observed fluxes.}
\end{figure*}
\end{center}

\newpage

\begin{center}
\begin{figure*}
\begin{center}
\includegraphics*[height=6cm]{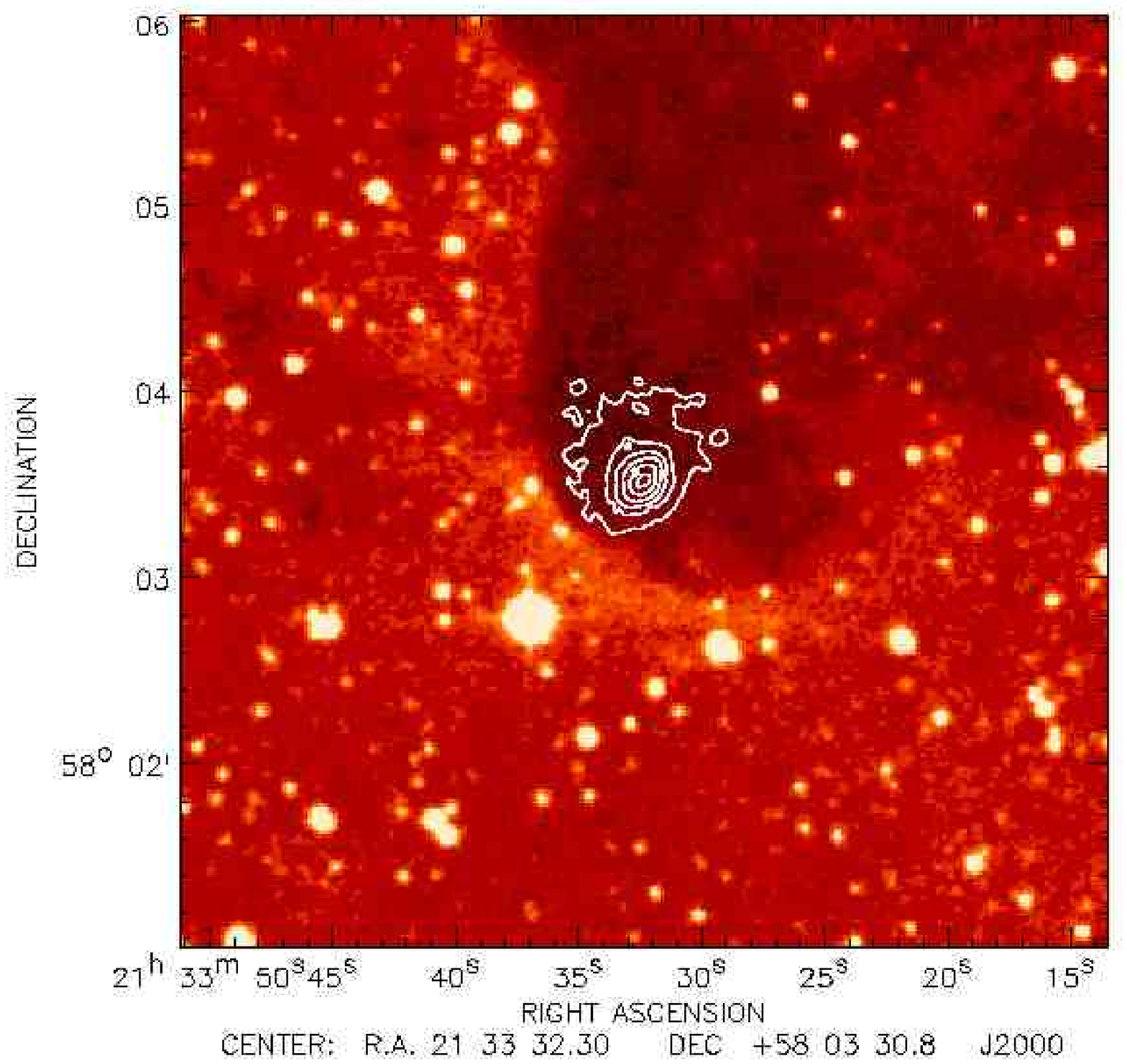}
\includegraphics*[height=6cm]{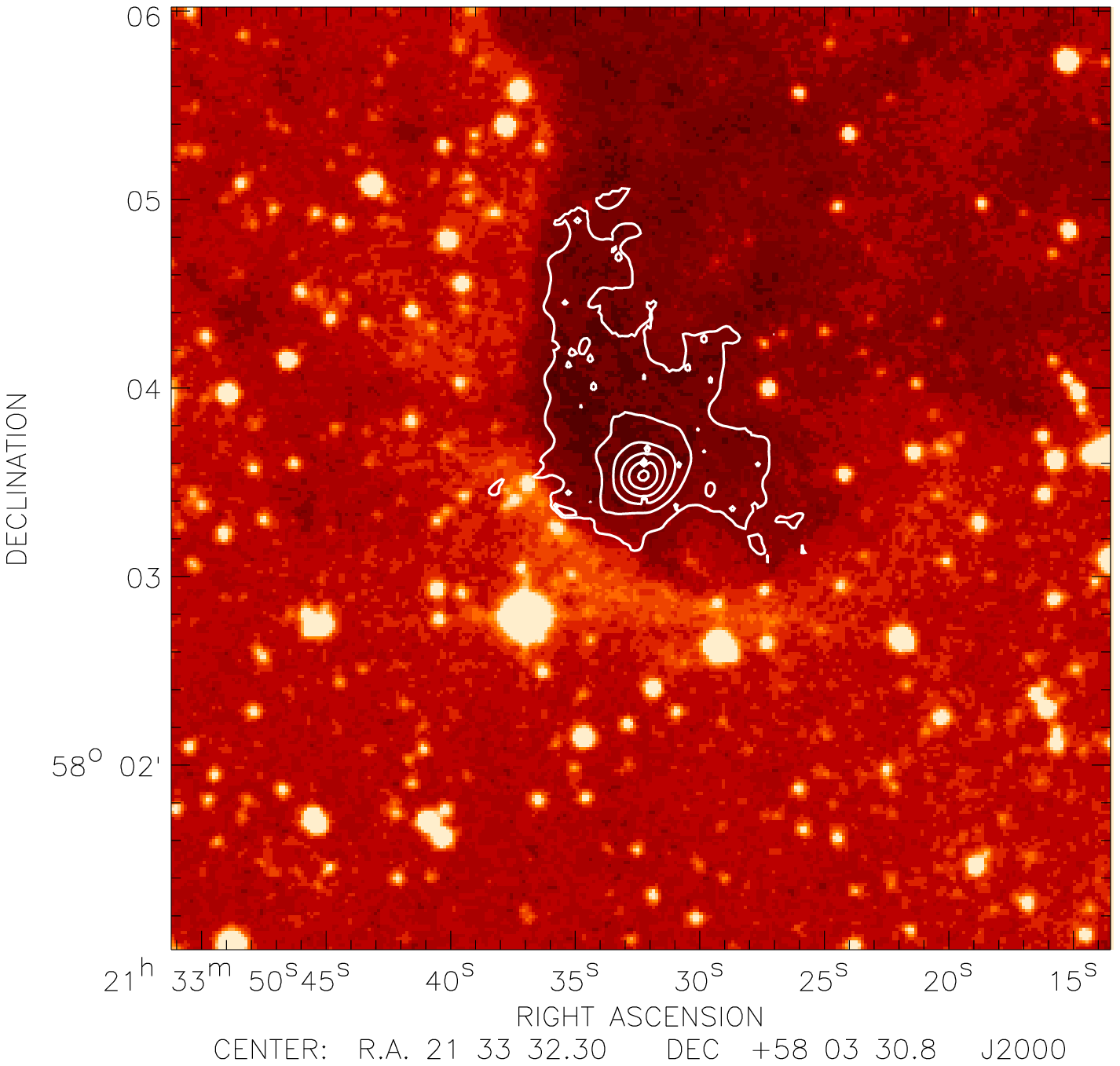}\\
\includegraphics*[height=6cm]{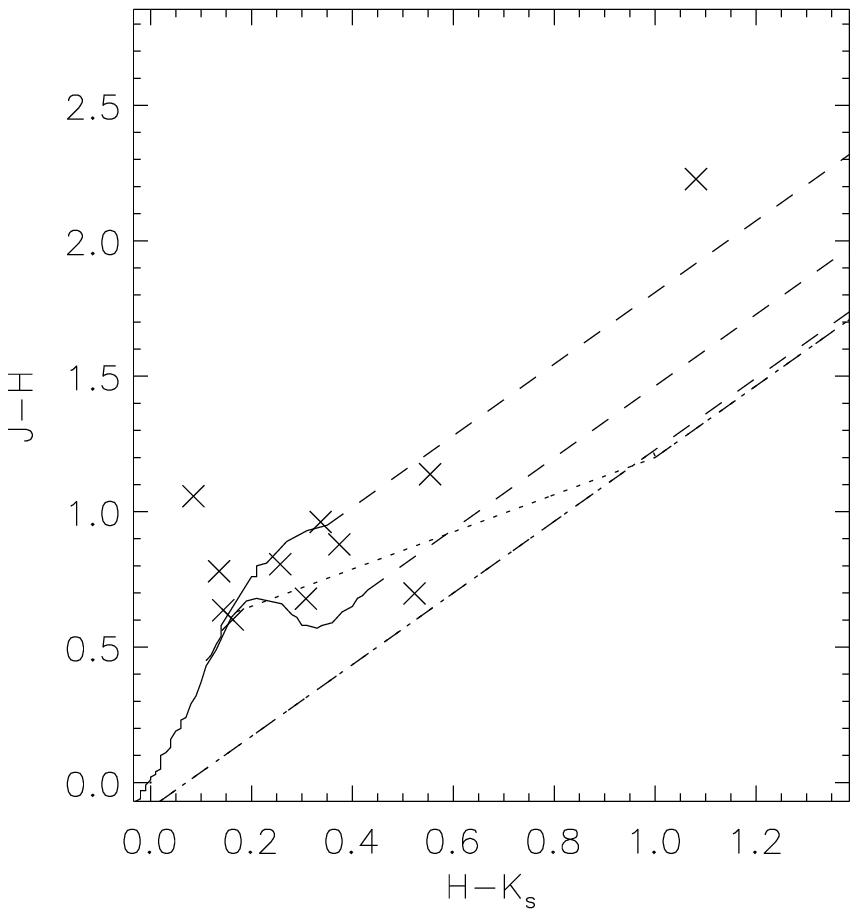}
\includegraphics*[height=6cm]{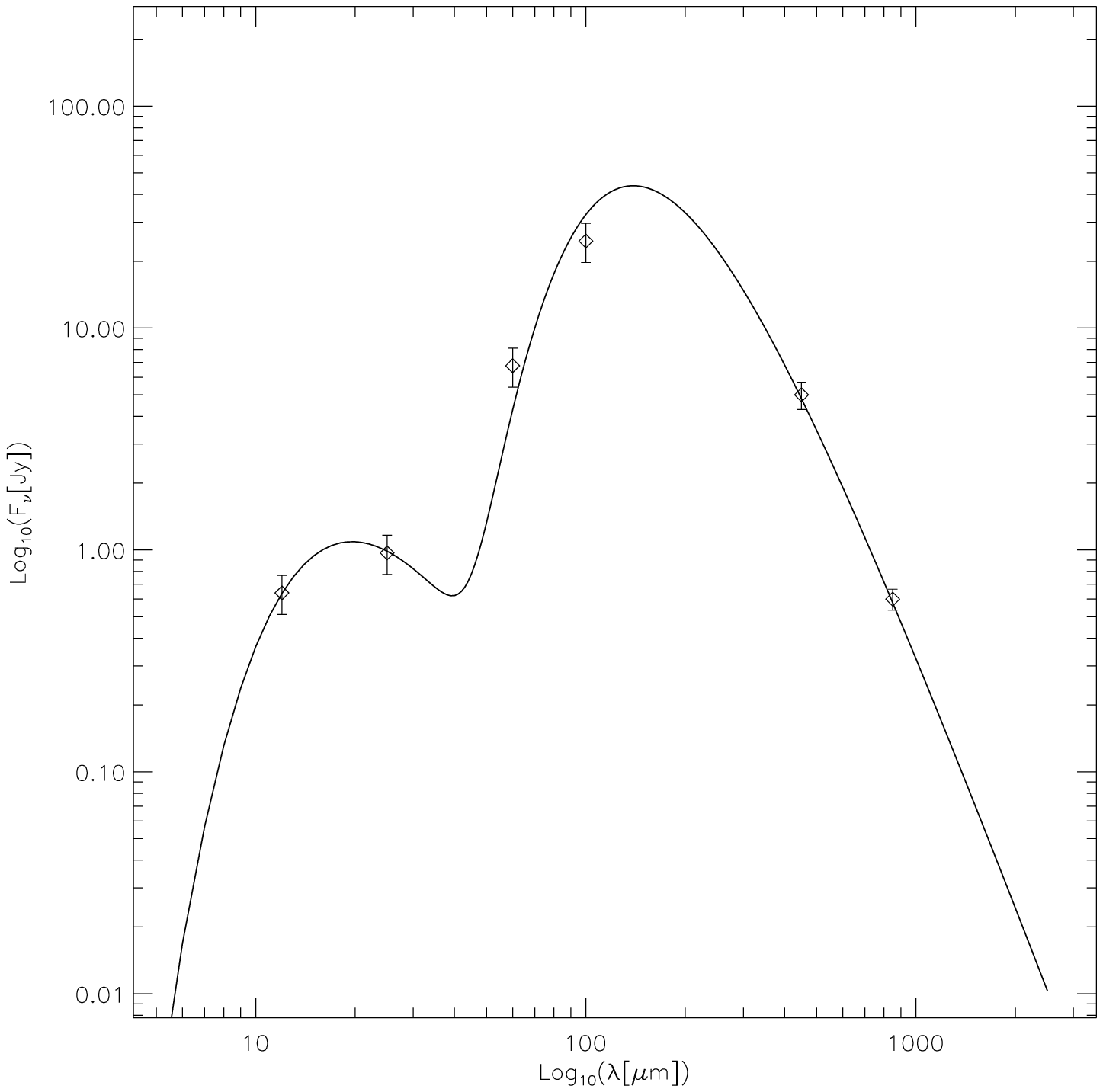}\\
\end{center}
\caption{Plots and images associated with the object SFO 34. The top images show SCUBA 450 \micron ~(left) and 850 \micron ~(right) contours overlaid on a DSS image, infrared sources from the 2MASS Point Source Catalogue \citep{Cutri2003} that have been identified as YSOs are shown as triangles.  850 \micron ~contours start at 6$\sigma$ and increase in increments of 20\% of the peak flux value, 450 \micron ~contours start at 3$\sigma$ and increase in increments of 20\% of the peak flux value.
\indent The bottom left plot shows the J-H versus H-K$_{\rm{s}}$ colours of the 2MASS sources associated with the cloud while the bottom right image shows the SED plot of the object composed from a best fit to various observed fluxes.}
\end{figure*}
\end{center}

\newpage

\begin{center}
\begin{figure*}
\begin{center}
\includegraphics*[height=6cm]{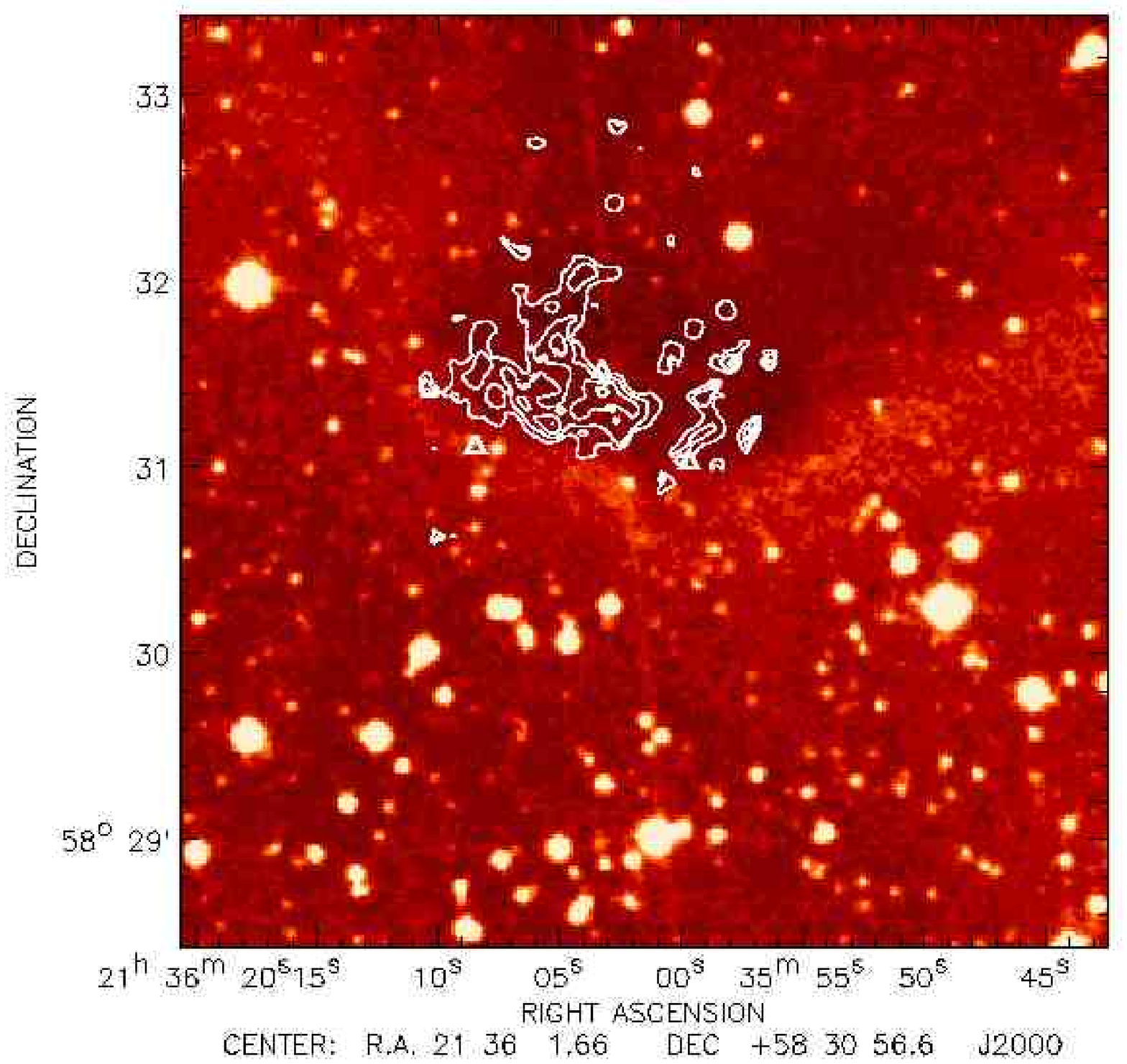}\\
\includegraphics*[height=6cm]{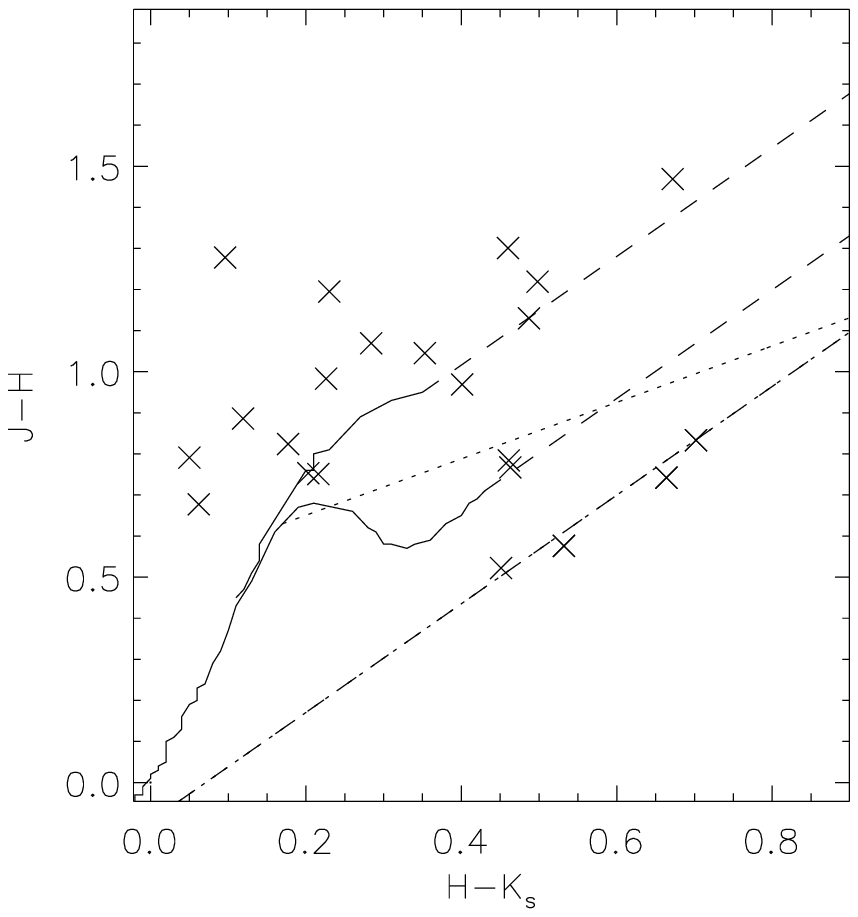}
\end{center}
\caption{Plots and images associated with the object SFO 35. The left image shows SCUBA 850 \micron ~contours overlaid on a DSS image, infrared sources from the 2MASS Point Source Catalogue \citep{Cutri2003} are shown as triangles.  850 \micron ~contours start at 6$\sigma$ and increase in increments of 33\ of the peak flux value.
\indent The right plot shows the J-H versus H-K$_{\rm{s}}$ colours of the 2MASS sources associated with the cloud.}
\end{figure*}
\end{center}

\newpage

\begin{center}
\begin{figure*}
\begin{center}
\includegraphics*[height=6cm]{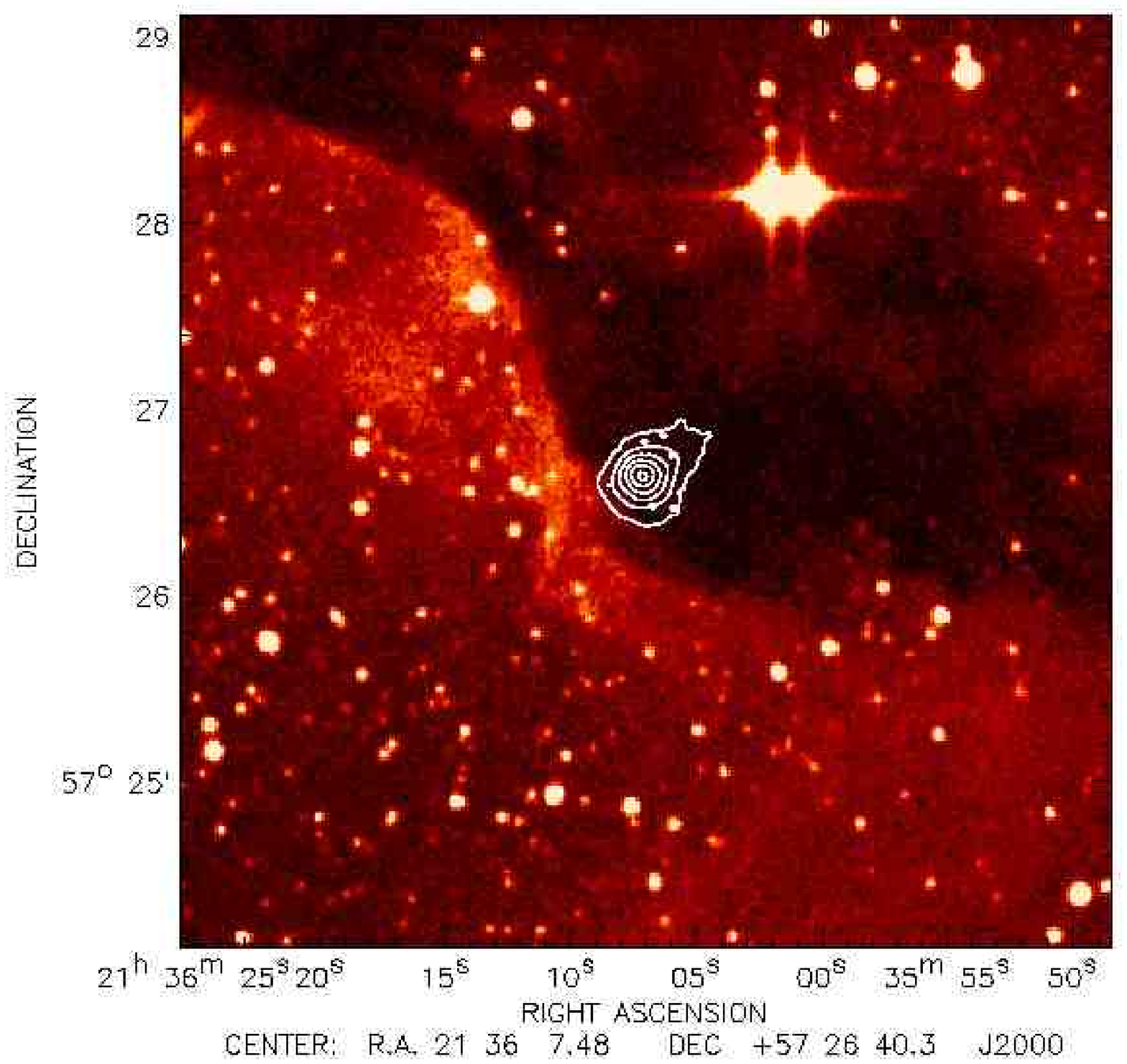}
\includegraphics*[height=6cm]{8104A27b.eps}\\
\includegraphics*[height=6cm]{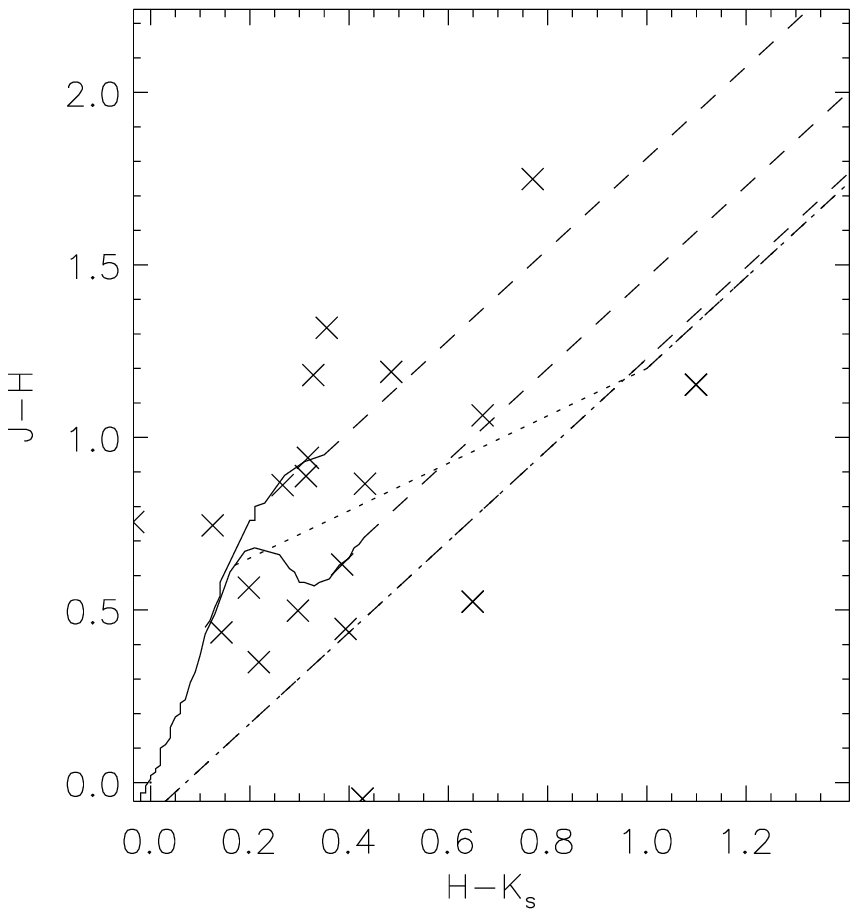}
\end{center}
\caption{Plots and images associated with the object SFO 36. The top images show SCUBA 450 \micron ~(left) and 850 \micron ~(right) contours overlaid on a DSS image, infrared sources from the 2MASS Point Source Catalogue \citep{Cutri2003} that have been identified as YSOs are shown as triangles.  850 \micron ~contours start at 9$\sigma$ and increase in increments of 20\% of the peak flux value, 450 \micron ~contours start at 9$\sigma$ and increase in increments of 20\% of the peak flux value.
\indent The bottom plot shows the J-H versus H-K$_{\rm{s}}$ colours of the 2MASS sources associated with the cloud.}
\end{figure*}
\end{center}

\newpage

\begin{center}
\begin{figure*}
\begin{center}
\includegraphics*[height=6cm]{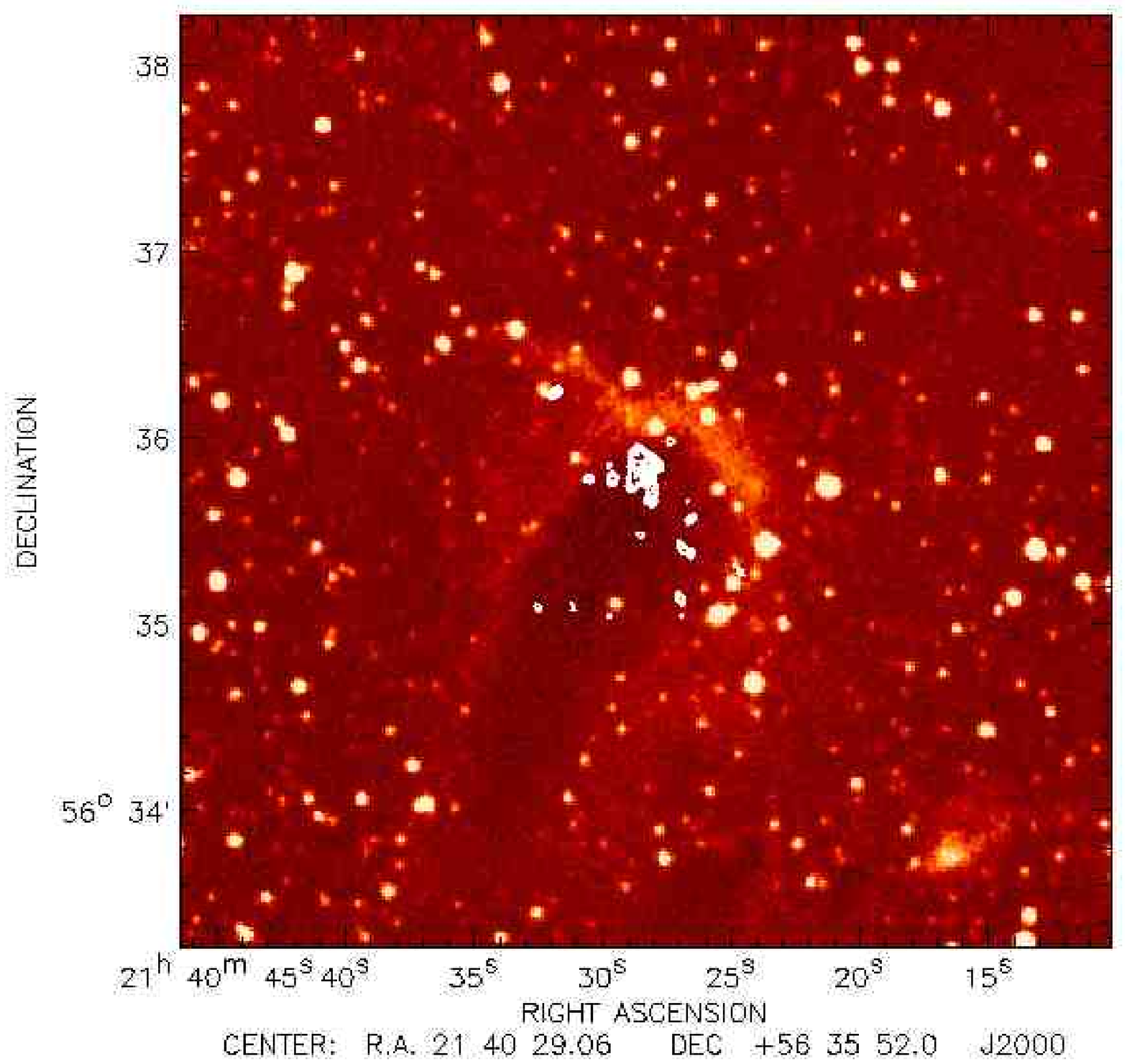}
\includegraphics*[height=6cm]{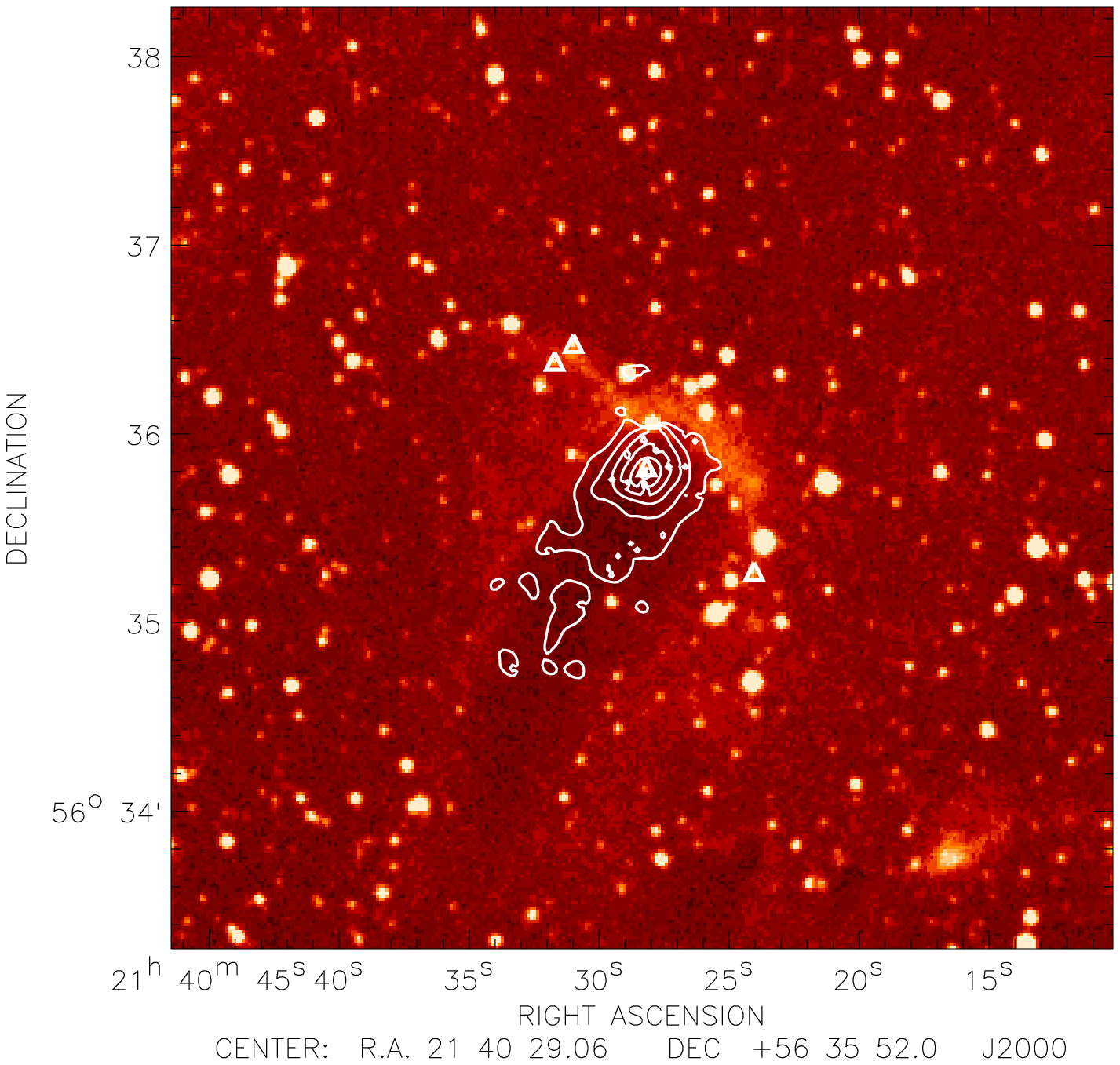}\\
\includegraphics*[height=6cm]{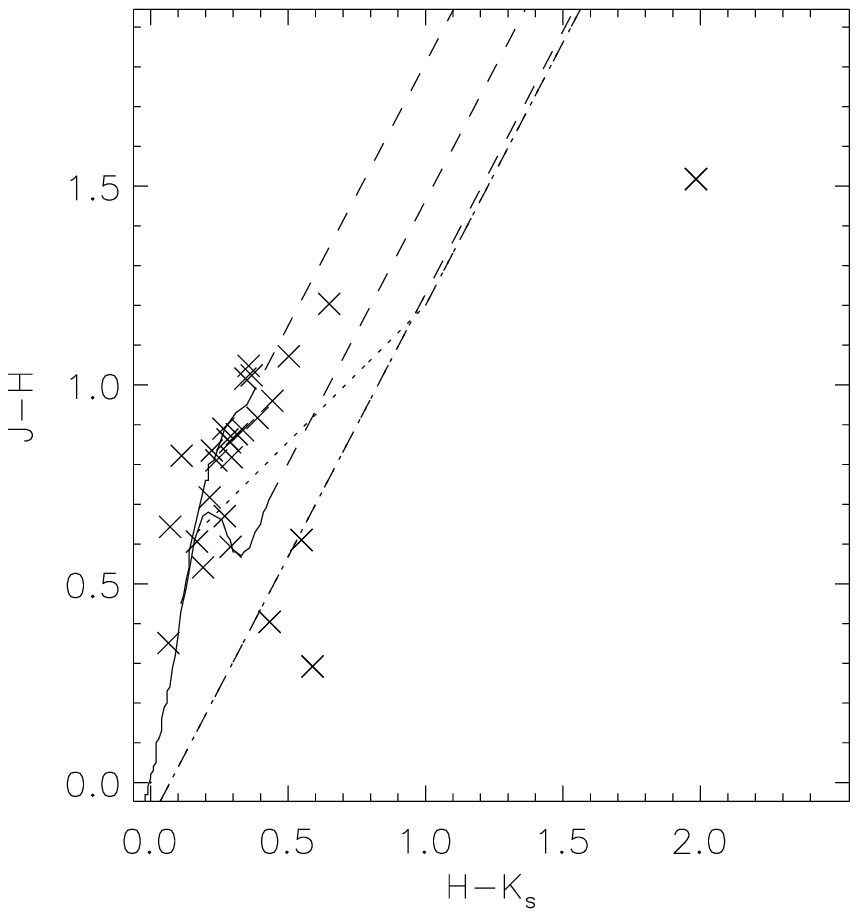}
\end{center}
\caption{Plots and images associated with the object SFO 37. The top images show SCUBA 450 \micron ~(left) and 850 \micron ~(right) contours overlaid on a DSS image, infrared sources from the 2MASS Point Source Catalogue \citep{Cutri2003} that have been identified as YSOs are shown as triangles.  850 \micron ~contours start at 3$\sigma$ and increase in increments of 20\% of the peak flux value, 450 \micron ~contours start at 2$\sigma$ and increase in increments of 20\% of the peak flux value.
\indent The bottom left plot shows the J-H versus H-K$_{\rm{s}}$ colours of the 2MASS sources associated with the cloud while the bottom right image shows the SED plot of the object composed from a best fit to various observed fluxes.}
\end{figure*}
\end{center}

\newpage

\begin{center}
\begin{figure*}
\begin{center}
\includegraphics*[height=6cm]{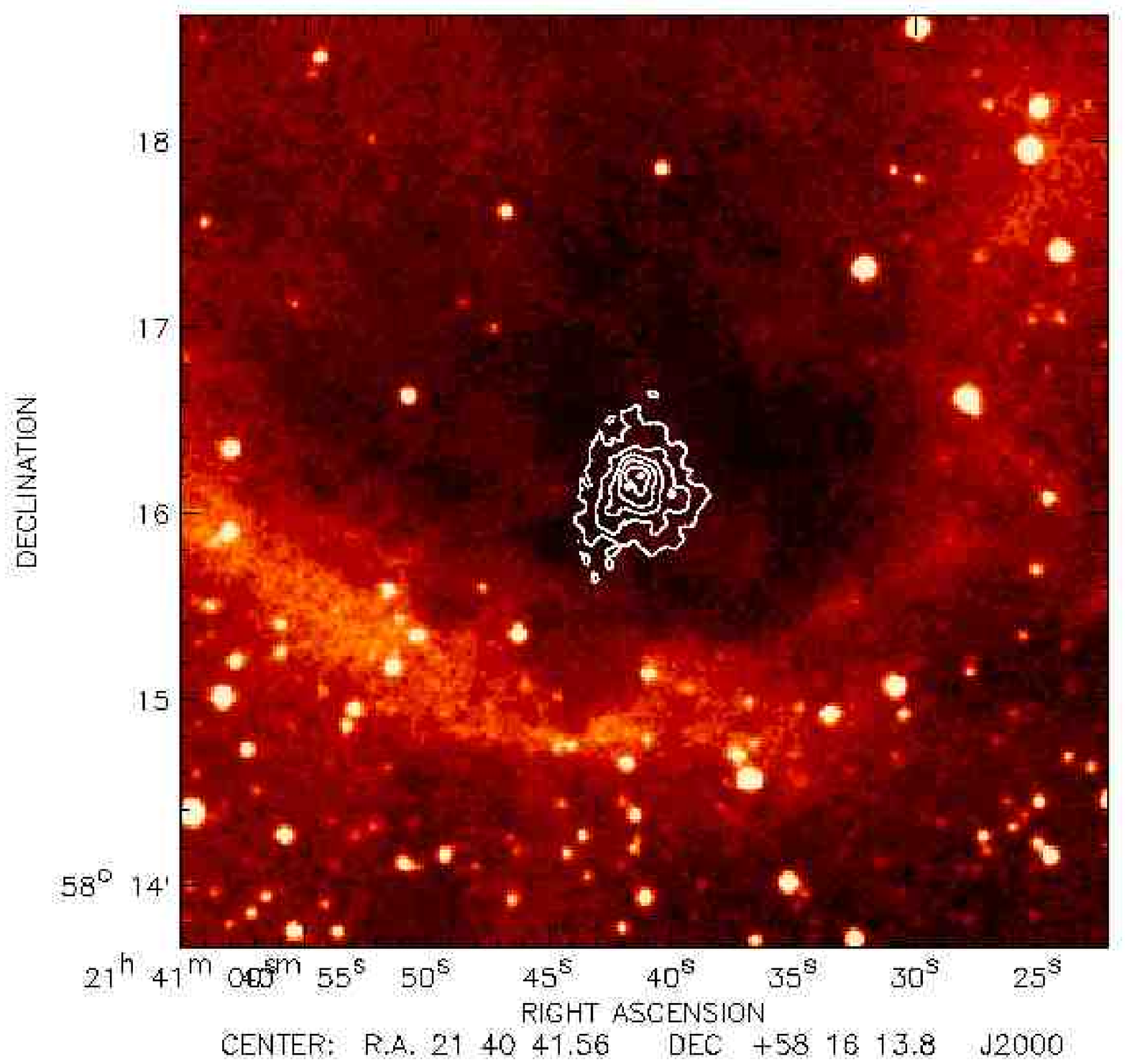}
\includegraphics*[height=6cm]{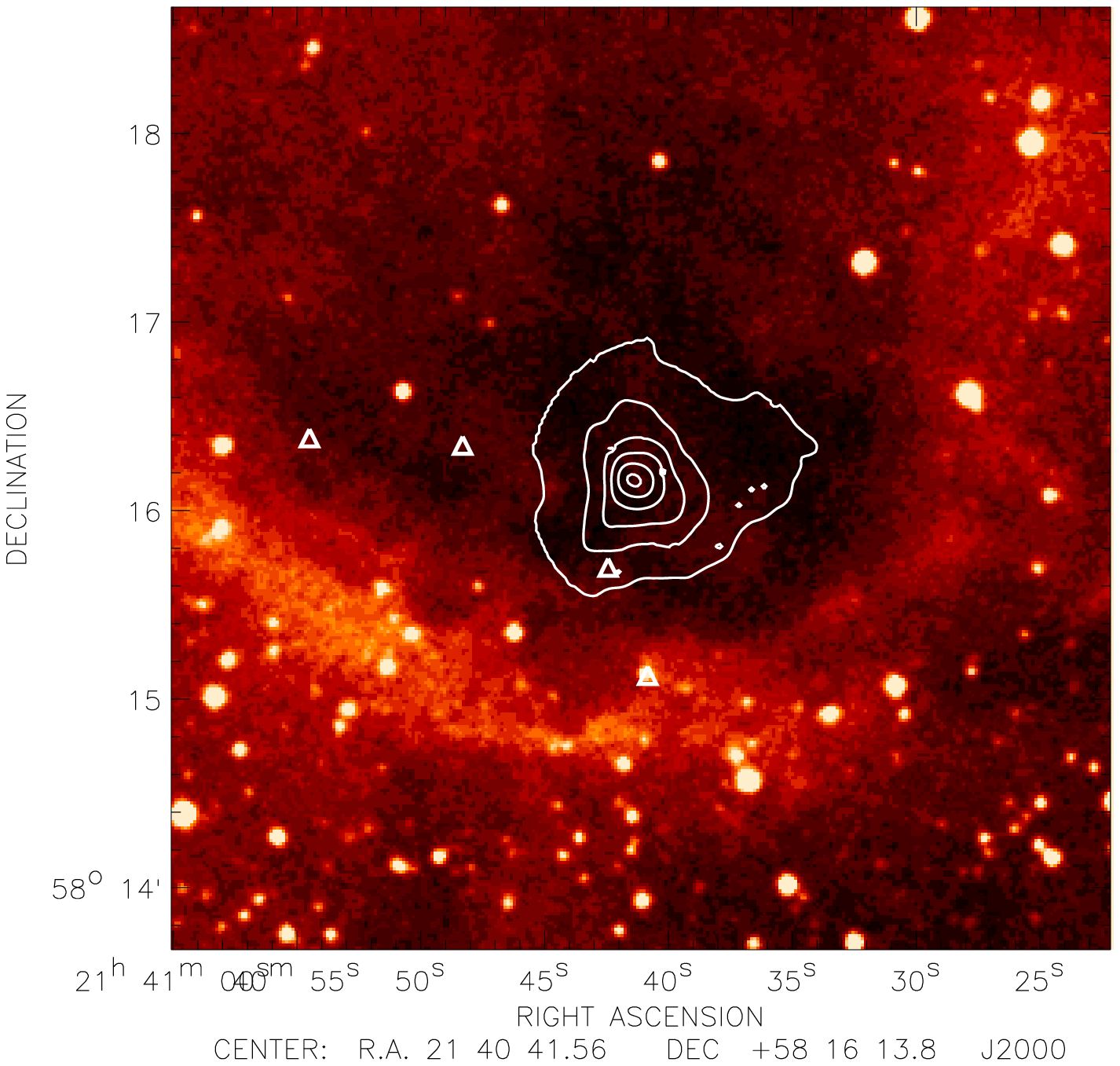}\\
\includegraphics*[height=6cm]{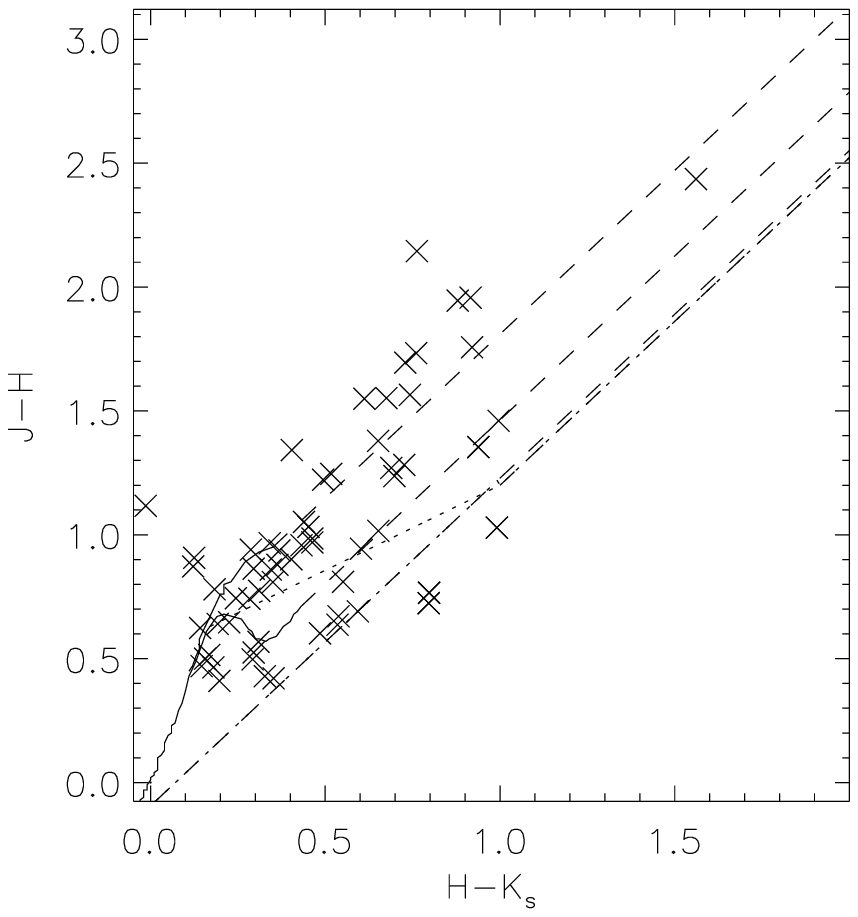}
\end{center}
\caption{Plots and images associated with the object SFO 38. The top images show SCUBA 450 \micron ~(left) and 850 \micron ~(right) contours overlaid on a DSS image, infrared sources from the 2MASS Point Source Catalogue \citep{Cutri2003} that have been identified as YSOs are shown as triangles.  850 \micron ~contours start at 9$\sigma$ and increase in increments of 20\% of the peak flux value, 450 \micron ~contours start at 6$\sigma$ and increase in increments of 20\% of the peak flux value.
\indent The bottom plot shows the J-H versus H-K$_{\rm{s}}$ colours of the 2MASS sources associated with the cloud.}
\end{figure*}
\end{center}

\newpage

\begin{center}
\begin{figure*}
\begin{center}
\includegraphics*[height=6cm]{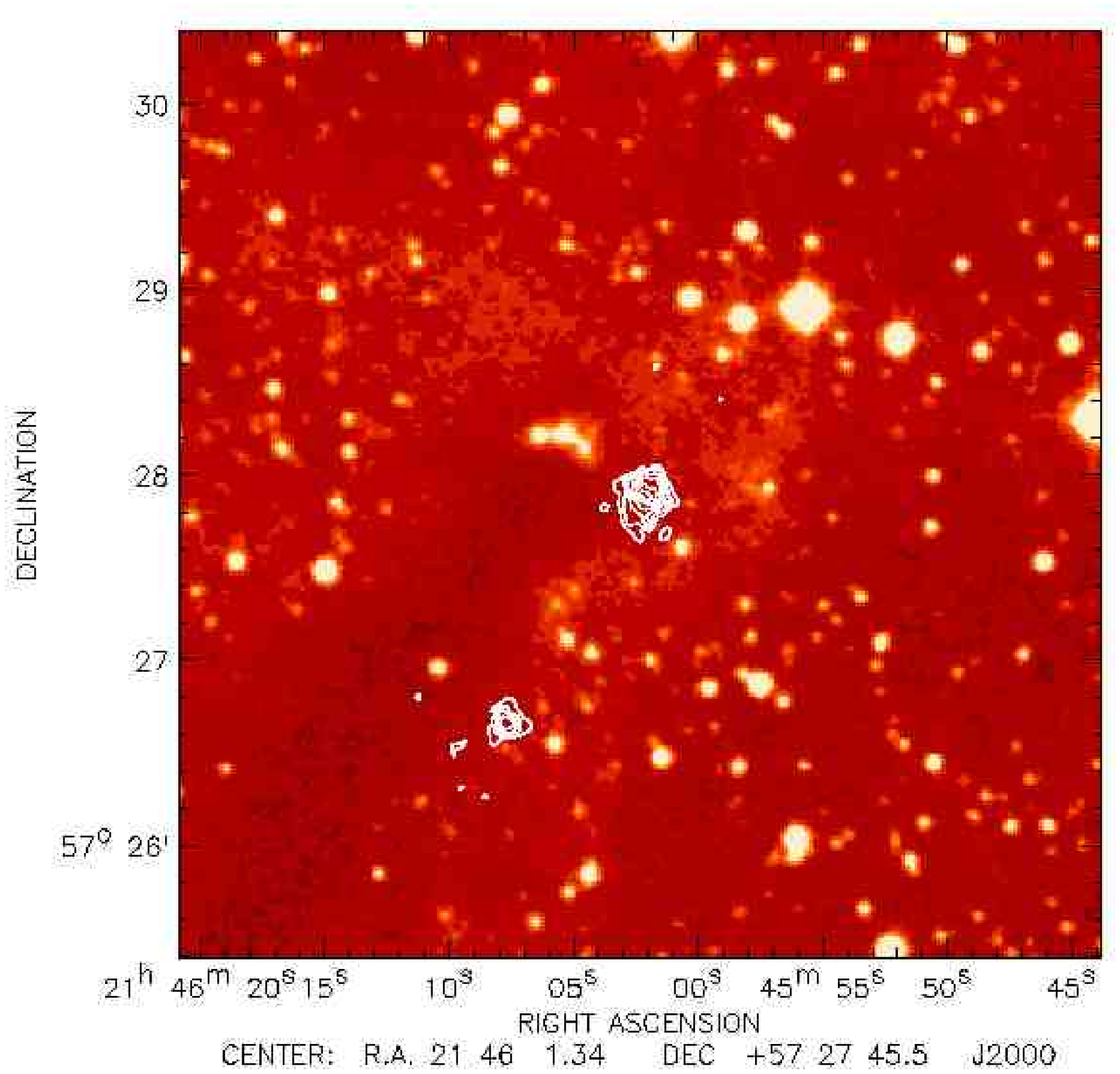}
\includegraphics*[height=6cm]{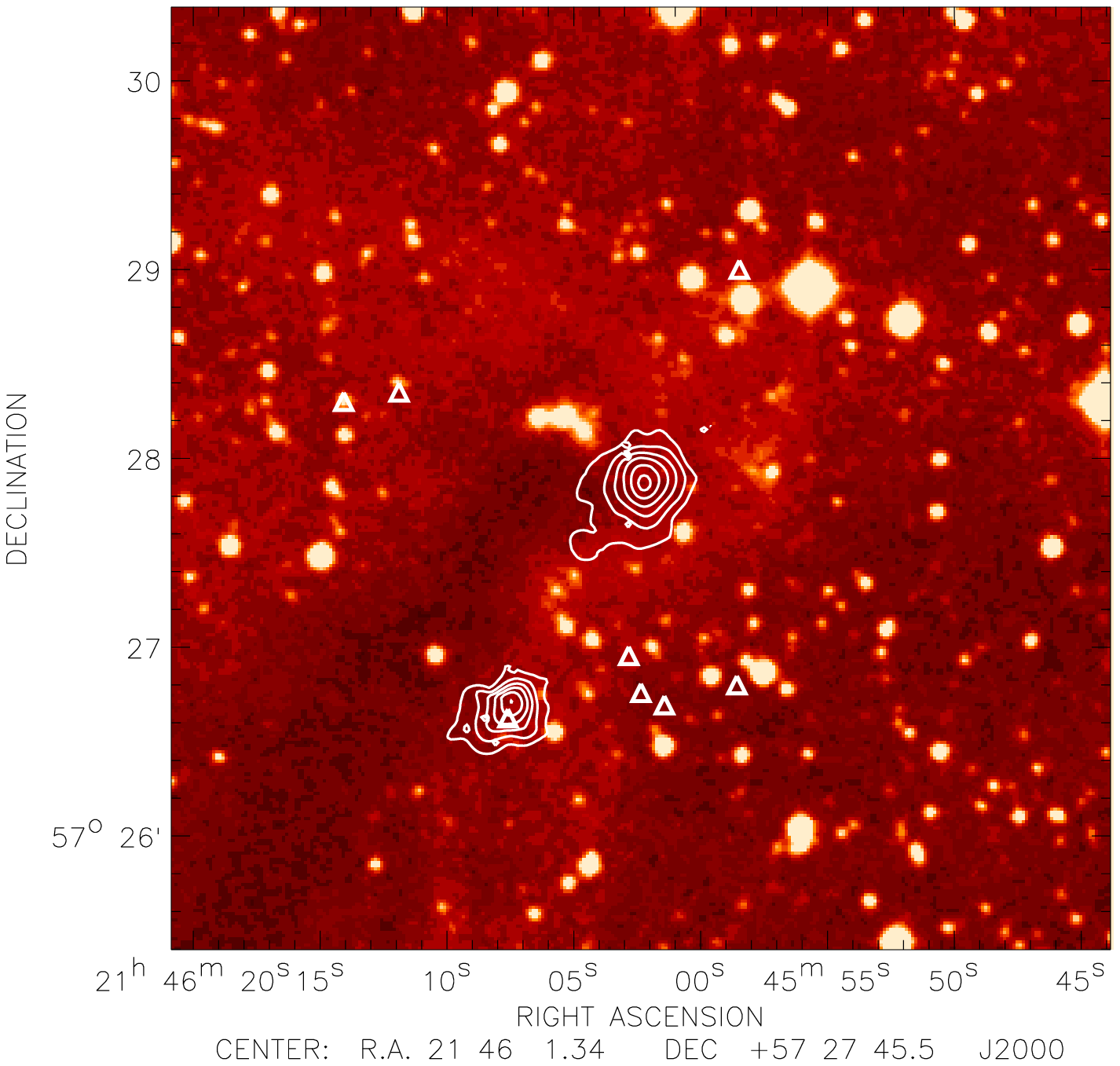}\\
\includegraphics*[height=6cm]{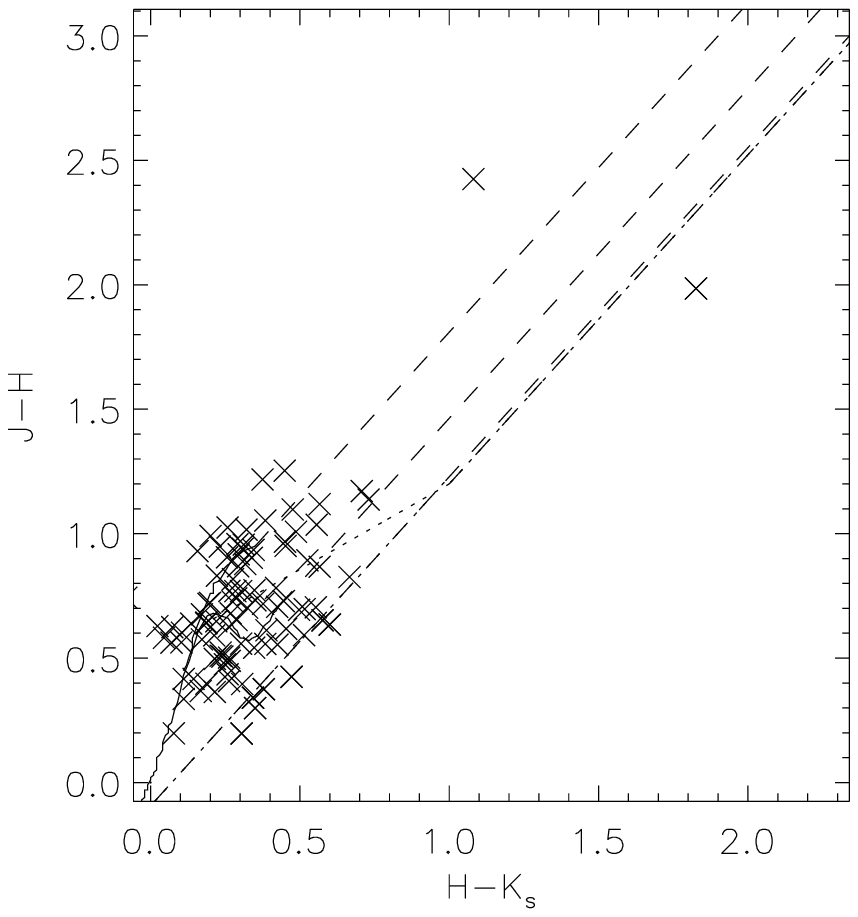}
\includegraphics*[height=6cm]{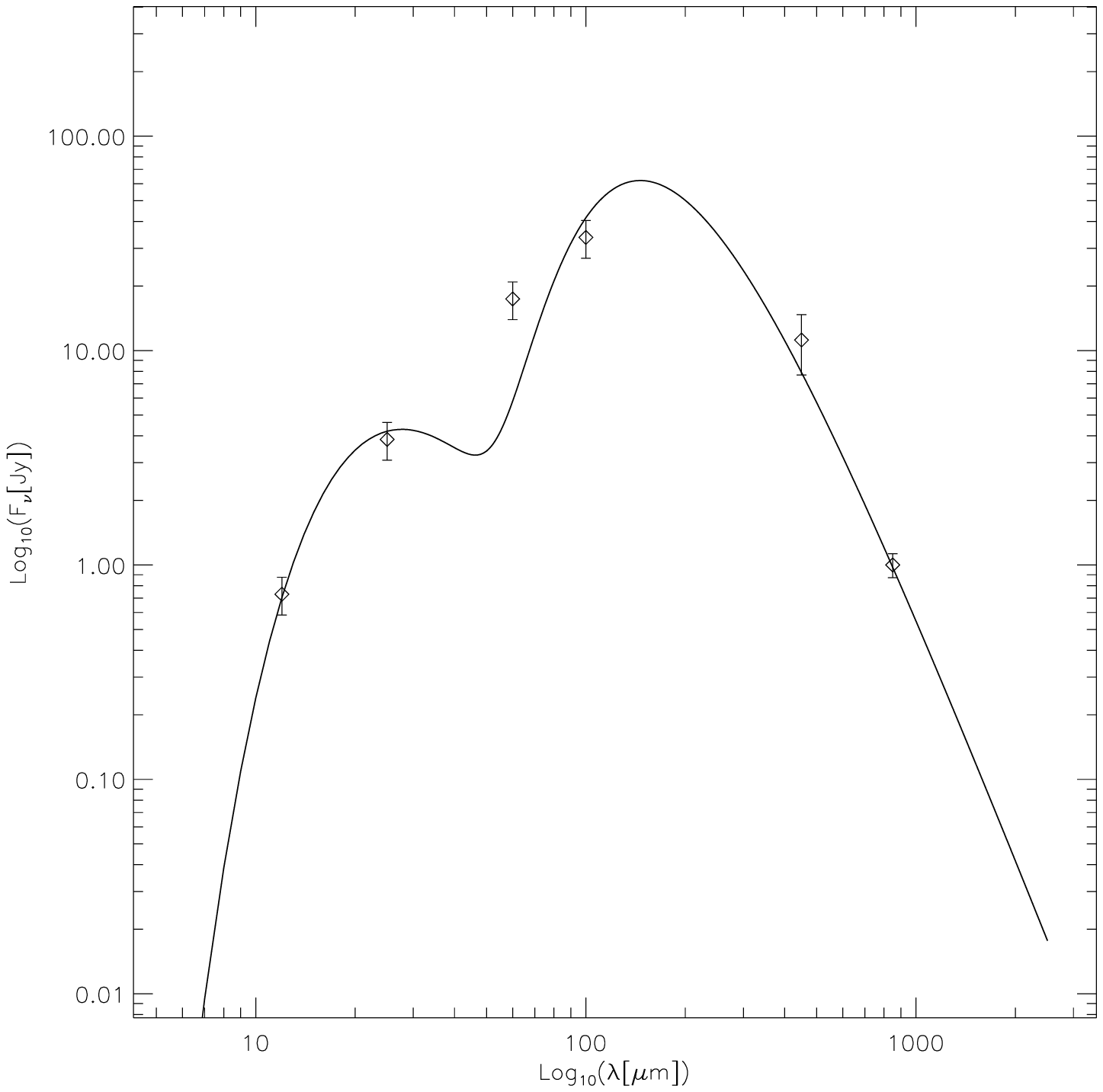}
\includegraphics*[height=6cm]{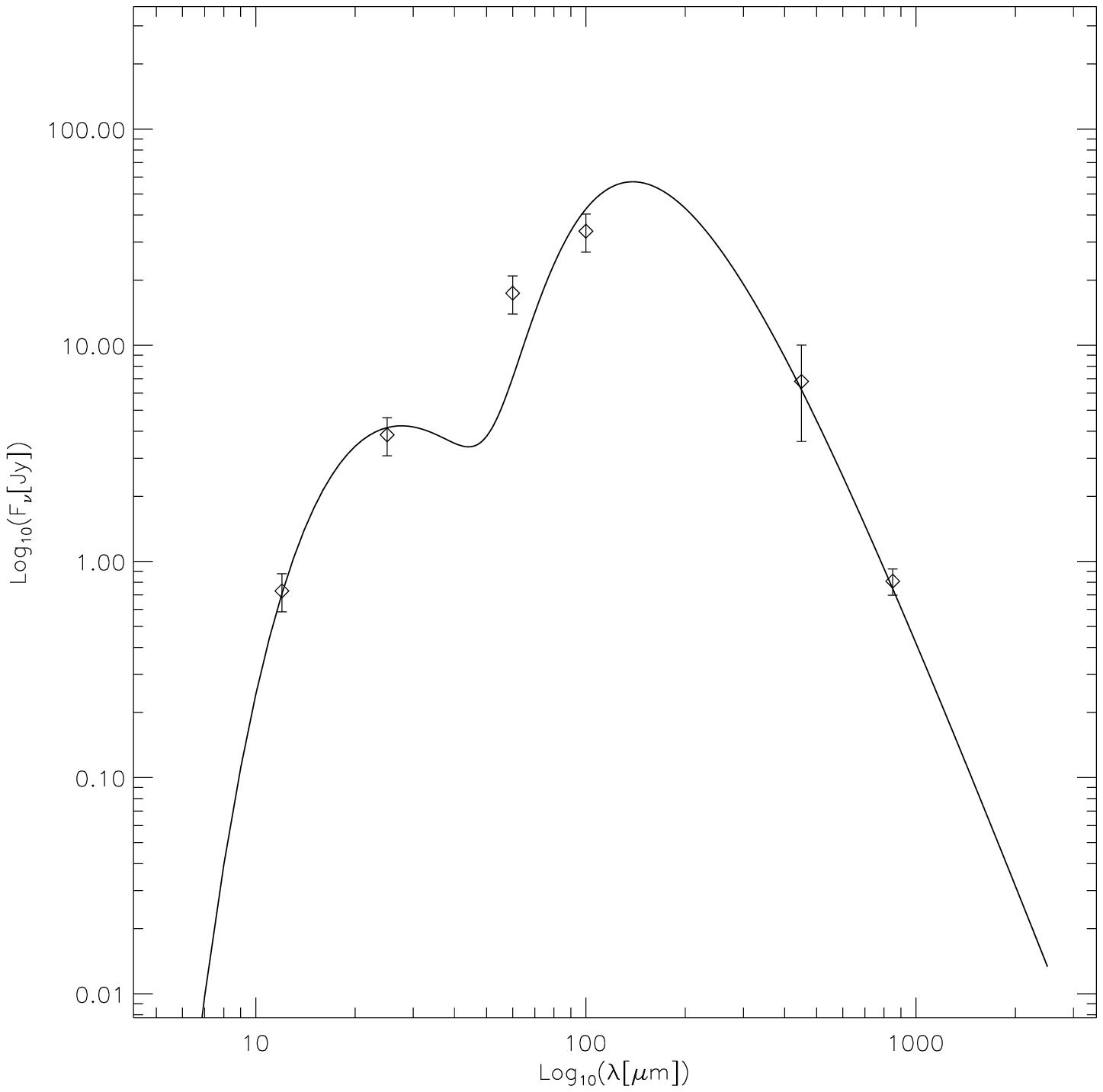}\\
\end{center}
\caption{Plots and images associated with the object SFO 39. The top images show SCUBA 450 \micron ~(left) and 850 \micron ~(right) contours overlaid on a DSS image, infrared sources from the 2MASS Point Source Catalogue \citep{Cutri2003} that have been identified as YSOs are shown as triangles.  850 \micron ~contours start at 8$\sigma$ and increase in increments of 20\% of the peak flux value, 450 \micron ~contours start at 3$\sigma$ and increase in increments of 20\% of the peak flux value.
\indent The central plot shows the J-H versus H-K$_{\rm{s}}$ colours of the 2MASS sources associated with the cloud while the bottom images show the SED plots of the separate core objects composed from a best fit to various observed fluxes.}
\end{figure*}
\end{center}

\newpage

\begin{center}
\begin{figure*}
\begin{center}
\includegraphics*[height=6cm]{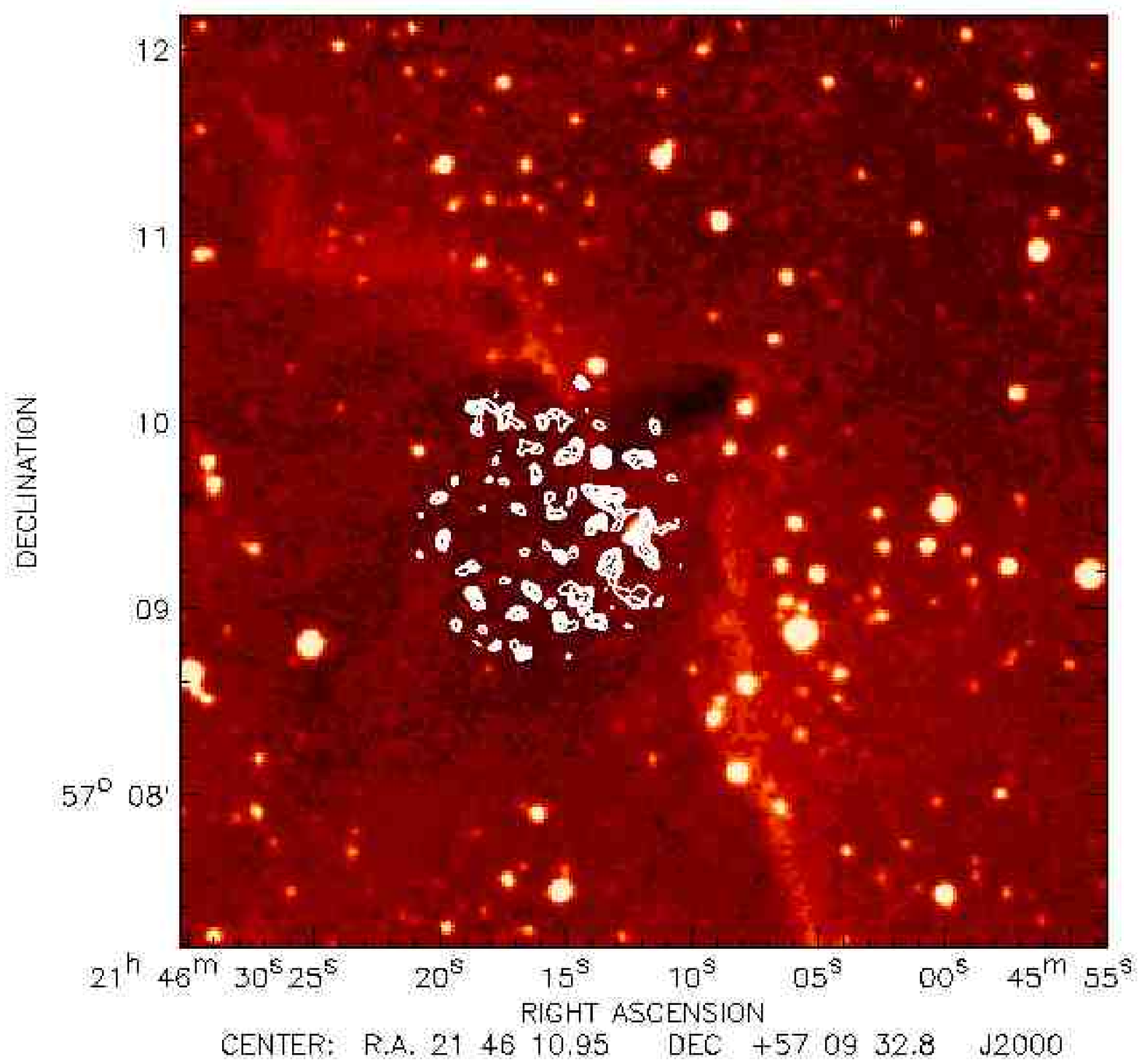}
\includegraphics*[height=6cm]{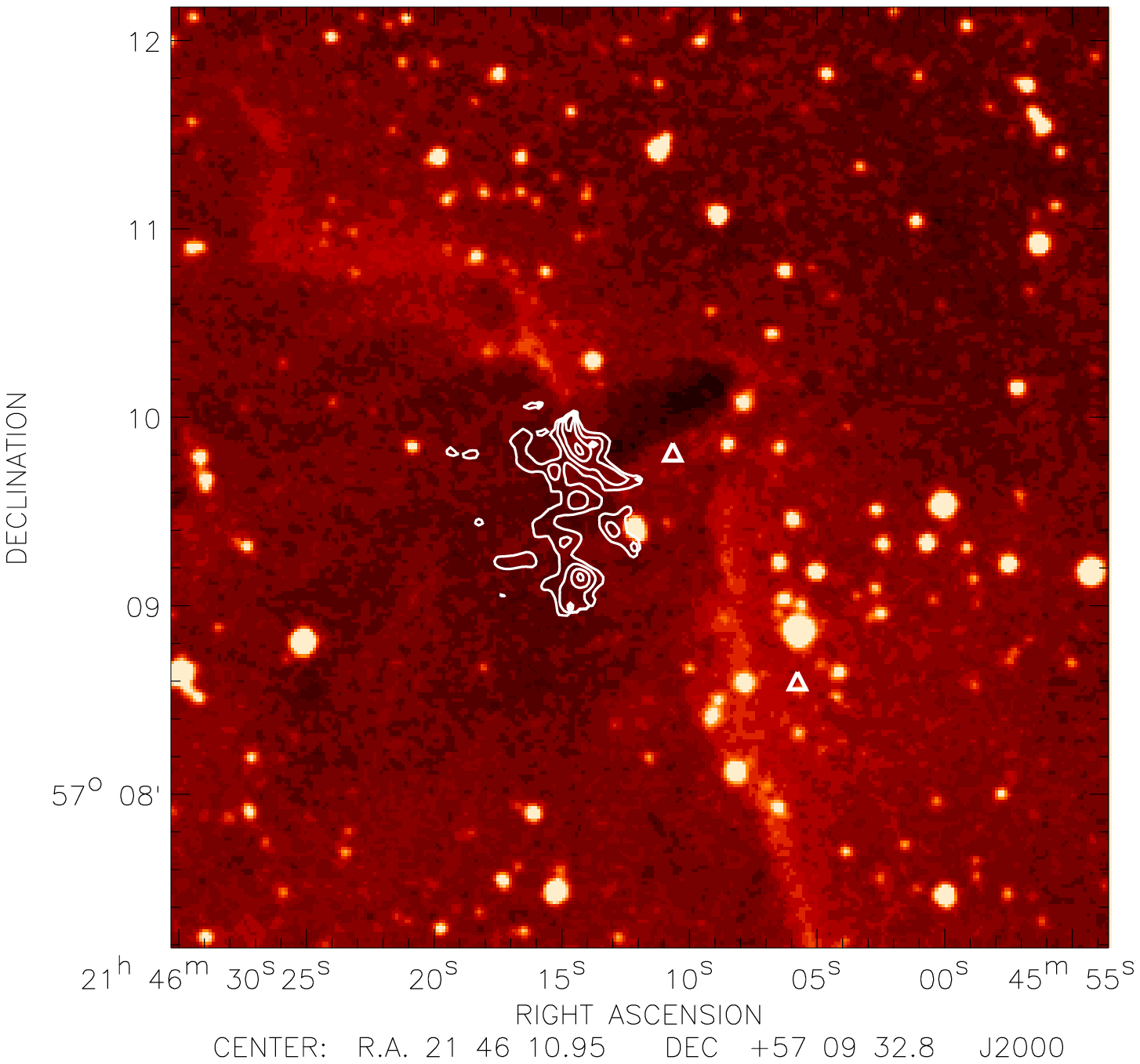}\\
\includegraphics*[height=6cm]{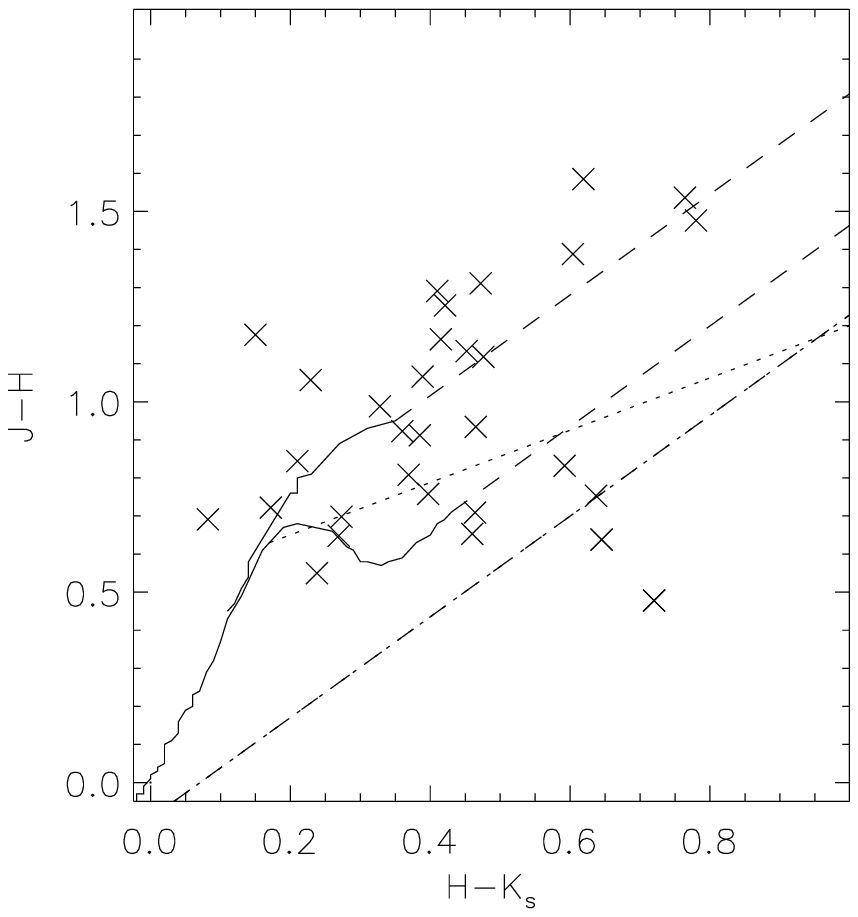}
\includegraphics*[height=6cm]{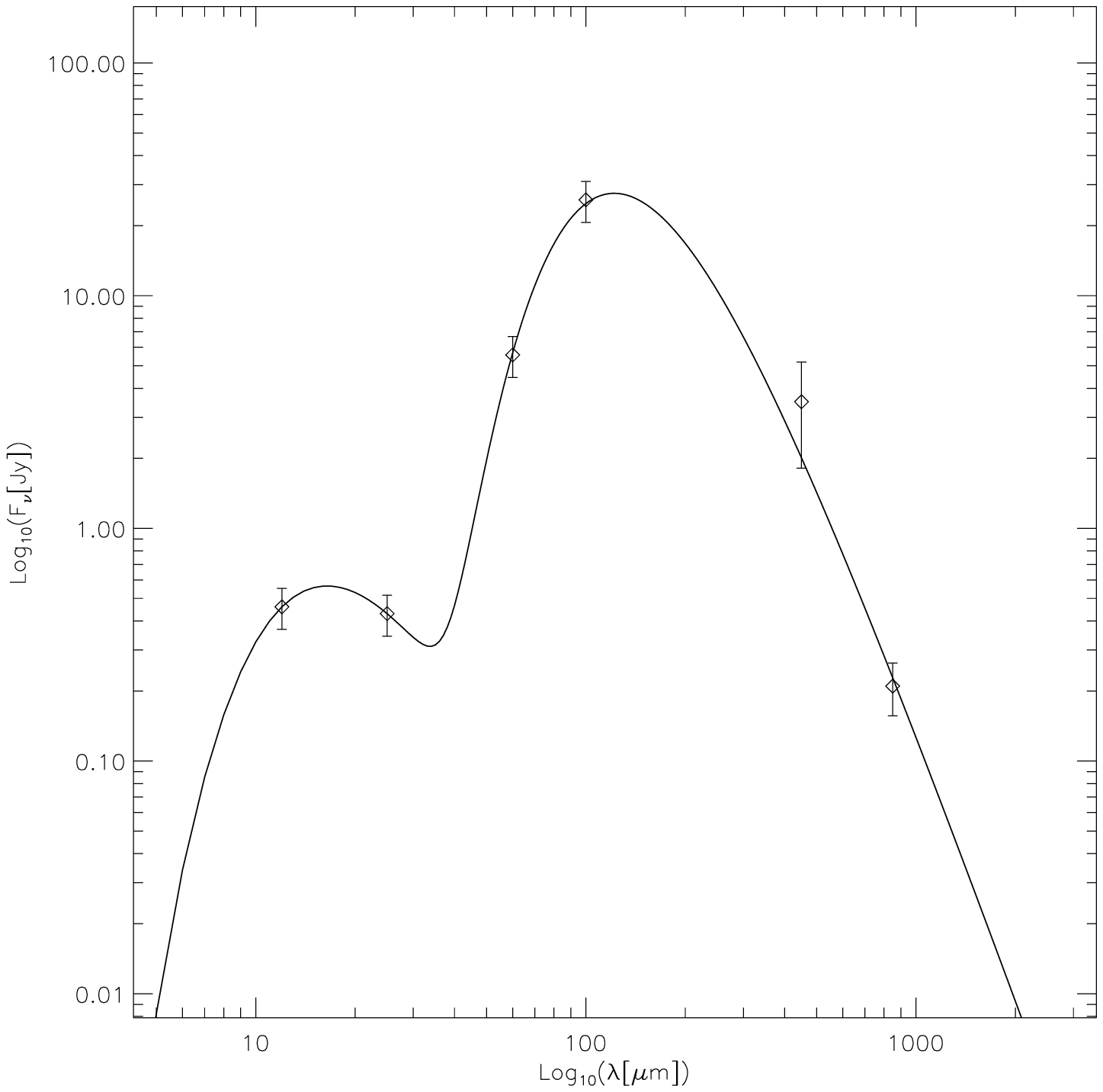}\\
\end{center}
\caption{Plots and images associated with the object SFO 40. The top images show SCUBA 450 \micron ~(left) and 850 \micron ~(right) contours overlaid on a DSS image, infrared sources from the 2MASS Point Source Catalogue \citep{Cutri2003} that have been identified as YSOs are shown as triangles.  850 \micron ~contours start at 4$\sigma$ and increase in increments of 33\ of the peak flux value, 450 \micron ~contours start at 3$\sigma$ and increase in increments of 20\% of the peak flux value.
\indent The bottom left plot shows the J-H versus H-K$_{\rm{s}}$ colours of the 2MASS sources associated with the cloud while the bottom right image shows the SED plot of the object composed from a best fit to various observed fluxes.}
\end{figure*}
\end{center}

\newpage

\begin{center}
\begin{figure*}
\begin{center}
\includegraphics*[height=6cm]{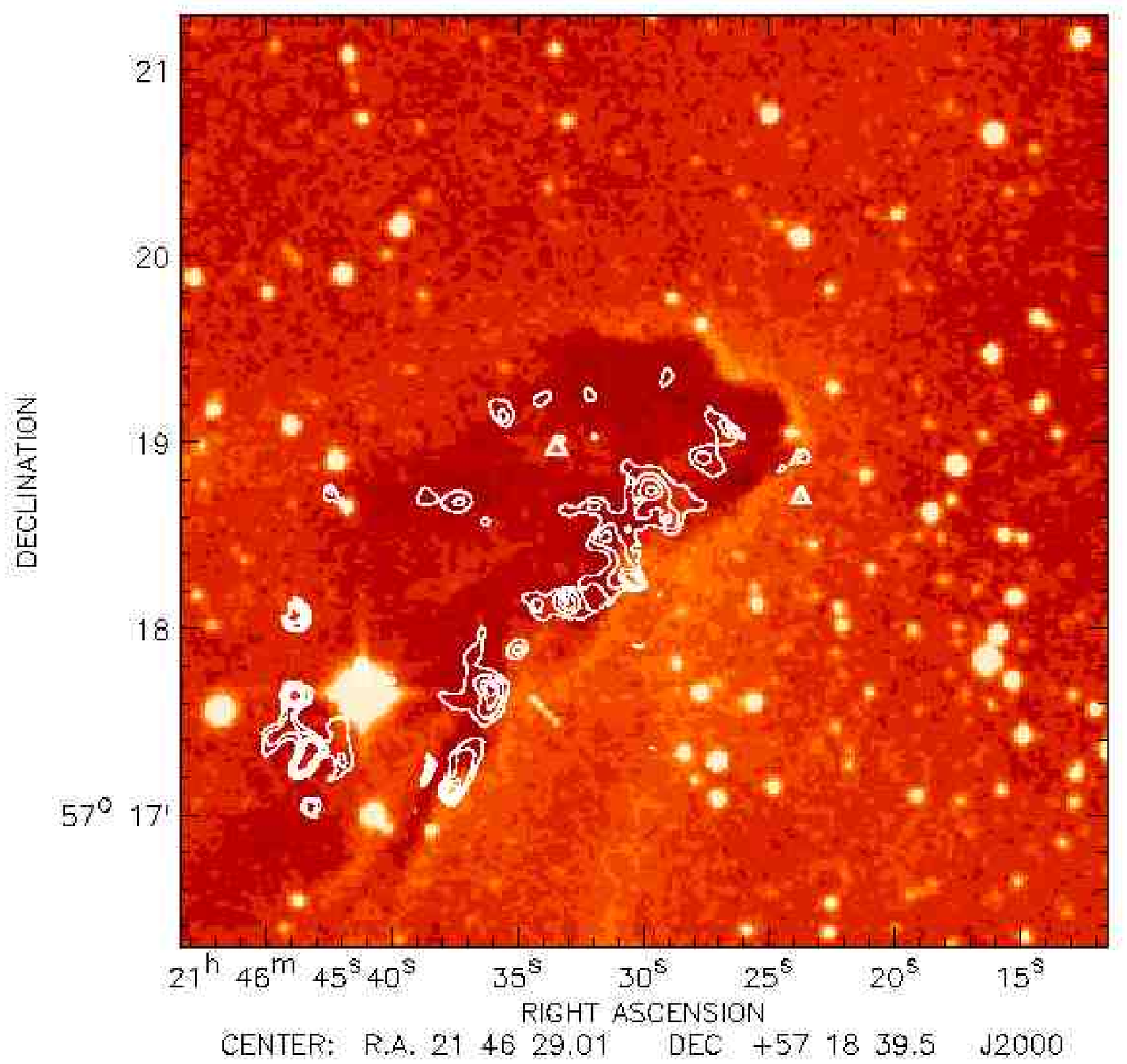}\\
\includegraphics*[height=6cm]{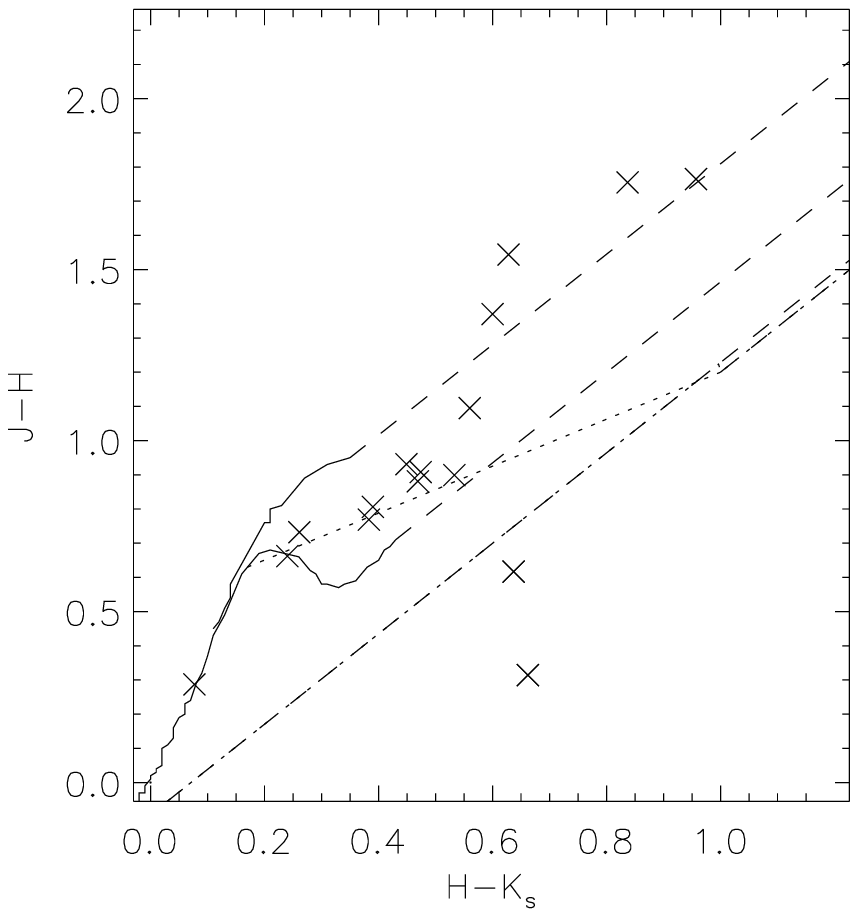}
\includegraphics*[height=6cm]{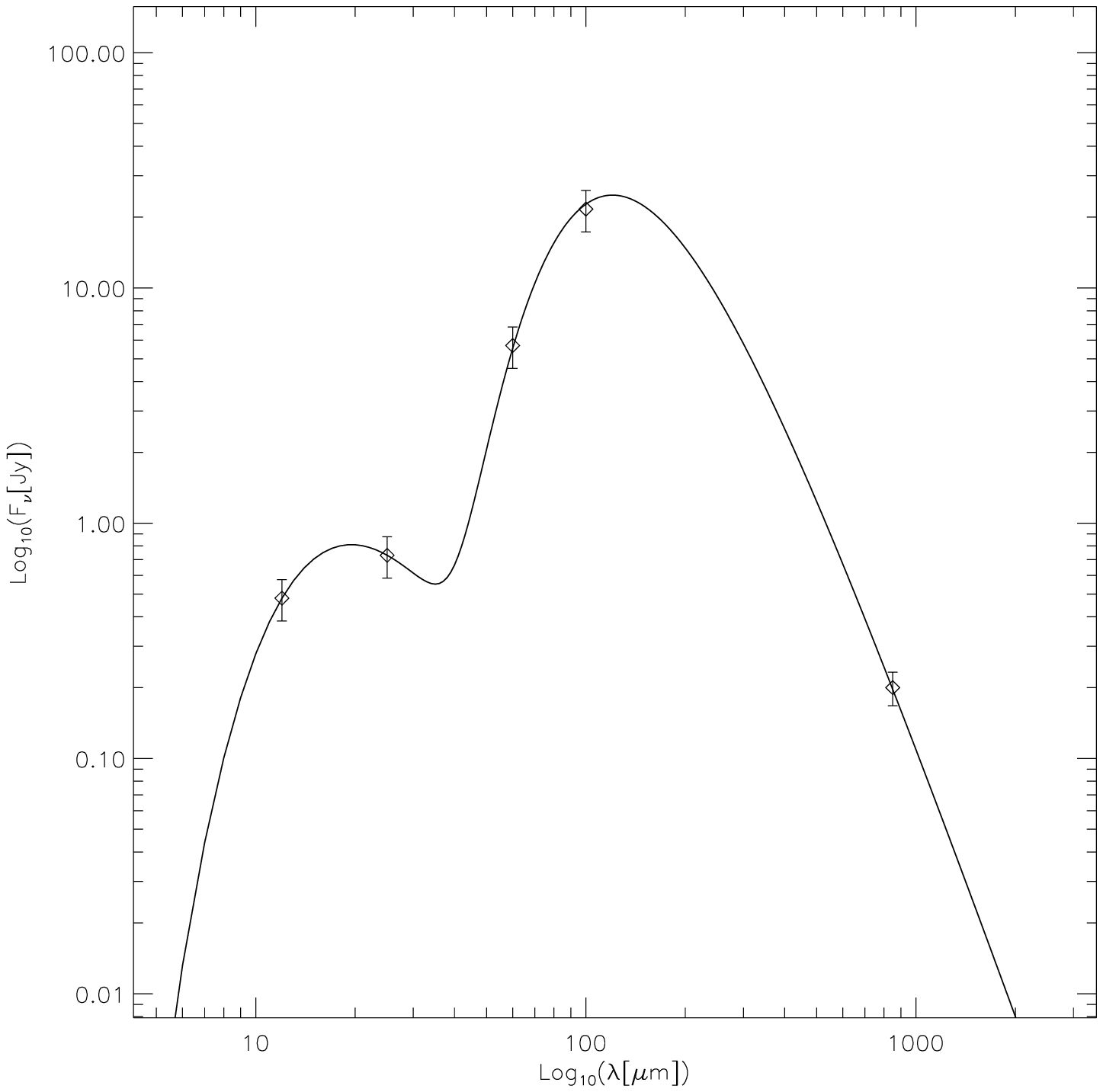}\\
\end{center}
\caption{Plots and images associated with the object SFO 41. The top image shows SCUBA 850 \micron ~contours overlaid on a DSS image, infrared sources from the 2MASS Point Source Catalogue \citep{Cutri2003} are shown as triangles.  850 \micron ~contours start at 7$\sigma$ and increase in increments of 33\ of the peak flux value.
\indent The bottom left plot shows the J-H versus H-K$_{\rm{s}}$ colours of the 2MASS sources associated with the cloud while the bottom right image shows the SED plot of the object composed from a best fit to various observed fluxes.}
\label{fig:images41}
\end{figure*}
\end{center}

\newpage

\begin{center}
\begin{figure*}
\begin{center}
\includegraphics*[height=6cm]{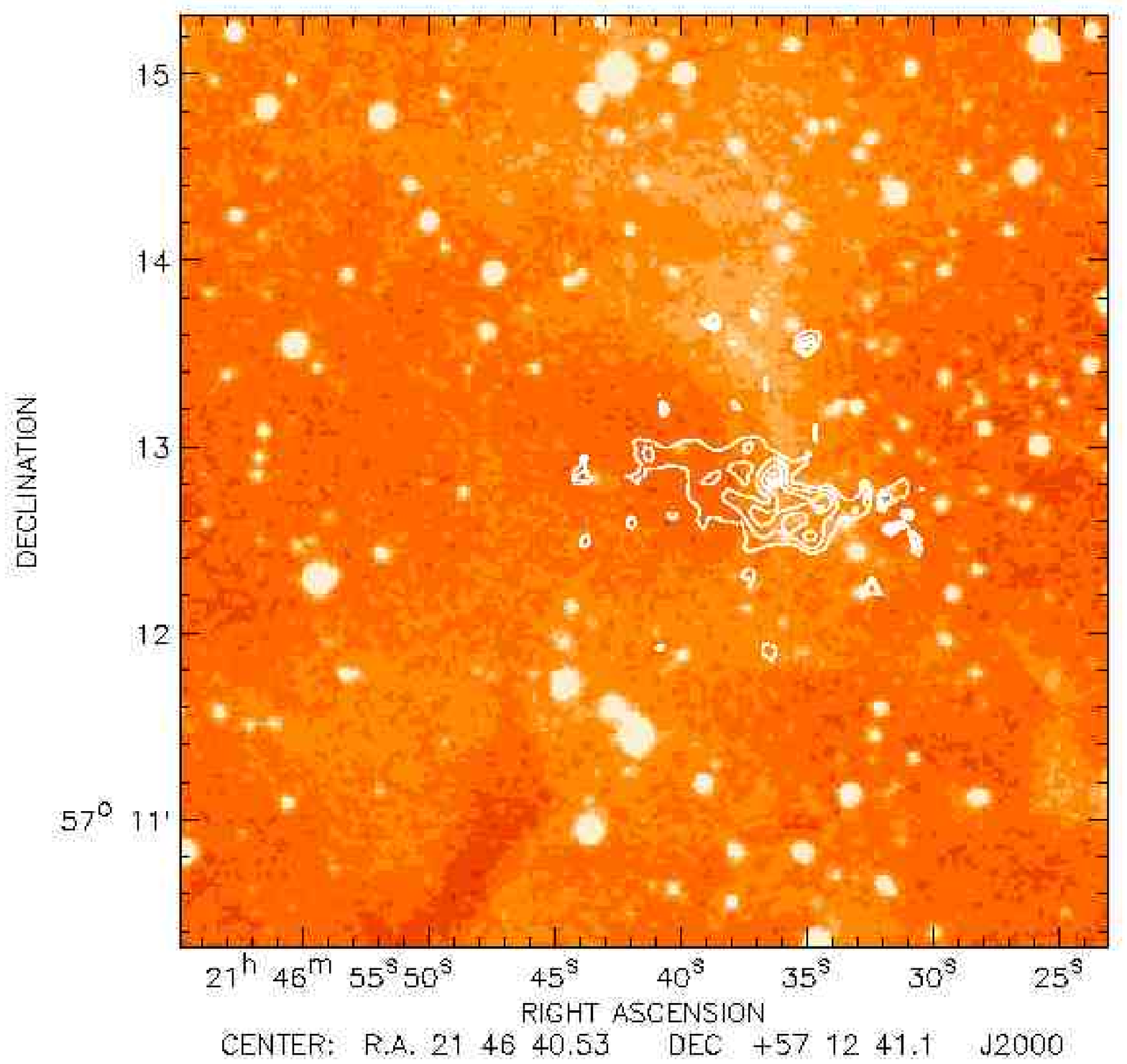}\\
\includegraphics*[height=6cm]{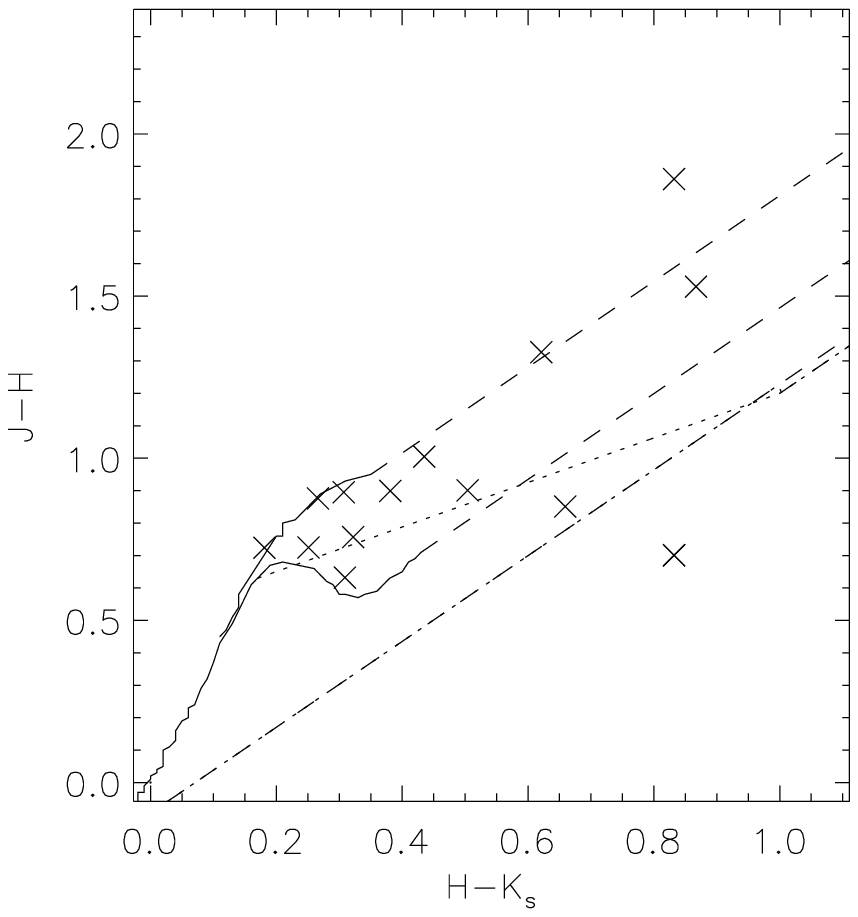}
\includegraphics*[height=6cm]{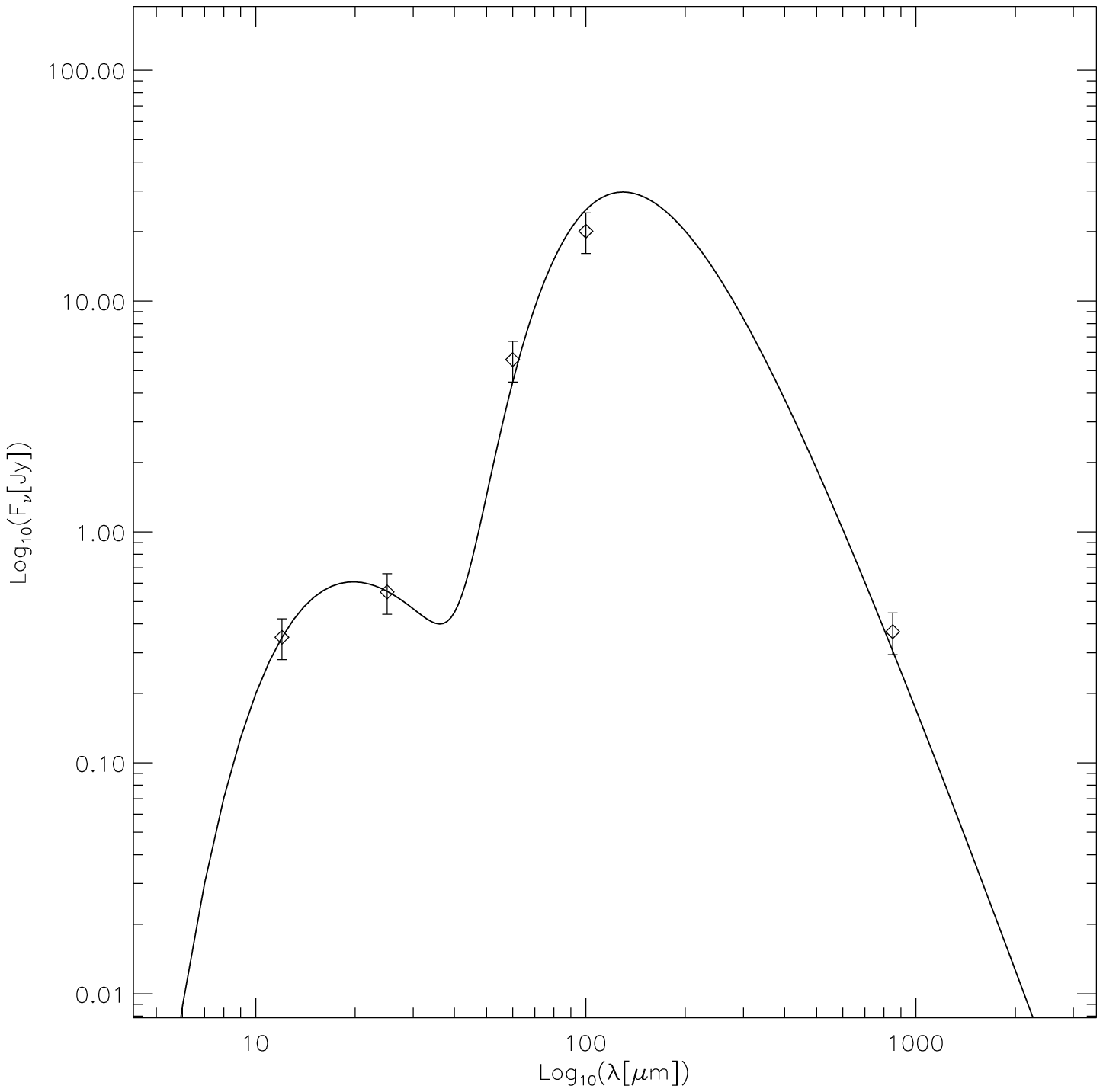}\\
\end{center}
\caption{Plots and images associated with the object SFO 42. The top image shows SCUBA 850 \micron ~contours overlaid on a DSS image, infrared sources from the 2MASS Point Source Catalogue \citep{Cutri2003} are shown as triangles.  850 \micron ~contours start at 5$\sigma$ and increase in increments of 33\ of the peak flux value.
\indent The bottom left plot shows the J-H versus H-K$_{\rm{s}}$ colours of the 2MASS sources associated with the cloud while the bottom right image shows the SED plot of the object composed from a best fit to various observed fluxes.}
\end{figure*}
\end{center}

\newpage

\begin{center}
\begin{figure*}
\begin{center}
\includegraphics*[height=6cm]{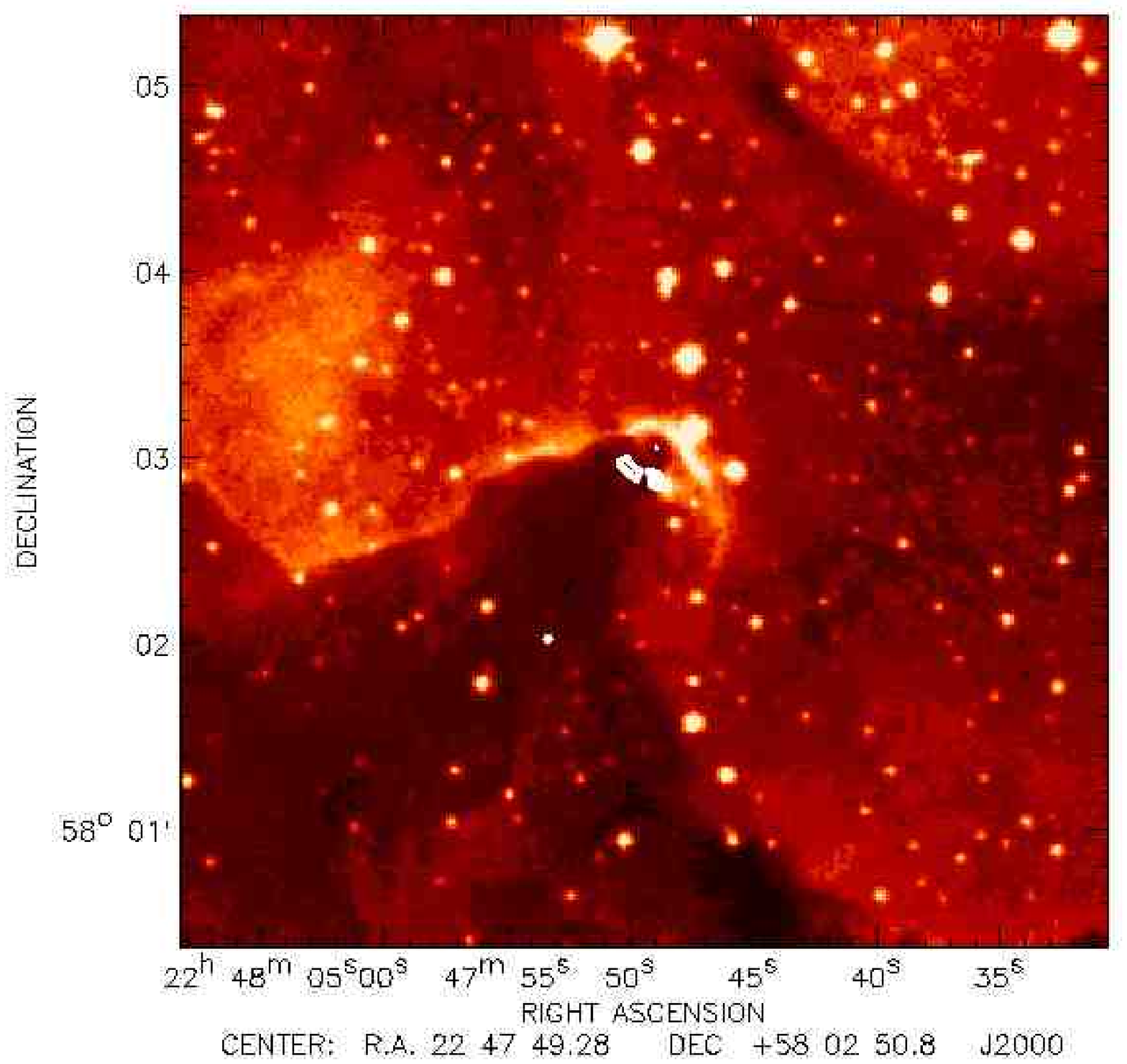}
\includegraphics*[height=6cm]{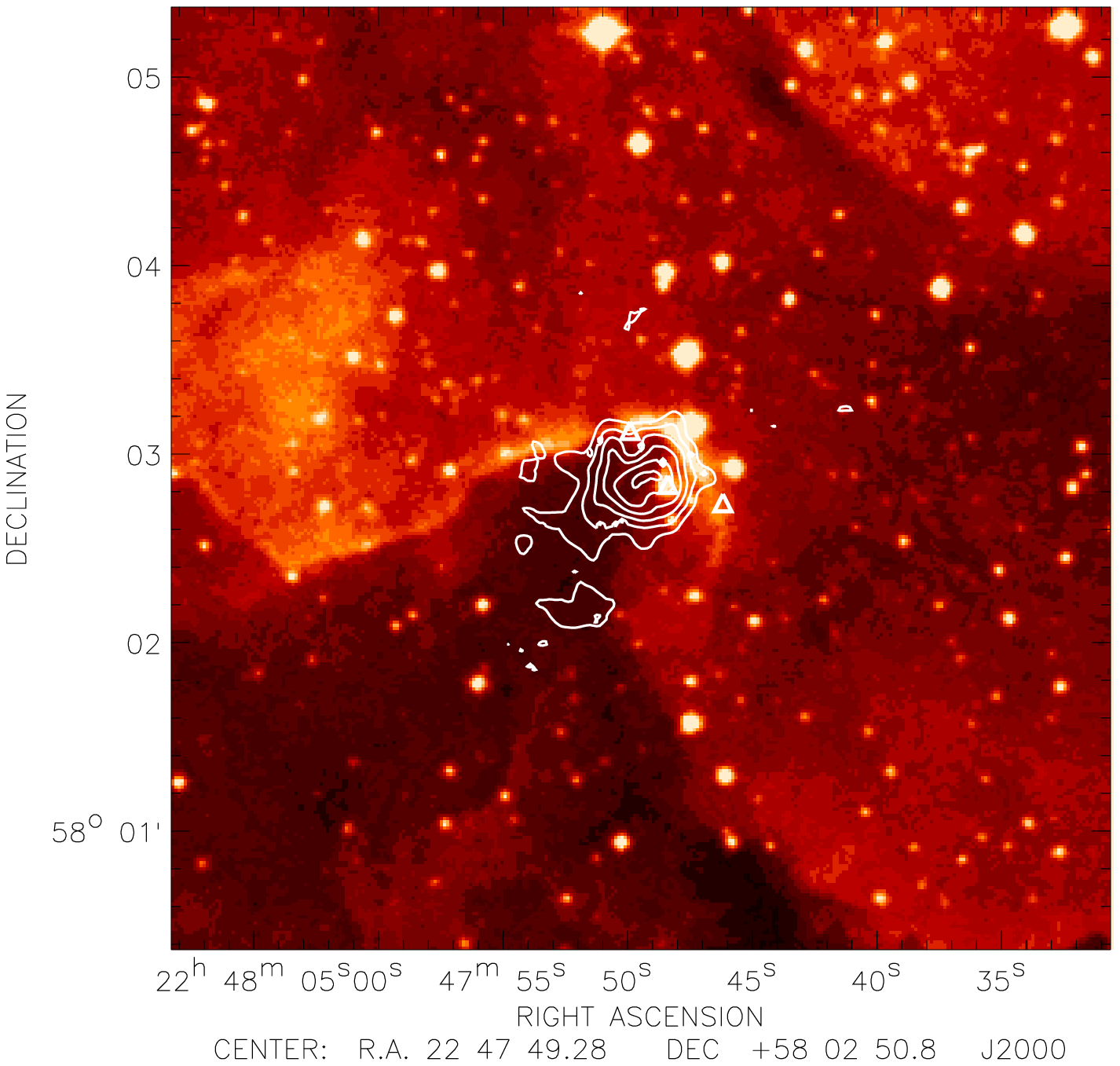}\\
\includegraphics*[height=6cm]{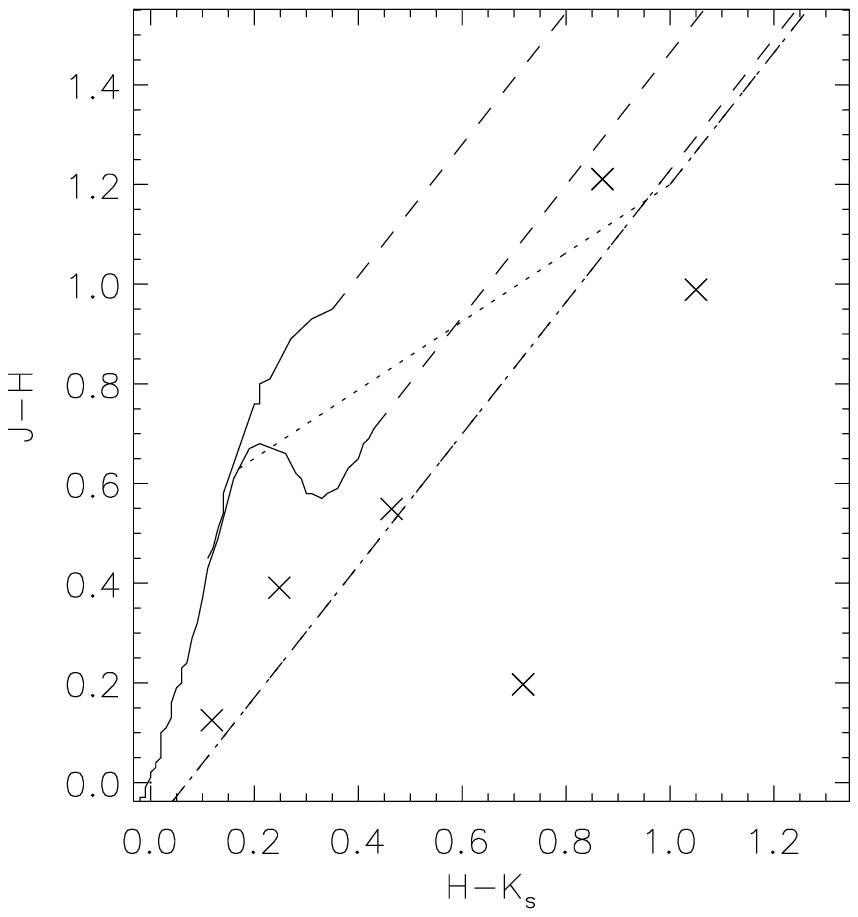}
\includegraphics*[height=6cm]{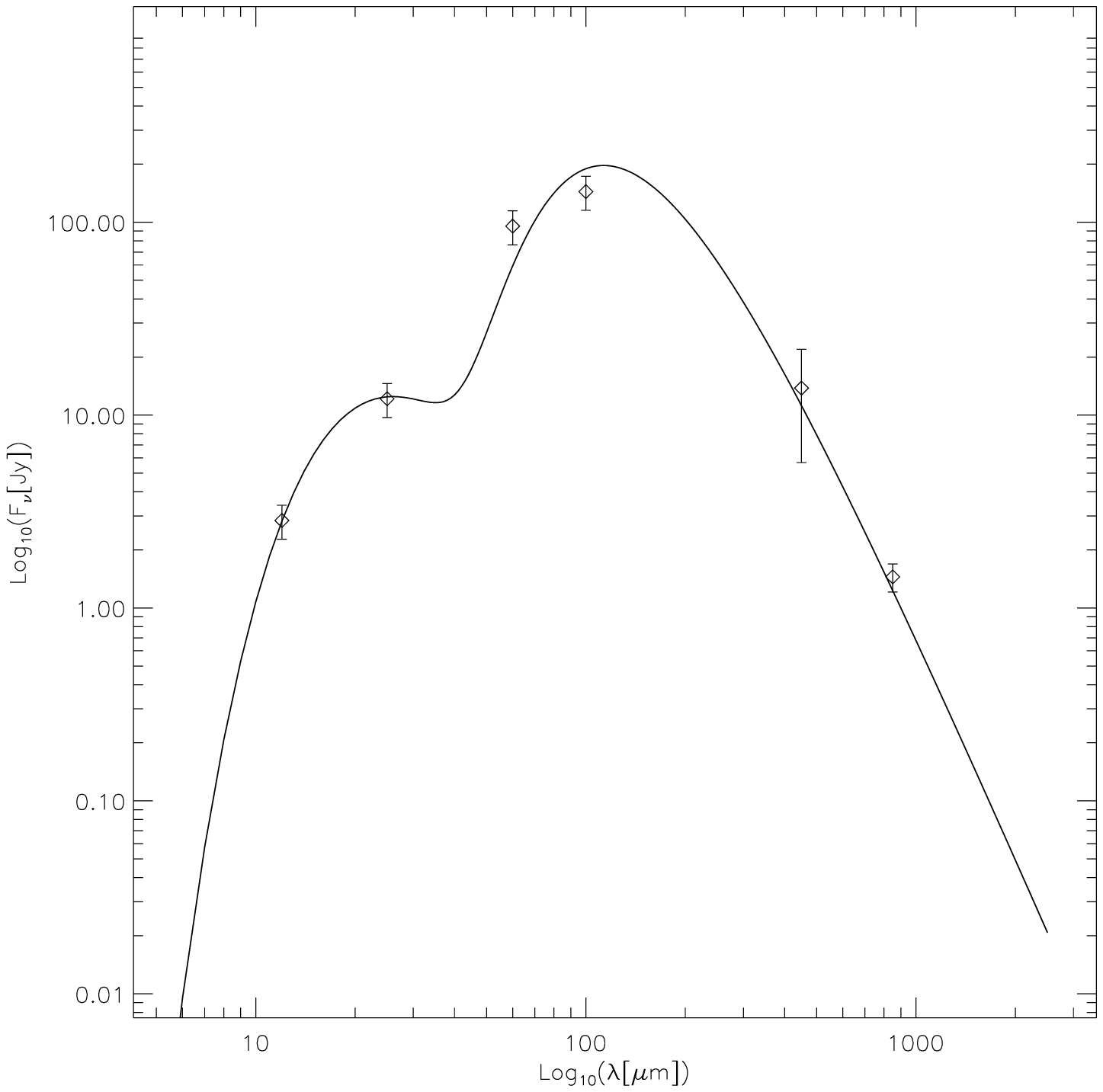}\\
\end{center}
\caption{Plots and images associated with the object SFO 43. The top images show SCUBA 450 \micron ~(left) and 850 \micron ~(right) contours overlaid on a DSS image, infrared sources from the 2MASS Point Source Catalogue \citep{Cutri2003} that have been identified as YSOs are shown as triangles.  850 \micron ~contours start at 3$\sigma$ and increase in increments of 20\% of the peak flux value, 450 \micron ~contours start at 3$\sigma$ and increase in increments of 20\% of the peak flux value.
\indent The bottom left plot shows the J-H versus H-K$_{\rm{s}}$ colours of the 2MASS sources associated with the cloud while the bottom right image shows the SED plot of the object composed from a best fit to various observed fluxes.}
\end{figure*}
\end{center}

\newpage

\begin{center}
\begin{figure*}
\begin{center}
\includegraphics*[height=6cm]{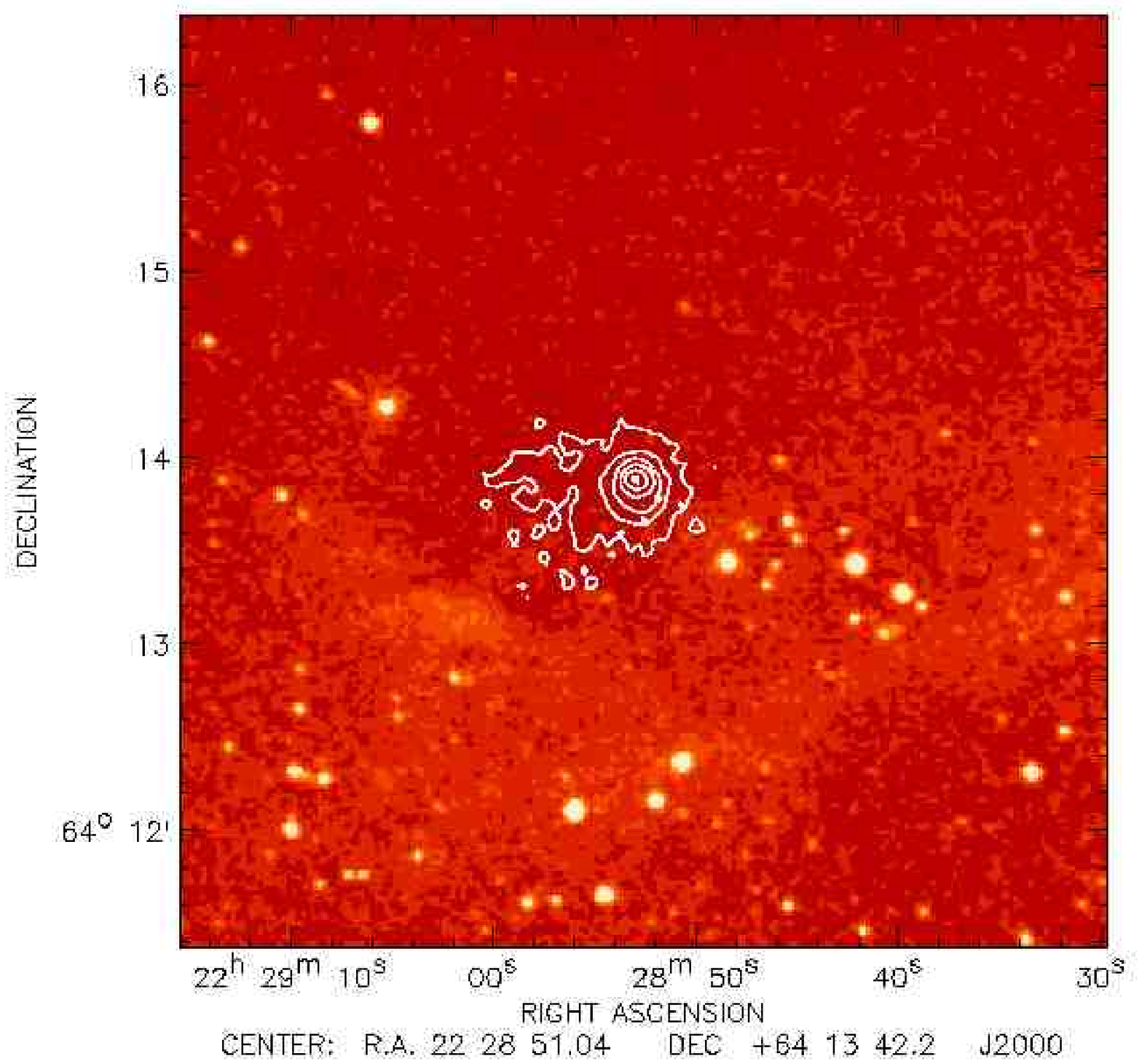}
\includegraphics*[height=6cm]{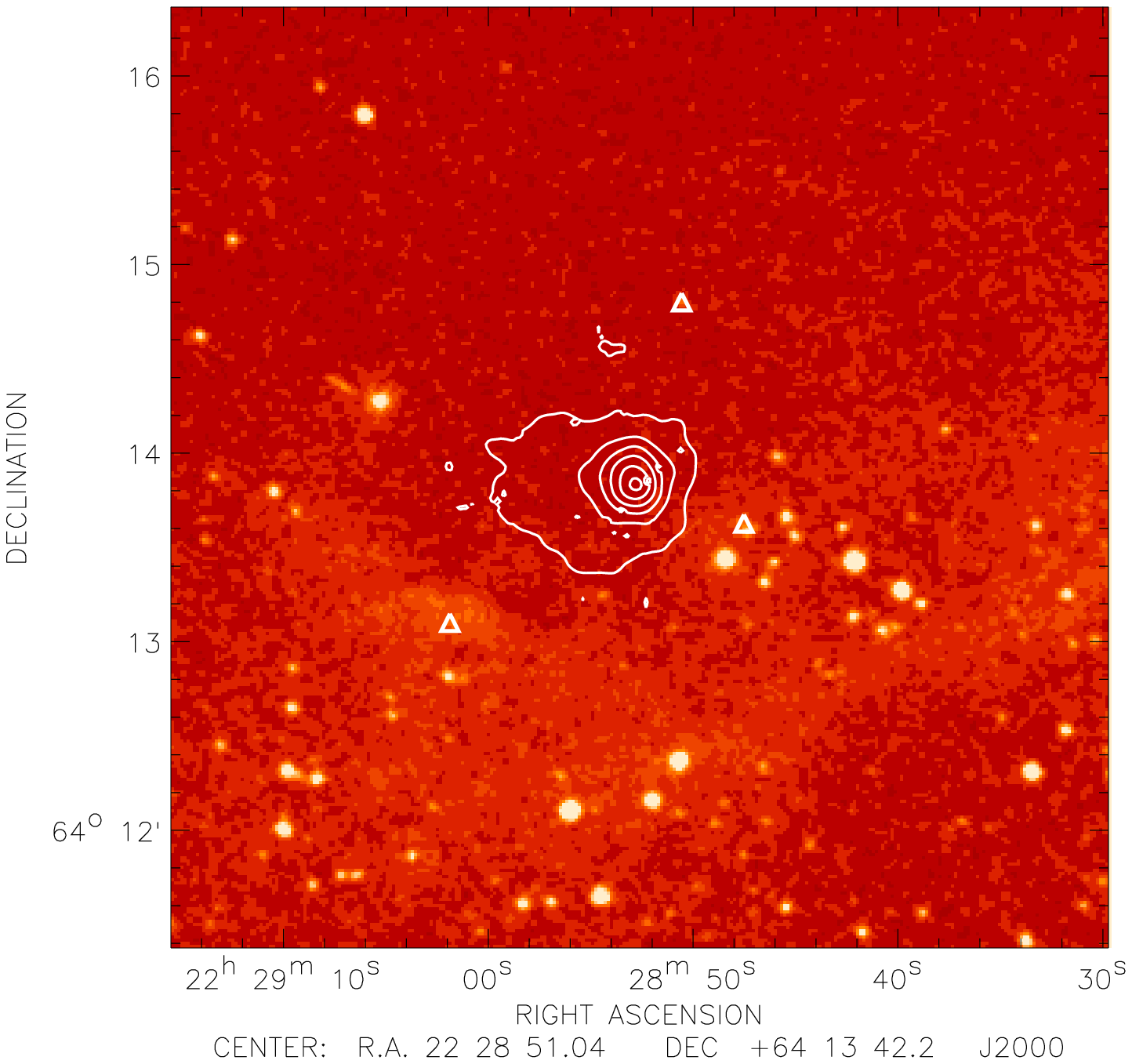}\\
\includegraphics*[height=6cm]{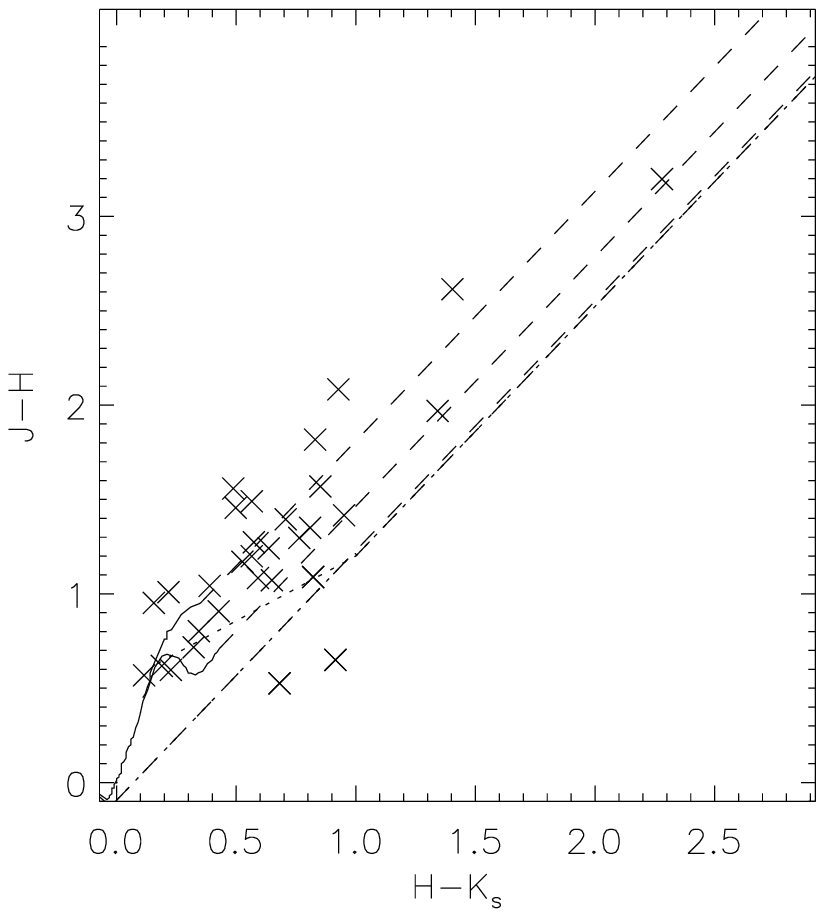}
\end{center}
\caption{Plots and images associated with the object SFO 44. The top images show SCUBA 450 \micron ~(left) and 850 \micron ~(right) contours overlaid on a DSS image, infrared sources from the 2MASS Point Source Catalogue \citep{Cutri2003} that have been identified as YSOs are shown as triangles.  850 \micron ~contours start at 6$\sigma$ and increase in increments of 20\% of the peak flux value, 450 \micron ~contours start at 3$\sigma$ and increase in increments of 20\% of the peak flux value.
\indent The bottom left plot shows the J-H versus H-K$_{\rm{s}}$ colours of the 2MASS sources associated with the cloud while the bottom right image shows the SED plot of the object composed from a best fit to various observed fluxes.}
\end{figure*}
\end{center}

\newpage

\begin{center}
\begin{figure*}
\begin{center}
\includegraphics*[height=6cm]{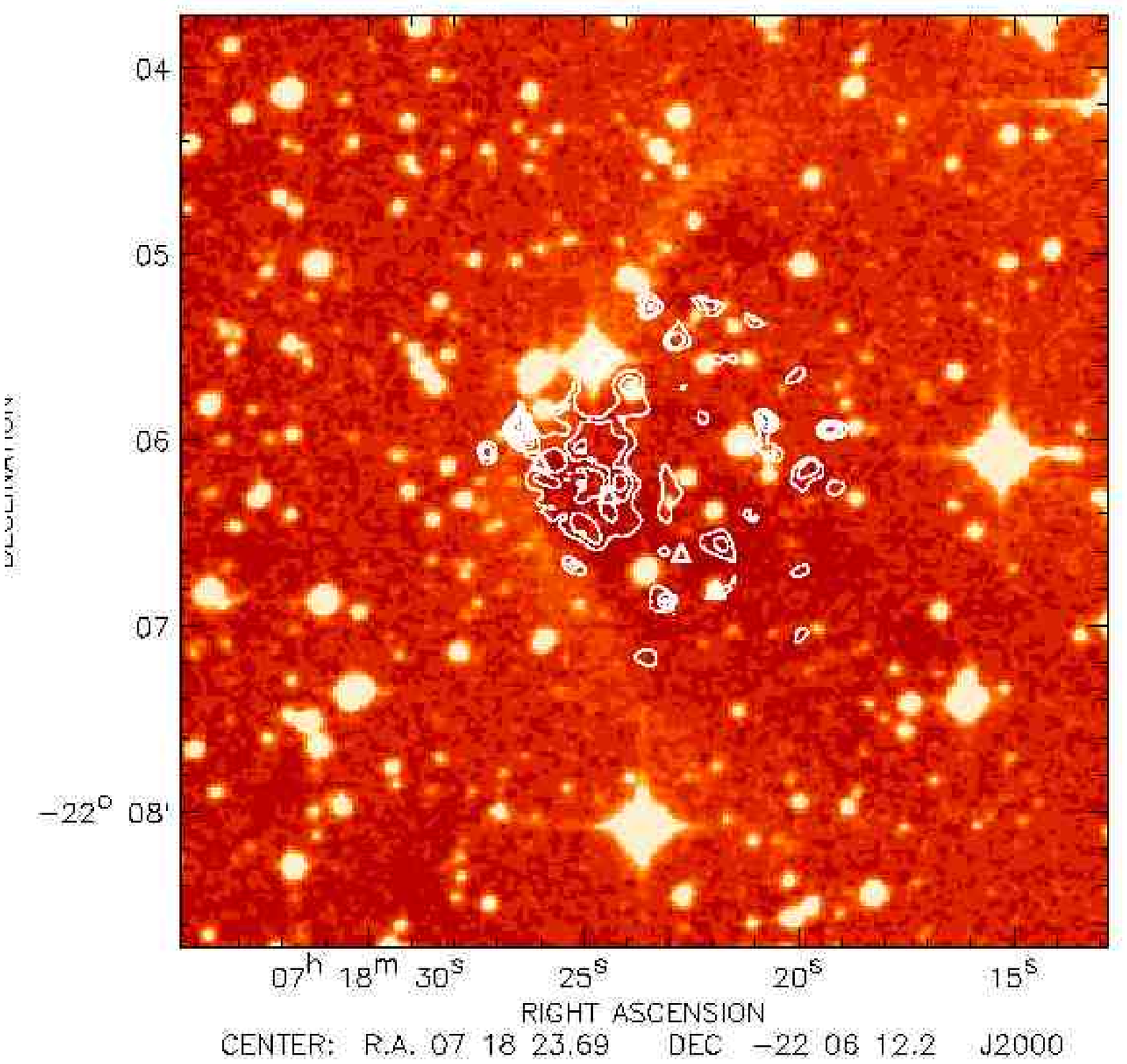}\\
\includegraphics*[height=6cm]{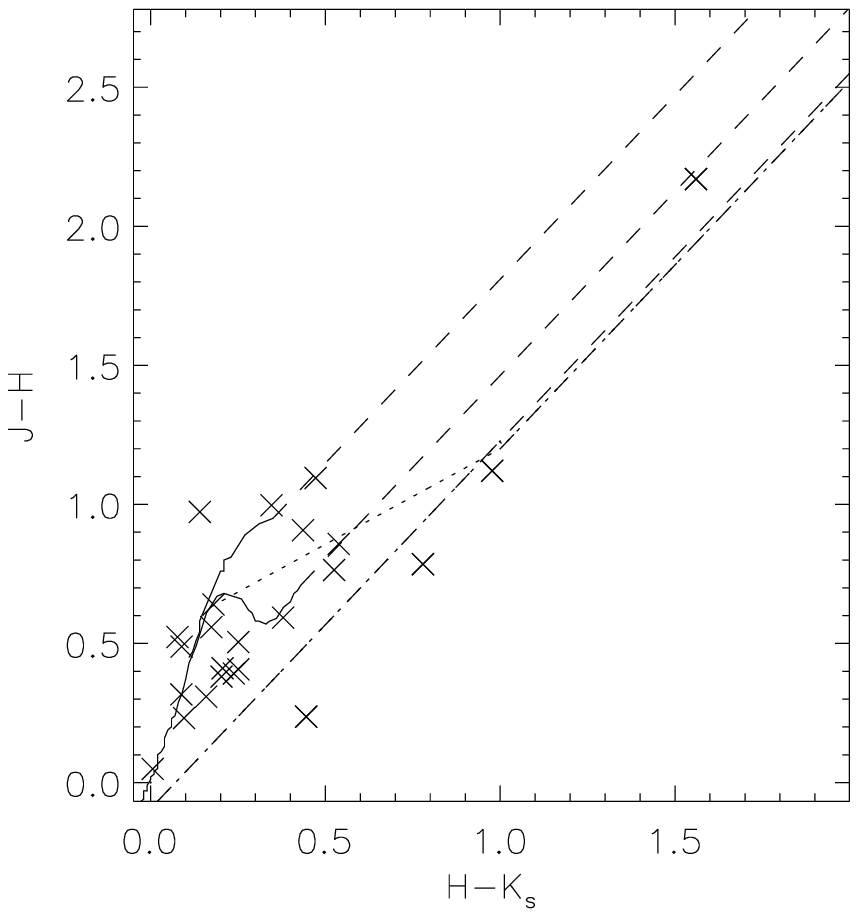}
\end{center}
\caption{Plots and images associated with the object SFO 45. The left image shows SCUBA 850 \micron ~contours overlaid on a DSS image, infrared sources from the 2MASS Point Source Catalogue \citep{Cutri2003} are shown as triangles.  850 \micron ~contours start at 3$\sigma$ and increase in increments of 20\% of the peak flux value.
\indent The right plot shows the J-H versus H-K$_{\rm{s}}$ colours of the 2MASS sources associated with the cloud.}
\end{figure*}
\end{center}

\newpage

\begin{center}
\begin{figure*}
\begin{center}
\includegraphics*[height=6cm]{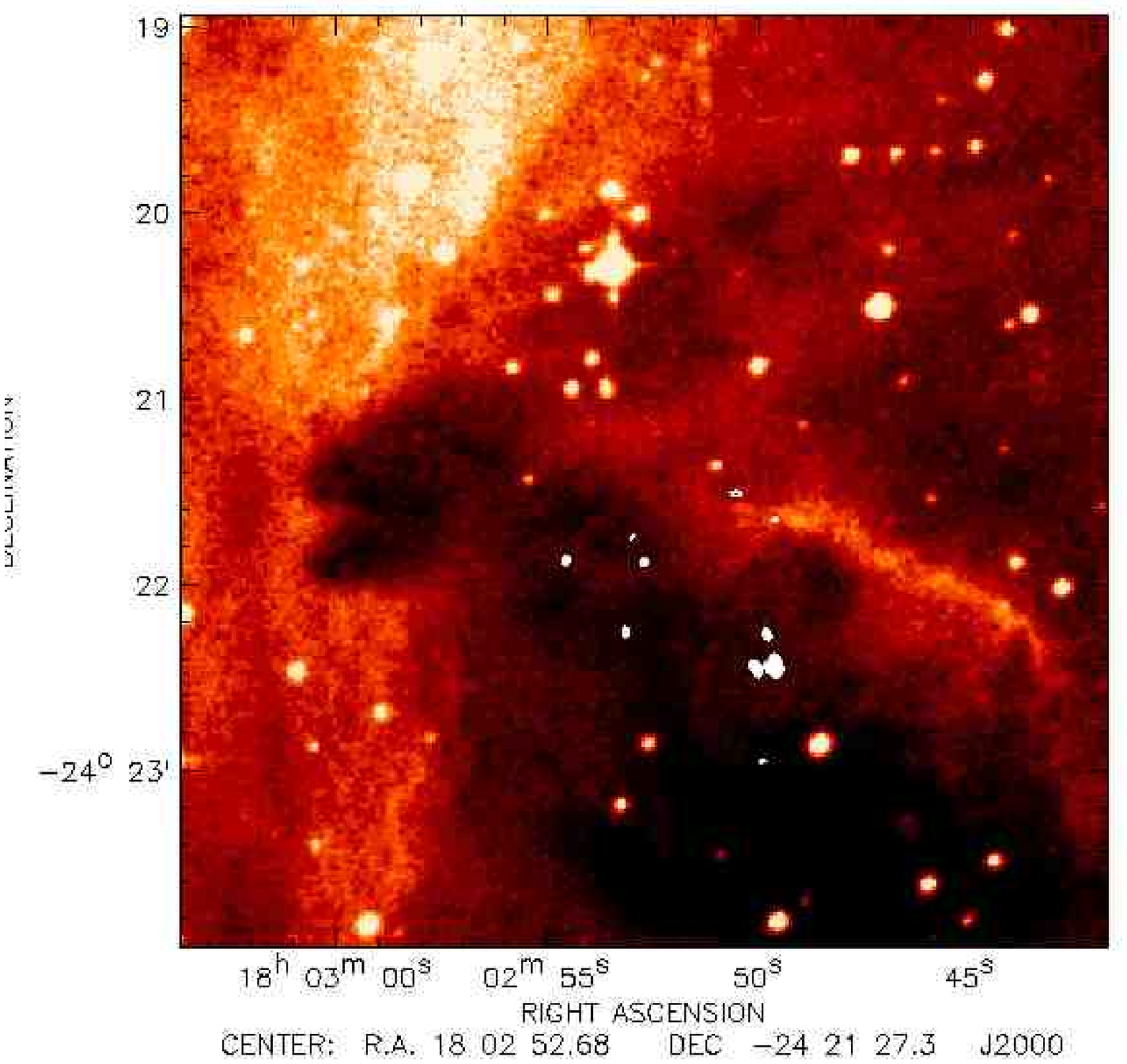}
\includegraphics*[height=6cm]{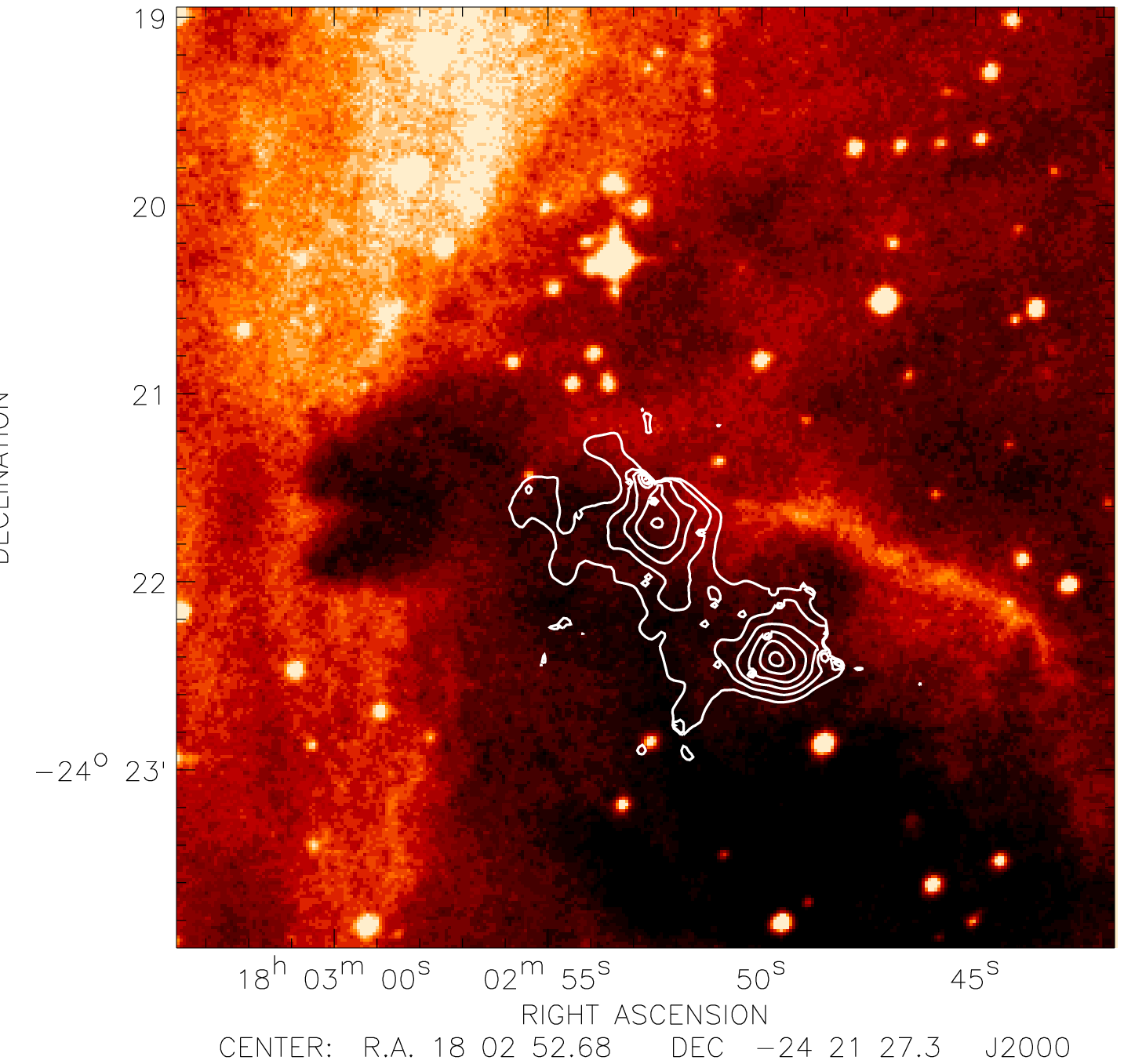}\\
\includegraphics*[height=6cm]{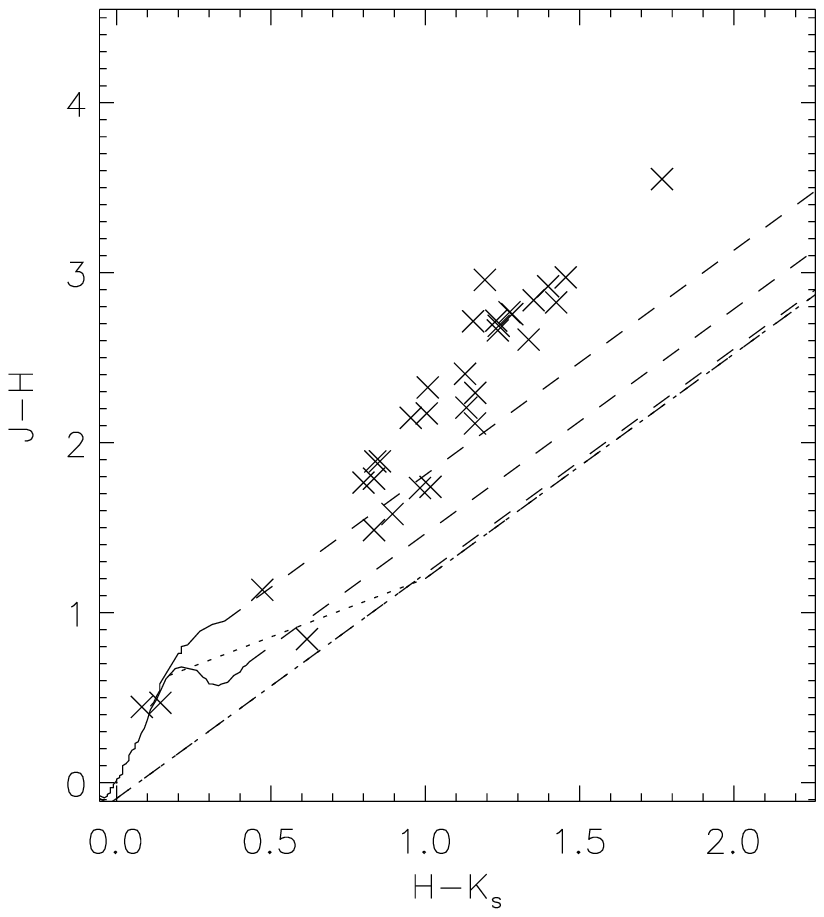}
\includegraphics*[height=6cm]{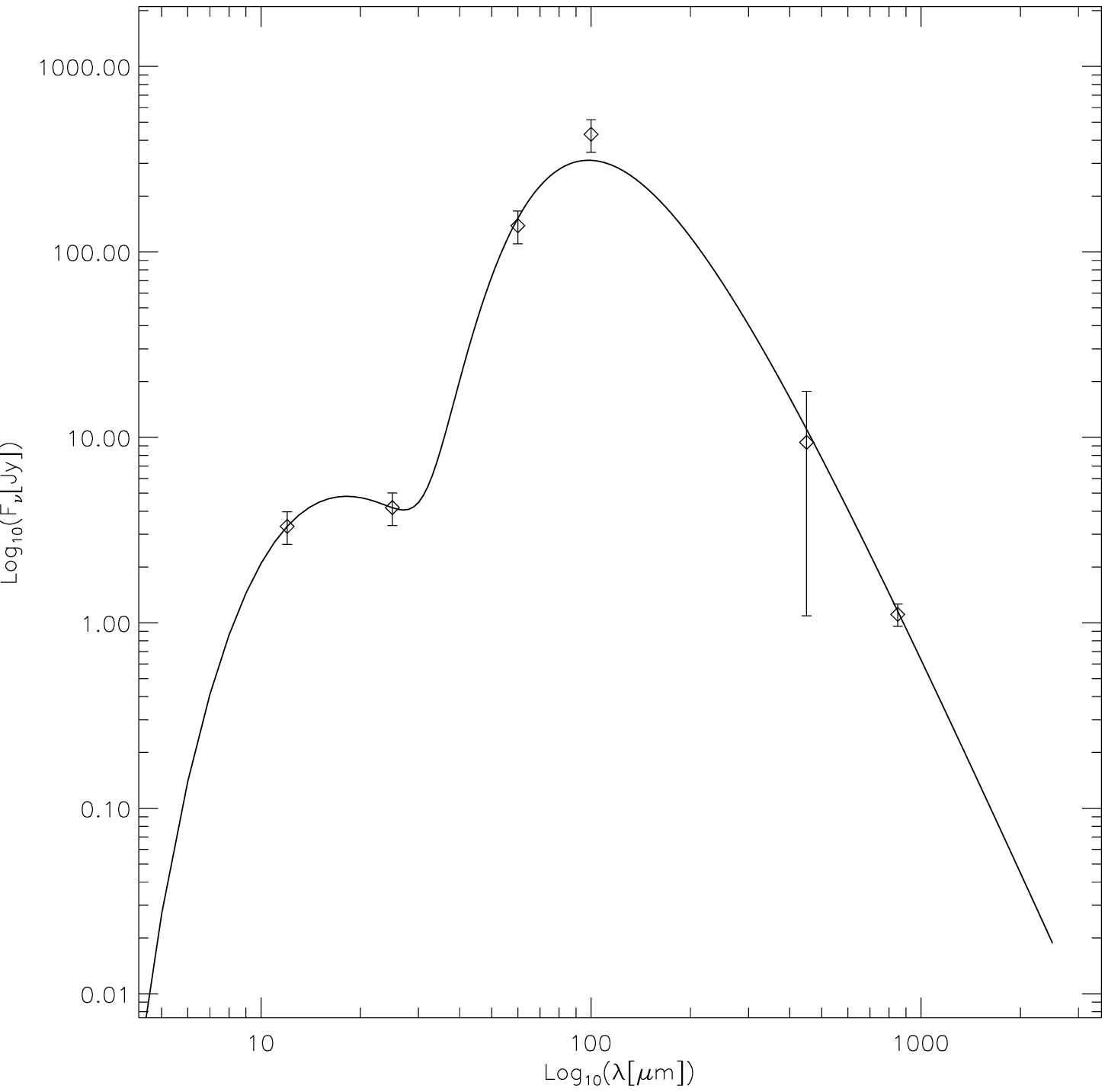}\\
\end{center}
\caption{Plots and images associated with the object SFO 87a. The top images show SCUBA 450 \micron ~(left) and 850 \micron ~(right) contours overlaid on a DSS image, infrared sources from the 2MASS Point Source Catalogue \citep{Cutri2003} that have been identified as YSOs are shown as triangles.  850 \micron ~contours start at 7$\sigma$ and increase in increments of 20\% of the peak flux value, 450 \micron ~contours start at 3$\sigma$ and increase in increments of 20\% of the peak flux value.
\indent The bottom left plot shows the J-H versus H-K$_{\rm{s}}$ colours of the 2MASS sources associated with the cloud while the bottom right image shows the SED plot of the object composed from a best fit to various observed fluxes.}
\end{figure*}
\end{center}

\newpage

\begin{center}
\begin{figure*}
\begin{center}
\includegraphics*[height=6cm]{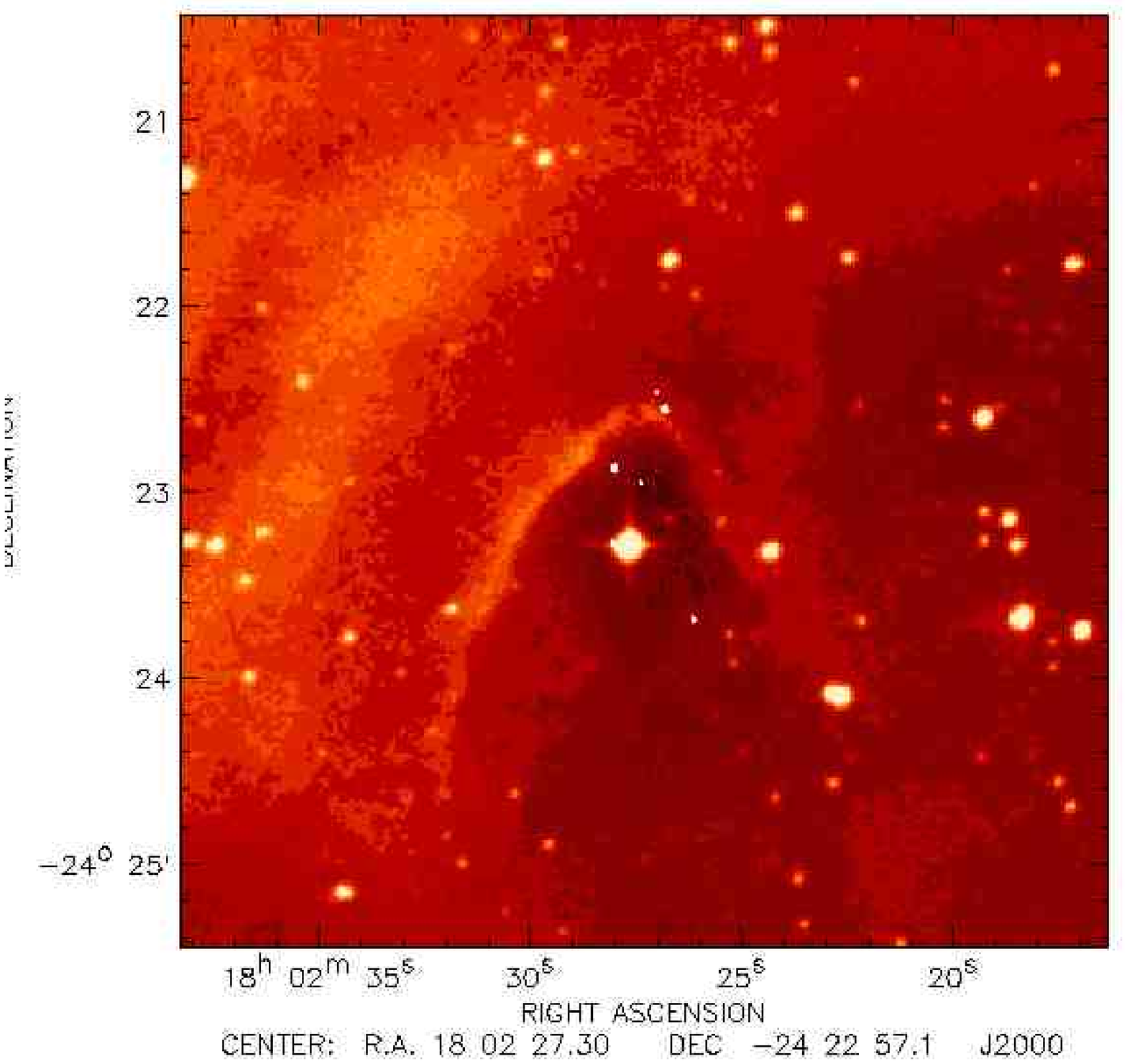}
\includegraphics*[height=6cm]{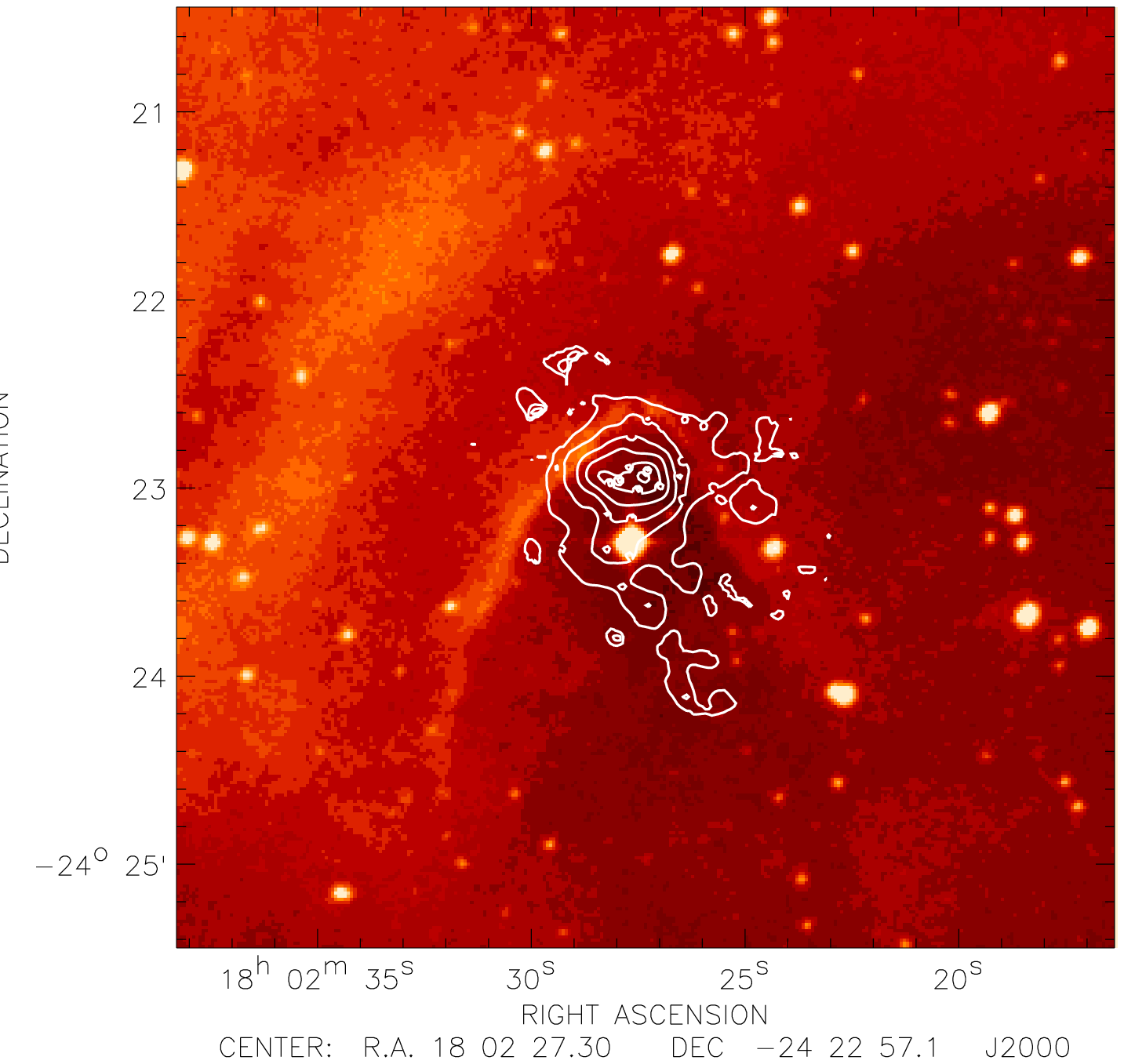}\\
\includegraphics*[height=6cm]{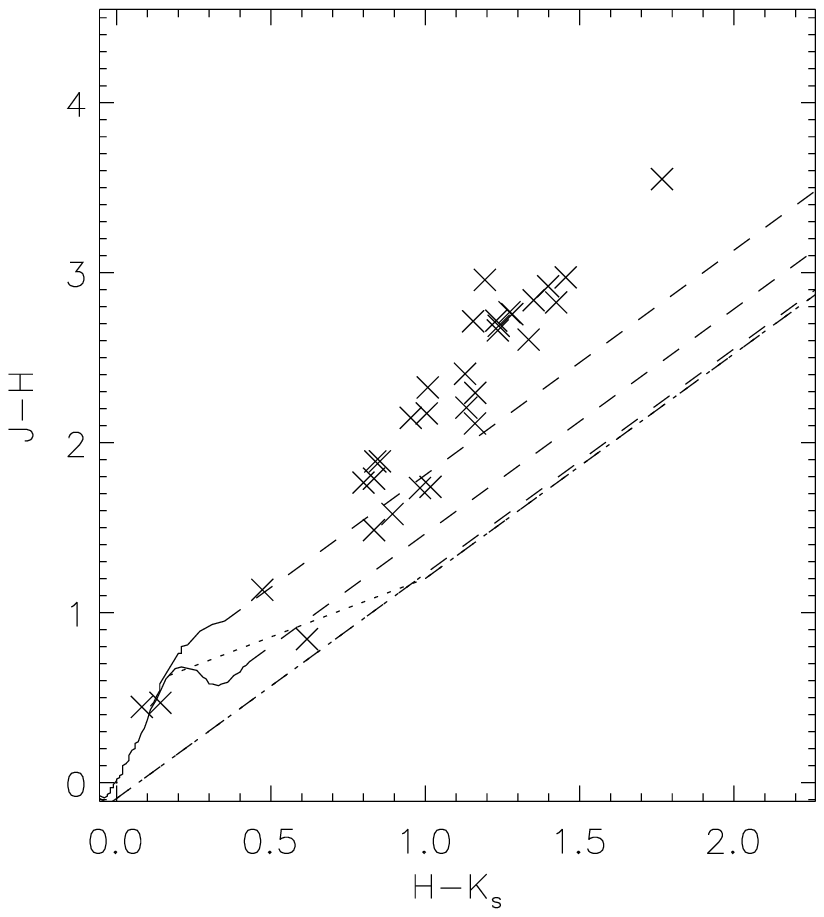}
\includegraphics*[height=6cm]{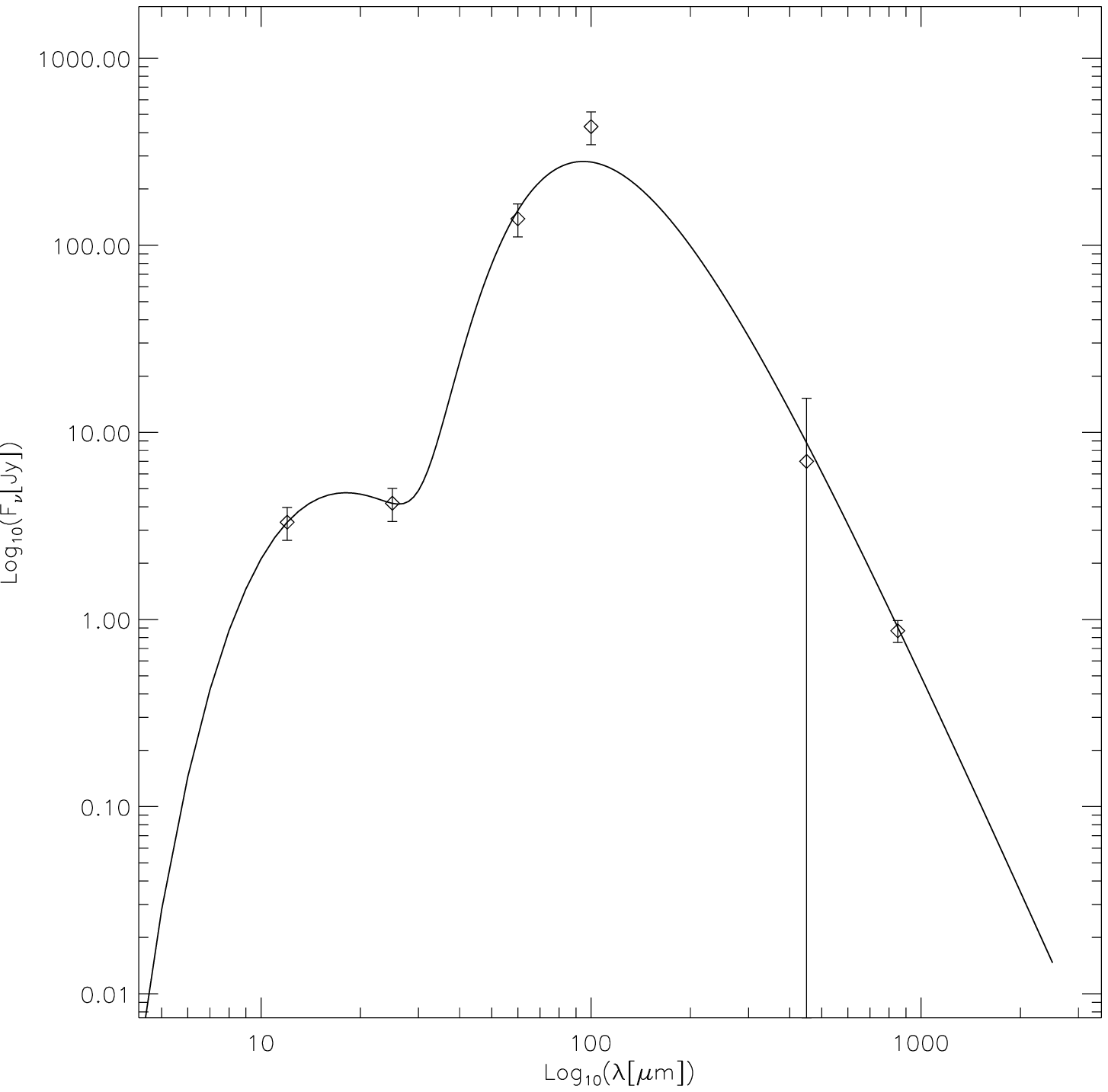}\\
\end{center}
\caption{Plots and images associated with the object SFO 87b. The top images show SCUBA 450 \micron ~(left) and 850 \micron ~(right) contours overlaid on a DSS image, infrared sources from the 2MASS Point Source Catalogue \citep{Cutri2003} that have been identified as YSOs are shown as triangles.  850 \micron ~contours start at 4$\sigma$ and increase in increments of 20\% of the peak flux value, 450 \micron ~contours start at 3$\sigma$ and increase in increments of 20\% of the peak flux value.
\indent The bottom left plot shows the J-H versus H-K$_{\rm{s}}$ colours of the 2MASS sources associated with the cloud while the bottom right image shows the SED plot of the object composed from a best fit to various observed fluxes.}
\end{figure*}
\end{center}

\newpage

\begin{center}
\begin{figure*}
\begin{center}
\includegraphics*[height=6cm]{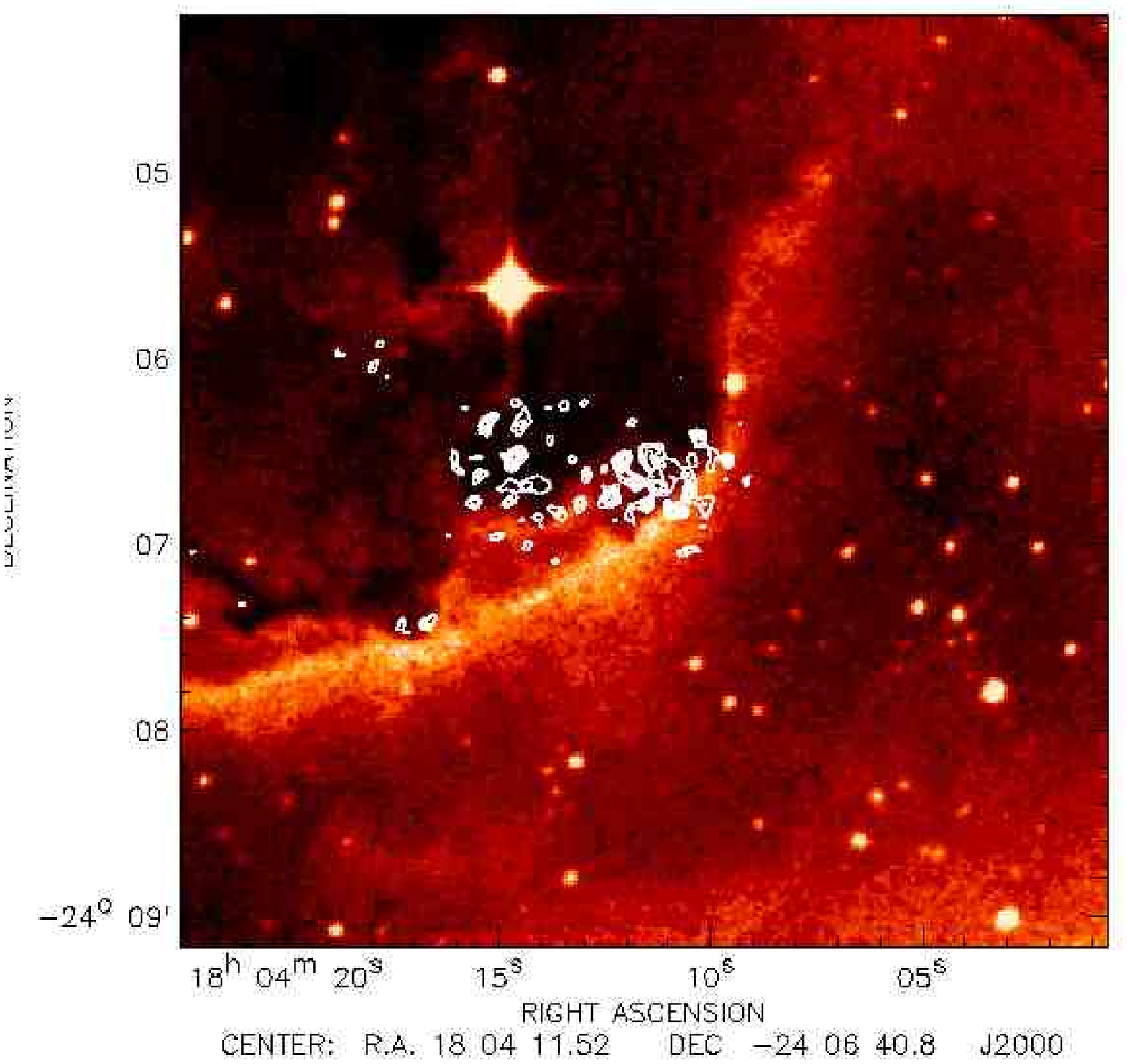}
\includegraphics*[height=6cm]{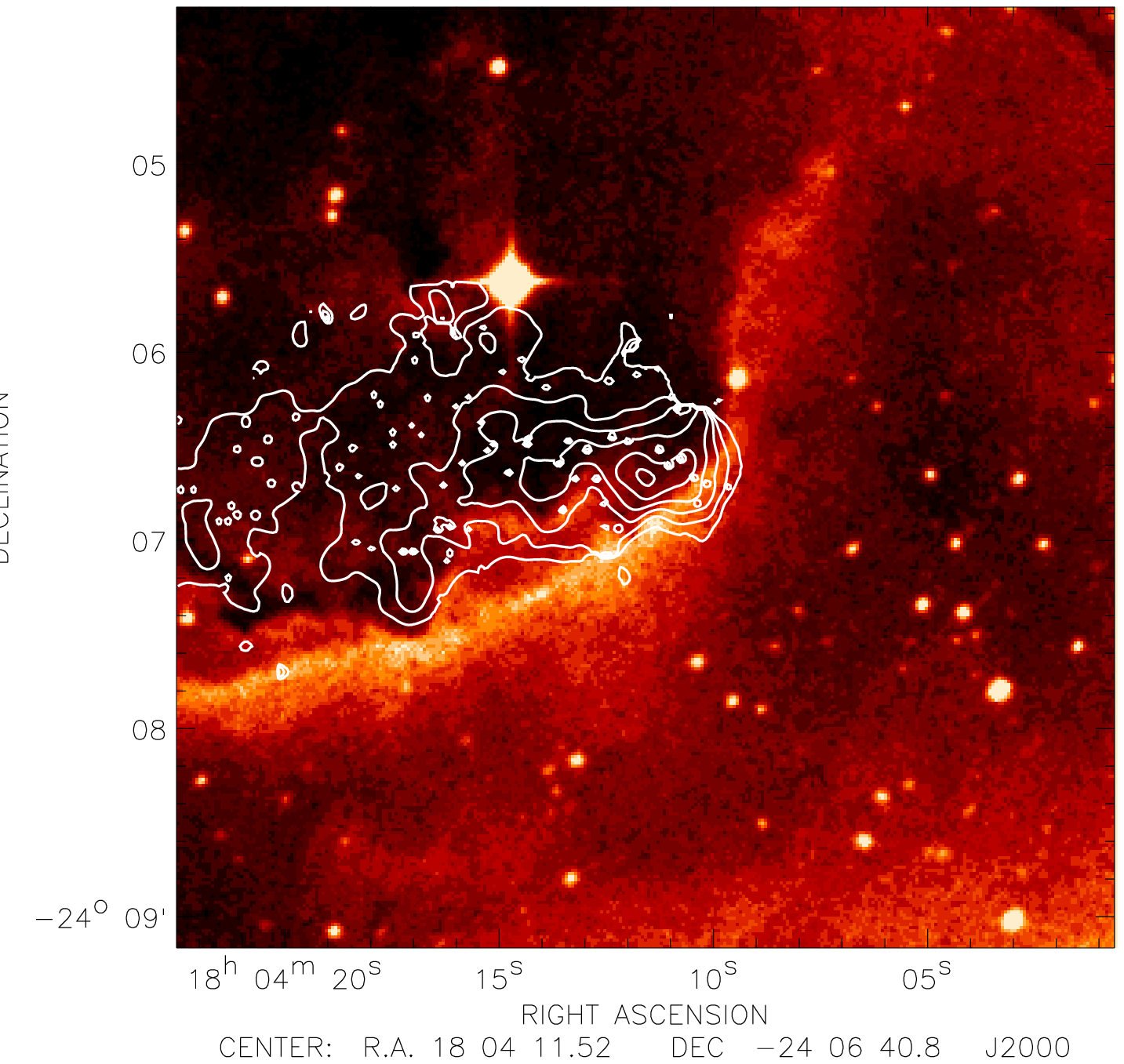}\\
\includegraphics*[height=6cm]{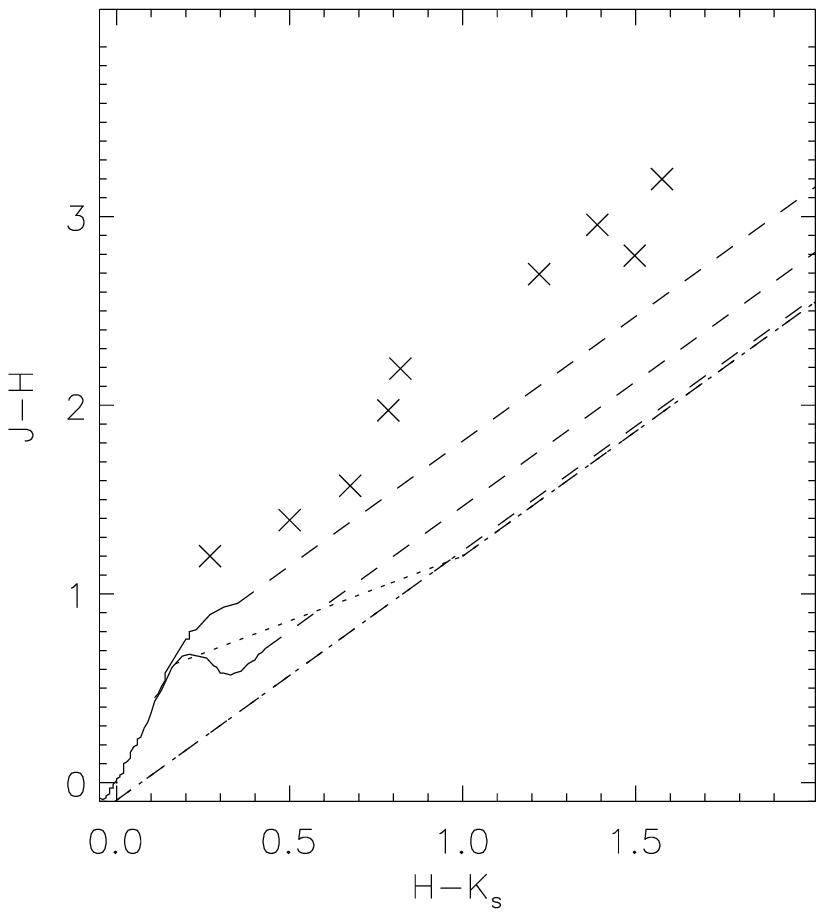}
\end{center}
\caption{Plots and images associated with the object SFO 88. The top images show SCUBA 450 \micron ~(left) and 850 \micron ~(right) contours overlaid on a DSS image, infrared sources from the 2MASS Point Source Catalogue \citep{Cutri2003} that have been identified as YSOs are shown as triangles.  850 \micron ~contours start at 9$\sigma$ and increase in increments of 20\% of the peak flux value, 450 \micron ~contours start at 4$\sigma$ and increase in increments of 20\% of the peak flux value.
\indent The bottom left plot shows the J-H versus H-K$_{\rm{s}}$ colours of the 2MASS sources associated with the cloud while the bottom right image shows the SED plot of the object composed from a best fit to various observed fluxes.}
\label{fig:images88}
\end{figure*}
\end{center}

\newpage

\begin{center}
\begin{figure*}
\begin{center}
\includegraphics*[height=6cm]{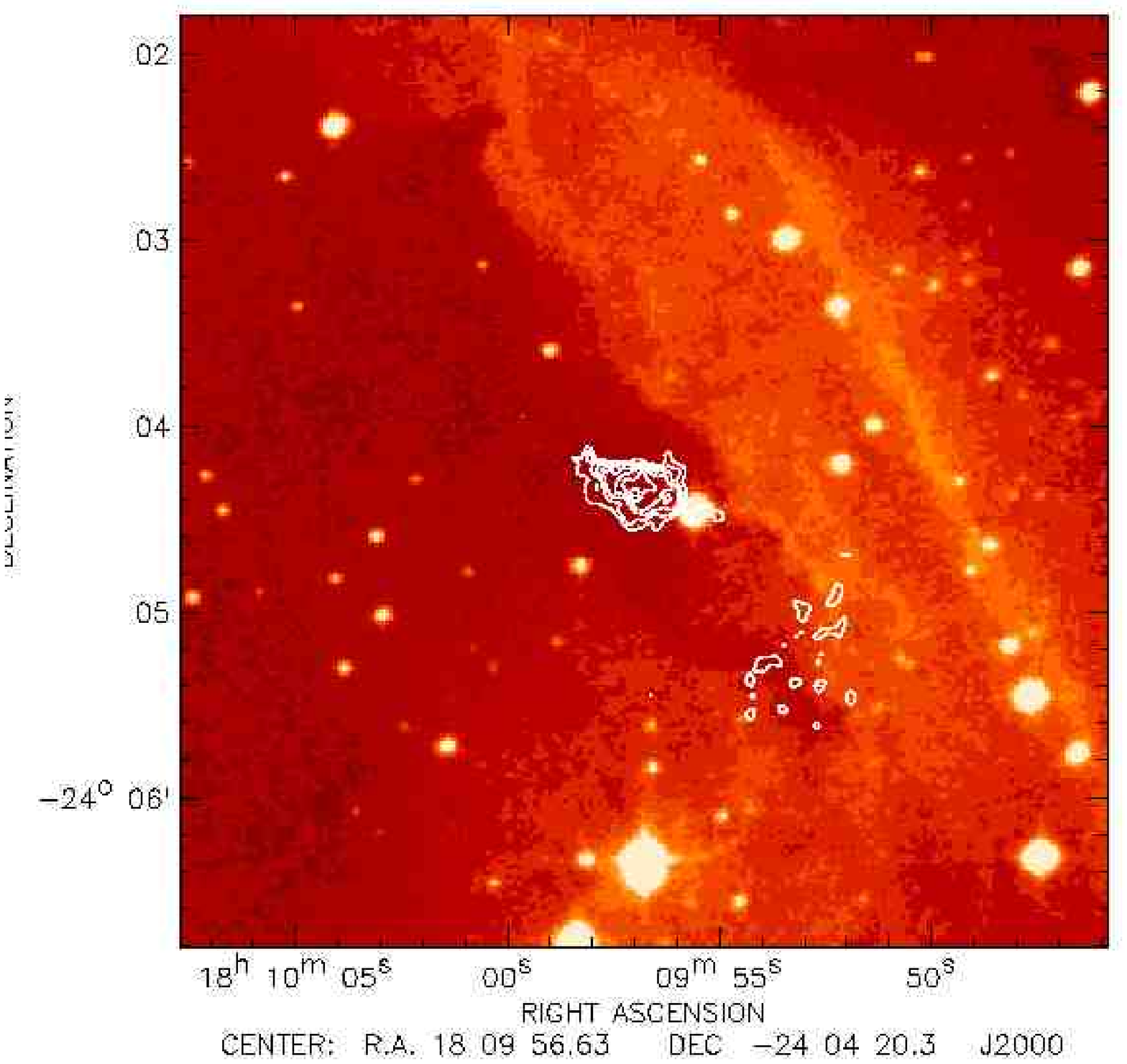}
\includegraphics*[height=6cm]{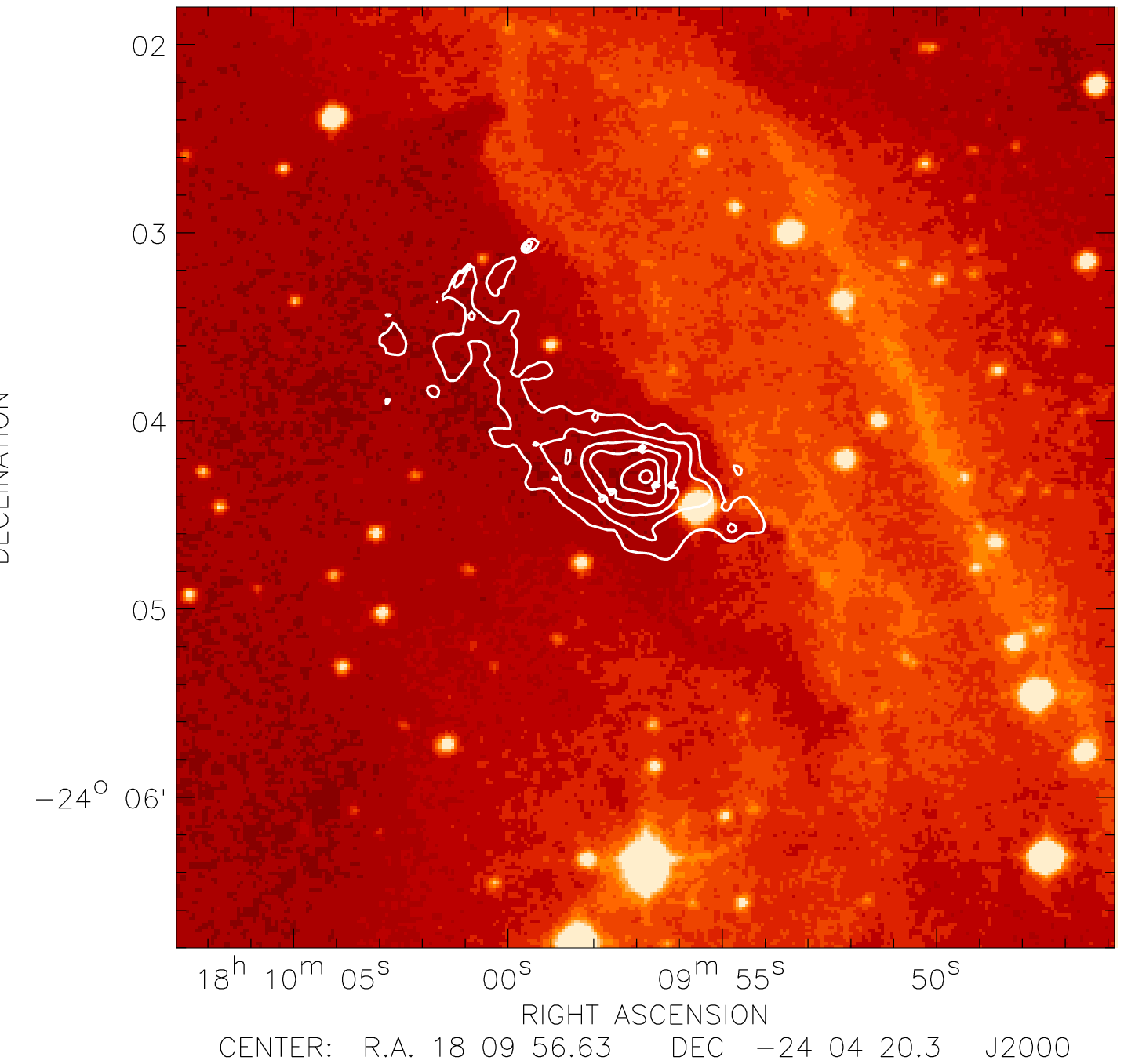}\\
\includegraphics*[height=6cm]{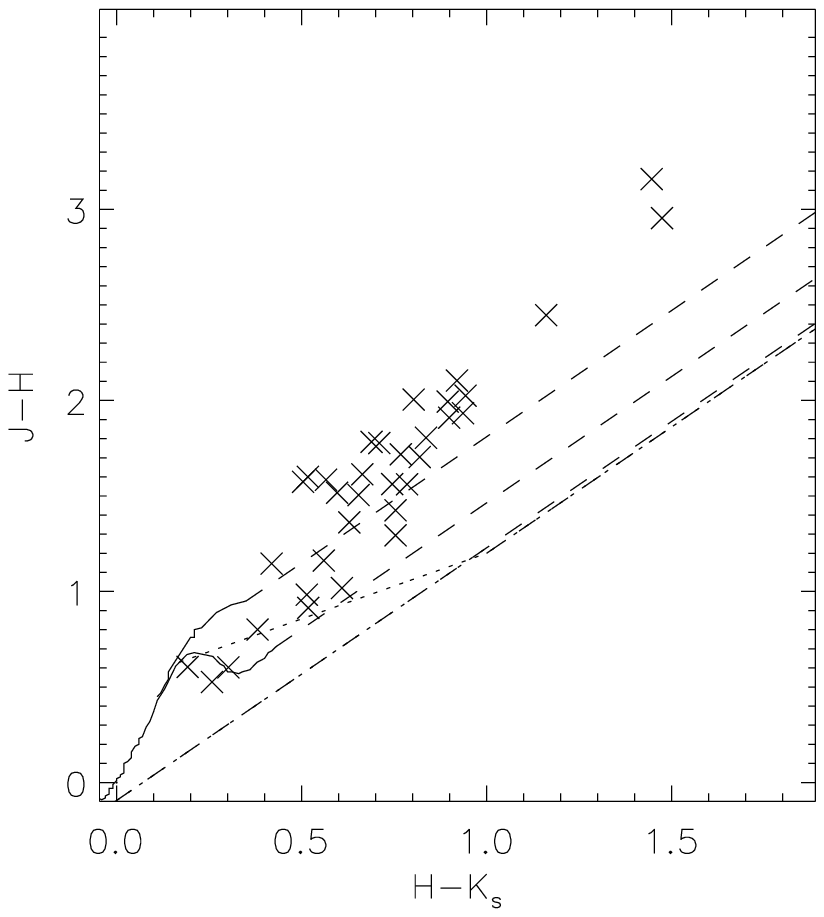}
\includegraphics*[height=6cm]{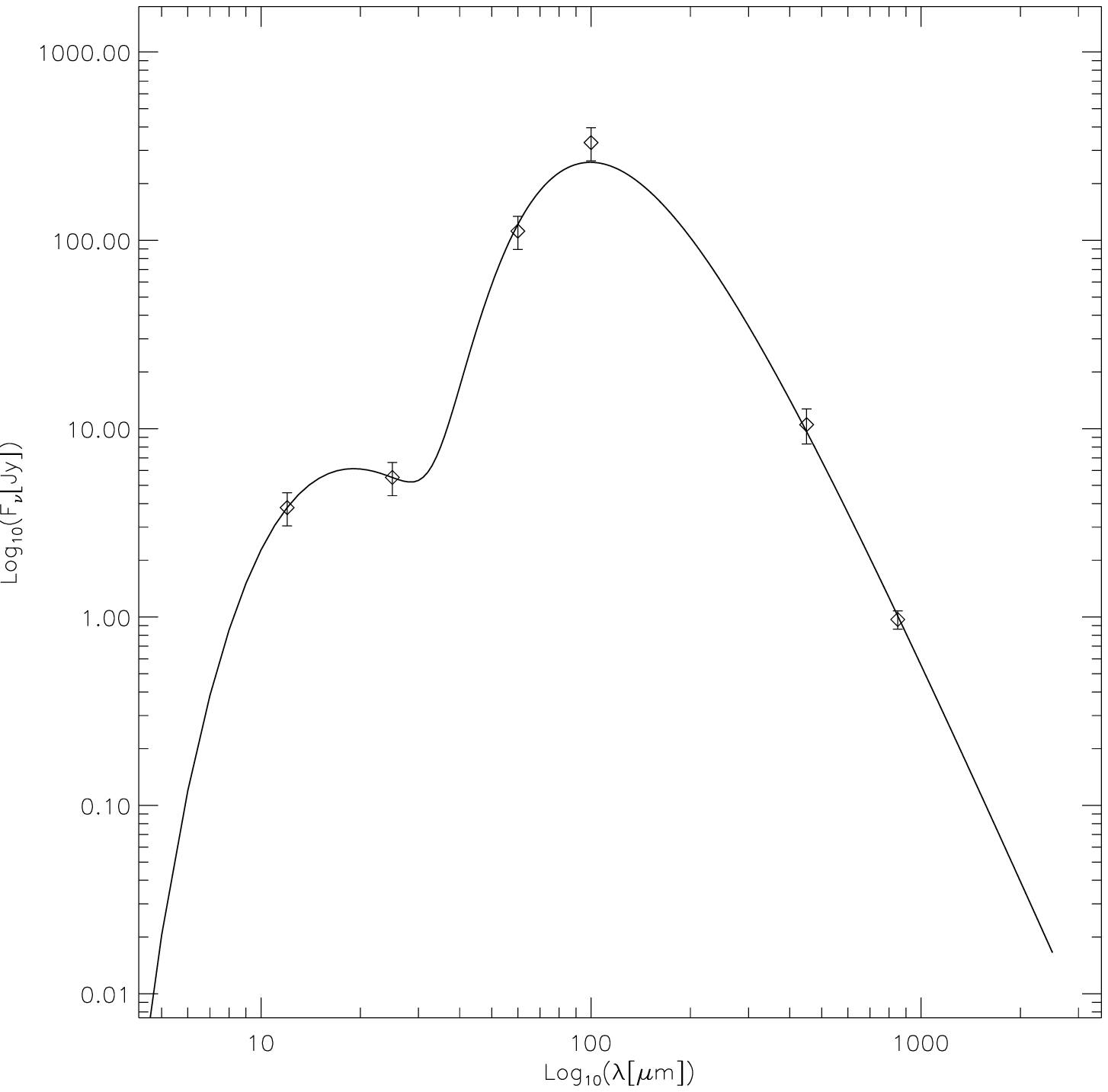}\\
\end{center}
\caption{Plots and images associated with the object SFO 89. The top images show SCUBA 450 \micron ~(left) and 850 \micron ~(right) contours overlaid on a DSS image, infrared sources from the 2MASS Point Source Catalogue \citep{Cutri2003} that have been identified as YSOs are shown as triangles.  850 \micron ~contours start at 9$\sigma$ and increase in increments of 20\% of the peak flux value, 450 \micron ~contours start at 4$\sigma$ and increase in increments of 20\% of the peak flux value.
\indent The bottom left plot shows the J-H versus H-K$_{\rm{s}}$ colours of the 2MASS sources associated with the cloud while the bottom right image shows the SED plot of the object composed from a best fit to various observed fluxes.}
\end{figure*}
\end{center}

\label{lastpage}

\end{document}